\documentclass{elsart}

\usepackage{amsmath}
\usepackage{amscd}   %
\usepackage{amssymb} %

\usepackage{psfrag}
\usepackage[dvips]{graphicx}

\usepackage{cite} %
\usepackage{array}     %
\usepackage{longtable}

\numberwithin{equation}{section}  
\numberwithin{figure}{section}  
\numberwithin{table}{section}  

\newcommand{\myparagraph}[1]{\subsubsection*{#1}}

\DeclareSymbolFont{rsfscript}{OMS}{rsfs}{m}{n}
\DeclareSymbolFontAlphabet{\mathrsfs}{rsfscript}
\renewcommand{\epsilon}{\varepsilon}
\renewcommand{\phi}{\varphi}
\renewcommand{\Re}{\mathop{\mathrm{Re}}\nolimits}
\renewcommand{\Im}{\mathop{\mathrm{Im}}\nolimits}
\newcommand{\q}{\quad}
\newcommand{\qq}{\qquad}
\newcommand{\nn}{\nonumber}
\newcommand{\nnn}{\nonumber\\}
\newcommand{\ts}{\textstyle}
\newcommand{\ds}{\displaystyle}
\newcommand{\mb}[1]{\mathbf{#1}}
\newcommand{\figurecaption}[1]{\caption{#1}} %
\newcommand{\mnote}[1]{}

\newcommand{\eref}[1]{(\ref{#1})}
\newcommand{\rme}{\mathrm{e}}
\newcommand{\rmi}{\mathrm{i}}
\newcommand{\rmd}{\mathrm{d}}
\newcommand{\tr}{\mathop{\mathrm{tr}}\nolimits}
\newcommand{\Or}{\mathord{\mathrm{O}}}
\newcommand{\opencircle}{\mbox{\Large$\circ\,$}}

\newcommand{\fullcircle}{\mbox{{\Large$\bullet\,$}}}

\newcommand{\sgn}{\mathop{\mathrm{sgn}}\nolimits}

\newcommand{\Ai}{\mathop{\mathrm{Ai}}\nolimits}
\newcommand{\Bi}{\mathop{\mathrm{Bi}}\nolimits}

\newcommand{\oFo}{\mathop{{_{1\,}{\!\rm F}_{\!1}}}\nolimits} %

\newcommand{\fpint}{=\hspace{-1.2em}\int}            %
\newcommand{\tfpint}{{\textstyle=\hspace{-1em}\int}} %
\newcommand{\bra}[1]{\langle#1|}
\newcommand{\ket}[1]{|#1\rangle}
\newcommand{\Bra}[1]{\Big\langle#1\Big|}
\newcommand{\Ket}[1]{\Big|#1\Big\rangle}
\newcommand{\braket}[2]{\langle #1|#2\rangle}%
\newcommand{\defas}{:=}
\newcommand{\Len}{\mathrsfs{L}}  %
\newcommand{\Area}{\mathrsfs{A}} %
\newcommand{\Domain}{\mathcal{D}} %
\newcommand{\openDomain}{\smash{\stackrel{_\circ}{\Domain}}}
\newcommand{\Boundary}{\Gamma}
\newcommand{\Complex}{{{\mathbb{C}}}}
\newcommand{\Real}{{{\mathbb{R}}}}
\newcommand{\Natural}{{{\mathbb{N}}}}
\newcommand{\Z}{{{\mathbb{Z}}}}
\newcommand{\Rtwo}{{\Real^2}}
\newcommand{\Nnull}{{\Natural_0}}
\newcommand{\Hilbert}{{{\mathcal{L}_2}}}
\newcommand{\Chi}{{\raisebox{.1ex}{\large$\chi$}}}
\newcommand{\grad}{{\boldsymbol{\nabla}}}
\newcommand{\mass}{m_{e}}
\newcommand{\Lag}{\mathcal{L}}   %
\newcommand{\Ham}{\mathrm{H}}
\newcommand{\Angle}{\raisebox{-.15ex}{\Large$\sphericalangle$}}
\newcommand{\llongrightarrow}{\,\begin{CD}@>>>\,\end{CD}} %
\newcommand{\epscomb}[1]{#1^{^{\scriptstyle\epsilon}}}
\newcommand{\PO}{\ .}   %
\newcommand{\CO}{\ ,}   %
\newcommand{\oh}{{\frac{1}{2}}} %
\newcommand{\piot}{\frac{\pi}{2}}
\newcommand{\piof}{\frac{\pi}{4}}
\newcommand{\rvec}{\mathbf{r}}
\newcommand{\tvec}{\hat{\mathbf{t}}}
\newcommand{\nvec}{\hat{\mathbf{n}}}
\newcommand{\cvec}{\mathbf{c}}
\newcommand{\pvec}{\mathbf{p}}
\newcommand{\vvec}{\mathbf{v}}
\newcommand{\vvech}{\hat{\mathbf{v}}}
\newcommand{\Avec}{\mathbf{A}}
\newcommand{\jvec}{\mathbf{j}}

\newcommand{\rhovec}{\boldsymbol{\rho}}
\newcommand{\xt}{\tilde{x}}
\newcommand{\yt}{\tilde{y}}
\newcommand{\pxt}{\tilde{p}_x}
\newcommand{\pyt}{\tilde{p}_y}
\newcommand{\rvect}{\tilde{\rvec}}
\newcommand{\cvect}{\tilde{\cvec}}
\newcommand{\pvect}{\tilde{\pvec}}
\newcommand{\vvect}{\tilde{\vvec}}
\newcommand{\Avect}{\tilde{\Avec}}

\newcommand{\rhovect}{\tilde{\boldsymbol{\rho}}}
\newcommand{\Chit}{\widetilde{\Chi}}
\newcommand{\chit}{\tilde{\chi}} %
\newcommand{\Lt}{\tilde{L}}
\newcommand{\Lagt}{\widetilde{\Lag}}
\newcommand{\Hamt}{\widetilde{\Ham}}
\newcommand{\rt}{{\tilde{r}}}
\newcommand{\Rt}{{\widetilde{R}}}
\newcommand{\tit}{{\tilde{t}}}

\newcommand{\PSet}{\mathcal{P}}
\newcommand{\wasgamma}{\zeta}
\newcommand{\SoL}{\eta}
\newcommand{\unfolded}[1]{\check{#1}} %
\newcommand{\mm}{{\rm m}} %
\newcommand{\mmt}{\widetilde{\mm}} %
\newcommand{\Mag}{\mathcal{M}} %
\newcommand{\Magt}{\widetilde{\mathcal{M}}} %
\newcommand{\magn}{{\scriptscriptstyle(\mathcal{M})}} %
\newcommand{\W}{W} %
\newcommand{\n}{N} %
\newcommand{\m}{M} %
\newcommand{\GF}{\mathrsfs{G}}  %
\newcommand{\oc}{\omega_{\rm c}}
\newcommand{\rmrn}{\rvec-\rvec_0}
\newcommand{\rmrnt}{\rvect-\rvect_0}
\newcommand{\rxrn}{\rvec\times\rvec_0}
\newcommand{\rxrnt}{\rvect\times\rvect_0}
\newcommand{\dedge}{d_{\rm edge}}
\newcommand{\dedgesm}{\overline{d}_{\rm edge}}
\newcommand{\dosc}{d_{\rm osc}}
\newcommand{\N}{{\rm N}}
\newcommand{\Nosc}{{\N}_{\rm osc}}
\newcommand{\Nsm}{\overline{\N}}
\newcommand{\Nedge}{{\N}_{\rm edge}}
\newcommand{\Nedgesm}{\Nsm_{\rm edge}}

\newcommand{\aconst}{\alpha_{\rm c}} 
\newcommand{\au}{\alpha_u} 
\newcommand{\qm}{q_m} 
\newcommand{\SL}{{\left(\begin{smallmatrix}{\rm S}\\{\rm L}\end{smallmatrix}\right)}}
\newcommand{\At}{\tilde{A}}
\newcommand{\Gn}{{\rm G}^0}
\newcommand{\Gnu}{{\rm G}_\nu}
\newcommand{\Gsc}{{\rm G}^{\rm (sc)}}
\newcommand{\Gnsc}{{\rm G}^{0({\rm sc})}}
\newcommand{\Gtn}{\widehat{\rm G}^0}
\newcommand{\Gtnsc}{\widehat{\rm G}^{0({\rm sc})}}

\newcommand{\dnb}{\partial_{n/b}}
\newcommand{\dnnb}{\partial_{n_0/b}}
\newcommand{\ohmnu}{{\ts\frac{1}{2}-\nu}}
\newcommand{\ohpnu}{{\ts\frac{1}{2}+\nu}}
\newcommand{\cpn}{\cos(\pi\nu)}

\newcommand{\pshift}{\alpha_{\Lambda}}
\newcommand{\Gam}{\Gamma_{\!\rm d}} %
\newcommand{\ga}{{\rm a}} %
\newcommand{\Ga}{{\rm A}} %
\newcommand{\aop}{\hat{a}}
\newcommand{\nR}{m}
\newcommand{\nL}{n}

\newcommand{\po}{{\gamma}}
\newcommand{\pon}{{\smash{\gamma^{(n)}}}}

\newcommand{\ipo}{{\gamma_{\rm\,int}}}
\newcommand{\ipon}{{\smash{\gamma_{\rm\,int}^{(n)}}}}
\newcommand{\ipons}{{\gamma_{\rm\,int}^{(n)}}}
\newcommand{\epo}{{\gamma_{\rm ext}}}
\newcommand{\epon}{{\smash{\gamma_{\rm ext}^{(n)}}}}

\newcommand{\dpo}{{\overline{\gamma}}}
\newcommand{\dpon}{{\smash{\overline{\gamma}^{(n)}}}}

\begin{document}

\begin{frontmatter}
\title{Magnetic Edge States}
\author[mpq,wis]{Klaus Hornberger} 
and 
\author[wis]{Uzy Smilansky}
\address[mpq]{Max-Planck-Institut f\"ur Physik komplexer Systeme,\\
N\"othnitzer Stra{\ss}e 38, 01187 Dresden, Germany}
\address[wis]{The Weizmann Institute of Science, 76100 Rehovot, Israel}
\date{mpipks/0112003}

\begin{abstract}
  Magnetic edge states are responsible for various phenomena of
  magneto-transport.  Their importance is due to the fact that, unlike
  the bulk of the eigenstates in a magnetic system, they carry
  electric current along the boundary of a confined domain.  Edge
  states can exist both as interior (quantum dot) and exterior
  (anti-dot) states.  In the present report we develop a consistent
  and practical spectral theory for the edge states encountered in
  magnetic billiards.  It provides an objective definition for the
  notion of edge states, is applicable for interior and exterior
  problems, facilitates efficient quantization schemes, and forms a
  convenient starting point for both the semiclassical description and
  the statistical analysis.  After elaborating these topics we use the
  semiclassical spectral theory to uncover nontrivial spectral
  correlations between the interior and the exterior edge states. We
  show that they are the quantum manifestation of a classical duality
  between the trajectories in an interior and an exterior magnetic
  billiard.


\end{abstract}

\end{frontmatter}

\newpage
\tableofcontents

\section{Introduction}

Magnetic edge states are formed if a confined two-dimensional electron
gas is penetrated by a strong magnetic field.  
Unlike the bulk of the electronic eigenstates, which approach the
Landau levels as the field is increased, these states 
localize at the edge of the
confinement region and carry a finite current along the boundary.  Due
to their quasi one-dimensional extension and the ability to mediate
transport the edge states play an important role in
various phenomena of semiconductor physics,
most notably the quantum Hall effect \cite{Halperin82}.

In the present review we shall \emph{not} be concerned with the
physics of interacting electrons in real semiconductor samples.
Rather, we study an idealized system: A single charged particle moving
ballistically in a plane which is subject to a homogeneous,
perpendicular magnetic field.  The confinement is caused by
impenetrable walls such that the quantum wave function vanishes
outside the considered region and the corresponding classical particle
is reflected specularly at the boundary.  This simple setup permits to
study in some detail the spectral properties of magnetic edge states
and their relation to the corresponding classical motion, which is
typically chaotic.  Thus, on the one hand, the present study extends
the field of quantum chaos 
\cite{OzoriodeAlmeida88,Gutzwiller90,Haake90,LesHouches91,%
BB97b,Gutzwiller98} 
to magnetic systems which
could not be accounted for so far.  On the other hand, we expect that
many of the results and insights obtained from the model system will
carry over to the analysis of the more realistic case of interacting
electrons in real samples.

Throughout this report the confining boundary will be a closed line
separating the plane into two parts -- a compact {\it interior} and an
unbounded {\it exterior}.  The particle can move in either of these
domains -- forming a quantum dot or the respective anti-dot.  In the
\emph{absence} of a magnetic field the interior system constitutes a
billiard problem whose classical and quantum properties are a paradigm
in the study of chaos and its quantum implications
\cite{Tabachnikov95,OzoriodeAlmeida88,Gutzwiller90,Haake90,%
LesHouches91,BB97b,Gutzwiller98}.  
It exhibits a discrete
quantum spectrum in the interior, while from the exterior the billiard
boundary 
acts as an obstacle of a scattering problem. It is well known that
there exists an intimate relation between the interior quantization
and the exterior scattering system  called the
interior-exterior duality \cite{DS93,EP95}.  The situation changes
if a finite magnetic field is present since now the exterior classical
motion is also bounded: The classical particle is either trapped on a
cyclotron orbit or it performs a skipping motion around the billiard
boundary.  Consequently, the quantum spectrum is purely discrete in
the exterior as well.  It is natural to ask whether any correlations
are to be expected between the interior and the exterior spectra, and
the investigation of this issue is one of the motivations for the
present work.

We shall show that a duality in the underlying classical dynamics of
the skipping trajectories leads to nontrivial cross-correlations
between the interior and the exterior spectra.  In order to observe
this relation it is crucially important to have a proper quantitative
definition for the notion of edge states at hand.  Although the
classical trajectories exhibit a clear partitioning into the skipping
type and the cyclotron orbits, such a sharp division is no longer
valid in the quantum treatment,
and one finds many eigenstates which interpolate between states in the
bulk and the proper edge states.  In the present work we offer a very
natural way of treating this gradual transition between the edge and
the bulk states.  It yields an objective and physically meaningful
definition for edge states
which permits a semiclassical description.

Apart from presenting our results on the properties and the dual
nature of the edge state spectra, the present report is aimed at
providing a consistent and self-contained formulation and exposition
of the following subjects:
\begin{enumerate}
\renewcommand{\labelenumi}{(\alph{enumi})}
\item the exact quantization of interior and exterior magnetic
  billiards based on  boundary integral equations,
\item the semiclassical quantization of interior and exterior magnetic
  billiards in terms of the classical dynamics,
\item a spectral measure for edge states and its semiclassical form, and
\item the relation between the interior and the exterior edge state
  spectra.
\end{enumerate}
\subsubsection*{Structure of the article}

In the next two chapters, we give a survey of the classical and
quantum dynamics in the free magnetic plane and in magnetic billiards,
respectively.  Although many of the statements in \emph{Chapter
  \ref{chap:plane}} are elementary, we shall present them in some
detail for the sake of completeness and to introduce a consistent set
of notations. These chapters include also the discussion of 
concepts, such as the scaling properties or the semiclassical
approximation, to which we refer to frequently in the remainder of the
report.  In the first part of \emph{Chapter \ref{chap:boundary}} the
classical interior-exterior duality is explained.  Turning to the
quantum problem, we introduce general boundary conditions and discuss
the asymptotic properties of magnetic spectra.  The introductory
chapters conclude with the definition of a scaled edge magnetization.

In \emph{Chapter \ref{chap:bim}} we solve the quantization problem in
the interior and exterior of arbitrary magnetic billiards by means of
a boundary integral method.  We explain why spurious solutions arise
initially and how they can be systematically avoided.  The application
of this method in numerical simulations, its accuracy and its
performance is demonstrated in \emph{Chapter \ref{chap:numres}}.  We
focus mainly on two issues: the computation of wave functions in the
extreme semiclassical regime and the extraction of large sequences of
eigenvalues.  The former serves to visualize the properties of edge
and bulk states and the latter enables the study of spectral
statistics and their relation to the underlying classical motion.

\emph{Chapter \ref{chap:trace}} is devoted to the derivation of the
semiclassical trace formula for hyperbolic and integrable magnetic
billiards by means of a surface-of-section method. We start from the
boundary integral operators and formulate the semiclassical
quantization condition in terms of map operators which are
semiclassically unitary and which refer to either the interior or the
exterior. These operators are related in a way which reflects the
underlying classical interior-exterior duality. The integrable disk
billiard is then quantized for a second time making use of its
separability.  In conjunction with the former results, it allows the
trace formula to be extended to general boundary conditions. This
chapter is rather technical but it lays the foundation for the
subsequent analysis.

The spectral density of edge states is introduced in \emph{Chapter
  \ref{chap:edge}}. It gives the concept of edge states a quantitative
meaning and is appropriate, both in the deep quantum and in the
semiclassical regime. As a matter of fact, we propose two different
methods to define the edge spectral densities and discuss their
relative advantages and connections. The new measures allow a spectral
analysis to be performed also in the exterior. The consistency with
random matrix theory is checked in \emph{Chapter \ref{chap:stat}} and
the quantum edge state densities are compared to the predictions of
the semiclassical trace formula.

In \emph{Chapter \ref{chap:cross}} we finally identify non-trivial
cross-correlations between interior and exterior edge state spectra.
We show that they are based on a classical duality of the periodic
orbits.  In order to observe the correlations the spectral density of
edge states or an equivalent measure, such as the edge magnetization,
is of crucial importance.  We conclude this report with a summary and
a list of open problems for further research.

Most of the material which we considered of technical nature is
deferred to the appendices. However, the reader may  (justifiably) become
impatient with some of the derivations deemed by us to be needed for
the coherent exposition.
The busy reader is encouraged to skip directly to Chapters
\ref{chap:bim}, \ref{chap:trace}, and \ref{chap:edge} and to go back
to the  earlier parts  whenever needed.  Note that a list of the
most frequently used symbols  can be found in the appendix.

\section{Motion in the free magnetic plane}
\label{chap:plane}

We start by collecting a number of elementary statements on the
classical and quantum motion in the magnetic plane. This allows to
introduce the notation used throughout the report, and to set the
stage for the discussion in
the following chapters.  
In particular,
the treatment of the quantum time evolution operator in Section
\ref{sec:propagator} yields the opportunity to discuss
the semiclassical approximation.  In Section \ref{sec:freeGreen} the
Green function of a particle in the free magnetic plane is derived in
both its semiclassical and its exact form.

\subsection{Classical motion}
\label{sec:classplane}

Consider the motion of a non-relativistic, spinless, charged
particle in the two-dimensional Euclidean plane,\footnote{%
  The motion on magnetic surfaces of \emph{finite} curvature received
  some attention in recent years both in the classical
  \cite{Comtet87,CGO93,Tasnadi98,Gutkin01} and the quantum treatment
  \cite{Comtet87,AKP92,CGO93,Dunne92,FV01}.  One motivation for
  introducing a non-vanishing curvature is the possibility to study
  the quantum spectrum of the free magnetic motion on a \emph{compact}
  domain (a modular domain in the case of constant negative
  curvature).  This has considerable mathematical advantages since
  the spectrum remains discrete in the limit of vanishing field. }
which is subject to a magnetic field.  Its Lagrangian has the form
\cite{LaLi2}
\begin{gather}
  \label{eq:Lagrangian}
  \Lag = \frac{\mass}{2} {\vvec}^2 + q \,  \vvec \Avec(\rvec)
\CO
\end{gather}
where $\mass$ and $q$ denote mass and charge, respectively.
The vectors $\rvec=(x,y)^{\rm T}$ and $\vvec=\dot{\rvec}$ give the
position and velocity of the particle. Both of them determine the
canonical momentum
\begin{gather}
  \label{eq:pdef}
  \pvec = \frac{\partial\Lag}{\partial\vvec}=\mass\vvec+q\Avec(\rvec)
  \PO
\end{gather}
The classical time evolution is given by
the Lagrangian equation of motion
\begin{gather}
  \label{eq:peom}
    \dot{\pvec }
    = q\,\grad(\vvec \Avec(\rvec))
    \PO
\end{gather}
Here, the magnetic field is described by the (time-independent)
two-dimensional vector potential $\Avec(\rvec)$.  
It
follows
immediately that the equation of motion for the velocity $\vvec$
depends only on the rotation $B=\grad\times\Avec$ of the vector
potential.  It reads
\begin{gather}
\label{eq:newtoneq}
 \mass \, \ddot{\rvec}
   = {qB}\,  \grad(\rvec\times\vvec)
\end{gather}
which is Newton's equation of motion under the action of the
(magnetic) Lorentz force. The latter acts perpendicularly to the
velocity and is proportional to the magnetic field $B$ (the magnetic
induction).

Throughout this report we are interested in the case of a \emph{homogeneous}
magnetic field $B$ (with $q\,B>0$).  Equation \eref{eq:newtoneq} is
then easily integrated, yielding the \emph{cyclotron motion}
\begin{align}
  \label{eq:rotrho}
  \rvec(t)&=\rvec(0) + \frac{1}{ \oc}
  \begin{pmatrix}
    \sin(\oc t) & 1-\cos(\oc t)
    \\
    -1+\cos(\oc t) &  \sin(\oc t)
  \end{pmatrix}
  \vvec(0)
  \\
  \tag{\ref{eq:rotrho}a} 
  \label{eq:rotrho2}
  &= \rvec(0) - \rhovec(0) +\rhovec(t)
\end{align}
with $\rvec(0)$ and $\vvec(0)$ the initial position and velocity,
respectively, and $\oc$=$q B/\mass$ the cyclotron frequency.
The particle moves clockwise on a circle with constant angular
velocity $\oc$.  Below, the velocity will be needed as a function of
the initial and the final position, $\rvec(0)$ and $\rvec(t)$. Apart
from the points in time which are multiples of the cyclotron period
$2\pi/\oc$ it is given by
\begin{gather} 
  \label{eq:vorrn}
   \vvec(t)=\frac{\oh\oc}{\sin(\oh\oc t)}
  \begin{pmatrix}
    \cos(\oh\oc t) & \sin(\oh\oc t)
    \\
    -\sin(\oh\oc t) &  \cos(\oh\oc t)
  \end{pmatrix}
  (\rvec(t)-\rvec(0)) 
  \PO
\end{gather}
The radius vector
\begin{gather} 
  \label{eq:rhovecdef}
  \rhovec(t)\defas\frac{1}{\oc}
  \begin{pmatrix}
    -v_y(t)\\v_x(t) 
  \end{pmatrix}\ ,
\end{gather}
points from the (instantaneous) center of motion to the particle
position.  Clearly, the position of the center
$\cvec(t)=\rvec(t)-\rhovec(t)$ is a constant of the motion.  To verify
this in a more formal way one may consider the classical Hamiltonian
\begin{align}
  \label{eq:Hamconv}
  \Ham=\pvec\dot{\rvec}-\Lag
  &=\frac{1}{2\mass}(\pvec-q\Avec(\rvec))^2
\end{align} 
as a function of the canonically conjugate variables $\rvec$ and
$\pvec$.  A short calculation shows that the Poisson bracket indeed
vanishes,
\begin{gather}
  \label{eq:ddtc}
  \frac{\rmd}{\rmd t} (\rvec-\rhovec)
  \equiv
  \frac{\rmd}{\rmd t} \cvec
  =
  \{\Ham,\cvec\}=0
\PO
\end{gather}
Similarly, the (kinetic) energy
$E\defas\Ham(\rvec,\pvec)=\frac{\mass}{2}\vvec^2$ is constant, as well
as the cyclotron radius $|\rhovec|$ and the kinetic angular momentum
with respect to the center of motion $\rhovec\times\vvec$, which are
functions thereof.  In contrast, the canonical momentum $\pvec$
itself is not a constant of the motion. In general, it does not have a
kinetic meaning since it depends on the vector potential, cf
\eref{eq:pdef}, which is not uniquely specified by the magnetic field.
Rather, the gradient of any scalar field $\Chi(\rvec)$, ie, any
``gauge field'', may be added to the vector potential without
affecting the classical equation of motion \eref{eq:newtoneq}.

We note that the general vector potential for homogenous magnetic
fields may be written in the form
\begin{gather}
  \label{eq:Aconv}
  \Avec(\rvec)=\frac{B}{2} {-y \choose x} + \grad \Chi(\rvec)
\PO
\end{gather}
The choice of $\Chi$ is a matter of convenience.  An important case is
the \emph{symmetric} gauge, $\Chi=0$, which distinguishes merely a
point in the plane (the origin).  Choosing $\Chi=-\frac{B}{2}xy$, on
the other hand, yields the \emph{Landau} gauge which distinguishes a
direction (the orientation of the $y$-axis).
These two gauges are particularly important because they turn
components of the canonical momentum into constants of the motion. In
the Landau case $p_x$ is given by the (constant) $y$-component of the center of
motion,
\begin{align}
  \label{eq:AconvL}
  \Avec = \Avec_{\rm Lan}\equiv B {-y \choose 0}
  &\q\Rightarrow\q
  p_x = -\mass\oc\,c_y
\CO
\intertext{%
while the symmetric gauge fixes the (canonical) angular momentum with
respect to the origin,
$L=\rvec\times\pvec$,}
  \label{eq:Aconvsym}
  \Avec=\Avec_{\rm sym}\equiv
  \frac{B}{2} {-y \choose x} 
  &\q\Rightarrow\q
  L\defas \rvec\times\pvec = \frac{\mass\oc}{2}\left(|\cvec|^2-|\rhovec|^2\right)
\PO  \!\!\!\!\! \!\!\!\!\!
\end{align}
It is determined by the distance $|\cvec|$ of the center of motion
from the origin, and the cyclotron radius $\rho=|\rhovec|$.

Below, it will be important at several points to state equations in a
\emph{manifestly} gauge invariant fashion.  This is done by keeping
$\Chi$ unspecified and verifying that the resulting expressions do not
depend on its choice.  As the only restriction, $\Chi$ will be assumed
to be a harmonic function, ie, $\grad^2\Chi=0$, throughout.  This
rules out conveniently the occurrence of singularities in $\Chi$ but
keeps the essential gauge freedom.  Moreover, it ensures that the
vector potential \eref{eq:Aconv} is a transverse field, ie, divergence
free, $\grad\Avec=0$, which facilitates a number of mathematical
transformations.

Turning to the quantum mechanical description, the quantum time
evolution will be treated in terms of the path integral formulation in
Section \ref{sec:propagator}. Before that we discuss the stationary
solutions of the Schr{\"o}dinger equation (in a specific gauge, to
prove the rule stated above). This permits to obtain the spectrum and
the scaling properties of the Hamiltonian in a straightforward manner.

\subsection{Quantization}
\label{sec:quantization}

In the quantum description the canonical variables $\rvec$ and $\pvec$
become observables expressed as operators in $\Hilbert(\Rtwo)$.  They
turn the Hamiltonian \eref{eq:Hamconv} into an operator
whose spectrum determines the energies $E$ of the stationary states.
In position representation,  $\pvec=-\rmi\hbar\grad$,
the stationary Schr{\"o}dinger equation reads
\begin{align}
  \label{eq:sse}
    \frac{1}{2\mass}(-\rmi\hbar\grad-q\Avec)^2
    \,\psi(\rvec)
    &=E\,\psi(\rvec) \ . 
\end{align}
In addition, the solution $\psi(\rvec)$ must be \emph{normalizable}
to qualify as a stationary quantum state.
The energy eigenstates in the magnetic plane were obtained not before
1930, when Landau published his article on orbital diamagnetism
\cite{Landau30}. 
Although he used the gauge \eref{eq:AconvL}, the symmetric vector
potential \eref{eq:Aconvsym} will prove more convenient in the
following. 
First, we introduce a (quantum) length scale
\begin{gather}
\label{eq:bdef}
  b \defas \left(\frac{2\hbar}{qB}\right)^\oh
\end{gather}
and call it the magnetic length, although it \emph{differs} from
Landau's definition\footnote{%
  Landau's definition of the magnetic length $\ell_B=b/\sqrt{2}$ is
  appropriate for the Landau gauge \eref{eq:AconvL}.
  The length $b$ (which is the suitable scale of the symmetric
  gauge)  proves more convenient since
  it avoids the appearance of the factor $2$ and $\sqrt{2}$ at various
  places.  It gives the radius of a disk the area $b^2\pi$ of which
  assumes the role of Planck's quantum, cf \eref{eq:INtota}.
  (The flux through the disk equates the ``flux quantum''
  $\Phi_0=h/q=B\, b^2\pi$).
  } by a factor of $\sqrt{2}$. It allows to transform position and
momentum operators into dimensionless quantities, denoted by a tilde,
\begin{gather}
  \label{eq:defrtpt}
  \rvect \defas \frac{\rvec}{b}
  \q\text{and}\q
  \pvect  \defas \frac{b}{\hbar}\pvec
\PO
\end{gather}
In the symmetric gauge the Hamiltonian \eref{eq:Hamconv} now assumes
a particularly simple form,
\begin{gather}
  \label{eq:Hdl}
  \Ham=
  \hbar\omega\,  
  \oh\left({\pvect^2}+{\rvect^2}\right)
  - \omega\, \hbar(\rvect\times\pvect)
  = \Ham_{\rm osc} - \omega\, {\rm L}.
\end{gather}
It is given by the energy of a two-dimensional harmonic oscillator
$\Ham_{\rm osc}$ minus its angular momentum ${\rm L}=\rvec\times\pvec$,
in quanta of the same size.  The oscillator eigen-frequency
{differs} from the cyclotron frequency by a factor of $2$. It is given
by
\begin{gather}
  \label{eq:larmordef}
  \omega\defas\frac{qB}{2\mass}=\frac{\oc}{2}
\CO
\end{gather}
and known from the precession of magnetic moments as the Larmor
frequency.  In order to construct the complete set of energy
eigenstates on the magnetic plane
 consider the annihilation operators of the
left- and right-circular quanta,
\begin{gather}
  \label{eq:adef} 
  \aop_{\rm R\choose L} = \oh(\xt\mp\rmi \yt+\rmi(\pxt\mp\rmi\pyt)) \ ,
\end{gather}
with $[ \aop_{\rm L},\aop_{\rm L}^\dagger]=[ \aop_{\rm R},\aop_{\rm R}^\dagger]=1$
as the only non-vanishing commutators. It is well known \cite{CTDLqm1} that
the simultaneous eigenstates of the left- and right-circular number
operators $( \aop_{\rm L}^\dagger \aop_{\rm L})$ and $( \aop_{\rm R}^\dagger
\aop_{\rm R})$ form a complete basis set of $\Hilbert(\Rtwo)$. An
oscillator eigenstate
corresponding to $\nL$ left-circular and $\nR$ right-circular quanta
is given by
\begin{gather}
  \label{eq:defcs}
  \ket{\nL,\nR}
  =
  \frac{1}{\sqrt{\nL!\,\nR!}}
   (\aop_{\rm L}^\dagger)^{\nL}
   (\aop_{\rm R}^\dagger)^{\nR}
   \ket{0,0}\ ,
\end{gather}
with $n,m\in\Nnull$.  Here, $\ket{0,0}$ denotes the harmonic oscillator
ground state, a Gaussian in position representation,
$ \braket{\rvec}{0,0} = \exp(-\oh \rvec^2/b^2)/\sqrt{b^2\pi}$.  Like
all the states \eref{eq:defcs} it is square-integrable and 
normalized.

Inverting equations \eref{eq:adef} the Hamiltonian of a particle
in the magnetic plane may be expressed in terms of the circular
operators.
It assumes a form
\begin{align}
  \label{eq:Hamco}
  \Ham &= 
  \Ham_{\rm osc} - \omega\, \Lt = 
  \hbar\omega\,(  
  \aop_{\rm R}^\dagger \aop_{\rm R}
  +
  \aop_{\rm L}^\dagger \aop_{\rm L}
  + 1
  )
  -  \hbar\omega\,(  
  \aop_{\rm R}^\dagger \aop_{\rm R}
  -
  \aop_{\rm L}^\dagger \aop_{\rm L}
  )
  \nnn
  &=
  \hbar{\oc}\, \left( \aop_{\rm L}^\dagger \aop_{\rm L} +\oh\right)
  \CO
\end{align}
which depends only on the number operator of the left-circular quanta.
It follows  that the states \eref{eq:defcs} form a complete
set of eigenstates of the magnetic plane. Their energies are
determined by the number $n$ of left-circular quanta, called the
Landau level,
\begin{gather}
  E = \hbar\oc\left(\nL+\oh\right)  \PO
\end{gather}
This proves that the spectrum 
of $\Ham$ is
{discrete} and equidistant.\footnote{ For mathematical literature on
  the spectral properties of magnetic Schr{\"o}dinger operators see
  \cite{AHS78,Helffer94}.  }  The fact that the energy does not depend
on $\nR$ shows that each Landau level is
infinitely {degenerate} (with a countable infinity).
This degeneracy is due to the energy independence of the position of
the center of motion.  To show that the latter is indeed determined by
the right-circular quanta alone we note the operators corresponding
to the classical radius vector \eref{eq:rhovecdef} and the center of
motion $\cvec=\rvec-\rhovec$, respectively,
\begin{gather}
  \label{eq:rhocqm}
  \rhovect \equiv
  \frac{\rhovec}{b}
  =\oh
  \begin{pmatrix}
    \aop_{\rm L}+\aop_{\rm L}^\dagger
    \\
    -\rmi(\aop_{\rm L}-\aop_{\rm  L}^\dagger)
  \end{pmatrix}
  \qq\text{and}\qq
  \cvect \equiv
  \frac{\cvec}{b}=
  \oh
  \begin{pmatrix}
    \aop_{\rm R}+\aop_{\rm  R}^\dagger 
    \\
    \rmi(\aop_{\rm R}-\aop_{\rm R}^\dagger)
  \end{pmatrix}
  \ .
\end{gather}
Here,  \eref{eq:pdef} was used to express the velocity in
terms of momentum and position. Clearly, $\cvec$ commutes with the
Hamiltonian like in the classical case.  The components $\rho_x$ and
$\rho_y$, on the other hand, are not constants of the motion, although
the cyclotron radius $|\rhovec|$ is again fixed and determined solely
by the energy.  This can be seen from the squared moduli of the
vectors,
\begin{gather}
  \label{eq:r2c2}
  |\rhovect|^2 = \aop_{\rm  L}^\dagger \aop_{\rm  L}+\oh
\qq\text{and}\qq
  |\cvect|^2 = \aop_{\rm  R}^\dagger \aop_{\rm  R}+\oh \CO
\end{gather}
which contain only the number operators of left- and right-circular
quanta.  Consequently, the states \eref{eq:defcs} with fixed $n$ and
$m$ are eigenstates of these operators. They are characterized by
definite expectation values for the cyclotron radius and for the distance
from the origin to the center of motion.  Moreover, these
stationary states are eigenvectors of the (canonical) angular momentum
given by the difference $|\cvect|^2-|\rhovect|^2 =L/\hbar$, in analogy
to the classical result \eref{eq:Aconvsym}.
The general eigenstate of $ |\rhovect|^2$ (with eigenvalue $n+\oh$) is
given by a superposition of states \eref{eq:defcs} with different
quantum numbers $m$. We will call any such stationary state a Landau
state in the Landau level $n$.

\subsubsection*{Coherent states}

Since
the states \eref{eq:defcs} are eigenstates of the radial components of
the operators $\rhovec$ and $\cvec$ their azimuthal components are
maximally uncertain. It is known from the two-dimensional harmonic
oscillator that the  common eigenvectors of $\aop_{\rm
  L}$ and $\aop_{\rm R}$ have the property to minimize the uncertainty
product \cite{CTDLqm1}.  These \emph{coherent} states are given by the
superposition
\begin{gather}
  \label{eq:coherent}
  \Ket{{{\alpha_{\rm L}};{\alpha_{\rm R}}}} 
  \defas
  \exp\left(\!-\,\frac{|{\alpha_{\rm L}}|^2+|{\alpha_{\rm R}}|^2}{2}\right)
  \sum_{\nL,\nR=0}^\infty
  \frac{({\alpha_{\rm L}})^{\nL}({\alpha_{\rm R}})^{\nR} }{\sqrt{\nL!\nR!}}
  \,
  \ket{\nL,\nR}
\CO
\end{gather}
with $\alpha_{\rm L}$, $\alpha_{\rm R}\in\Complex$ the associated
eigenvalues.  If considered in the magnetic plane, 
the expectation values of
$\rhovec$ and $\cvec$ are  determined  directly
by these eigenvalues,
\begin{align}
  \Bra{{{\alpha_{\rm L}};{\alpha_{\rm R}}}} 
  \rhovec
  \Ket{{{\alpha_{\rm L}};{\alpha_{\rm R}}}} 
  &= b
  \begin{pmatrix}
    \Re({\alpha_{\rm L}}) \\   \Im({\alpha_{\rm L}})
  \end{pmatrix}
  \nnn
  \Bra{{{\alpha_{\rm L}};{\alpha_{\rm R}}}} 
  \cvec
  \Ket{{{\alpha_{\rm L}};{\alpha_{\rm R}}}} 
  &= b
  \begin{pmatrix}
    \Re({\alpha_{\rm R}}) \\   -\Im({\alpha_{\rm R}})
  \end{pmatrix}
  \CO
\end{align}
as can be found immediately from equation \eref{eq:rhocqm}.
The corresponding uncertainties $\Delta\rho_x=\Delta\rho_y= \Delta
c_x=\Delta c_y=b/2$ are minimal, indeed. 
Furthermore, the  wave functions \eref{eq:coherent}
remain of the coherent type as they evolve in time. From \eref{eq:Hamco}
one observes  that the state at time $t$,
\begin{gather}
  \label{eq:coherev}
  \rme^{-\rmi\Ham t/\hbar}
  \,
  \Ket{{{\alpha_{\rm L}};{\alpha_{\rm    R}}}}
  =
  \rme^{-\rmi\oc t/2}
  \Ket{  \rme^{-\rmi\oc t}{{\alpha_{\rm L}};{\alpha_{\rm    R}}}}
  \CO
\end{gather}
is merely characterized by a different phase of ${\alpha_{\rm L}}$. 
It is a localized wave packet rotating with cyclotron frequency $\oc$
around the constant center of motion $\cvec$.  As such it embodies the
closest quantum analogy \cite{Schroedinger26}
 to the classical motion discussed in Section
\ref{sec:classplane}.

\subsubsection*{Gauge invariance}

So far, the quantum problem was discussed for the symmetric gauge
\eref{eq:Aconvsym} only.  We will now admit an arbitrary gauge again
and consider the consequences of a finite choice of $\Chi$.  Although
the canonical momentum is gauge dependent, its representation as a
differential operator, $\pvec=-\rmi\hbar\grad$, contains no dependence
on the vector potential.  This can be understood by the observation
that the velocity operator
\begin{gather}
  \vvec =
  \frac{1}{\mass}(\pvec-q\Avec)
  =
  \frac{\rmi}{\hbar}[\Ham,\rvec]
\end{gather}
undergoes a unitary transformation
as one changes  the gauge: 
\begin{gather}
  \label{eq:vunit}
  \frac{1}{\mass}\big(
  -\rmi\hbar\grad-q \Avec(\rvec)
  \big)
  =
  \rme^{\rmi q \Chi(\rvec)/\hbar}
  \,
  \frac{1}{\mass}\big(
    -\rmi\hbar\grad-q \Avec_{\rm sym}(\rvec)
  \big)
  \,
  \rme^{-\rmi q \Chi(\rvec)/\hbar}
\end{gather}
Consequently, in order to preserve the gauge independence of the
velocity expectation value also the wave functions must be
transformed unitarily as the gauge is changed.
This is found immediately by applying \eref{eq:vunit} twice to the
time dependent Schr{\"o}dinger equation at arbitrary gauge,
\begin{align}
  \label{eq:tdse}
  \rmi\hbar\,\partial_t\ket{\psi_\chi}
  &=
    \frac{1}{2\mass}(-\rmi\hbar\grad-q\Avec)^2
    \,\ket{\psi_\chi}
  \nnn
    &=
  \rme^{\rmi q \Chi(\rvec)/\hbar}
  \,
  \frac{1}{2\mass}(-\rmi\hbar\grad-q\Avec_{\rm sym})^2
  \,
  \rme^{-\rmi q \Chi(\rvec)/\hbar}
  \,\ket{\psi_\chi}
  \PO
\intertext{Comparing the wave function with the one of  the symmetric gauge,}
  \rmi\hbar\,\partial_t\ket{\psi_0}
  &=
  \frac{1}{2\mass}(-\rmi\hbar\grad-q\Avec_{\rm sym})^2
  \,\ket{\psi_0}
  \CO
\end{align}
we see that they are  related by a local, unitary transformation
\begin{gather}
  \label{eq:gtrafowf}
  \ket{\psi_\chi} = \rme^{\rmi q\Chi(\rvec)/\hbar}  \ket{\psi_0}
  \equiv  \rme^{\rmi \Chit(\rvect)}  \ket{\psi_0}
\end{gather}
which is determined by the gauge field $\Chi$ (in dimensionless units
$\Chit(\rvect)\defas 2 \Chi(\rvec)/(B b^2)$).
It follows that the velocity expectation value is gauge invariant.
The same holds for all observables which commute with $\rvec$, due to
the local nature of the transformation \eref{eq:gtrafowf}.  
As an immediate consequence, the probability
density $ |\psi|^2(\rvec)$ and the probability flux,
$\jvec(\rvec)$ are also gauge-invariant. The latter may
be identified from the continuity equation
$\grad\jvec=-\partial_t|\psi|^2$, which follows from
\eref{eq:tdse}, as
\begin{align}
  \label{eq:defj}
  \jvec%
  \defas 
  \Re(\psi^*\vvec\psi)
  = \frac{\hbar}{\mass}\Im(\psi^*\grad\psi)-\frac{q}{\mass}\Avec|\psi|^2
  \PO
\end{align}
Like all observables which include the gradient in position
representation it contains the vector potential explicitely to
account for the gauge-dependent phase of the wave function.

\subsection{The scaling property}
\label{sec:scaling}

The magnetic Schr{\"o}dinger operator conventionally contains the four
parameters $\hbar$, $\mass$, $q$, $B$, along with the energy $E$ as
the spectral variable.  Due to the homogeneity of the vector potential
\eref{eq:Aconvsym} it is possible to reduce those to the two
principal length scales which we encountered in the previous sections.
Those are the cyclotron radius $\rho$  \eref{eq:rhovecdef} and the
magnetic length $b$ \eref{eq:bdef},
respectively, given by
\begin{gather}
  \label{eq:rhobdef}
  \rho^2\defas\frac{2\mass E}{q^2 B^2}
  \qq\text{and}\qq
  b^2\defas\frac{2\hbar}{q B}
  \PO
\end{gather}
The cyclotron radius is a quantity of classical mechanics. The
magnetic length, in contrast, has a pure quantum meaning. As discussed
above, it determines the mean extension of a minimum uncertainty
state, and vanishes as $\hbar\to 0$.

In the preceding section the dimensionless variables $\rvect=\rvec/b$
and $ \pvect=b\pvec/\hbar$ were introduced.  In fact, the homogeneity
of the potential \eref{eq:Aconvsym}, in conjunction with the
requirement $[\xt,\pxt]=[\yt,\pyt]=i$, leads necessarily to the
magnetic length as the appropriate scale. The only freedom is a
numerical factor in the definition of $b$. We took it such that the
induced time scale $\tit=\omega t$ is given by the (classical) Larmor
frequency $\omega$  \eref{eq:larmordef}.  It is appropriate to
measure time in terms of the Larmor period $T=2\pi/\omega$,
rather than the cyclotron period $T_{\rm cyc}=\oh T$, because the
former is the fundamental time scale of the quantum problem:
It takes \emph{two} cyclotron periods, as one observes from equation
\eref{eq:coherev} (and more generally from the propagator
\eref{eq:propagator}), before a wave packet returns to its initial state
with correct parity.

The respective dimensionless Lagrangian, furnished with a tilde like all scaled
units, reads
\begin{gather}
  \label{eq:Lagtdef}
  \Lagt=\frac{\Lag}{\hbar\omega}
  =  \oh\vvect^2+\rvect\times\vvect+\vvect\,\grad_{\rvect}\Chit
  = \oh\vvect^2+\vvect\Avect(\rvect)
  \PO
\end{gather}
It contains no parameters any more, but for the definition of the
scaled gauge field,
\begin{gather}
  \label{eq:Chitdef}
  \Chit(\rvect)\defas \frac{2}{B b^2}  \Chi(b\rvect)
\end{gather}
(which is not necessarily homogeneous of order two). This implies the
definition of the general scaled vector potential 
${\Avect(\rvect)={2}\Avec(b\rvect)/({B b})}$.
The scaled Hamiltonian, given by
\begin{gather}
\label{eq:Hamtdef}
  {\Hamt}=\frac{\Ham}{\hbar\omega}=
  \oh(\pvect-\Avect)^2
  \CO
\end{gather}
shows that the proper, scaled energy reads
$\widetilde{E}=E/(\hbar\omega)=2\rho^2/b^2$.  We will 
state the energy in terms of the spacing between Landau levels, 
\begin{gather}
  \label{eq:defnu}
  \nu \defas 
  \frac{E}{\hbar\oc} =
  \frac{E}{2\hbar\omega} = \frac{\rho^2}{b^2}
  \CO
\end{gather}
and call $\nu=\tilde{E}/2$ the scaled energy, nonetheless. This way we
conform with the popular convention that the Landau levels start at
one half, rather than at one.

Below, it will be important to distinguish between the two independent
short-wave limits of magnetic dynamics. From expression
\eref{eq:defnu} one observes that the spectral variable $\nu$ can be
increased by either increasing $\rho$ at constant magnetic length $b$,
or by decreasing $b$ at fixed cyclotron radius $\rho$.
The former direction is realized by raising the conventional energy at
constant magnetic field. It is  the standard \emph{high-energy} limit.
Here, the curvature of the classical trajectory tends to zero, which
shows that in this limit the dynamical effect of the magnetic field
vanishes.
On the other hand, one may increase both the conventional energy and
the field at a fixed ratio of $E/B^2$, thereby keeping the cyclotron
radius fixed. This way the underlying classical phase space is kept
invariant, while the magnetic length tends to zero. It is a
realization of the \emph{semiclassical} limit since $b^2$ plays the
role of $\hbar$ as the semiclassically small parameter.%

In order to be able to consider both limits most equations will
\emph{not} be written in scaled variables, since they might depend on
the choice of the independent variable.  Rather, the formulas will be
stated in terms of combinations like $\rvec/b$ so that they can be
immediately replaced by scaled variables.  This includes the scaled
gradient, $\grad_{\rvect}\equiv b\grad_\rvec$,  written as
\begin{gather}
  \grad_{\rvec/b}\defas b\grad_\rvec 
\CO
\end{gather}
which is an admittedly unusual but consistent notation.  The spectral
variable is always stated as $\nu$.

\subsection{The free quantum propagator}
\label{sec:propagator}

We return to the Lagrangian formulation of mechanics in order to
calculate the time evolution operator $ {{\rm U}(t;0)}:=
\exp[-{\rmi}\Ham t/{\hbar}]$ for arbitrary gauge.  According to
\mbox{Feynman} its position representation (for $t>0$) is given by the path
integral \cite{FH65,Schulman81,GS98}
\begin{gather}
  \label{eq:pathint}
   {{\rm U}}(t,\rvec;0,\rvec_0)
   =
   \overset{\mathbf{q}(t)=\rvec}{\underset{\mathbf{q}(0)=\rvec_0}{\int}}
   \!\!\!\!
   \mathcal{D}[\mathbf{q}]
   \,
   \exp\left(\frac{\rmi}{\hbar}\W[\mathbf{q}]\right)
   \ .
\end{gather}
Here, the functional $\W$ attributes a classical
action
\begin{gather}
  \label{eq:acq}
  \W[\mathbf{q}] \defas \int_0^{t} 
   \!\!
  \Lag(\mathbf{q}(t'),\dot{\mathbf{q}}(t')) 
 \, \rmd t'
\end{gather}
to all paths $\mathbf{q}(t')$ going from
$\rvec_0$ to $\rvec$  in the given time $t$.
(All equations are stated for a time independent Lagrangian, and the
zero indicating the initial time will be omitted in the
following.)

The formulation in terms of a path integral permits the calculation of
the time evolution operator in a straightforward manner. Its most
important advantage is that the \emph{semiclassical} approximant of the
propagator can be obtained in a transparent way.
The situation is called semiclassical if $\hbar$ is small
compared to the actions \eref{eq:acq}.  In this case the dominant
contributions to the path integral are represented by those paths for
which the phase in \eref{eq:pathint} is stationary. They are solutions
of the variational problem $\delta \W[\mathbf{q}]=0$ with fixed initial
and final position and time. According to Hamilton's principle these
are classical trajectories. The integral is then evaluated by 
expanding the variations of  \eref{eq:acq} to second order. 
Provided the trajectories are isolated one obtains the asymptotic
expression of the propagator to leading order in $\hbar$
\cite{Gutzwiller67}.
\begin{gather}
  \label{eq:vanVleck}
  {\rm U}(t,\rvec;\rvec_0)
  =
  \frac{1}{2\pi\rmi\hbar}
  \sum_{\mathbf{q}_{\rm cl}}
  \left|
    \det
    \bigg(\!\!-
      \frac{\partial^2{\W[\mathbf{q}_{\rm cl}]}}{\partial\rvec\partial\rvec_0}
    \bigg)
  \right|^\oh
  \exp\left(\frac{\rmi}{\hbar}{\W}[\mathbf{q}_{\rm cl}]
    -\rmi\piot\nu_{\mathbf{q}_{\rm cl}}\right)
  \,\,(1+\Or(\hbar))
\end{gather}
It is a sum over all classical trajectories $\mathbf{q}_{\rm cl}$ going
from $\rvec_0$ to $\rvec$, in the given time $t$.  The only quantum
ingredient is the finite size of $\hbar$, which sets the scale of the
associated classical action in the phase factor. 
The additional phase shift is determined by
the number
$\nu_{\mathbf{q}_{\rm cl}}$ of negative
eigenvalues of the matrix 
$\big(-{\partial^2{\W[\mathbf{q}_{\rm
      cl}]}}/({\partial\rvec\partial\rvec_0})\big)$
\cite{Gutzwiller67}.  The latter has a dynamical meaning
\cite{Schulman81,GS98}, it is the inverse of the Jacobi field of
$\mathbf{q}_{\rm cl}$, which describes the linearized deviation of
classical trajectories with different initial momenta. The points on
$\mathbf{q}_{\rm cl}$ where classical trajectories coalesce are called
focal or \emph{conjugate}.  They determine 
$\nu_{\mathbf{q}_{\rm cl}}$ geometrically by virtue of the Morse
theorem \cite{Morse73}:
The value of $\nu_{\mathbf{q}_{\rm cl}}$ is equal to the number of
conjugate points the particle encounters on its journey (counted with
their multiplicities \cite{Morse73}) and is called the Morse index.

We are now in a position to derive the time evolution operator in the
free magnetic plane.  The quadratic dependence of the Lagrangian
\eref{eq:Lagrangian} on position and velocity renders the expression
\eref{eq:vanVleck} for the time evolution operator \emph{exact} rather
than asymptotic.  First, the (scaled) action of a trajectory is needed as a
function of the initial and the final position, $\rvec_0$ and $\rvec$,
and the time of flight $t$.  From the classical solution in Section
\ref{sec:classplane} one obtains 
\begin{align}
  \label{eq:Wt}
  \frac{1}{\hbar}{\W}(t,\rvec;\rvec_0)
  &
  = \widetilde{\W}(\tit,\rvect;\rvect_0)
  =\int_0^\tit   \!\! 
  \left(\oh \vvect(\tit')^2
    +\vvect(\tit')\Avect(\rvect(\tit'))
    \right)
    \,\rmd \tit'  
  \nnn
  &= \oh \int_0^\tit \vvect^2(\tit') \rmd \tit' 
  +\int_{\partial Q} \!\!
  \Avect(\rvect')\rmd \rvect' 
  +\underset{\rvect_0\to\rvect}{\int} \!\! \Avect(\rvect')
  \,\rmd  \rvect'
  \nnn
  &= \oh{(\rmrnt)^2}\cot(\tit) - \rxrnt+\Chit(\rvect)-\Chit(\rvect_0)
  \PO
\end{align}
Here, the action integral was split into three parts:
\begin{align}
  \label{eq:vnull}
   &\int_0^\tit \!\! \vvect^2(\tit',\rvect;\rvect_0)  \,\rmd \tit'  
   =
   \frac{(\rmrnt)^2}{\sin^2(\tit)} \,\frac{\tit}{2}
   \\
   \label{eq:areaint}
   &\int_{\partial Q} \!\!  \Avect(\rvect')\,\rmd \rvect' 
    =
    - \frac{(\rmrnt)^2}{\sin^2(\tit)}\,\frac{\tit}{2}
    + \frac{(\rmrnt)^2}{2}\cot(\tit)
   \\
   \label{eq:strint}
   &\underset{\rvect_0\to\rvect}{\int} \!\! \Avect(\rvect')\,\rmd \rvect' 
   = -\rxrnt+\Chit(\rvect)-\Chit(\rvect_0)
\end{align}
In the first, the modulus of the velocity is constant. Its value
\eref{eq:vnull} follows from \eref{eq:vorrn}.  The second part was
made a closed line integral, encircling a domain $Q$, which is confined
by the trajectory and the straight line from $\rvect$ back to
$\rvect_0$.  
By Stokes' theorem
one obtains \eref{eq:areaint}, with the minus sign due to the
clockwise, ie, negative sense of integration.  The remaining part
\eref{eq:strint} is a line integral along the straight path from
$\rvect_0$ to $\rvect$.  Unlike the other contributions, it depends on
$\rvect$ and $\rvect_0$ individually and carries the gauge
dependence.

In principle, more than one classical trajectory could connect the two
points $\rvect$ and $\rvect_0$ in a given time. However, since the
determinant of the matrix in \eref{eq:vorrn} is non-zero for $\tit\ne
n\pi, n=1,2,\ldots$, the initial velocity is uniquely specified for
those times. At integer multiples of the cyclotron period, in
contrast, any trajectory returns to its starting point.  Excluding
these instances for the time being, the time evolution operator is
determined by only one trajectory.  For the matrix of second
derivatives one obtains 
\begin{gather}
  \label{eq:Rsecdif}
  \det\left(
    \frac{\partial^2\widetilde{\W}}{\partial\rvect\partial\rvect_0}
    \right)
  =
  \frac{1}{\sin^2(\tit)}
  \PO
\end{gather}
The determinant of its inverse has doubly degenerate zeros at
$\tit=n\pi$. Hence, the Morse index reads $\nu_{\mathbf{q}_{\rm cl}} = 2[\tit/\pi]$
(with $[\cdot]$ the integer part),
and one arrives immediately at the time evolution operator in the free
magnetic plane
\begin{align}
  \label{eq:propagator}
  {\rm U}(t,\rvec;\rvec_0)
  =&
  \frac{1}{2\pi\rmi b^2}
  \frac{1}{\sin(\omega t)}
  \exp
  \left[\rmi
    \frac{(\rmrn)^2}{2b^2}\cot(\omega t) 
    -\rmi\frac{\rxrn}{b^2}
  \right]  
  \nnn
  &\times\exp
  \left[\rmi
    \Big(
    \Chit\Big(\frac{\rvec}{b}\Big)-\Chit\Big(\frac{\rvec_0}{b}\Big)
    \Big)
  \right]  
\PO
\end{align}
As noted above, this expression is 
identical to the exact path integral
\cite{Glasser64,FH65,LS77,Cheng84}.  It is valid except for the times
equal to integer multiples of the cyclotron period. At these instances
the propagator is just a unit operator,
\begin{align}
  \label{eq:deltaprop}
  \lim_{\omega t\to{n\pi}} 
  {\rm U}(t,\rvec;\rvec_0)
  &=
  \lim_{\epsilon\to 0}\,
  \frac{1}{2\pi\rmi b^2}\,
  \frac{(-)^n}{\sin(\epsilon)}
  \exp
  \left[\rmi
    \frac{(\rmrn)^2}{2b^2}\cot(\epsilon) 
  \right]  
  \nnn
  &\phantom{=\lim_{\epsilon\to 0}\,}
  \times
  \exp\left[
    -\rmi\frac{\rxrn}{b^2}
     +\rmi\Chit\Big(\frac{\rvec}{b}\Big)-\rmi\Chit\Big(\frac{\rvec_0}{b}\Big)
    \right]  
  \nnn
  &=  (-)^n \,\delta\left(\frac{\rmrn}{b}\right) 
\CO
\end{align}
with a sign which is positive only after even multiples of the
cyclotron period. This means that any wave function which is
propagated by multiples of the Larmor period $T=2\pi/\omega=2T_{\rm
  cyc}$ returns precisely to its initial state.
Equation \eref{eq:deltaprop} follows from a special representation of
the two-dimensional $\delta$-function which is given in the
appendix, see \eref{eq:delta2d}.
Note  that the propagator \eref{eq:propagator} was derived
for positive times $t>0$ only. It is valid for all times, nonetheless,
since it clearly obeys the unitarity relation 
$ {\rm  U}(-t,\rvec;\rvec_0) = [{\rm U}(t,\rvec_0;\rvec)]^*$.  
Furthermore,
it is given for arbitrary vector potentials.  The dependence on $\Chi$
shows how the propagator transforms as the gauge is changed.  It is
consistent with the gauge dependence of the wave functions
\eref{eq:gtrafowf} discussed in Section \ref{sec:quantization}.

\subsection{The free Green function}
\label{sec:freeGreen}

We are now in a position to calculate the Green function of the free
magnetic plane. It will be an important ingredient in the theory of
the exact and semiclassical quantization of magnetic billiards.  We
define the Green function to be the Fourier transform of the free
propagator
\begin{gather}
  \label{eq:fourint}
  {\rm G}(E,\rvec;\rvec_0)
  \defas \frac{\hbar}{2\rmi \mass }
  \lim_{\epsilon\downarrow 0}
  \int_0^\infty
  {{\rm U}}(t,\rvec;\rvec_0)
  \rme^{\rmi (E+\rmi\epsilon)t/\hbar}
  \, \rmd t 
  \ .
\end{gather}
As such, it is a resolvent of the Hamiltonian obeying the
inhomogeneous Schr{\"o}dinger equation
\begin{gather}
  \label{eq:seconvinh}
  (\Ham-E) {\rm G}(E,\rvec;\rvec_0)=-\frac{\hbar^2}{2 \mass} \delta(\rmrn)
  \PO
\end{gather}
For later reference we note that there exists a second, independent
solution of \eref{eq:seconvinh} which differs appreciably from ${\rm G}$.
We shall call it the unphysical or \emph{irregular} Green function
${\rm G}^{\rm (irr)}$.  

One 
procedure to obtain the Green function is based on the observation
that the differential equation \eref{eq:seconvinh} separates in polar
coordinates if the symmetric gauge is used.  This way one is led to an
angular momentum decomposition of ${\rm G}$, which is of little use
for our purposes. It was derived (with some errors) in
\cite{KR92,TCA97} and is summarized in Appendix \ref{app:nullfield}.
Here, we perform the Fourier integral \eref{eq:fourint}
directly. It yields the Green function 
in a clear-cut fashion,
in Cartesian representation and arbitrary gauge.
Substituting  scaled variables the integral \eref{eq:fourint} reads
\begin{align}
  \label{eq:Gintexp}
  &G_\nu(\rvec;\rvec_0) \defas  {\rm G}(2\hbar\omega\nu,\rvec;\rvec_0)
  \nnn
  &=  
  \frac{-1}{4\pi}
  \int_0^\infty \!\!
  \frac{\rmd\tit}{\sin(\tit)}
  \exp\left[
    \rmi
    \left(
      \frac{(\rmrnt)^2}{2}\cot(\tit)-\rxrnt+\Chit-\Chit_0+2\nu\tit
    \right)
  \right]
\end{align}
with the abbreviations $\Chit\defas\Chit(\rvect),
\Chit_0\defas\Chit(\rvect_0) $. (The energy $\nu$ is assumed to have
an infinitesimally small positive imaginary part.)

Like in the case of the propagator, stating the Green function as an
integral has the advantage that its semiclassical approximation can
be obtained in a straightforward way.  This is shown in the following. The
exact integration will be carried out afterwards.

\subsubsection{The semiclassical  Green function}
\label{sec:Gsc}

The semiclassical approximation to the Green function $\Gsc_\nu$ is
obtained by performing the Fourier transform in the stationary phase
approximation, which is summarized in Appendix \ref{app:statphase}.
It yields an asymptotic expansion to leading order in the
semiclassically large parameter $1/b^2$.
Requiring the integrand of the Fourier integral \eref{eq:Gintexp} to
have a stationary phase leads to a condition
\begin{gather}
  \label{eq:spc}
  |\sin(\tit)|
  \stackrel{!}{=}\frac{|\rmrnt|}{2\sqrt{\nu}}
  \equiv  \frac{|\rmrn|}{2\rho}
\CO
\end{gather}
which selects the times of flight of classical trajectories connecting
the initial position $\rvec_0$ with the final point $\rvec$ at fixed
energy $\nu$.  It can be satisfied only if the distance between the
two points is smaller than the cyclotron diameter.  If this is the
case, the time derivative of the phase in \eref{eq:Gintexp} vanishes
at an infinite number of (discrete) times,
\begin{align}
  \label{eq:times}
  \tit_{\rm S}^{\,(n)} &= \arcsin(\wasgamma)+n\pi
  \CO
\nnn
  \tit_{\rm L}^{\,(n)} &=  \pi-\arcsin(\wasgamma)+n\pi
  \CO
  \qq\text{with $n=0,1,\ldots$}
\end{align}
{Here,}
\begin{align}
\label{eq:wasgammadef}
  \wasgamma
  \defas
  \frac{|\rmrn|}{2\rho}
\end{align}
measures the distance between the initial and the final  point
relative to the classical cyclotron diameter.
The two times of flight $\tit_{\rm S}^{\,(0)}$ and $\tit_{\rm
  L}^{\,(0)}$ belong to the two distinct trajectories which connect
the initial and the final point directly. They are ``short'' and ``long''
arcs, respectively, ie, span an angle smaller and larger than
$\pi$ (cf Fig.~\ref{fig:alphabeta}).
For $n>0$ the trajectories perform in addition $n$ complete
cyclotron orbits.
After the Fourier transform the trajectories entering the
semiclassical Green function exhibit an action
$\tilde{S}=\tilde{\W}+2\nu \tit$ which is a function of energy
$\nu=\widetilde{E}/2$ rather than time.
As specified by \eref{eq:times} the actions read
\begin{gather}
  \tilde{S}^{\,(n)}_\SL
   =  2\pi\nu \left(\ga_\SL+n\right) +\Chit-\Chit_0
   \PO
\end{gather}
Here, we introduced the notation
\begin{align}
  \label{eq:deftsl}
  \ga_{\rm S}(\rvec;\rvec_0)
  &\defas \frac{1}{\pi}  
  \left(\arcsin(\wasgamma)+\wasgamma\sqrt{1-\wasgamma^2} 
    - \frac{\rvec\times\rvec_0}{2\rho^2}
  \right)
 \q\text{and}
  \nnn
  \ga_{\rm L}(\rvec;\rvec_0)
  &\defas \frac{1}{\pi}  
  \left(\pi-\arcsin(\wasgamma)-\wasgamma\sqrt{1-\wasgamma^2}
    - \frac{\rvec\times\rvec_0}{2\rho^2}
  \right)
\end{align}
for the geometric part of the action.  
Note that $\ga_{\rm S}$ and $\ga_{\rm L}$ depend on the initial
and the final point individually, due to the term $\rxrn$, which means
that they are \emph{not} translationally invariant.  However, one observes the
relation $ \ga_{\rm S}(\rvec;\rvec_0)+\ga_{\rm
  L}(\rvec_0;\rvec)=1$. It follows that the (scaled) action of a
closed cyclotron orbit -- a short arc followed by a long one -- is
given by $2\pi\nu$.

To compute the stationary phase approximation \eref{eq:stphase1d} we
also need the second derivative of the phase in \eref{eq:Gintexp}. It
is given by $(\rmrnt)^2\cos(\tit)/\sin^3(\tit)$ and at times
\eref{eq:times} assumes the values
$\pm4\nu\sqrt{1-\wasgamma^2}/\wasgamma$ (where the positive sign
stands for trajectories of the short type).
It follows that in the semiclassical approximation an infinite number
of trajectories contributes to the Fourier integral.
\begin{align}
  \Gsc_\nu(\rvec;\rvec_0)
  =
  \frac{-1}{4\pi}
  \sum_{n=0}^\infty
  (-)^n  
  \left(
    \frac{\frac{\pi}{2\nu}}{\wasgamma\sqrt{1-\wasgamma^2}}
  \right)^{\oh}
  \Big\{
  &\exp\Big({2\pi\rmi\nu(\ga_{\rm S}+n)+\rmi\chit-\rmi\chit_0 
    +\rmi\piof}\Big)
  \nnn
    +
    &\exp\Big({2\pi\rmi\nu(\ga_{\rm L}+n)+\rmi\chit-\rmi\chit_0  
      -\rmi\piof}\Big)
  \Big\}
\end{align}
The sum over the repetitive cyclotron orbits $n$ converges since
$\nu$ was assumed to have a small positive imaginary part. It adds a
factor $(1+\rme^{2\pi\rmi\nu})^{-1}$ which is singular at the energies
of the Landau levels.  The semiclassical Green function is therefore
given by a sum of two contributions, belonging to the short and the
long arc trajectory --- the principal classical trajectories
connecting $\rvec_0$ and $\rvec$:
\begin{align}
  \label{eq:Gsctsl}
    \Gsc_\nu(\rvec;\rvec_0)
  =
  \frac{1}{2(1+\rme^{2\pi\rmi\nu})}
  \frac{1}{(2\pi\rmi)^\oh}
  \frac{\frac{1}{2\rmi\sqrt{\nu}}  }
  {\left(\wasgamma\sqrt{1-\wasgamma^2}\right)^\oh}
    \left\{
    \rme^{2\pi\rmi\nu \ga_{\rm S}}
    +
    \rme^{-\rmi\piot}
    \rme^{2\pi\rmi\nu \ga_{\rm L}}
  \right\}
  \rme^{\rmi(\chit-\chit_0)}
\end{align}
This form will be used in Chapter \ref{chap:trace} for periodic
orbit theory.  Alternatively, one can combine the short and long arc
contributions pulling out that part of the phase which was time
independent in \eref{eq:Gintexp}.  This leads to the expression
\begin{align}
  \label{eq:Gsccos}
    \Gsc_\nu(\rvec;\rvec_0)
  &=
  \exp\left[-\rmi\left(\frac{\rxrn}{b^2}-\chit+\chit_0\right)\right]
  \;   \Gnsc_\nu\left(\frac{(\rvec-\rvec_0)^2}{b^2}\right)
\CO
\end{align}
{with}
\begin{align}
\label{eq:Gscn}
   \Gnsc_\nu(z)&\defas 
   \frac{-1}{4\pi}\,\frac{({2\pi})^\oh}{\cpn}\,
  \frac{1}
  { \big[z\,(4\nu-z)\big]^\frac{1}{4}}
\nnn
&\times  {\cos
    \left(%
      2\nu
      \left[
        \arcsin\!\left(\left(\frac{z}{4\nu}\right)^\oh\right)
        +\left(\frac{z}{4\nu}\,\left(1-\frac{z}{4\nu}\right)\right)^\oh
        -\piot
     \right]  +\piof
    \right)}
\PO
\end{align}
It shows that the semiclassical Green function is given by a phase factor which
contains the gauge dependence and a \emph{real} function
$\Gnsc_\nu$ which depends only on the distance between the initial
and the final point.  The exact Green function has the same property,
as manifest in \eref{eq:Gintexp}.

Note that the expressions \eref{eq:Gsctsl} and \eref{eq:Gsccos} are
defined only for separations smaller than the cyclotron diameter
$|\rmrn|<2\rho$. For larger distances, the semiclassical Green function
vanishes by definition, since the stationary phase condition
\eref{eq:spc} has no solution. As the distance between the initial and
the final points approaches the cyclotron diameter,
the short and long arcs coalesce and are therefore no longer isolated.
In this case the approximation \eref{eq:stphase1d} fails, which is
indicated by the diverging prefactor of $\Gsc$, as $\wasgamma\to 1$.
If a semiclassical expression is needed for the domain
$|\rmrn|\gtrapprox 2\rho$, eg to describe tunneling effects, uniform
approximations  \cite{BM72} must be employed as discussed in
Appendix \ref{app:uniform}.

\subsubsection{The exact  Green function}
\label{sec:Gexact}

When evaluating the exact Green function we may separate the part of
the phase in \eref{eq:Gintexp} which is not explicitely time
dependent, like in the semiclassical case.
\begin{align}
  \label{eq:GreenPhase}
  {\rm G}_\nu(\rvec;\rvec_0)=
  \exp\left[-\rmi\left(\frac{\rxrn}{b^2}-\chit+\chit_0\right)\right]
  \;   \Gn_\nu\left(\frac{(\rmrn)}{b^2}\right)
\end{align}
Now, the integral can be performed exactly by contour integration \cite{HS00a}
\begin{align}
\label{eq:contint}
 {\rm G}^0_\nu(z) &= \frac{-1}{4\pi} 
 \int_0^\infty\!\! \frac{{\rm d}\tit}{\sin(\tit)}\,
  \exp\left[\rmi\left(\frac{z}{2}\cot(\tit)+2\nu\tit\right)\right]
 \nnn
&= 
 \frac{-1}{4\pi}\,
 \Gamma(\tfrac{1}{2}-\nu)\, z^{-\frac{1}{2}}\, W_{\nu,0}(z)
\end{align}
Here, $W_{\nu,0}$ is the  (real valued)
\emph{irregular Whittaker function} \cite[eq (13.1.34)]{AS65}.
This expression was also obtained \cite{Ueta92} using the separability
of \eref{eq:seconvinh} in the symmetric gauge.

Both, the function \eref{eq:contint} and its semiclassical
approximant \eref{eq:Gscn} exhibit simple poles as the energy $\nu$
approaches the Landau levels. It is often convenient to remove these
poles by considering the \emph{regularized} version of $\Gn_\nu$,
\begin{align}
  \label{eq:Gregdef}
  \Gtn_\nu(z) \defas
  \lim_{\mu\to\nu}\cos(\pi\mu)\,{\rm G}^0_{\mu}(z)
  \PO
\end{align}
We finally state the regularized Green function in terms of the
\emph{irregular confluent hypergeometric function} $U$ \cite{AS65}
which is more common than the Whittaker function:
\begin{eqnarray}
  \label{eq:Gregdef2}
  \Gtn_\nu(z) 
  = \frac{-1}{4\pi}\,\frac{\pi}{\Gamma(\nu+\tfrac{1}{2})}\,
  \rme^{\ts-z/2} \, {\rm U}(\ohmnu,1;z)
\end{eqnarray}

\begin{figure}[tbh]
  \begin{center}%
  \psfrag{sqrt(z)}{$\sqrt{z}$}
  \psfrag{G}{\hspace*{-2em}$\Gtn_\nu$, $\Gtnsc_\nu$ }
  \psfrag{g}{\hspace*{-2em}$|\Gtn_\nu-\Gtnsc_\nu|$   }
  \includegraphics[width=\linewidth] {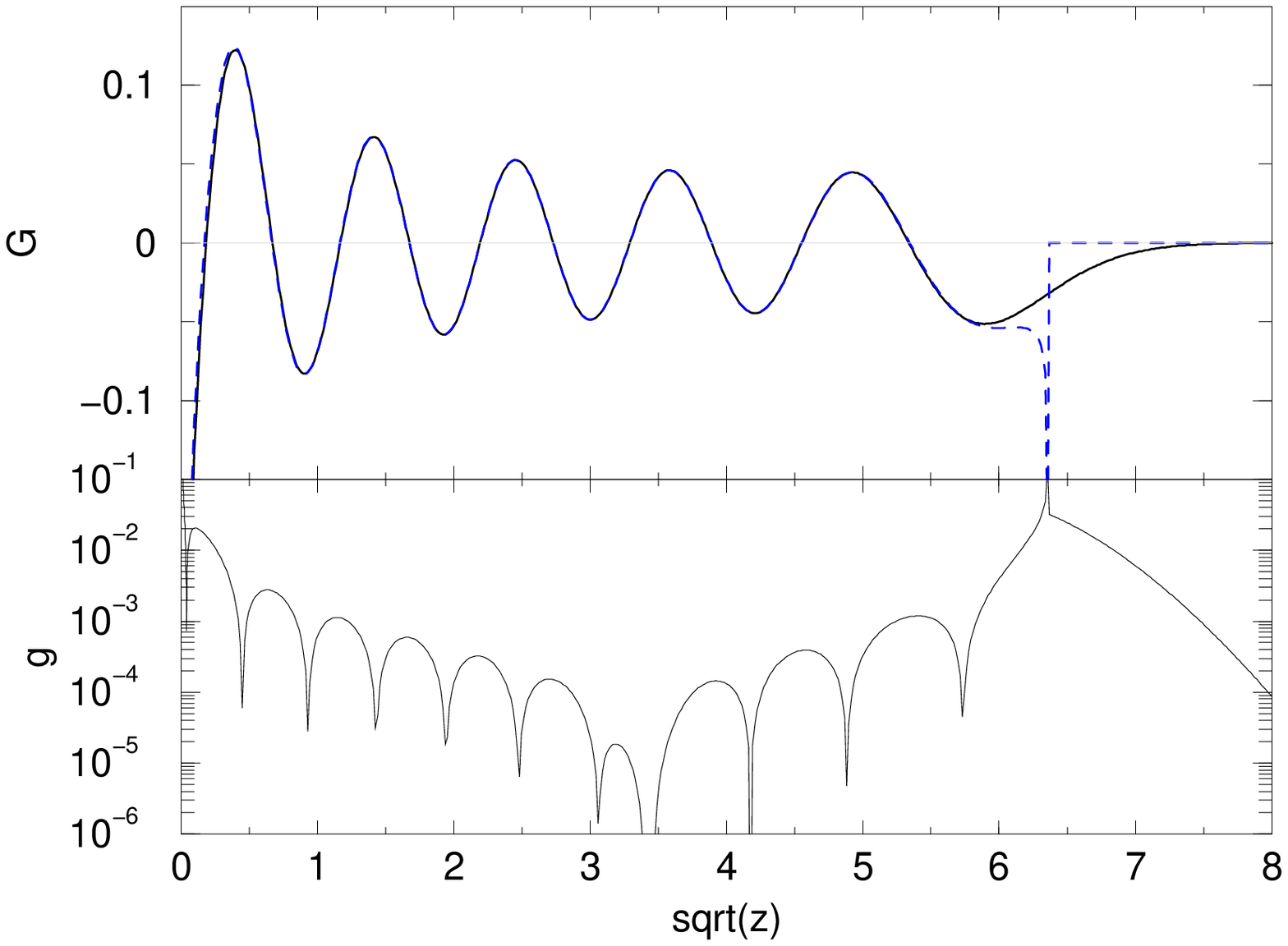}
  \figurecaption{%
Regularized gauge-independent part of the free Green function.
Top: Exact (solid line) and semiclassical (dashed line)
functions $\Gtn_\nu$ at $\nu=10.1$. Bottom: Error of the
semiclassical approximation. Even at this moderate value of $\nu$
strong deviations occur only at the classical turning point
$\sqrt{z}=2\sqrt{\nu}\approx 6.36$ and at small distances.  
(The deviations arise since the
semiclassical approximation does not account for the logarithmic
singularity at $z=0$ and the tunneling into distances larger than
the cyclotron diameter.)
}%
  \label{fig:G}
  \end{center}
\end{figure}

\subsubsection{Properties of the free Green function}
\label{sec:Gprop}

Figure \ref{fig:G} displays the gauge-independent, regularized part of
the exact and semiclassical Green function.  As one expects, the exact
Green function decays exponentially once the points are separated by a
distance, $|\rmrn|>2\rho$, (ie $z>4\nu$) which cannot be traversed
classically.\footnote{%
  The above mentioned irregular solution of \eref{eq:Gdef} grows
  exponentially beyond the classically allowed region. Its derivation
  is sketched in Appendix \ref{app:Gang}.
}
At small distances, $\rvec\to\rvec_0$, it displays a logarithmic
singularity, similar the (complex valued) field-free Green function
\cite{MF53}.  We find
\begin{align}
  \Gtn_\nu(z) 
  =& \, \frac{\cpn}{4\pi}
  \big(\log(z)+{\Psi}(\ohpnu)-2{\Psi}(1) \big)
  - \frac{\sin(\pi\nu)}{4}
+ {\rm O}(z\log z)
\end{align}
as $z\to 0$,
with $\Psi(z)$  the Digamma function \cite{AS65}.
Our method to evaluate the free Green function
numerically with high precision and efficiency is discussed in
\cite{HS00a}.

The gauge invariant part of the Green function has the remarkable
feature that its derivatives can be expressed by the function itself,
at a different energy. For the regularized version one finds
\begin{align}
\label{eq:zdG}
  z\frac{\rmd}{\rmd z}\Gtn_\nu(z) 
  =& -(\ohmnu)\big(\Gtn_\nu+\Gtn_{\nu-1}\big)-\frac{z}{2}\Gtn_\nu
  \\
\label{eq:zzddG}
  z^2\frac{\rmd^2}{\rmd z^2}\Gtn_\nu(z) 
  =& (\tfrac{3}{2}-\nu)(\ohmnu)\big(\Gtn_\nu+2\Gtn_{\nu-1}+\Gtn_{\nu-2}\big)
\nnn
  &+z (\ohmnu) \big(\Gtn_\nu+\Gtn_{\nu-1}\big)
  +\frac{z^2}{4}\Gtn_\nu
  \PO
\end{align}
These formulas were obtained by employing the differential properties
of the confluent hypergeometric function \cite{AS65}.  Their
asymptotic behavior reads
\begin{align}
  z\frac{\rmd}{\rmd z}\Gtn_\nu(z) 
  &= \frac{\cpn}{4\pi}
    \big[ 
       1 -z\,\nu\,
         \big(\log(z)+{\Psi}(\ohmnu)-2{\Psi}(1)-1\big)
    \big] 
  + {\rm O}(z^2\log z)
\CO
\\
  z^2\frac{\rmd^2}{\rmd z^2}\Gtn_\nu(z) 
  &= - \frac{\cpn}{4\pi}   + {\rm O}(z\log z)\CO \q\q\mbox{as $z\to 0$.}
\end{align}
It can be deduced from the logarithmic representation of U in terms of
the regular Kummer function \cite[eq. (13.6.1)]{AS65} and will be
needed below.

\section{Introducing a boundary}

\label{chap:boundary}

The motion in the magnetic plane turns into a non-trivial problem once
the particle is restricted to a
bounded domain.  

\subsection{Motion in a restricted domain}
\label{sec:restr}

Let us assume that the particle is confined to move in a compact and
simply connected domain $\Domain \subset \mathbb{R}^2$ with smooth
boundary $\Boundary=\partial\Domain$.  The classical equation of
motion \eref{eq:newtoneq} applies in the interior of the domain
$\openDomain$. Here, the particle moves on arcs of constant curvature,
which may at some point impinge on the boundary.
At these instances the trajectories must obey the law of
\emph{specular reflection} to qualify as a classical solution.  This
follows directly from Hamilton's principle, as will be shown in
Sect.~\ref{sec:sp}.  Clearly, any trajectory which was reflected once
must run into the boundary again.  It follows that the phase space is
in general split up into two disjunct parts.  One part consists of
\emph{skipping orbits}.  Their classical motion is no longer described
by a continuous Hamiltonian flow (but by a discrete map) and may
range from regular (integrable) to completely chaotic (hyperbolic).
We will briefly review this classical billiard problem below, in Sect.
\ref{sec:cbilliard}.
The remaining part of phase space describes the trivial motion on
closed \emph{cyclotron orbits}. It has a finite volume whenever the
cyclotron radius is small enough to enable a disk of radius $\rho$
to fit into the domain.  We will call the magnetic field
\emph{strong}, accordingly, if the cyclotron radius is comparable to
or smaller than the size of the billiard -- a criterion which is
purely classical.

In the corresponding quantum problem the eigenfunctions are required
to satisfy the Schr{\"o}dinger equation in the open domain
$\openDomain$, together with a boundary condition on the border line
$\Boundary$ (as discussed in Sect. \ref{sec:qbilliard}).
One observes that, at strong fields, the spectrum reflects the
partitioning of the classical phase space.  There are eigenstates
which hardly touch the boundary and have energies very close to the
Landau levels.  They are called \emph{bulk states} because in the
limit of strong fields they constitute the major part of the
spectrum.  We will see that these states are based on that part of
phase space which is given by the unperturbed cyclotron motion.
At the same time, one finds eigenstates which are localized at the
boundary. These \emph{edge states} correspond to the skipping
trajectories and are expected to reflect the underlying billiard
motion. Albeit being an effect of the boundary they may be quite
significant. For instance, they typically exhibit a directed
probability flux causing a large magnetic moment. This way they
balance the magnetic moments
of the bulk leading to a vanishing mean magnetization, as discussed in
Section \ref{sec:mag}.

The separation into edge and bulk states is intuitively clear and
often used.  Early studies
were concerned with the surface electron states inside metals
\cite{NP67,KMF69}, and
after the discovery of the Quantum Hall Effect \cite{KDP80,Laughlin81}
the notion of edge states was employed to explain this phenomenon
\cite{Halperin82,Buttiker88,Shizuya94,MMP99,FGW00,SBKR00}. (In the
latter problem the Hamiltonian must include an additional impurity
potential.)  However, the above characterization of edge states is not
precise and
we are not aware of a general quantitative definition in the literature.
In due course, we will introduce a spectral measure, which permits to
quantify the edge character of a state \cite{HS02a}.  Having a
meaningful spectral density of edge states at our disposal, it will be
worthwhile to consider the quantum problem also in the exterior.

\subsubsection*{Motion in the exterior}

The \emph{exterior} billiard problem is obtained by restricting the
particle to the domain $\mathbb{R}^2 \setminus \Domain$ -- henceforth
called the exterior domain.
From the classical point of view there is little difference between
the interior and the exterior dynamics.  
A particle impinging on the boundary from outside is reflected
specularly and performs a skipping motion around the billiard.  Like in
the interior the skipping trajectories cover a finite volume in phase
space and are described by a discrete billiard bounce map.  Complete
cyclotron orbits, on the other hand, now exist for any $\rho$.  The
corresponding phase space volume is unbounded because the cyclotron
center may be located at an arbitrarily large distance from the
billiard.

The fact that a  ``free particle'' cannot escape to
infinity but is trapped on a cyclotron orbit is reflected by the exterior
quantum spectrum. It is \emph{discrete}, in marked contrast to
the field-free scattering situation.  
The exterior quantum problem requires the stationary wave function to
satisfy the Schr{\"o}dinger equation in $\mathbb{R}^2 \setminus
\openDomain$, again with a boundary condition on $\Gamma$.  In addition,
the normalization condition 
implies that the wave functions must vanish at infinity.
In the absence of a boundary the spectrum would be given by a discrete
set of Landau energies, each infinitely degenerate, as shown in the
preceeding chapter. The presence of a billiard lifts this degeneracy
turning each Landau level into a \emph{spectral accumulation point}.
This means that there are infinitely many discrete eigenenergies in
the vicinity of each Landau energy.

We shall address the general quantum problem in
Section~\ref{sec:qbilliard}.  There, the main concern will be on the
boundary conditions and the average spectral behavior, whereas the
actual quantization is performed in Chapter \ref{chap:bim}.
To prepare for the semiclassical quantization in Chapter
\ref{chap:trace}
let us first take a closer look at the classical problem.

\subsection{The classical billiard}
\label{sec:cbilliard}

Classical magnetic billiards were first examined by Robnik and Berry
\cite{RB85} and are still the subject of active research
\cite{MBG93,Kleberetal96,BK96,Tasnadi96,Tasnadi97,Kovacs97,AL97,DA00,Gutkin01}.
In this section we collect basic results, limiting the discussion to
those aspects which will be needed later on.

The classical dynamics is completely  specified by the size of the
cyclotron radius $\rho$ and by the
shape of the  billiard.
Throughout this work, the billiard boundary $\Gamma$ is assumed to be
smooth, so that its normals $\nvec$ exist everywhere.  We
define them to point outwards (ie, into $\mathbb{R}^2 \setminus
\Domain$).  Keeping their orientation fixed will allow to distinguish
the interior from the exterior problem.
The boundary is parameterized  by the arc length $s$,
\begin{align}
  \label{eq:parametr}
  \Boundary: s \in [0;\Len] \mapsto \rvec(s)\in \Rtwo
  \CO
\end{align}
such that the derivative yields the normalized tangent
\begin{align}
  \label{eq:parametr2}
  \frac{{\rm d}\rvec(s)}{{\rm d}s} 
  \defas \tvec(s) 
  = {-n_y(s)\choose n_x(s)}
  \PO
\end{align}
We define the local curvature
\begin{align}
  \label{eq:parametr3}
  \kappa(s_0)
  \defas
  2\lim_{s\to s_0}
  \frac{\big(\rvec(s)-\rvec(s_0)\big)\,\nvec(s)}
  {\big(\rvec(s)-\rvec(s_0)\big)^2}
\end{align}
to be positive for convex domains.  The area of the domain is denoted  by
$\Area$, and  $\Len$ represents its circumference.

\subsubsection{The billiard bounce map}

As mentioned above, the particle's skipping motion may be described by
the mapping of a Poincar\'e surface of section onto itself.  Like in
the case of field-free billiards
\cite{Sinai76,Bunimovich74,Tabachnikov95,Smilansky95} it is
natural to use the Birkhoff coordinates $(s,p_s)$ to define the
surface of section. They are given by the position on the boundary $s$
(the curvilinear abscissa) and the (normalized) tangential component
of the reflected velocity $p_s=\vvech_0(s)\,\tvec(s)$ at the point of
reflection. The variables $s$ and $p_s$ are canonically conjugate in
the sense described below.  It is worth noting, therefore, that $p_s$
is defined as a component of the \emph{velocity} vector,
rather than the (gauge-dependent) canonical momentum.

A point $(s,p_s)$ in the Birkhoff phase space describes the position
of incidence, and the direction of the velocity after reflection (once
it is agreed on whether to consider the interior or exterior problem).
Tracking the classical trajectory until its first intersection with
the boundary specifies the next point of reflection $s'$ uniquely, and
$p_s'$ follows from the law of specular reflection.  Since any
reflected trajectory is included this way the complete billiard
dynamics is described by the bounce map
\begin{gather}
  \label{eq:Birmap}
  \mathcal{B}: (s,p) \mapsto (s',p')
\end{gather}
which maps the Poincar\'e surface of section $ [0;\Len]\times(-1;1)$
onto itself.
In  order to see that the map generates a discrete Hamiltonian evolution,
one may look for a generating function $\GF(s,s')$, which yields the
(canonically) conjugate coordinates by differentiation,
\begin{align}
  \label{eq:dGF}
  p_s=-\frac{\rmd \GF(s,s')}{\rmd s}
  \qq\text{and}\qq
  p_s'=\frac{\rmd \GF(s,s')}{\rmd s'}
  \PO
\end{align}
The relation \eref{eq:dGF} is the
discrete analogue to the case of continuous Hamiltonian dynamics, where
the canonical momenta are similarly given by the derivative of the
action.
If the mixed second derivative of $\GF$ has a definite sign
the equations \eref{eq:dGF} may be globally inverted
\cite{Smilansky95}, yielding the bounce map \eref{eq:Birmap}.

\begin{figure}[tbp]%
  \begin{center}%
    \includegraphics[width=\linewidth] {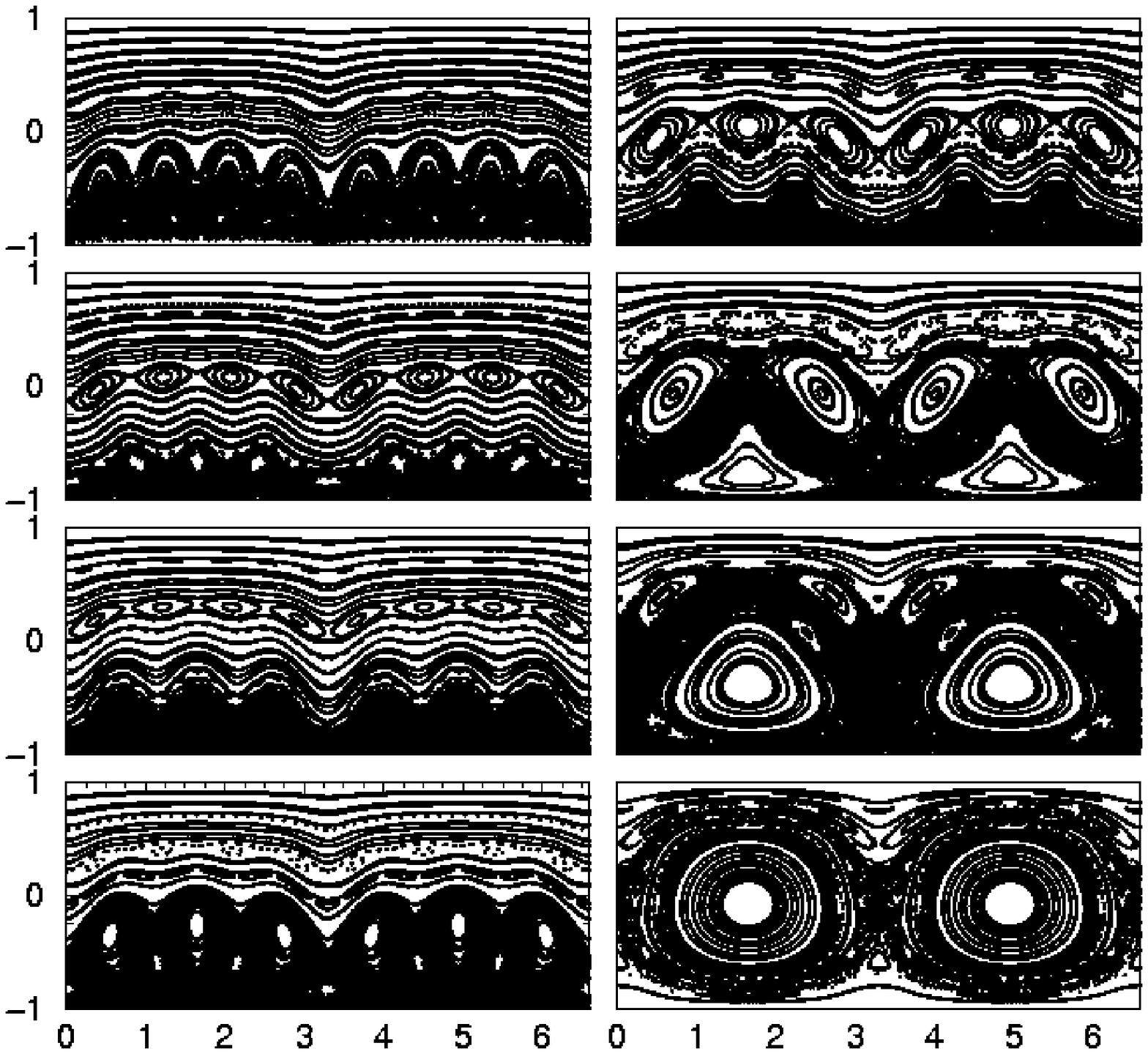}
    \figurecaption{%
      Birkhoff phase space portraits of the interior ellipse (strong
      eccentricity $0.8$, area $\Area=\pi$), for different values of
      the cyclotron radius $\rho=0.40, 0.44,
      0.50, 0.54$ (left column, top to bottom) and $\rho=0.6, 1.0,
      2.0, 10.0$ (right column, top to bottom). The motion turns
      (more) regular as the limit of a strong field, $\rho\to 0$, and
      a vanishing field, $\rho\to\infty$, is approached. 
[figure quality reduced]
       }%
    \label{fig:pport}%
  \end{center}
\end{figure}

The billiard dynamics may now be studied conveniently by investigating
the properties of the map.  
In Fig.~\ref{fig:pport} we show surface of section plots of an
interior ellipse at different values of the the cyclotron radius.  One
observes the standard picture of mixed chaotic dynamics
\cite{Berry78,LL83,Reichl92}.  The trajectories either lie on
invariant curves (characterizing regular motion) or cover a whole area
in the surface of section (chaotic motion). Stable periodic orbits, in
particular,
are characterized by surrounding invariant lines (``stability islands'').

\subsubsection{Integrable and hyperbolic billiards}

In the field free case the ellipse is known to be the only smooth and
simply connected billiard with two integrals of motion (including the
circle as a special case).  At finite magnetic fields, the ellipse
turns chaotic, as we have just seen, except for the circle billiard.
The latter exhibits the canonical angular momentum \eref{eq:Aconvsym}
as the second integral of the motion (provided the circle is centered at the
origin of the symmetric gauge).  This suggests that circular shapes,
ie, the disk and the annular billiard, are the only boundaries which
yield integrable motion in the magnetic field.

The other extreme type of motion is called \emph{hyperbolic}, or displaying
\emph{hard chaos}. It is present if the stable part of phase space has
zero measure rendering almost all trajectories unstable.  Hyperbolic
billiards are popular, although they form a small class. Early
examples of  field-free billiards displaying hard chaos were given by Sinai
\cite{Sinai76}
and Bunimovich \cite{Bunimovich74}.  
Conditions for the instability of orbits in magnetic billiards are
discussed in \cite{Tasnadi96,Tasnadi97,Kovacs97}.  In his recent work
\cite{Gutkin01} Gutkin applied a general hyperbolicity criterion
\cite{GSG99} to construct classes of hyperbolic magnetic billiards.
The critical parameter in these sets is given by the sum of the
reciprocal cyclotron radius and the (local) curvature of the boundary.
Hard chaos is guaranteed in these cases only for cyclotron radii above
a certain minimal value.
Most of the billiards studied numerically in this report are
hyperbolic at zero field, but assume a mixed chaotic phase space at
any finite cyclotron radius.
An example of a
billiard shape which generates truly hyperbolic motion even at fairly
strong fields is given in the right part of Fig.~\ref{fig:shapes}.

Since the above statements apply equally to the interior and the exterior
dynamics  there was no need to distinguish between them. We now
turn to the question of how the classical interior and exterior
problems are related.

\begin{figure}[tb]%
  \begin{center}%
    \includegraphics[width=0.8\linewidth] {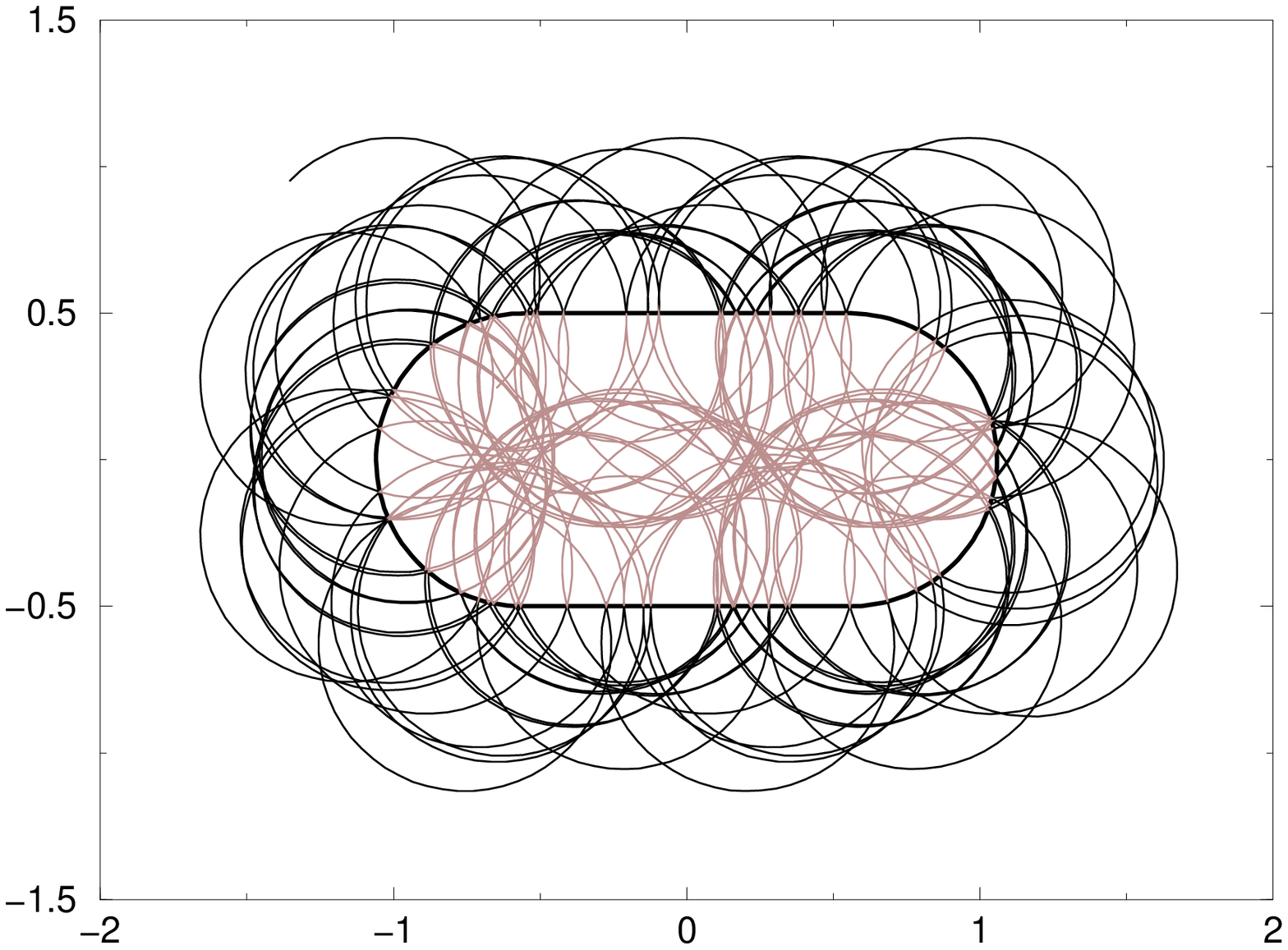}
    \figurecaption{%
      Parts of dual trajectories in the interior and exterior of a
      stadium-billiard at $\rho=0.5$ (sequence of 75 reflections).
      The billiard shape is defined in Fig.~\ref{fig:s14shape}.
       }%
    \label{fig:corbs}%
  \end{center}%
\end{figure}

\subsubsection{The classical interior-exterior duality}
\label{sec:duality}

\begin{figure}[tb]%
  \begin{center}%
    \includegraphics[width=0.6\linewidth] {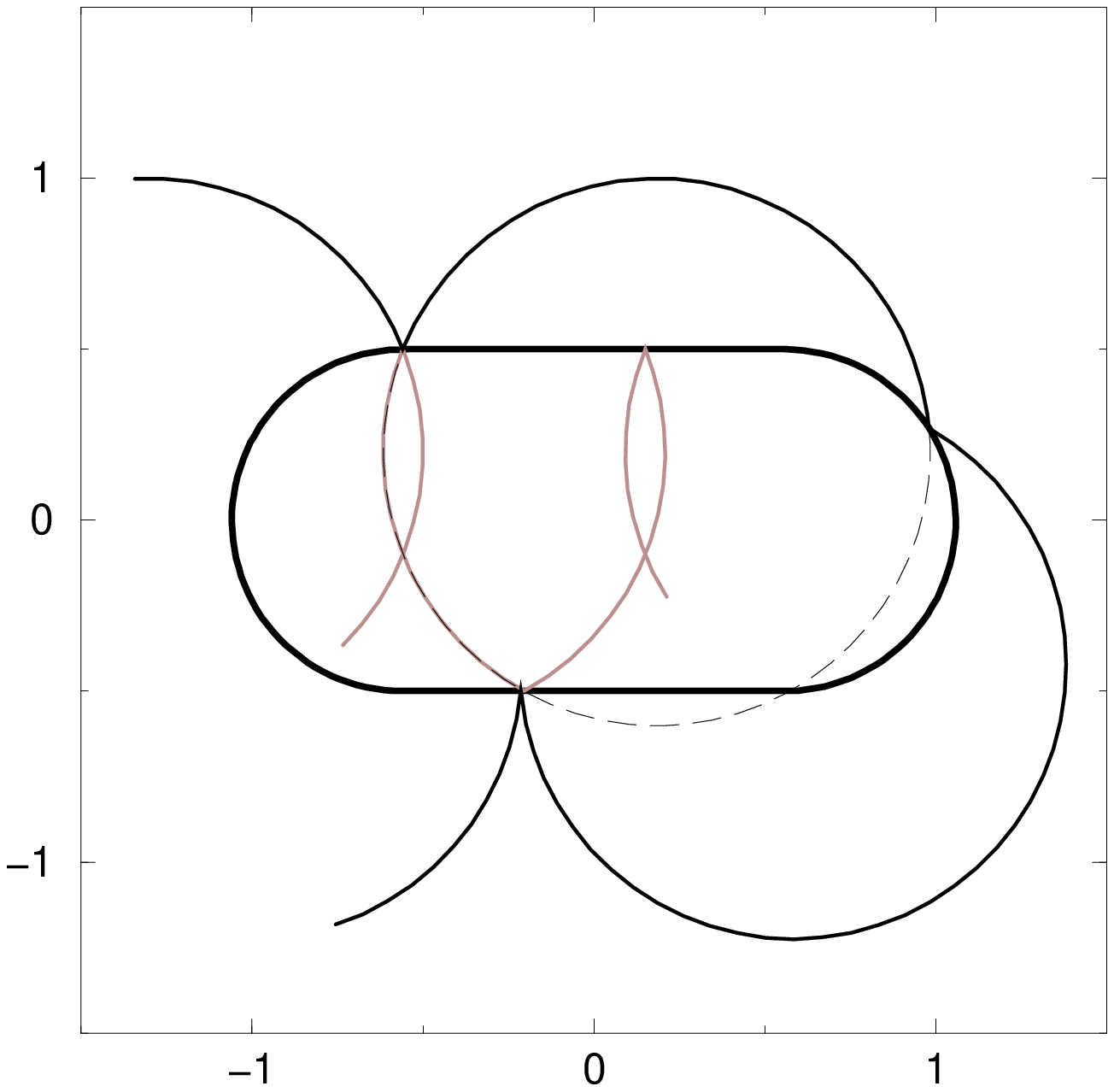}
    \figurecaption{%
      Breakdown of the duality in  segments of partially corresponding
      trajectories (stadium of Fig.~\ref{fig:corbs} at $\rho=0.8$.)
      Only the two left (top) arcs in the interior (exterior) meet
      with a dual partner. The breakdown occurs because a cyclotron
      orbit, which is obtained by continuing the arcs (dashed line)
      intersects the boundary more than twice.
       }%
    \label{fig:ncorbs}%
  \end{center}%
\end{figure}

When comparing the interior and the exterior motion the size of the
cyclotron radius $\rho$ plays a crucial role.
An important  situation is encountered
if the cyclotron radius and the billiard shape are such that any circle 
with radius $\rho$ intersects the boundary at most twice.
For convex domains, 
a sufficient condition is the cyclotron radius being greater than the
maximum radius of curvature, or less than the minimum radius of
curvature.  However, convexity is not necessary for the above
condition --- which we shall assume to hold for the moment.

Now consider a segment of an interior trajectory going from $\rvec(s)$
to $\rvec(s')$. The same two points are connected by a valid exterior
trajectory which travels backwards in time. Necessarily, the two arcs
form a complete circle of radius $\rho$.  (They do not intersect with
the boundary, except at the points $\rvec(s)$ and $\rvec(s')$, because the
above criterion was assumed to hold.)
The interior trajectory is reflected specularly and finally runs into
the boundary at $\rvec(s'')$. Clearly, the time-reversed exterior
trajectory obeys the same law of specular reflection, leading to
the same boundary point  $\rvec(s'')$.
It follows that the interior dynamics and the time-reversed exterior
one are described by the same Poicar\'e surface of section.  Every
interior trajectory is linked with a \emph{dual} exterior trajectory,
which travels backwards in time.  We call this property the
\emph{classical duality} of interior and exterior motion.  Pairs of
dual trajectories are displayed in Figure \ref{fig:corbs} and
\ref{fig:copair}.

As an immediate consequence of the classical duality one finds for
any given interior \emph{periodic} orbit a dual periodic orbit in the
exterior, and vice versa.  Being periodic, both may now be thought of
as running forward in time, but then with opposite orders in the
sequence of reflection points.  Clearly, these dual partners are
intimately related. We will see that they have the same stability
properties and that the sum of their actions is an integer multiple
of the action of a full cyclotron orbit (with the integer given by the
number of reflections). Examples of dual periodic orbits are given in
Figure \ref{fig:copair}.

\begin{figure}[tbp]%
  \begin{center}%
    \psfrag{rho}{$\rho$}
    \includegraphics[width=0.8\linewidth] {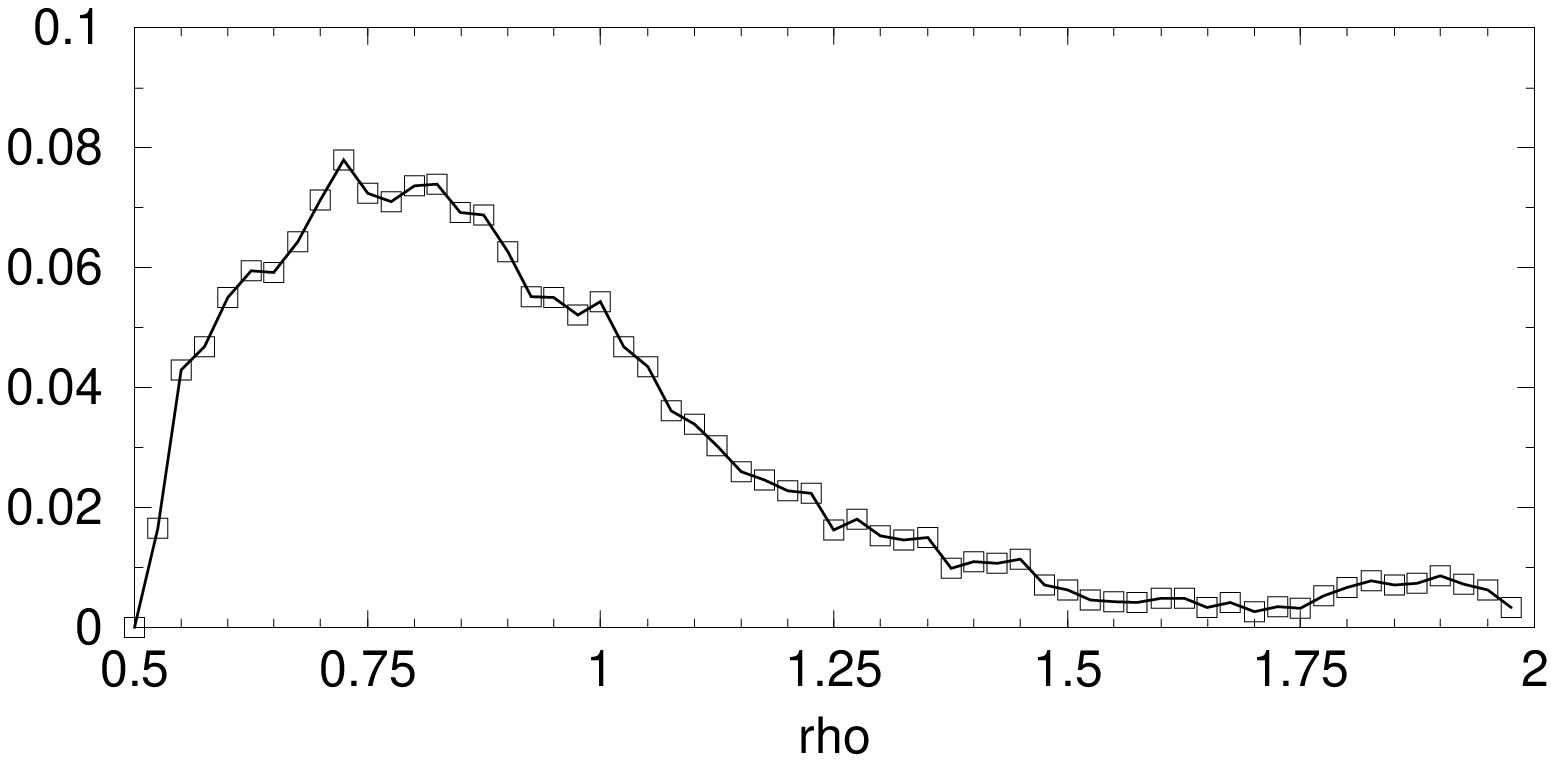}
    \figurecaption{%
      Fraction of the complete interior phase space belonging to arcs
      which violate the duality criterion, as a function of the
      cyclotron radius $\rho$. (Calculated for the stadium billiard in
      Fig.~\ref{fig:corbs}; the squares indicate the 
      error of the statistical sampling.)
       }%
    \label{fig:ncfrac}%
  \end{center}%
\end{figure}

Figure \ref{fig:ncorbs} shows that the duality breaks down once the
\emph{duality condition} that ``any circle of radius $\rho$ intersects
the boundary at most twice'' is no longer fulfilled.
Typically, only a small fraction of the phase space corresponds to
arcs which violate the duality condition. Fig \ref{fig:ncfrac} gives
an impression of the fraction of phase space belonging to arcs whose
extension intersects the boundary more than twice.  

\subsection{Quantum billiards}
\label{sec:qbilliard}

An early study of a magnetic quantum billiard was carried out by
Nakamura and Thomas \cite{NT88} (see \cite{PAKGEC94} for a
correction).  Later works are concerned with the spectral implications
of the absence of time-reversal invariance \cite{BGOdAS95,YH95,JB95}.
Special geometries, such as the disk \cite{Rensink69,BB97a} or, more
recently, the square \cite{RUJ96,NPZ00}, received attention as well.
All these studies were limited to the first few hundred eigenvalues,
and only to the interior problem.

\subsubsection{General boundary conditions}
\label{sec:bcond}

The mentioned works use Dirichlet boundary conditions, ie, demand the
wave function to vanish on the boundary.  It is the natural choice
from a physical point of view, which takes
 the boundary as due to an infinite potential wall.
However, it will prove worthwhile to consider slightly more general,
``mixed'' boundary conditions which include the Dirichlet choice as a
special case. They are defined by the equation
\begin{gather}
  \label{eq:bcond}
  \psi(\rvec) =  \pm\frac{\lambda}{b} 
  \big(
  \dnb\psi(\rvec)-\rmi\At_n(\rvec)\psi(\rvec)
  \big),
  \q\rvec\in\Boundary
  \PO
\end{gather}
The lower sign stands for exterior problem and the symbols $\dnb\defas
b\nvec(\rvec)\grad_{\!r}$ and $\At_n=\nvec(\rvec)\tilde{\mb{A}}$
denote the scaled normal derivative and the normal component of the
scaled vector potential, respectively.

The ``mixing'' parameter $\lambda$ interpolates between the two
extremes, \emph{Dirichlet}, $\lambda=0$, and \emph{Neumann} boundary
conditions, $\lambda^{-1}=0$.  In principle, $\lambda$ may be a
function of the position on the boundary, but will be assumed constant
throughout.  At non-vanishing $\lambda$ our boundary conditions
\eref{eq:bcond} are the gauge-invariant generalization of the mixed
boundary conditions known for the Helmholtz problem
\cite{KSG64,BB70,SPSUS95}.  They imply that the normal component of
the current density $\tilde{\jmath}_n={\rm
  Im}(\psi^*\dnb\psi)-\At_n|\psi|^2$ vanishes for any $\lambda$.
(Take the imaginary part after multiplying \eref{eq:bcond} with
$\psi^*$.)  The resulting conservation of the probability density
explains why the condition \eref{eq:bcond} keeps the problem
self-adjoint for any $\lambda$.
The explicit appearance of the vector potential in \eref{eq:bcond} is
needed to ensure the gauge-invariance of the boundary conditions.  The
fact that the definition does not depend on the gauge freedom $\chi$
is easily seen observing the gauge dependence of a general wave
function \eref{eq:gtrafowf}.
Finally, note that $\lambda$ has the dimension of a length, cancelling
the dimensionality introduced by the normal derivative. The magnitude
of the latter depends on the modulus $k=\sqrt{2 \mass E}/\hbar$ of the
wave vector.
To account for this trivial energy dependence of the eigenstates on
the boundary condition it will be convenient (later in the
semiclassical treatment) to use the dimensionless mixing parameter
\begin{align}
  \label{eq:Lambdadef}
  \Lambda \defas 
  k\lambda =  2\sqrt{\nu}\,\frac{\lambda}{b}
\PO
\end{align}
We did not state the definition \eref{eq:bcond} of the boundary
condition in terms of
$\Lambda$ because its dependence on the spectral variable $\nu$ would
destroy the self-adjointness of the problem rendering different
eigenstates non-orthogonal.

A quite different type of boundary conditions for magnetic billiards
was proposed recently by Akkermans \etal\ \cite{AANS98}.  It was
designed specifically to be sensitive on the ``chirality'' of the wave
functions.
For the special situation of a separable problem (disk billiard) they
allow to split the interior eigenspace into two subspaces with
definite chirality.  We will see that this is quite close to the
desired separation into bulk and edge states.  However, it cannot be
generalized to billiards with arbitrary shapes, and the resulting
spectrum has no relation to the standard Dirichlet problem.  Below,
we take a different approach to separate edge and bulk, by adjusting
the spectral measure according to our needs, rather than modifying the
spectrum.

\subsubsection{The quantum spectrum}
\label{sec:qspec}

Unlike their field-free relatives, magnetic quantum billiards offer
two independent external parameters -- the cyclotron radius and the
magnetic length.  As discussed in Section~\ref{sec:scaling}, one must
specify which one is to be fixed in order to define a quantum
spectrum.
In the main part of this report the formulas for spectral
densities are constructed
at a fixed magnetic length $b$. This is done to avoid clumsy notation
(and to minimize the danger of confusion).  A summary of formulas for
spectra defined in the semiclassical direction is given in Appendix
\ref{app:rho}.  Still, some of the numerical investigations presented
below are carried out on spectra defined in the semiclassical
direction, which will be clearly indicated.

The simplest function to characterize a spectrum is the \emph{spectral
  staircase} (or \emph{number counting function}) which gives the
number of spectral points below the specified energy.  For a set
of eigenvalues $\{\nu_n\}$ it is formally defined as a sum
\begin{gather}
\label{eq:Ndef}
  \N(\nu) \defas \sum_{n=1}^\infty \Theta\big(\nu-\nu_n\big)
\end{gather}
over Heaviside step functions $\Theta$.  Note that $\N(\nu)$ is a
well-defined function only for the interior problem, due to the
infinite number of exterior bulk states close to each Landau level.
The spectral density is conveniently defined as the energy derivative
of the counting function,
\begin{gather}
\label{eq:ddef}
  d(\nu) \defas \frac{\rmd}{\rmd\nu}  \N(\nu)
  =
 \sum_{n=1}^\infty \delta(\nu-\nu_n)
\CO
\end{gather}
and should be understood in the sense of distributions.  Formally, such
a sum of Dirac $\delta$-functions could be defined for the exterior
problem as well. However, this density would be meaningful at most in
a local sense since the convolution with a test function would
diverge at all the Landau energies.  Therefore, the following
discussion of the smooth, asymptotic properties of magnetic spectra
must be restricted to the interior problem.

\subsubsection{Asymptotic counting functions}
\label{sec:acf}

The spectral staircase is described asymptotically by the
\emph{mean number counting function} $\Nsm(\nu)$, which is
uniquely defined \cite{BH78}.
For Dirichlet boundary conditions it is given by the asymptotic
expression \cite{MMP97}
\begin{align}
  \label{eq:Nsmooth}
  \Nsm(\nu)
  =& \frac{\Area}{b^2\pi} \, \nu
  - \frac{\Len}{2\pi b} \, \nu^\oh  + \frac{1}{6}
  +\Or\big(\nu^{-\oh}\big)
  \PO
\end{align}
The expression includes only geometric quantities and the
conventional wave vector $\sqrt{2 \mass E}/\hbar=2\sqrt{\nu}/b$, which are
all \emph{independent} of the magnetic field.
The field independence of the leading order term follows immediately
from Weyl's law, as discussed below. However, it is not obvious that
the next two orders are identical to the field free case as well.
This was proved only recently in \cite{MMP97}, and for circular
billiards in \cite{NS99}.

Note the hierarchy of the geometric quantities appearing in
\eref{eq:Nsmooth}.  The leading and the second term are proportional to the
area and the circumference, respectively.  The constant is
determined\footnote{%
  The constant term in \eref{eq:Nsmooth} is modified if there are
  corners in the boundary \cite{MMP97}.
  } by the mean curvature $\int_\Gamma \kappa(s) \rmd s=2\pi$.
Moreover, the higher order terms are typically proportional to
higher moments of the curvature \cite{BH94}.  This hierarchy
reflects the systematic method to derive the boundary corrections to
asymptotic quantities (see eg. \cite{BB70}): The boundary is locally
approximated first by a by a straight line, then a circular arc, and
so on.

\begin{figure}[tbp]%
  \begin{center}%
    \includegraphics[width=0.8\linewidth] {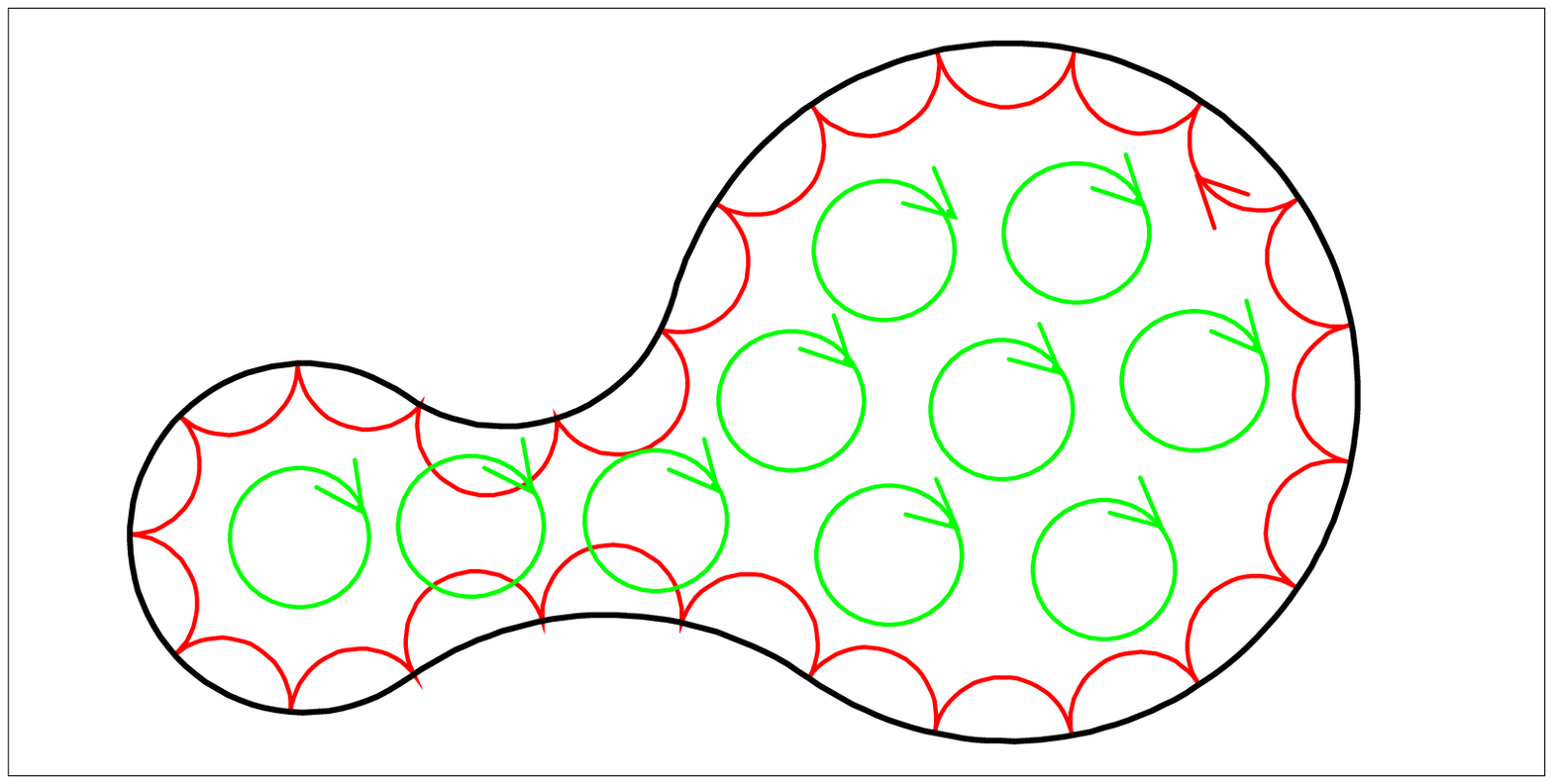}
    \figurecaption{%
      At strong magnetic fields, $\rho\ll\Len$,
      the major part of the available
      phase space consists of complete cyclotron orbits. The skipping
      orbits give rise to a net current along the boundary.  It has a
      counter-clockwise sense of orientation, in contrast to the
      cyclotron orbits.
       }%
    \label{fig:skipcyc}%
  \end{center}%
\end{figure}

\myparagraph{Weyl's law revisited}

\begin{figure}[tbp]%
  \begin{center}%
    \includegraphics[width=0.9\linewidth] {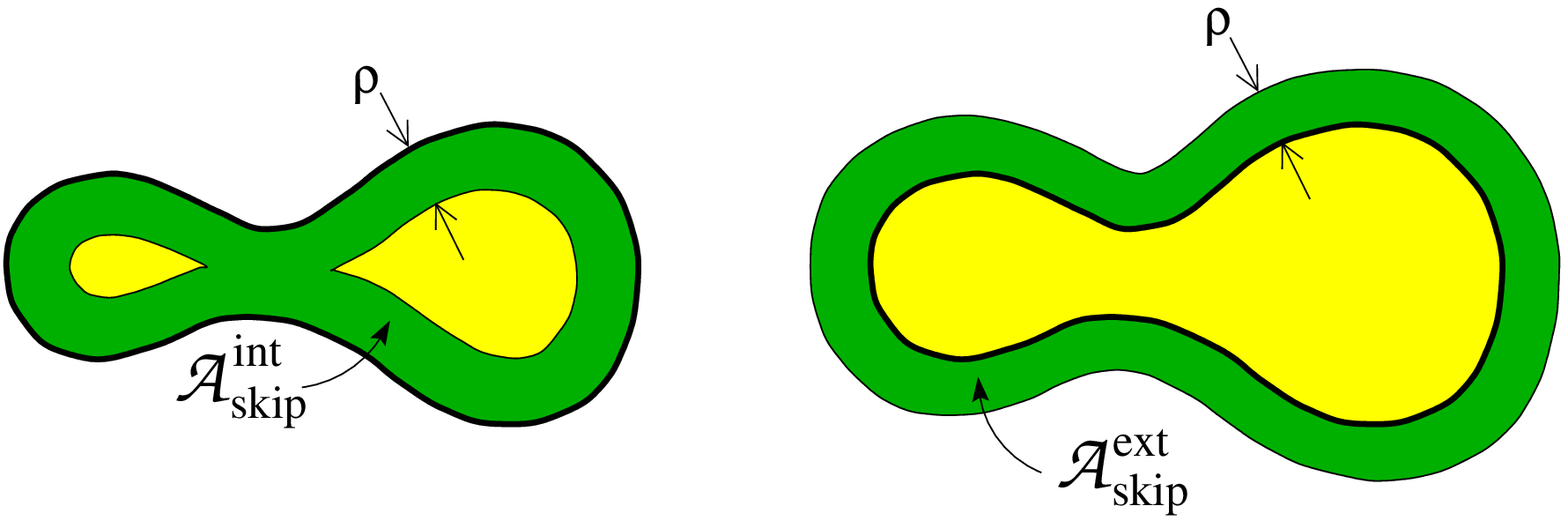}
    \figurecaption{%
      The dark shaded region indicates the area determining the phase
      space volume of interior (left) and exterior (right)
      skipping orbits. It is given by those points of the interior and
      exterior domain, respectively, which have a distance
      less than \emph{one} cyclotron radius $\rho$ to the boundary.
       }%
    \label{fig:area}%
  \end{center}%
\end{figure}

Let us consider Weyl's law more explicitely. It states that the
number of quantum states below a given energy is determined, to
leading order, by the volume of phase space, within the
energy shell, divided by (a power of) Planck's quantum
\begin{align}
  \label{eq:INtot}
  \Nsm_{\rm tot}(\rho^2, b^2)
  &=\frac{1}{(2\pi\hbar)^2}\int\!\!\!\int\Theta(E-\Ham) 
  \,\rmd^2\rvec\,  \rmd^2\pvec
  \\
  &=
  \label{eq:INtota}
  \tag{\ref{eq:INtot}a} \frac{1}{(b^2\pi)^2} \int\!\!\!\int
  \Theta\big(\rho^2 - |\rhovec'|^2 \big) \, \rmd^2\cvec\,
  \rmd^2\!\rhovec' \PO
\end{align}
This is the first term in the asymptotic expansion \eref{eq:Nsmooth}.
Changing the integration of the canonical momentum to the velocity
vector in the first line renders the phase space integral
\emph{independent} of the magnetic field (since the Jacobian is
constant \cite{Peierls79}).  This shows immediately that the leading
order term of the counting function (like any quantity which may be
written as a phase space integral of position and velocity) cannot
depend on the field strength.
In \eref{eq:INtota}, however,  we transformed the variables of integration
to the radius vector $\rhovec'$, cf eq \eref{eq:rhovecdef}, and the
cyclotron center $\cvec=\rvec-\rhovec'$, which \emph{do} depend on the
magnetic field.  As a result, the role of Planck's quantum is now
played by the area $b^2\pi$.
This second form of the phase space integral has the advantage that
it permits to separate the volumes of skipping
and cyclotron motion.
The center $\cvec$ is a constant of the motion for all cyclotron
orbits. Hence, integrating only the cyclotron part of the centers
one obtains the area $\Area_{\rm cyc}(\rho)$ of the set of points in
$\Domain$ with a distance from the boundary greater than $\rho$.
Consequently, the number of quantum states which correspond to
cyclotron motion is given, to leading order, by the integral
\begin{align}
   \Nsm_{\rm cyc}(\rho^2, b^2) =
  \frac{2\pi}{(b^2\pi)^2}  \int_0^{\rho}
  \Area_{\rm cyc}(\rho')
  \rho' \rmd\rho'
\PO
\end{align}
We note from \eref{eq:Nsmooth} that the total number of states reads
to leading order,
\begin{align}
   \Nsm_{\rm tot}(\rho^2, b^2) =
  \frac{2\pi}{(b^2\pi)^2} \frac{\rho^2\Area}{2}
  \PO
\end{align}
Hence, the number of states associated with the skipping part of phase space
can be written as an integral
\begin{gather}
\label{eq:Nsmskip}
   \Nsm_{\rm skip}(\rho^2, b^2) =\Nsm_{\rm tot} 
   -  \Nsm_{\rm cyc} =
  \frac{2\pi}{(b^2\pi)^2}  \int_0^{\rho}
  \Area_{\rm skip}(\rho')
  \rho' \rmd\rho'
\CO
\end{gather} 
{involving the area}
$   \Area_{\rm skip}(\rho) \defas \Area -  \Area_{\rm cyc}(\rho)$.
By definition, this area is given by those points in the interior domain which
are closer to the boundary than the cyclotron \emph{radius},
cf Fig.~\ref{fig:area}.  It determines the mean density of those states,
which correspond to the skipping part of phase space.
\begin{align}
  \label{eq:dskip}
  \overline{d}_{\rm skip}(\nu)
  =\frac{\rmd}{\rmd\nu}\Nsm_{\rm skip}(\nu\,b^2,b^2)
  = \frac{ \Area_{\rm skip}( b\sqrt{\nu})}{b^2\pi}
\end{align}
This is a remarkably simple and intuitive formula. 
It should be made clear, however, that we do not yet have a criterion
at our disposal, which provides a clear distinction of edge and bulk
states.  Clearly, a reasonable definition should pass the
requirement of being consistent with \eref{eq:dskip}.

Furthermore, a proper ``density of edge states'' will have to be
well-defined also in the \emph{exterior}. Let us therefore comment on
the expected mean number of {exterior} states which correspond to
skipping motion. By symmetry, it should be determined by the area
$\Area_{\rm skip}^{\rm ext}$ of those points in the exterior domain
which are closer to the boundary than $\rho$.
This can be confirmed for the circular geometry, where the integral over
the skipping part of phase space in \eref{eq:INtota} can be performed
explicitely.  For a disk of radius $R$ one obtains
\begin{align}
\label{eq:Nskipsmdi}
  \Nsm^{\rm int}_{\rm skip}
  &= 
  \begin{cases}
    \displaystyle
    \frac{4}{3}\,\frac{\Len}{2\pi b}\, \nu^{\frac{3}{2}} -\oh \nu^2
    & \text{if $\sqrt{\nu}\,b<R$}
    \\[1.5ex] %
    \displaystyle
    \frac{\Area}{b^2\pi}\nu
    & \text{if $\sqrt{\nu}\,b>R$}
  \end{cases}
  \\[1.3ex]
\label{eq:Nskipsmde}
  \Nsm^{\rm ext}_{\rm skip}
  &= \;\;\;\frac{4}{3}\,\frac{\Len}{2\pi b}\, \nu^{\frac{3}{2}} +\oh \nu^2
  \CO
\end{align}
for the interior and the exterior problem, respectively.  Note that
the interior number is determined by the area $\Area$ of the domain
once the cyclotron radius $\rho=\sqrt{\nu}\,b$ exceeds the radius $R$
of the disk, preventing any cyclotron orbits in the interior.

At very strong fields, $b\ll\rho\ll\Len$, in contrast, it is the
circumference term which dominates.
Since in this case we may neglect the mean curvature, the average
number of skipping states is approximately given by
\begin{align}
\label{eq:Nskipsm}
 \Nsm_{\rm skip}=
\frac{4}{3}\,\frac{\Len}{2\pi b}\, \nu^{\frac{3}{2}}
\PO
\end{align}
This expression coincides with the phase space estimate for a straight
line with periodic boundary conditions (see Appendix \ref{app:line}).

Let us turn to another quantity which serves to characterize interior
magnetic billiards --- the orbital magnetism which measures the
response of the spectrum to changes in the magnetic field. Its
asymptotic properties may be related to a phase space integral as
well.

\subsection{Orbital magnetism}
\label{sec:mag}

Employing the notion of \emph{orbital magnetism} we slightly abuse a
thermodynamic concept for our one-particle problem.  Nonetheless, it
is worthwhile to ask for the magnetic response of the billiard
dynamics in the sense of statistical mechanics.  We consider only
micro-canonical ensembles (since we are not concerned with effects of
finite temperature) which means that averages are performed on the
energy shell in phase space, ie, among all orbits of a given cyclotron
radius.

Let us first consider the classical motion along  a single 
\emph{periodic}\footnote{%
  We may confine the discussion to periodic orbits because the set of
  periodic orbits is known to be dense in phase space.
} trajectory. Being charged the particle constitutes an electric
current which in turn induces a magnetic moment. Will it serve to
strengthen or to weaken the applied magnetic field?  Clearly, the
latter is expected in the case of a cyclotron orbit. Here, the
(scaled) magnetic moment turns negative,
\begin{gather}
  \label{eq:mcyccl}
  \oh
  \int_0^{\widetilde{T}_{\rm cyc}}
  \! \rvect(\tit)\times\vvect(\tit)\, \rmd\tit= -\nu
  \CO
\end{gather}
which shows that the cyclotron part of phase space is
\emph{diamagnetic}.
The skipping orbits, on the other hand, will in general give rise to
both signs.
At strong fields (if the cyclotron radius is shorter than the minimum
diameter of the billiard) skipping trajectories carry a net current
along the boundary. It is orientated clockwise, ie, opposite to the
cyclotron orbits (see Fig.~\ref{fig:skipcyc}).  A detailed analysis
 \cite{Leeuwen21} shows that, in any case, a subtle cancellation
mechanism
between cyclotron and skipping orbits is at work, which guarantees
that classically there is \emph{no} net orbital magnetization.  This
is  the van Leeuwen theorem
\cite{Leeuwen21,Peierls79}.

The statement is proved immediately by evoking the thermodynamic
definition of the magnetization as
the derivative of a thermodynamic potential (the free energy or the
grand canonical potential) with respect to the magnetic field. The
potentials are determined by the partition sum, which is a phase space
integral in the classical case. As such it cannot  depend on the
magnetic field for the reasons given in the preceding section
\cite{Peierls79}.

Before we turn to the precise quantum definition it should be
emphasized that orbital magnetism in its proper sense is an effect of
many particles at finite temperature.
Assuming the temperature to be much larger than the spacing between
Landau levels,   $T\gg\hbar\omega_c/k_B$,
Landau showed \cite{Landau30} that 
a degenerate Fermi gas exhibits a small\footnote{%
  The effect is one third of the Pauli spin paramagnetism \cite{AM76}.}  net
diamagnetic response. This Landau diamagnetism is an effect of the
bulk.
Asymptotic corrections due to the
existence of a boundary are discussed in
\cite{vRvL93,Kunz94,JS95,MMP97,NSA98,NS99}.
Recently, the effect met some renewed interest since the
\emph{geometry} of mesoscopic devices may greatly enhance orbital
magnetism.  Semiclassical treatments in terms of periodic orbit theory
may be found in \cite{vonOppen94,Agam94,URJ95,RUJ96,RM98}. In 
these works the magnetic field was assumed to be very weak such that
the bending of the trajectories could be neglected. An exception is
the study of the quantum and semiclassical magnetization of the
magnetic disk in \cite{Tanaka98}.
A comprehensive review on the subject of orbital magnetism is given in
\cite{Richter00}.

In the following we shall use the concept of orbital magnetization
merely as a means of characterizing magnetic billiards.
We shall argue that it is advantageous to adopt a modified definition
of orbital magnetization.  In order to motivate this we start with
the conventional one.

\myparagraph{Conventional magnetization}

Given the spectrum $\{E_n\}$ at finite magnetic field $B$ one may
conventionally define the magnetization as
\begin{align}
  \label{eq:Macconv}
  \Mag_{\rm conv}(E,B)
  \defas &
  -\!\!\!\sum_{E_n \le E} \frac{\rmd E_n}{\rmd B}
  =  
  \int_0^E \!\!\! \mm(E';B) \,\rmd E' 
\PO
\end{align}
This 
is the one-particle and zero-temperature limit of the standard
thermodynamic definition.  By means of equation \eref{eq:Macconv}
the function $\mm(E,B)$ is introduced which we  call the magnetization
density,
\begin{align}
  \label{eq:macconv}
  \mm(E,B) 
  \defas& 
  \frac{\rmd {\N}_{\rm tot}}{\rmd B}(E,B) 
  =\, - \sum_n \frac{\rmd E_n}{\rmd B}\,\delta(E-E_n)
\PO
\end{align}
The relation of $\mm(E,B)$ to the electrodynamic interpretation of the
magnetization is seen once we note the derivative of the Hamilton
operator \eref{eq:Hamconv} with respect to the magnetic field,
\begin{align}
   \frac{\rmd \Ham }{\rmd B} 
   = -\frac{q}{2}\,(\rvec\times\vvec)_{\rm  sym}
   \PO
\end{align}
It is the operator of the magnetic moment, where $(\cdot)_{\rm sym}$
indicates the symmetrized form.  It follows that the energy derivatives
${\rmd E_n}/{\rmd B}$ in eq \eref{eq:macconv} are given by the
corresponding expectation values of the magnetic moment, ie, the
magnetization density \eref{eq:macconv}
reads
\begin{gather}
  \label{eq:macconva}
  \mm(E,B)=
  \sum_n \frac{q}{2}   \bra{\psi_n}(\rvec\times\vvec)_{\rm
  sym}\ket{\psi_n}\,\delta(E-E_n)
\PO
\end{gather}

The fact that the \emph{mean} magnetization (density) vanishes follows
immediately from the field-independence of $\Nsm$ \eref{eq:Nsmooth},
as noted above.  At strong fields the negative moments of (many) bulk
states are balanced, consequently, by the large, positive magnetic
moments of relatively few edge states.  This is seen much more clearly
once we modify the definition of the magnetization such that is
complies with the scaling properties of the system.

\myparagraph{Bulk and edge magnetization}

We proceed to define a scaled magnetization which has considerable
advantages compared to the conventional one.
According to \eref{eq:defnu} the spectrum $\{\nu_n\}$ depends
parametrically on the magnetic length, $\nu_n=\nu_n(b^2)$.  It is
natural to define the \emph{scaled magnetization density} such that
it yields the density of the {scaled} magnetic moment \eref{eq:mag},
in analogy to \eref{eq:macconva}.  Hence, one is led to the definition
\begin{align}
  \mmt(\nu,b^2) 
  \defas
  \sum_n \oh\bra{\psi_n}(\rvect\times\vvect)_{\rm sym}\ket{\psi_n}
  \,\delta(\nu-\nu_n)
\PO
\end{align}
From the explicit form of the scaled Hamiltonian one can easily show that
\begin{align}
\label{eq:mag}
 \oh\bra{\psi_n}(\rvect\times\vvect)_{\rm sym}\ket{\psi_n}
 =  b^2\frac{\rmd \nu_n}{\rmd b^2} - \nu_n
\PO
\end{align}
Thus, the scaled magnetization density can be written in terms of
derivatives of the number counting function,
\begin{align}
  \label{eq:magdef}
  \mmt(\nu,b^2) 
  =  \sum_n \Big( 
    b^2\frac{\rmd \nu_n}{\rmd b^2} - \nu_n \Big) 
  \,\delta(\nu-\nu_n) =
  - b^2 \frac{\partial \N}{\partial b^2}
  - \nu \frac{\partial \N}{\partial \nu}
\PO
\end{align}
The scaled magnetization follows by integrating the density.
\begin{align}
  \label{eq:Mdef}
  \Magt (\nu,b^2)
  \defas&\, 
  \int_0^\nu \!\! \mmt(\nu',b^2)
  \,\rmd \nu'
  =
  \sum_{\nu_n\le\nu} \Big( b^2\frac{\rmd \nu_n}{\rmd b^2} - \nu_n \Big)
  \\
  \label{eq:Mdefa}
  \tag{\ref{eq:Mdef}a}
  =&\,\Magt_{\rm edge} +  \Magt_{\rm bulk}  
\end{align}
As indicated in the second line the scaled magnetization splits up
naturally into two parts which we call, respectively, the \emph{edge}
magnetization,
\begin{align}
  \label{eq:Medge}
  \Magt_{\rm edge}  (\nu,b^2) &\defas  
  \sum_{\nu_n\le\nu} b^2\frac{\rmd \nu_n}{\rmd b^2}
  =-\int_0^\nu b^2\frac{\rmd}{\rmd b^2}\N(\nu',b^2)\,\rmd\nu'
  \CO
\intertext{and the \emph{bulk} magnetization,}
  \label{eq:Mbulk}
  \Magt_{\rm bulk}  (\nu,b^2)  &\defas -  \sum_{\nu_n\le\nu}  \nu_n
  =-\int_0^\nu \nu'\frac{\rmd}{\rmd \nu'}\N(\nu',b^2)\,\rmd\nu'
  \PO
\end{align}
This naming is appropriate since any Landau state  \eref{eq:defcs}
exhibits a scaled magnetic moment 
$ \bra{n,m}\oh(\rvect\times\vvect)_{\rm sym}\ket{n,m}=-(n+\oh)=-\nu$,
like the classical cyclotron orbit \eref{eq:mcyccl}.
Each eigenstate contributes to both
magnetization densities,
\begin{align}
  \label{eq:mmedge}
  \mmt_{\rm edge}(\nu,b^2)
  &=  \sum_n  b^2\frac{\rmd \nu_n}{\rmd b^2}\,\delta(\nu-\nu_n)
\intertext{and}
  \mmt_{\rm bulk}(\nu,b^2)
  &=  - \sum_n   \nu_n\, \delta(\nu-\nu_n)
\PO
\end{align}
The energies of bulk states lie close to the  Landau levels
and the nearer they get to the  level the less they depend on
$b^2$ (since the Landau energy is independent of $b^2$).
Hence, they give rise to a negligible edge contribution.
Edge states, in contrast, contribute to the edge magnetization much
stronger than to the bulk. This follows from the mean values of the
magnetization.
For the smooth  edge magnetization density one finds, cf \eref{eq:Nsmooth},
\begin{align}
  \label{eq:mmedgesm}
   \overline{\mm}_{\rm edge} (\nu,b^2) 
   = -b^2  \frac{\partial \Nsm}{\partial
     b^2}
   = \frac{\Area}{b^2\pi}\nu - \oh \frac{\Len}{2\pi b} \nu^\oh
\PO
\end{align}
Remarkably, the bulk mean value assumes a form,
\begin{align}
   \overline{\mm}_{\rm bulk} (\nu,b^2) 
   = -\nu  \frac{\partial \Nsm}{\partial
     \nu}(\nu,b^2) 
   = -  \frac{\Area}{b^2\pi}\nu + \oh \frac{\Len}{2\pi b} \nu^\oh
\CO
\end{align}
which cancels the mean edge magnetization identically.  Hence, the mean
(total) magnetization, $ \overline{\Mag}= \overline{\Mag}_{\rm edge}+
\overline{\Mag}_{\rm bulk}$
vanishes  like in the conventional
case.
This holds strictly for any field, independently of whether or not
there is a classical separation into skipping and cyclotron orbits.

The edge magnetization \eref{eq:Medge} defined in this section
embodies a first quantity which allows to distinguish edge states
quantitatively.  It gives the excess magnetization of the states which
arises due to the existence of a boundary, ie, as compared to the
expected diamagnetism of a state in the infinite plane.

\section[Quantization in the interior and the exterior]
{Quantization in the interior and the exterior: The boundary integral method}

\label{chap:bim}

In the present chapter, we show how to solve the quantization problem
for interior and exterior magnetic billiards by means of a boundary
integral method.  It provides the spectra and wave functions of
arbitrarily shaped billiard domains, and includes the general boundary
conditions discussed in Section \ref{sec:bcond}.  Moreover, the
boundary integral formalism constitutes the basis for the
semiclassical theory discussed in Chapter \ref{chap:trace}.

\subsection{Boundary methods}

As compared to the field free case, it is surprisingly difficult 
to obtain the quantum spectra of magnetic billiards.
So far, numerical studies were restricted to the interior problem and
performed almost exclusively by diagonalizing the
Hamiltonian\cite{NT88,BGOdAS95,YH95,JB95,deAguiar96}.  This requires
the choice and truncation of a basis, which is problematic for general
billiards, where no natural magnetic basis set exists.  Consequently,
results were limited to the first few hundred eigenvalues.

In the case of field free billiards
quantum spectra  are usually obtained by
transforming the eigenvalue problem into an integral equation of lower
dimension. The corresponding integral operator is defined in terms
of the free Green function, and depends only on the boundary
\cite{KR74,Riddell79,BW84,Boasman94,KS97,LRH98}.  
This method is known to be more efficient than diagonalization by an
order of magnitude \cite{MK88,CLH01}. 
We proceed to extend these ideas to magnetic billiards. A step in this
 direction was taken by Tiago \etal \cite{TCA97}, who
 essentially propose a null-field method\footnote{The authors of
 \cite{TCA97} inaccurately call their scheme a ``boundary integral
 method''. }  \cite{Martin82} for (interior) magnetic billiards.  It
 involves the irregular Green function
\eref{eq:Girrangular} in angular
momentum decomposition.  A drawback of the approach is that this function
must be known for large angular momenta, which turns out to be 
numerically impractical.
Moreover, the method does not apply for the exterior problem. 

In the following we derive the boundary integral method for magnetic
billiards.  Like in the field free case, it involves the regular
Green function in position space representation. We present the method
for the interior and the exterior problem, and general boundary
conditions.

\subsection{The boundary integral equations}
\label{sec:bie}

\subsubsection{Single and double layer equations}

The stationary eigenfunction 
of a magnetic billiard at
energy $\nu$ is defined by the differential equation
\begin{gather}
\label{eq:Schreq2}
\left({\tfrac{1}{2}}\big(-\rmi\grad_{\!r/b} -
\tilde{\mb{A}}(\rvec)\big)^2 
-2\nu\right)
\psi(\rvec)
=0\CO
\end{gather}
and a specification of the wave function on the billiard boundary
$\Boundary$.  
The free Green function, $\Gnu$, was shown
to satisfy the {inhomogeneous} Schr{\"o}dinger equation
\begin{align}
\label{eq:Gdef}
\left(\tfrac{1}{2}\big(-\rmi\grad_{\!r/b} 
  - \tilde{\mb{A}}(\rvec)\big)^2 -2\nu\right)
\Gnu(\rvec;\rvec_0)
= - \tfrac{1}{2}\,\delta\!\left(\frac{\rvec-\rvec_0}{b}\right)
\PO
\end{align}

Our goal is to cast the quantization problem into an integral
equation defined on the billiard boundary.
To that end,
we take the complex conjugate of \eref{eq:Schreq2} and
multiply it (from the left) with $\Gnu$.  Similarly, equation
\eref{eq:Gdef} is multiplied with $\psi^*$ and subtracted from the
former expression. One obtains an equation
\begin{gather}
  \label{eq:deleq}
  \psi^*\, \grad_{\!r/b}^2 \Gnu
  - \Gnu\,\grad_{\!r/b}^2 \psi^*
  -2 \rmi\, \grad_{\!r/b} \big( \tilde{\mb{A}}\psi^*\Gnu \big) 
  = \psi^*\,  \delta\!\left(\frac{\rvec-\rvec_0}{b}\right)\CO
\end{gather}
which has a form suitable for the Green and Gau{ss} integral theorems.
It holds everywhere in the plane, except for the boundary $\Boundary$,
where the boundary condition \eref{eq:bcond} introduces a
discontinuity in the derivative of $\psi$.

We start by considering the {interior} problem and sketch the treatment of
the {exterior} case afterwards.
Choosing the initial point of the Green function away from the
boundary, $\rvec_0\in \mathbb{R}^2\setminus\Boundary$, the integral of
\eref{eq:deleq} over the (interior) domain $\Domain$ may be transformed to a
line  integral,
\begin{multline}
\label{eq:split}
\int_{\Boundary}
\big[
\psi^*\, (\dnb\Gnu-\rmi\,\At_n\,\Gnu)
-\Gnu\,(\dnb\psi^*+\rmi\,\At_n\,\psi^*)
\big] \frac{{\rm d}\Boundary}{b}
\\
=
\begin{cases}
\psi^*(\rvec_0)&\text{if $\rvec_0\in \openDomain $ }
\\
0&\text{if $\rvec_0\in \mathbb{R}^2 \setminus{\Domain}$. }
\end{cases}
\end{multline}
It is defined on the boundary $\Boundary$ (with the normal components
of the vector potential and the gradient denoted as
$\At_n=\nvec(\rvec)\tilde{\mb{A}}$ and $\dnb\defas
b\nvec(\rvec)\grad_{\!r}$, respectively).
Note that the vector potential part of the integrand was split
which is necessary for a gauge invariant formulation of the
integral equations. 

\myparagraph{The single layer equations}

We choose $\rvec_0 \in \Boundary$ and define
$\rvec_0^\pm\defas\rvec_0\pm\epsilon \nvec_0$, for small $\epsilon>0$.
By adding the two equations in \eref{eq:split}, one obtains
\begin{gather}
\label{eq:SLepsin}
\int_{\Boundary}
\big[
\psi^*\, ({\partial}_{n/b}\epscomb{\Gnu} -\rmi\, \At_n\, \epscomb{\Gnu} )
- \epscomb{\Gnu}\, (\dnb\psi^*+\rmi\,\At_n\,\psi^*) )
\big] 
\frac{{\rm d}\Boundary}{b}
=
{\tfrac{1}{2}}\psi^*(\rvec_0^-)\PO
\end{gather}
Here, we used the abbreviation
$\epscomb{\Gnu}={\tfrac{1}{2}}\Gnu(\rvec;\rvec_0^+)
+{\tfrac{1}{2}}\Gnu(\rvec;\rvec_0^-)$.
Equation \eref{eq:SLepsin} holds for all (sufficiently small)
$\epsilon>0$, hence the limit
$\epsilon\to0$ exists.  Moreover, observing the asymptotic
properties of the Green function (cf Sect. \ref{sec:Gprop}), it can be
shown, that the integration and the limit $\epscomb{\Gnu}\to\Gnu$,
${\partial}_{n/b}\epscomb{\Gnu}\to\partial_{n/b}\Gnu$ may be
interchanged.  Inserting the boundary condition \eref{eq:bcond} we
obtain, after renaming the limiting function
$u=\dnb\psi^*+\rmi\At_n\psi^*$, $u_0\defas u(\rvec_0)$,
\begin{gather}
  \label{eq:SingleLayerin}
\int_{\Boundary}
\big[
\,\Gnu-\frac{\lambda}{b}
(\dnb\Gnu -\rmi\, \At_n\, \Gnu)
\big]\, u \,\frac{{\rm d}\Boundary}{b}
=
\frac{\lambda}{b}\, (- {\tfrac{1}{2}}u_0)
\CO
\end{gather}
an integral equation defined on the boundary $\Boundary$ \cite{HS00a}.

In order to derive the corresponding equation for the exterior
problem, consider a large disk $\mathcal{K}_p\supset\Domain$ of radius
$p$, and integrate \eref{eq:deleq} over
$\mathcal{K}_p\cap\openDomain$.  Once $\rvec_0$ lies in the vicinity
of $\Boundary$, the contribution of $\partial\mathcal{K}_p$ to the
boundary integral vanishes as $p\to\infty$, due to the exponential
decay of the regular Green function $\Gnu$.
Similar to eq \eref{eq:SLepsin} one  obtains an equation
which permits the limit $\epsilon\to 0$ to be taken before
performing the integration.  The resulting boundary integral equation
differs from \eref{eq:SingleLayerin} only by a sign.  In the
following, we shall treat both cases simultaneously, with the
convention that the upper sign stands for the interior problem, and
the lower sign for the exterior one,
\begin{gather}
  \label{eq:SingleLayer}
\int_{\Boundary}
\big[\,
\Gnu\mp\frac{\lambda}{b}
(\dnb\Gnu -\rmi\, \At_n\, \Gnu)
\big]\, u\, \frac{{\rm d}\Boundary}{b}
=
\frac{\lambda}{b}\, (- {\tfrac{1}{2}}u_0)
\PO
\end{gather}
In analogy to the Helmholtz problem  \cite{KR74}, 
we will refer to these
equations as the \emph{single layer equations} for the interior and
the exterior domain.

\myparagraph{The double layer equations}

A second kind of boundary integral equations can be derived by applying
the differential operator $(\dnnb +\rmi
\At_{n_0})\defas\nvec(\rvec_0)(\grad_{\!r_0/b}+\rmi\tilde{\mb{A}}(\rvec_0)$
on equation \eref{eq:SLepsin},
\setlength{\multlinegap}{0.1\textwidth}
\begin{multline}
  \label{eq:DLeps}
  \int_{\Boundary}
  \psi^*\,
  ({\partial}_{n_0/b}+\rmi\, \At_{n_0})
  ({\partial}_{n/b}\epscomb{\Gnu}-\rmi\, \At_n\,\epscomb{\Gnu} ) 
  \frac{{\rm d}\Boundary}{b}
  \\
  - 
  \int_{\Boundary}
  ({\partial}_{n_0/b}\epscomb{\Gnu} +\rmi\, \At_{n_0}\, \epscomb{\Gnu} )
  (\dnb\psi^*+\rmi\At_n\psi^*)
  \frac{{\rm d}\Boundary}{b}
  \\
  =
  \pm{\tfrac{1}{2}}
  (\dnnb+\rmi\,\At_{n_0})\psi^*(\mb{r^\mp_0})\PO
\end{multline}
This equation is true for all $\epsilon >0$, which means that the
limit $\epsilon\to 0$ exists. As for the second integral, we may again
permute the limit and the integration which yields a proper integral.
Consequently, the limit of the first integral is finite, too.
However, in the first integral we are not allowed to exchange the
integration with taking the limit because the limiting integrand
\eref{eq:QexpslN} has a $1/(\rvec-\mb{r_0})^2$-singularity which is
not integrable (see below).

Integral operators of this kind are named \emph{hypersingular}
\cite{Guiggiani98}. Similar to a Cauchy principal value integral, they
are defined by taking a special limit. However, compared to the
principal value the singularity is stronger by one order in the
present case. Below, in
Section~\ref{sec:bops}, 
we define which limit is to be taken.  It is
denoted by $\tfpint$ and should be read ``finite part of the
integral''. With this concept and equation \eref{eq:bcond}, we obtain the
\emph{double~ layer} equations,
\begin{multline}
  \label{eq:DoubleLayer}
  \int_{\Boundary}
   (\dnnb\Gnu +\rmi\, \At_{n_0}\,\Gnu  )
   \,u\,
   \frac{{\rm d}\Boundary}{b}
   \\
  \mp
  \frac{\lambda}{b}
   \fpint_{\Boundary}
   ({\partial}_{n_0/b} +\rmi\, \At_{n_0})
   ({\partial}_{n/b}\Gnu -\rmi\, \At_{n}\,\Gnu)
   \,u\, 
   \frac{{\rm d}\Boundary}{b}
  = 
  \mp \tfrac{1}{2}u_0
\CO
\end{multline}
which are again integral equations defined on the boundary
$\Boundary$.

\myparagraph{The spectral determinants}

It is useful to introduce a set of integral operators (whose labels D
and N indicate correspondence to pure Dirichlet or Neumann
conditions):
\begin{align}
  \label{eq:Opdef}
  \mathsf{Q}^{\rm D}_{\rm sl} [u] &= 
  \int_{\Boundary}\!
  {\rm d}\Boundary
  \,\Gnu\, u
  \\
  \label{eq:Opdef2}
  \mathsf{Q}^{\rm N}_{\rm sl} [u] &= 
  \int_{\Boundary}\!
  \frac{{\rm d}\Boundary}{b}
  (\dnb\Gnu -\rmi\, \At_n\, \Gnu)
  \,u
  \\
  \label{eq:Opdef3}
  \mathsf{Q}^{\rm D}_{\rm dl} [u] &= 
  \int_{\Boundary}\!
  \frac{{\rm d}\Boundary}{b}
   (\dnnb\Gnu   +\rmi\, \At_{n_0}\,\Gnu   )
   \,u
   \\
  \label{eq:Opdef4}
  \mathsf{Q}^{\rm N}_{\rm dl} [u] &=  \,\,
   \fpint_{\Boundary}\!
   \frac{{\rm d}\Boundary}{b^2}
   ({\partial}_{n_0/b} +\rmi \At_{n_0})
   ({\partial}_{n/b}\Gnu -\rmi\, \At_{n}\,\Gnu)
    \,u
\end{align}
They act in the space of square-integrable periodic functions,
$u\in\Hilbert(\Boundary)$, with the period given by the circumference
$\Len$.

Nontrivial solutions of the single  layer equations  \eref{eq:SingleLayer}
and double layer equations \eref{eq:DoubleLayer} exist
for energies where the corresponding Fredholm determinants vanish,
\begin{align}
  \label{eq:detsl}
  \det\left[
    \mathsf{Q}^{\rm D}_{\rm sl}
    \mp \lambda
    \mathsf{Q}^{\rm N}_{\rm sl}
    +  \frac{\lambda}{2} \mathsf{id}
  \right] 
  &= 0
  \q\q\mbox{(single layer)}
\\
  \label{eq:detdl}
  \det\left[
    \mathsf{Q}^{\rm D}_{\rm dl}
    \mp \lambda
    \mathsf{Q}^{\rm N}_{\rm dl}
    \pm \frac{1}{2} \mathsf{id}
  \right] 
  &= 0
  \q\q\mbox{(double layer).}
\end{align}
Hence,
these are secular equations although the explicit dependence on the
spectral variable is not shown in our abbreviated notation.  
However, each of the determinants \eref{eq:detsl}
and \eref{eq:detdl} may have roots, which do \emph{not} correspond to
solutions of the original eigenvalue problem given by
\eref{eq:Schreq2} and \eref{eq:bcond}.  For finite $\epsilon$, the
equations \eref{eq:SLepsin} and \eref{eq:DLeps}
are still equivalent to the latter. They acquire additional spurious
solutions only as they are transformed to boundary integral equations
by the limit $\epsilon\to0$.

\subsubsection{Spurious solutions and the combined operator}
\label{ssec:spurious}

The physical origin of the redundant zeros is apparent in our gauge
invariant formulation:  They are proper solutions for the domain
\emph{complementary} to the one considered.  This is obvious for the
single layer equation with Dirichlet boundary conditions ($\lambda=0$),
where the spectral determinant does not depend on the orientation of
the normals. The same is true for the double layer equation with
Neumann boundary conditions ($\lambda^{-1}=0$).

In general, the character of the spurious solutions may be summarized
as follows: Independently of the boundary conditions, the
\emph{single layer} equation includes the \emph{Dirichlet} solutions
of that domain which is complementary to the one considered.
Likewise, the \emph{double layer} equation is polluted by the
\emph{Neumann} solutions of the complementary domain, irrespective of
the boundary conditions employed.
This statement is easily proved by observing that the
single-layer-Neumann operator and the double-layer-Dirichlet operator
are \emph{adjoint} to each other, \mbox{$\mathsf{Q}^{\rm N}_{\rm
sl}=(\mathsf{Q}^{\rm D}_{\rm dl})^\dagger$}, while the operators
$\mathsf{Q}^{\rm D}_{\rm sl}$ and $\mathsf{Q}^{\rm N}_{\rm dl}$ are
self-adjoint (see below).
Now assume that $u$ is a complementary Dirichlet solution. In Dirac
notation,
\begin{align}
  \mathsf{Q}^{\rm D}_{\rm sl} \ket{u} =0&
  \q\wedge\q
  \mathsf{Q}^{\rm D}_{\rm dl} \ket{u} \mp \tfrac{1}{2} \ket{u} =0
\\
  {\Rightarrow}\qq
  \bra{u}\mathsf{Q}^{\rm D}_{\rm sl} =0&
  \q\wedge\q
  \bra{u}\mathsf{Q}^{\rm N}_{\rm sl} \mp \tfrac{1}{2} \bra{u} =0\PO
  \nn
\end{align}
Applying the dual of $u$ to the single layer operator yields
\begin{align}
  \bra{u} \mathsf{Q}^{\rm D}_{\rm sl}  
  \mp \lambda \Big\{\bra{u}\mathsf{Q}^{\rm N}_{\rm sl}  
                    \mp \tfrac{1}{2} \bra{u}\Big\}
  = 0\CO
\end{align}
which implies that the Fredholm determinant of the single layer
operator vanishes.  Similarly, if $u$ is a complementary Neumann
solution,
\begin{align}
  &\pm\mathsf{Q}^{\rm N}_{\rm sl} \ket{u} + \tfrac{1}{2} \ket{u} =0
  \q\wedge\q
  \mathsf{Q}^{\rm N}_{\rm dl} \ket{u} =0
\\
  {\Rightarrow}\qq
  &\pm\bra{u}\mathsf{Q}^{\rm D}_{\rm dl} + \tfrac{1}{2} \bra{u} =0
  \q\wedge\q
  \bra{u}\mathsf{Q}^{\rm N}_{\rm dl}  =0
  \nn
\end{align}
then its dual satisfies the double layer equation, again for any $\lambda$,
\begin{align}
  \pm\Big\{\pm\bra{u} \mathsf{Q}^{\rm D}_{\rm dl} + \tfrac{1}{2} \bra{u}\Big\}
  \mp \lambda \bra{u}\mathsf{Q}^{\rm N}_{\rm dl}
  = 0\PO
\end{align}
Since the spurious solutions are never of the same type, it is possible
to dispose of them by requiring that both, the single and the double
layer equations, should be satisfied by the \emph{same} solution $u$.
Therefore, one obtains a necessary and sufficient condition for the
definition of the spectrum by considering a \emph{combined} operator
\begin{gather}
  \label{eq:combdef}
  \mathsf{Q}_{\rm c}^\pm
  \defas
  \left(
    \mathsf{Q}^{\rm D}_{\rm dl}
    \mp \lambda
    \mathsf{Q}^{\rm N}_{\rm dl}
    \pm \frac{1}{2} \mathsf{id}
  \right)
  + \rmi \aconst    
  \left(
    \mathsf{Q}^{\rm D}_{\rm sl}
    \mp \lambda
    \mathsf{Q}^{\rm N}_{\rm sl}
    +  \frac{\lambda}{2} \mathsf{id}
  \right)\CO
\end{gather}
with an arbitrary constant  $\aconst$.
It has a zero eigenvalue only if both, single and double layer
operators do.  In practice, the spectrum is
obtained by finding the roots of the \emph{spectral function}
\begin{gather}
  \label{eq:xibim}
  \xi(\nu) = \det\left[\mathsf{Q}_{\rm c}^\pm\right]
\PO
\end{gather}

It is worthwhile noting that (for the interior problem) spurious
solutions will not appear if one uses the irregular Green function.
The reason is that the gauge-independent part of this function is
\emph{complex}, which destroys the mutual adjointness of the
operators.  This is why the irregular Green function had to be chosen
for the null-field method \cite{TCA97}.  For the boundary integral
method, the option to use this exponentially divergent solution of
\eref{eq:Gdef} is excluded, since the corresponding operator would get
arbitrarily ill-conditioned once the diameter of the domain $\Domain$
exceeds the cyclotron diameter.  The exterior problem cannot even
formally be solved using ${\rm G}^{\rm (irr)}_\nu$ (due to an
essential singularity at the origin).

A last remark is concerned with the important case of Dirichlet
boundary conditions.  Here, one could as well derive a pair of
boundary integral equations that are \emph{not} gauge-invariant.
(Just set $\psi^*=0$ in \eref{eq:split} and consider $u=\dnb\psi^*$.)
Of course, these equations would yield all the proper gauge-invariant
eigen-energies of the problem. However, the energies of the additional
spurious solutions would depend on the chosen gauge, and a
characterization of the latter in terms of solutions of a
complementary problem would not be possible.

The fact that the spurious solutions can be removed by
considering a \emph{combined} integral operator is of great practical
importance for numerical calculations \cite{HS00a,Hornberger01}.
An individual spurious
solution of the single or the double layer operator may 
be identified as well after  evaluating the corresponding wave
functions by observing in which domain it vanishes. 

\subsubsection{Wave functions}

The eigenfunctions at points off the boundary,
$\psi(\rvec_0\notin\Boundary)$, are determined by the null vectors $u$ 
corresponding to the roots of the spectral determinant.
From equation \eref{eq:split} we obtain immediately an integral
representation of the (un-normalized) wave function,
\begin{align}
  \label{eq:psiint}
  \psi(\rvec_0)
  =\,\pm
  \bigg[
  \int_{\Boundary}
  \!
  \frac{{\rm d}\Boundary}{b}
  \big[
  \pm\frac{\lambda}{b} (\dnb\Gnu-\rmi\At_n\Gnu)-\Gnu
  \big] u
  \bigg]^*
\CO
\end{align}
for $\rvec_0\notin\Boundary$.
According 
to eq. \eref{eq:split} the integral vanishes identically either
in the interior or in the exterior. This is indeed 
confirmed by our numerical calculations which are reported  in the
next chapter. 

In order to calculate the current density \eref{eq:defj} one needs the gauge
invariant gradient of the wave function. An integral formula
is obtained from equation \eref{eq:split}, after applying the
differential operator $\grad_{r_0/b}+\rmi\Avect_0$,
\begin{align}
  \label{eq:gradint}
  \grad_{r_0/b}\psi(\rvec_0)-\rmi\Avect(\rvec_0)\psi(\rvec_0)
  =\,\pm
  \bigg[
  \int_{\Boundary}
  \!
  \frac{{\rm d}\Boundary}{b}
  \begin{aligned}[t]
    \big[&
    \pm\frac{\lambda}{b} (\grad_{r_0/b}+\rmi\Avect_0)(\dnb\Gnu-\rmi\At_n\Gnu)
    \\
    &-(\grad_{r_0/b}\Gnu+\rmi\Avect_0\Gnu)
    \big] u 
    \bigg]^*  
    \PO
  \end{aligned}
  \nn
  \\
\end{align}
The densities of other observables can
be obtained by similar boundary integrals.

\subsection{The boundary operators}
\label{sec:bops}

In the following, we give explicit expressions for the boundary
integrals.  This allows to define the ``finite part integral'' appearing
in the double layer equation \eref{eq:DoubleLayer}.

The integral operators \eref{eq:Opdef} -- \eref{eq:Opdef4},
\begin{gather}
  \big(\mathsf{Q}[u]\big)(\rvec_0) 
  = \int_\Boundary\! \rmd\Boundary\, {\rm q}(\rvec;\rvec_0) u(\rvec)
\CO
\end{gather}
are defined by their  integral kernels ${\rm q}(\rvec;\rvec_0)$. 
The form of the Green function \eref{eq:GreenPhase} leads to the
expressions
\begin{align}
\label{eq:QexpslD}
  {\rm q}_{\rm sl}^{\rm D}(\rvec;\rvec_0)
  =&
\,{\rm E}(\rvec;\rvec_0)
  \,  \Gn_\nu(z)
\\
\label{eq:QexpslN}
  {\rm q}_{\rm sl}^{\rm N}(\rvec;\rvec_0)
  =&
\,{\rm E}(\rvec;\rvec_0)
  \,
  \bigg\{
    -\rmi \frac{(\rmrn)\times\nvec}{b^2} 
    \, \Gn_\nu(z)
    + 2 \frac{(\rmrn)\,\nvec}{(\rmrn)^2} 
    \,
    z\frac{\rmd}{\rmd z}\Gn_\nu(z)
  \bigg\}
\\
\label{eq:QexpdlD}
  {\rm q}_{\rm dl}^{\rm D}(\rvec;\rvec_0)
  =&
\,{\rm E}(\rvec;\rvec_0)
  \,
  \bigg\{
    -\rmi \frac{(\rmrn)\times\mb{n_0}}{b^2} 
    \, \Gn_\nu(z)
    - 2 \frac{(\rmrn)\,\nvec_0}{(\rmrn)^2} 
    \,
    z\frac{\rmd}{\rmd z}\Gn_\nu(z)
  \bigg\}
\\
\label{eq:QexpdlN}
  {\rm q}_{\rm dl}^{\rm N}(\rvec;\rvec_0)
  =&
\,{\rm E}(\rvec;\rvec_0)
  \,
  \bigg\{
    \left(-\frac{((\rmrn)\times\nvec_0)((\rmrn)\times\nvec)}{b^4}
      -\rmi \frac{\nvec\times\nvec_0}{b^2}
      \right)   \Gn_\nu(z)
  \nnn
    & \phantom{\,{\rm E}(\rvec;\rvec_0) \,\bigg\{}
        +\left(- 2\rmi \frac{\nvec\times\nvec_0}{b^2}
        - 2  \frac{\nvec\,\nvec_0}{(\rmrn)^2}       \right)    
       \,
       z\frac{\rmd}{\rmd z}\Gn_\nu(z)  
  \nnn & \phantom{\,{\rm E}(\rvec;\rvec_0) \,\bigg\{}
    - 4   \frac{((\rmrn)\nvec)((\rmrn)\nvec_0)}{(\rmrn)^4}
       \;       
       z^2\frac{\rmd^2}{\rmd z^2}\Gn_\nu(z)  
  \bigg\}
\CO
\intertext{with $\nvec=\nvec(\rvec)$, $\nvec_0=\nvec(\rvec_0)$, 
$z\defas(\rmrn)^2/b^2$, and  the abbreviation}
{\rm E}(\rvec;\rvec_0)\defas& \exp\left[{-\rmi\left(
  \frac{\rvec\times\rvec_0}{b^2}
  -\Chit(\rvec)+\Chit(\mb{r_0})\right)}\right]
\end{align}
for the gauge dependent part.  Note that the gauge freedom $\Chi$ has
cancelled in the pre\-factors and appears in the phase only. It can be
absorbed by the substitution $u(\rvec) \to \exp(+\rmi\Chi(\rvec))
u(\rvec)$, proving the manifest gauge invariance of the boundary
integral equations \eref{eq:SingleLayer}, \eref{eq:DoubleLayer}.  Note
also that expressions \eref{eq:QexpslN} and \eref{eq:QexpdlD} are related
by a permutation of $\rvec$ and $\rvec_0$ with subsequent complex
conjugation (since $\Gn_\nu$ is real), hence the operators are the
adjoints of each other. The self-adjoint nature of \eref{eq:QexpslD}
and \eref{eq:QexpdlN} follows likewise.

The derivatives appearing in \eref{eq:QexpslN} -- \eref{eq:QexpdlN}
may be stated in terms of the gauge independent part of the Green
function, $\Gn_\nu$, itself, at different energies $\nu$. 
They are given 
in Section \ref{sec:Gprop} together with their asymptotic
properties. $\Gn_\nu$ displays a logarithmic singularity as
$\rvec\to\rvec_0$, while the differential expressions are bounded.
In that limit, most of the quotients vanish for a smooth boundary,
others tend to the curvature \eref{eq:parametr3}
at the boundary point $\rvec_0$.
As a consequence, all the terms in \eref{eq:QexpslD} --
\eref{eq:QexpdlN} are integrable --- but for the one containing the
$(\nvec\,\nvec_0)/(\rmrn)^2$-singularity.  The latter gives rise to
the need for a finite part integral.

\subsubsection*{The hypersingular integral operator}

For finite $\lambda$ the double-layer equation contains a
hypersingular integral defined as
\begin{align}
\label{eq:fpi}
  \mathsf{Q}^{\rm N}_{\rm dl} [u] 
  =&\,\, 
  \fpint_{\Boundary}
   \frac{{\rm d}\Boundary}{b^2}
   ({\partial}_{n_0/b} +\rmi \At_{n_0})
   ({\partial}_{n/b}\Gnu -\rmi \At_{n}\Gnu)
   u 
\nnn
  \defas& 
  \lim_{\epsilon\to0}
  \int_{\Boundary}
   \frac{{\rm d}\Boundary}{b^2}
   ({\partial}_{n_0/b} +\rmi \At_{n_0})
   ({\partial}_{n/b}\epscomb{\Gnu} -\rmi \At_{n}\epscomb{\Gnu})
    u\PO
\end{align}
We want to replace the integrand by its limiting form.  To this end
the boundary is split into the part $\gamma_{c\epsilon}$, which lies
within a $(c\epsilon)$-vicinity around $\rvec_0$ (with arbitrary
constant $c$), and the remaining part $\Boundary_{\!c\epsilon}$,
\begin{align}
  \label{eq:fpieps}
 =  \lim_{\epsilon\to0} \!
  \bigg[&
  \int_{\Boundary_{\!\scriptstyle c\epsilon}}
    \!\! \frac{{\rm d}\Boundary}{b^2}
    (\partial_{n_0/b} +\rmi \At_{n_0})
    ({\partial}_{n/b}\epscomb{\Gnu} -\rmi \At_{n}\epscomb{\Gnu})
     u
\nnn
   +&  \int_{\gamma_{\scriptstyle c\epsilon}}
    \!\! \frac{{\rm d}\Boundary}{b^2}
    (\partial_{n_0/b} +\rmi \At_{n_0})
    ({\partial}_{n/b}\epscomb{\Gnu} -\rmi \At_{n}\epscomb{\Gnu})
     (u-u_0)
\\
    +&
    u_0
  \int_{\gamma_{\scriptstyle c\epsilon}}
    \!\! \frac{{\rm d}\Boundary}{b^2}
    (\partial_{n_0/b} +\rmi \At_{n_0})
    ({\partial}_{n/b}\epscomb{\Gnu} -\rmi \At_{n}\epscomb{\Gnu})
  \bigg]
\CO
\nn
\end{align}
with $u_0 \defas u(\rvec_0)$.  For sufficiently small $\epsilon$ the
boundary piece $\gamma_{c\epsilon}$ may be replaced by its
tangent\footnote{We emphasize  that we assume the boundary to be
  smooth throughout.} and the Green function by its asymptotic
expression, cf Sect. \ref{sec:Gprop}.  This way the third integral in
\eref{eq:fpieps} may be evaluated to its contributing order,
\begin{align}
  \label{eq:fpix3}
  &\int_{\gamma_{\scriptstyle c\epsilon}}
    \!\! \frac{{\rm d}\Boundary}{b^2}
    (\partial_{n_0/b} +\rmi \At_{n_0})
    ({\partial}_{n/b}\epscomb{\Gnu} -\rmi \At_{n}\epscomb{\Gnu})
\nnn
  =\,&
  \frac{1}{4\pi}
  \int_{-c\epsilon}^{c\epsilon}
    \!\!\!    \!\!\!
    \cos\! \Big(\frac{\rvec_0\nvec_0}{b^2}s\Big)
    \cos\!\left[\epsilon\left(\frac{\nvec_0\times\rvec_0}{b^2}-s\right)\right]
    \bigg(
      \frac{-2}{s^2+\epsilon^2}
      +4\frac{\epsilon^2}{(s^2+\epsilon^2)^2}
    \bigg)
    \,    \rmd s
    \nnn
  &+ \Or(\epsilon^2\log\epsilon)  
\nnn
 =\,&
  \frac{1}{2\pi}
  \int_{-c\epsilon}^{c\epsilon}
    \!\!\!    \!\!\!
    \rmd s
      \frac{\epsilon^2-s^2}{(s^2+\epsilon^2)^2}
    \,\,
  + \Or(\epsilon^2\log\epsilon)  
 =
  \frac{1}{\pi}   \frac{1}{c\epsilon}
  \frac{c^2}{c^2+1}
  + \Or(\epsilon^2\log\epsilon)
\nnn
 \approx&\,
  \frac{1}{\pi}   \frac{1}{c\epsilon}
  + \Or(\epsilon^2\log\epsilon)\PO  
\end{align}
Here, the explicit form of the integrand was obtained from
\eref{eq:QexpdlN} by the replacement $\rvec_0\to\rvec_0^\pm$.
The last approximation in \eref{eq:fpix3} holds because $c$ may be
chosen arbitrarily large.  In a similar fashion it can be shown that
the second integral in \eref{eq:fpieps} is of order
$\Or(\epsilon)$. In the first integral we may replace (again for large
$c$) the integrand by its limit, because $\epsilon$ is small compared
to $\min(|\rvec-\rvec_0|)=c\;\epsilon$.  Therefore, the limit in
\eref{eq:fpi} may be expressed as
\begin{multline}
\label{eq:fpilim}
  \fpint_{\Boundary}
   \frac{{\rm d}\Boundary}{b^2}
   ({\partial}_{n_0/b} +\rmi \At_{n_0})
   ({\partial}_{n/b} -\rmi \At_{n})
   \Gnu u 
   \\
  =  
 \lim_{\epsilon\to0}
 \bigg[
  \int_{\Boundary_{\!\scriptstyle \epsilon}}
    \!\! \frac{{\rm d}\Boundary}{b^2}
    (\partial_{n_0/b} +\rmi \At_{n_0})
    ({\partial}_{n/b}\Gnu -\rmi \At_{n}\Gnu)
     u
   \,\,
   + u_0 \frac{1}{\pi\epsilon}
 \bigg]\CO
\end{multline}
where we replaced $c\epsilon$ by $\epsilon$. This equation
defines the finite part integral.  It completes the derivation
of the boundary integral equations.

\subsection{Solving the integral equations}

As discussed above, the integral equations \eref{eq:SingleLayer} and
\eref{eq:DoubleLayer} 
determine the spectra and wave functions of arbitrary interior and
exterior magnetic billiards.  
In the stated form the equations are not yet suitable for numerical
evaluation, though, since the integral kernels display (integrable)
singularities.

Fortunately, 
it is possible to treat the singular behavior
analytically  which renders a highly accurate and efficient numerical
scheme.
In brief, the boundary integral equations are regularized using the
known asymptotic behavior of the Green function and its
derivatives, cf Sect.~\ref{sec:Gprop}.
Representing the periodic boundary functions in a Fourier basis then
leads to an exponential localization of the integral kernels.  This
permits a well controlled truncation of the corresponding matrix. The
roots of the (Fredholm) determinant are accurately obtained by
singular value decomposition.
We refer the reader to our recent publication \cite{HS00a} for the
technical details 
and a convergence analysis.\footnote{
Note that the equations in \cite{HS00a} are stated in complex
conjugated form since the focus is there on the wave functions rather
than the Green function.}

\section{Results of the boundary integral method}
\label{chap:numres}

The numerical implementation of the boundary integral method provides
thousandths of eigenfunctions at high accuracy with moderate
computational effort.  This includes the bulk states as long as the
small energy difference to the Landau level can be represented
numerically.  In the following we demonstrate the performance of the
boundary integral method by exhibiting numerical results on magnetic
billiards which have been inaccessible by other methods \cite{HS00a}.

\subsection{Spectral statistics}
\label{sec:specstat}

We start by applying some of the standard tools of spectral statistics
to large data sets of interior spectral points.  The spectra are
expected to reproduce the features of random matrix theory (RMT) if
the underlying classical motion displays hard chaos \cite{Bohigas91}.
In this section we define the spectra in the semiclassical direction
$b\to0$, keeping the cyclotron radius $\rho$ constant.  This way we
can ensure that the classical dynamics are completely chaotic
throughout the spectral intervals considered.

\begin{figure}[tbp]%
  \begin{center}%
    \includegraphics[width=\linewidth] 
    {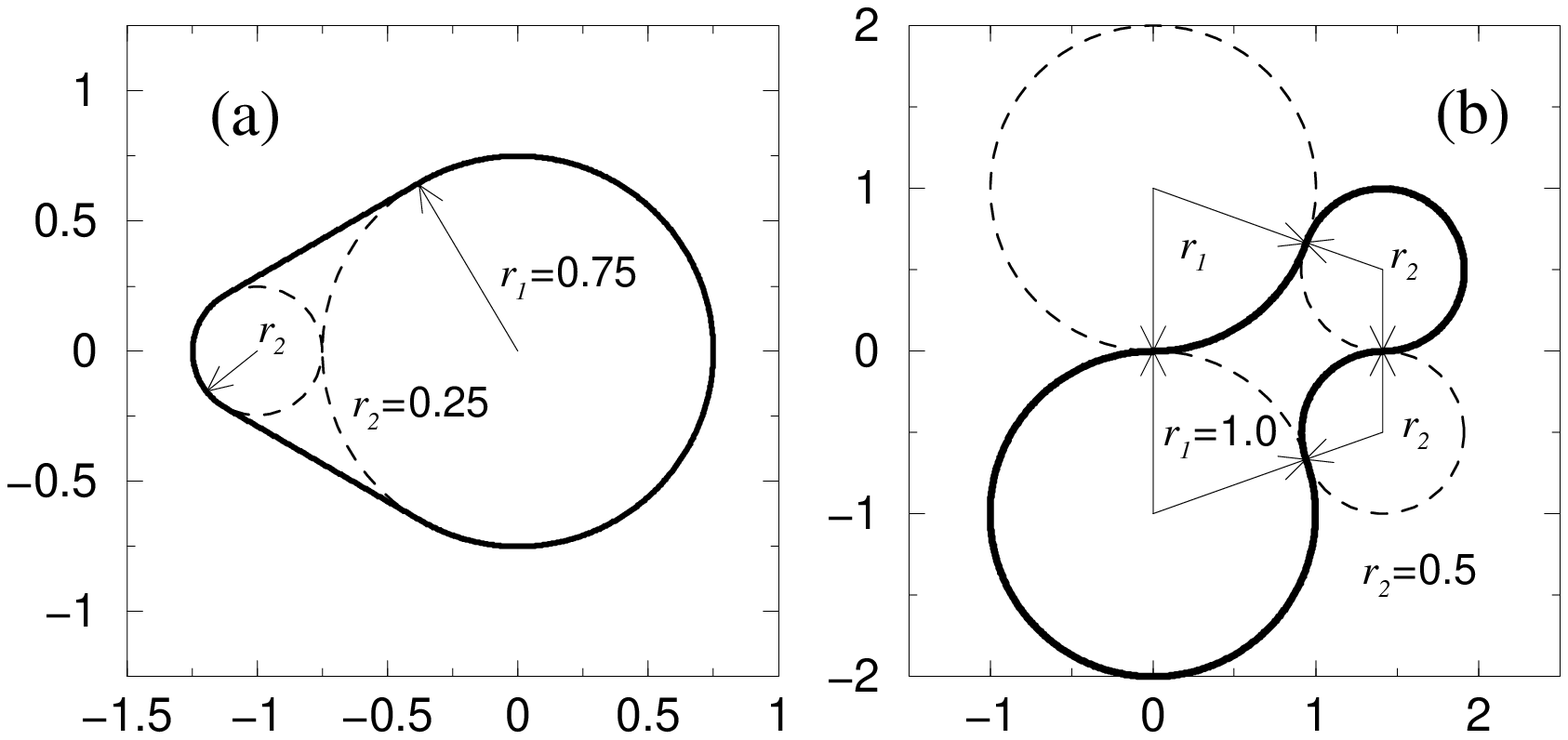}
    \figurecaption{%
      Definition of  domain boundaries considered in Chapter
      \ref{chap:numres}. The magnetic dynamics  in
           the asymmetric stadium (left) exhibits an anti-unitary
    symmetry, but no unitary  one.
      In contrast, the skittle shaped domain (right) is free
      of any symmetry. It generates \emph{hyperbolic} classical motion
      for $\rho>2$.       }%
    \label{fig:shapes}%
  \end{center}%
\end{figure}

We consider the two domains described in Fig.~\ref{fig:shapes}. One is
an asymmetric version of the Bunimovich stadium billiard ($r_1=0.75$,
$r_2=0.25$, $\Area=2.10957$, $\Len=5.39724$). In the magnetic field
its dynamics is free of unitary symmetries but exhibits an
anti-unitary one (time reversal and reflection at $y=0$). The skittle
shape, in contrast, (made up of the arcs of four symmetrically
touching circles, $r_1=1.0$, $r_2=0.5$, $\Area=4.33969$,
$\Len=9.42478$) does not display any symmetry.  It generates
hyperbolic classical motion even for
fairly strong magnetic fields
\cite{Gutkin01}. (The asymmetric stadium is not strictly hyperbolic,
but any possibly regular part in phase space is much smaller than the
uncertainty product $(b^2\pi)^2$ throughout the considered spectral
interval.)

\begin{figure}[tbp]%
  \begin{center}%
   \psfrag{x}{$\nu$}
   \psfrag{y}{\hspace*{-0.75em}$\N_{osc}^{(\rho)}(\nu)$}
    \includegraphics[width=\linewidth] {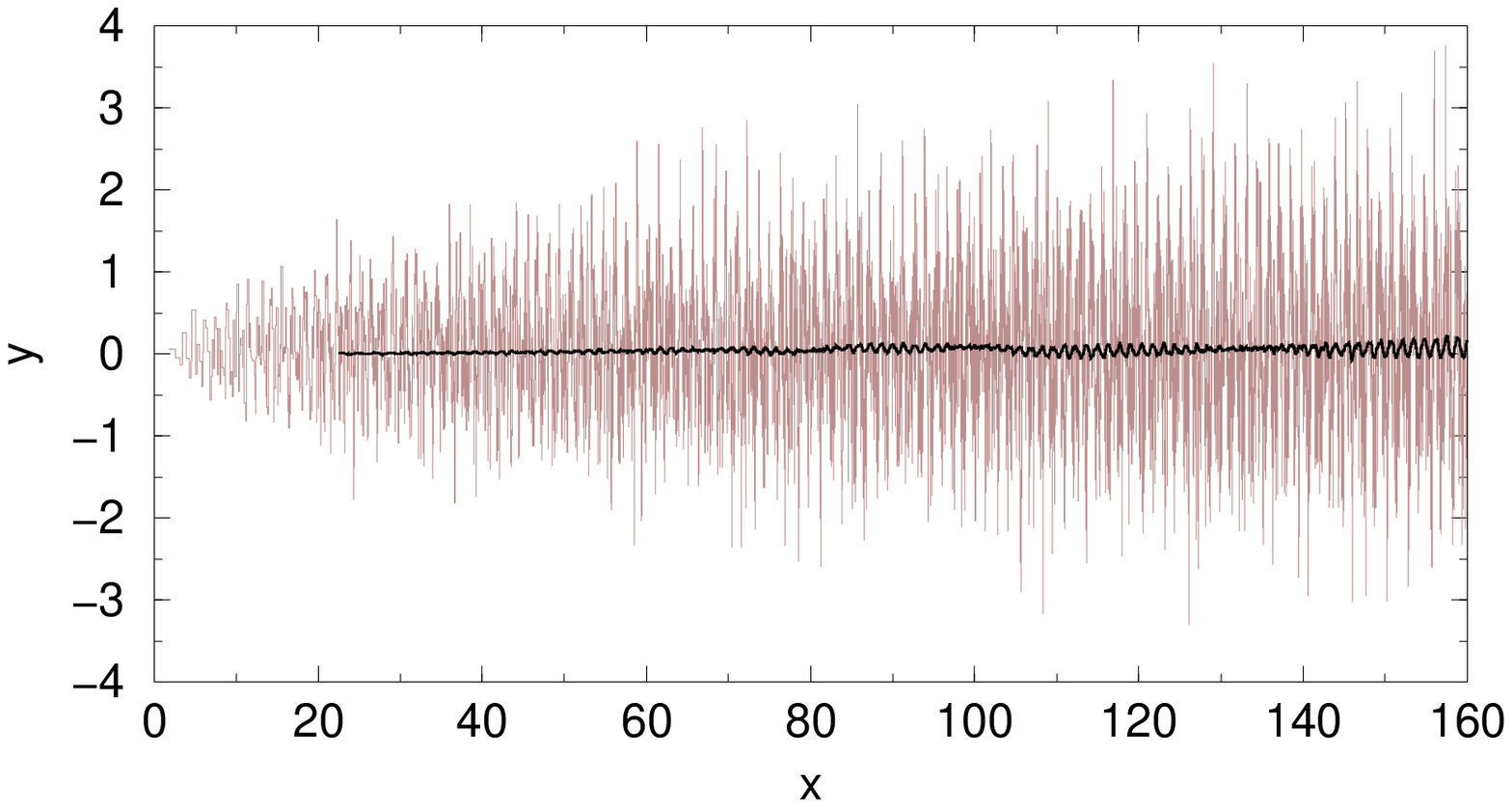}
    \figurecaption{%
      Fluctuating part of the spectral staircase
      in the asymmetric stadium at $\rho=1.2$.  The displayed range
      contains the first $12000$ points of the interior spectrum, with
      the heavy line a running average over 500 neighboring points.
      A missing spectral point  would show up as
      a distinct step by one.  }%
    \label{fig:Nosc}%
  \end{center}%
\end{figure}

We calculated 12300 and 7300 consecutive interior Dirichlet
eigenvalues at $\rho=1.2$ for the asymmetric stadium and the skittle
shaped domain, respectively.  Using the boundary integral method it is
possible to converge states even with much greater quantum numbers
\cite{HS00a}.  The time consuming task
is really to find \emph{all} energies, including the near-degenerate
ones, in a given interval.

\begin{figure}[tbp]%
  \begin{center}%
   \psfrag{x}{$s$}
   \psfrag{y}{\hspace*{-1.75em}$P(s),\, I(s)$}
    \includegraphics[width=\linewidth]{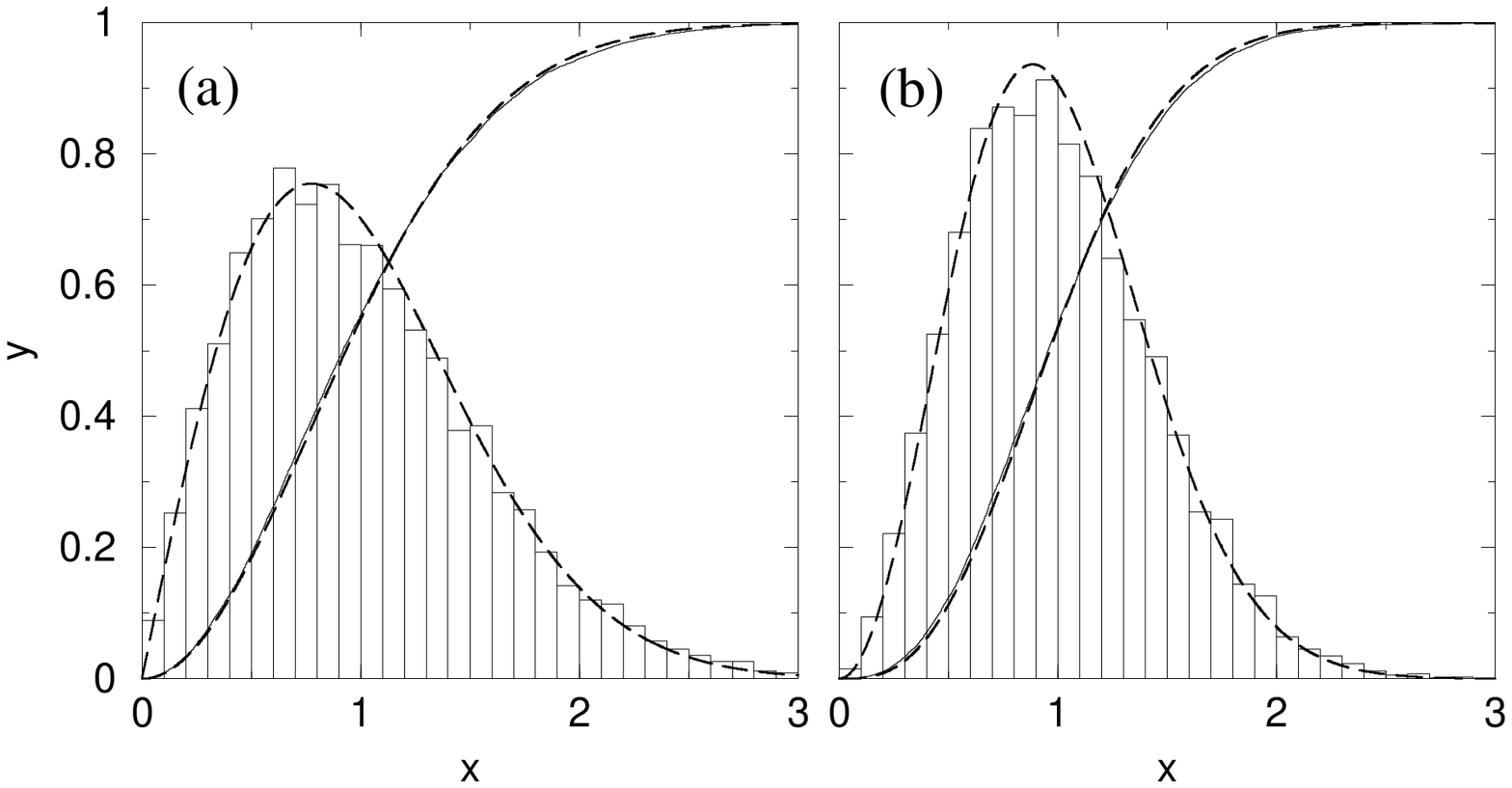}
    \figurecaption{%
      Nearest neighbor spacing distributions of (a) the asymmetric stadium 
      and (b) the  skittle shaped domain (right), at $\rho=1.2$.  The
      histograms should be compared to GOE and GUE predictions of
      random matrix theory, respectively (dashed lines.)  
      The corresponding cumulative probabilities are given by the
      monotonic lines. Here, the differences between data and RMT 
      are of the order of the error of Wigner's surmise.}%
    \label{fig:NN}%
  \end{center}%
\end{figure}

The integrity of the obtained spectrum 
may be checked by calculating
the fluctuating part ${\rm
\N_{osc}^{(\rho)}}(\nu)=\N^{(\rho)}(\nu)-\Nsm^{(\rho)}(\nu) $ of the
spectral counting function.
This quantity must average to zero indicating 
whether spectral points were missed. It is defined in terms of the
mean staircase (given in equation \eref{eq:Nrhosm} for fixed $\rho$).
Figure \ref{fig:Nosc} displays ${\rm \N_{osc}^{(\rho)}}$ for the
asymmetric stadium.  The strongly fluctuating function indeed vanishes
on average which indicates the completeness of the spectrum. This can be seen
from the heavy line which gives a running average over 500 neighboring
points. The oscillations of the running average can be related
semiclassically to the existence of bouncing ball modes, which are
discussed below.
A very similar result like Fig.~\ref{fig:Nosc} is obtained for the skittle
shaped domain (not shown).

The large spectral intervals at hand permit the direct calculation of
some of the popular statistical functions used to characterize
spectra.  Due to the underlying classical chaos and the symmetry
properties mentioned above one expects the statistics of the
Gau{ss}ian Orthogonal Ensemble (GOE) for the asymmetric stadium, and
of the Gau{ss}ian Unitary Ensemble (GUE) for the skittle.  Figure
\ref{fig:NN} shows the distributions of nearest neighbor spacings $P(s)$ of
the unfolded\footnote{The spectra are transformed to unit density; see
also the discussion in Sect.~\ref{sec:auto}.} spectra.  Indeed, one
finds excellent agreement with random matrix theory. The differences
between the numerical and the RMT cumulative functions $I(s)=\int_0^s
P(s^\prime)
\rmd s^\prime$ stay below $2\%$ (ie, below the error of Wigner's
surmise \cite{Bohigas91}).

\begin{figure}[tbp]%
  \begin{center}%
   \psfrag{x}{$\tau$}
   \psfrag{y}{\hspace*{-0.75em}$K(\tau)$}
    \includegraphics[width=\linewidth]{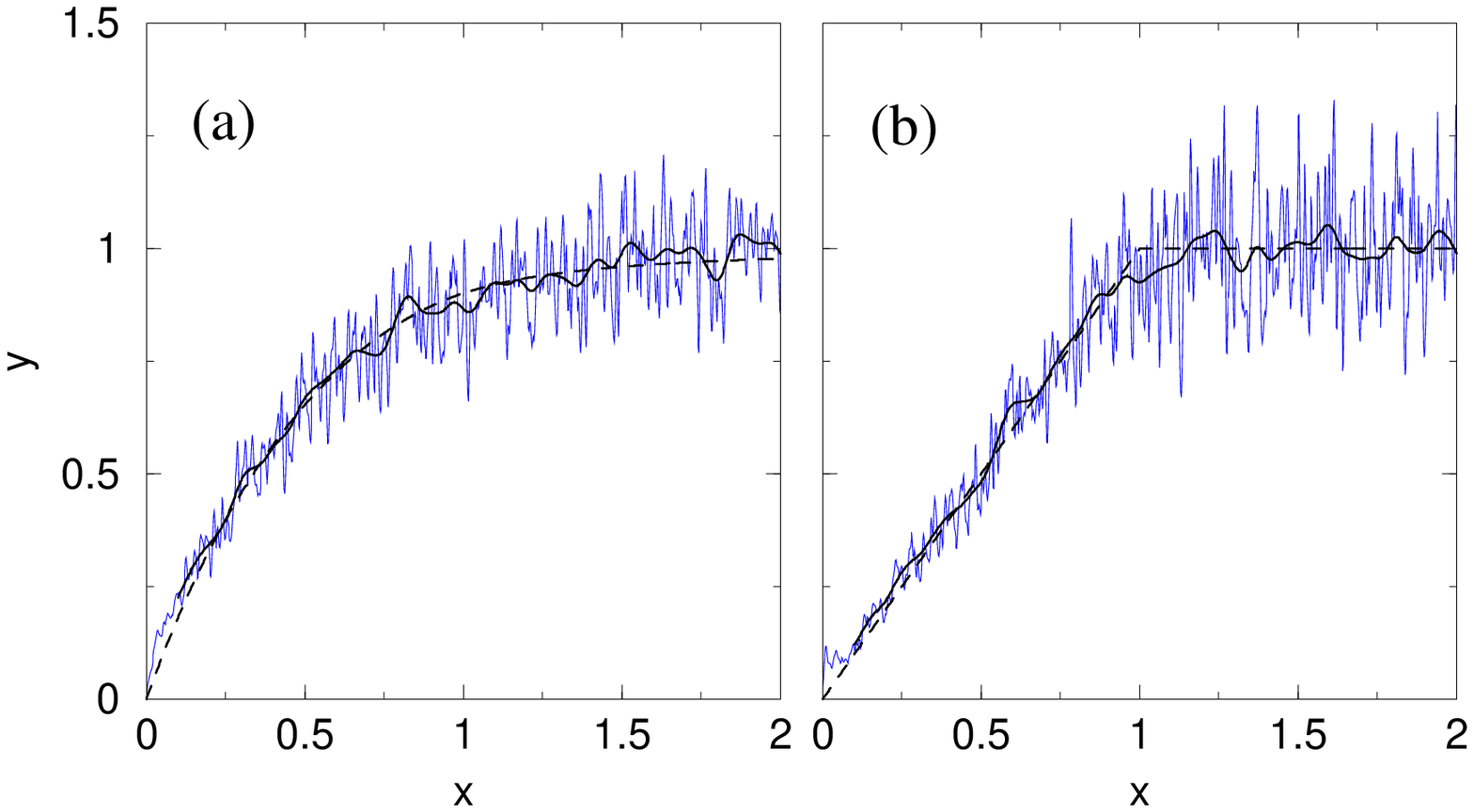}
    \figurecaption{%
      Spectral form factor of (a) the asymmetric stadium and (b) the
      skittle shaped domain based on 12300 and 7300 spectral
      points, respectively.  The heavy lines belong to the same data
      after stronger spectral averaging.
      One observes good agreement with
      the random matrix prediction of
      the Gau{ss}ian Orthogonal and the Gau{ss}ian Unitary Ensemble,
      respectively,  indicated by the dashed lines.
      }%
    \label{fig:formfactor}%
  \end{center}%
\end{figure}

In order to characterize the 
spectrum more sensitively
one often considers the form factor $K(\tau)$, ie, the (spectrally averaged)
Fourier transform of the two-point autocorrelation function of the
spectral density \cite{Berry85,AIS93}.
Figure \ref{fig:formfactor} gives the spectral form factors of the
asymmetric stadium and the skittle spectra. The thin and heavy lines
correspond to different degrees of averaging, using a spectral window
of width 3 and 30, respectively. The RMT predictions are shown as
dashed lines, and one observes again very good agreement.
Since most of the other popular spectral functions, such as Dyson's
$\Delta_3$ statistic, are transformations of the form factor we do not
present them here.

We emphasize that the good agreement with RMT is not only a
consequence of the large spectral intervals the statistics are based
on. It is as important to have the spectra defined at fixed classical
dynamics.  Had we calculated the spectra at fixed field, they would
have been based on a classical phase space that transforms from a
near-integrable, time-invariance-broken structure to a hyperbolic
time-invariant one as $\rho$ increases with energy.  This
transformation of spectral statistics from GOE to GUE as the field is
increased was studied in \cite{BGOdAS95,YH95,JB95}.

\subsection{Wave functions in the interior and in the exterior}
\label{sec:wavefunctions}

\subsubsection*{The skittle}

To get an overview of the various types of eigenstates one may
encounter in magnetic billiards we proceed to present a selection of
stationary wave functions. We focus on the semiclassical regime of
large scaled energies at cyclotron radii small enough to observe
strong effects of the magnetic field.  We start with the skittle
shaped domain choosing again $\rho=1.2$, such that the corresponding
classical skipping motion is chaotic in the interior as
well as in the exterior.

\begin{figure}[tbp]%
  \begin{center}%
    \includegraphics[scale=0.70] 
{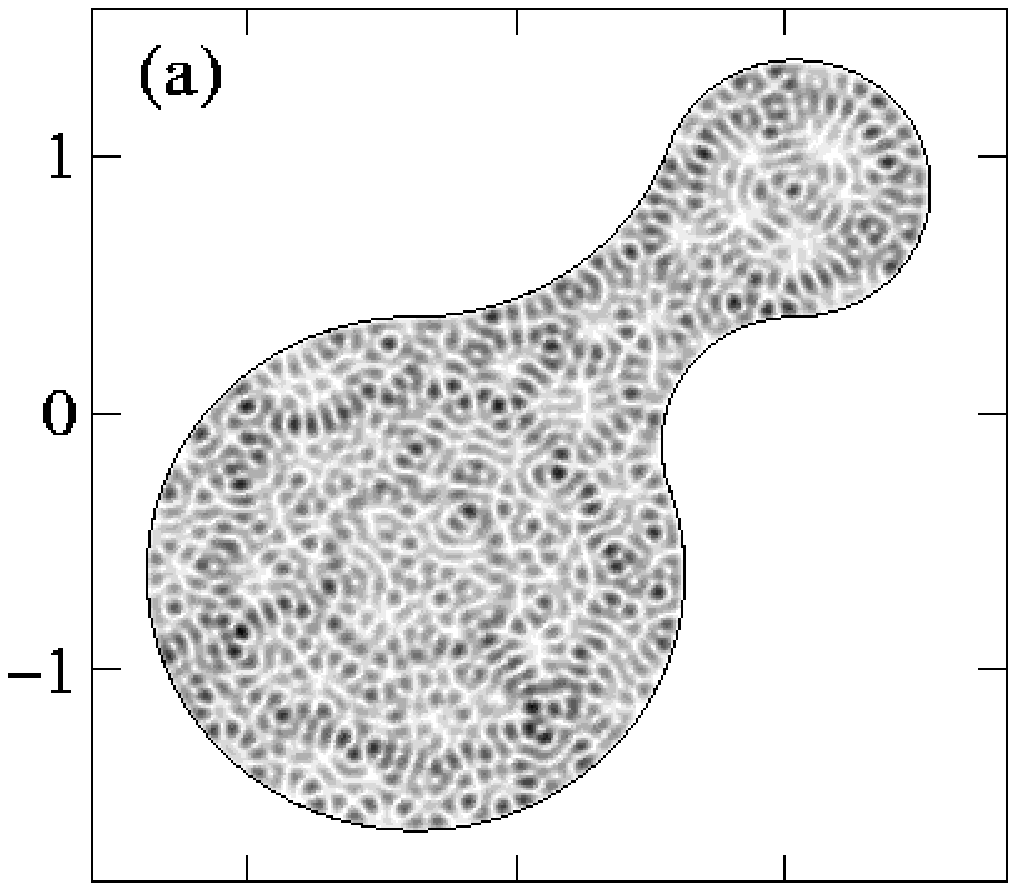}
    \includegraphics[scale=0.70] 
{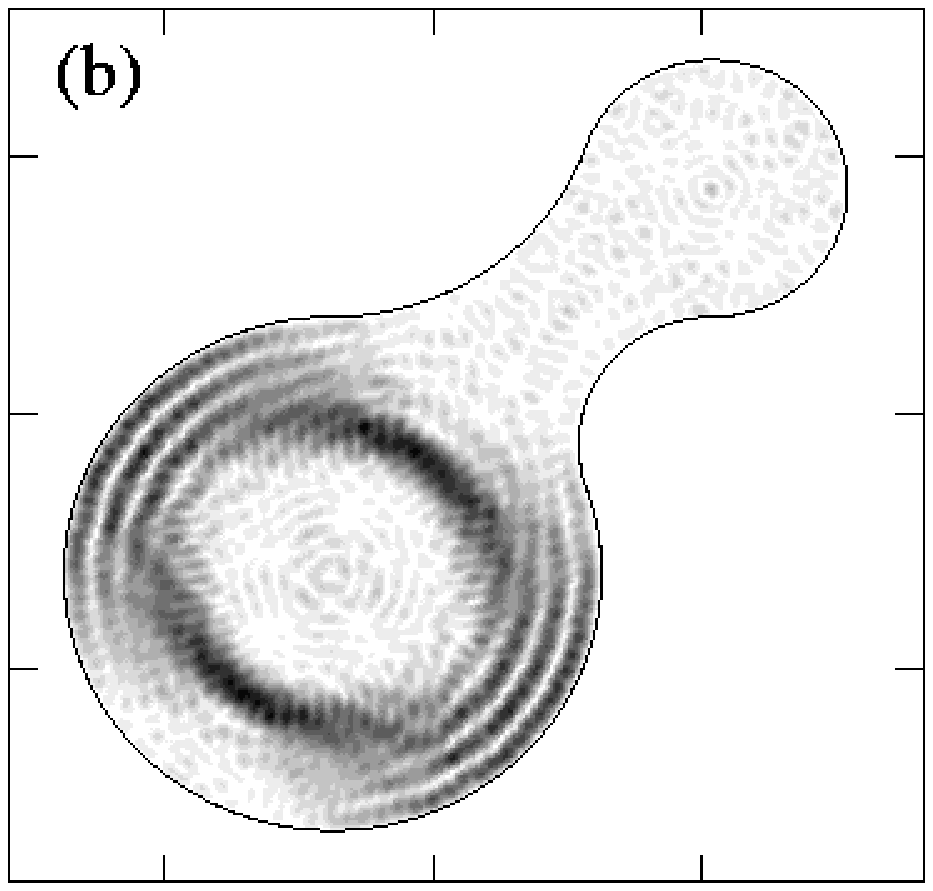}
    \\[1mm]
    \includegraphics[scale=0.70] 
{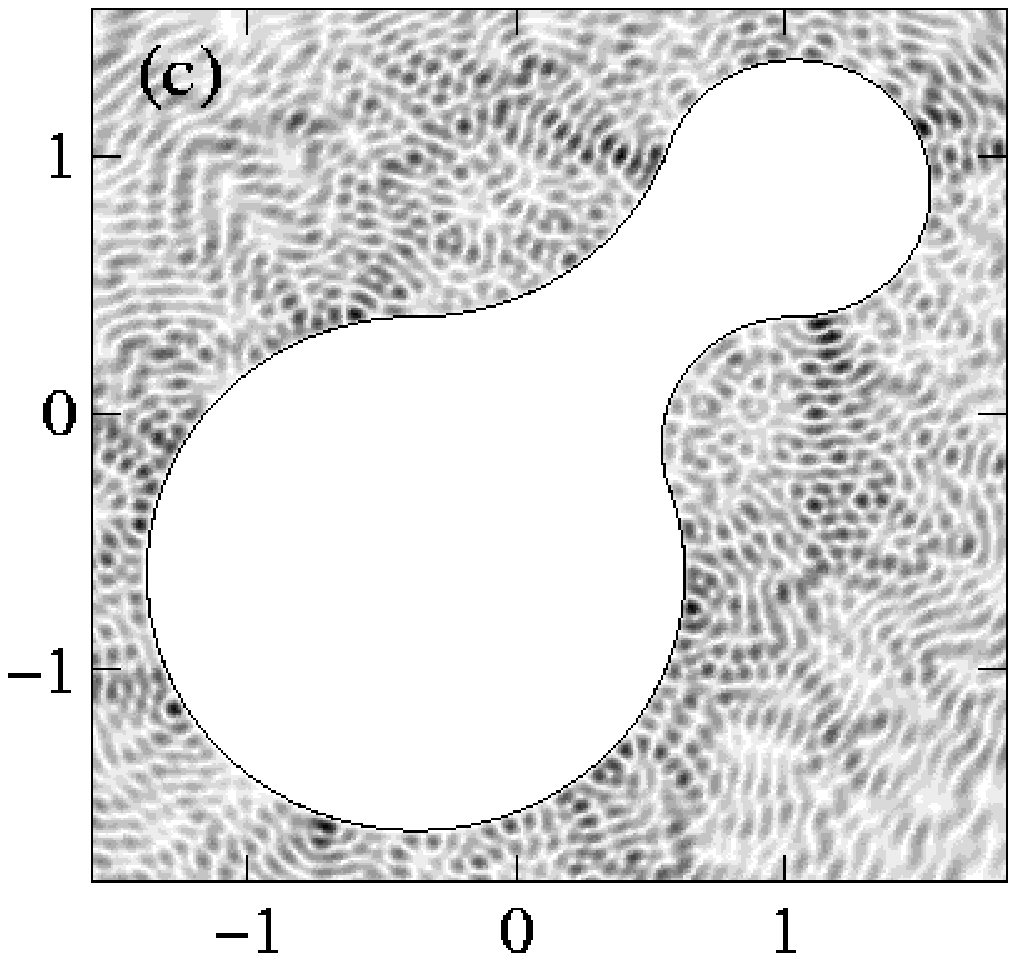} 
    \includegraphics[scale=0.70] 
{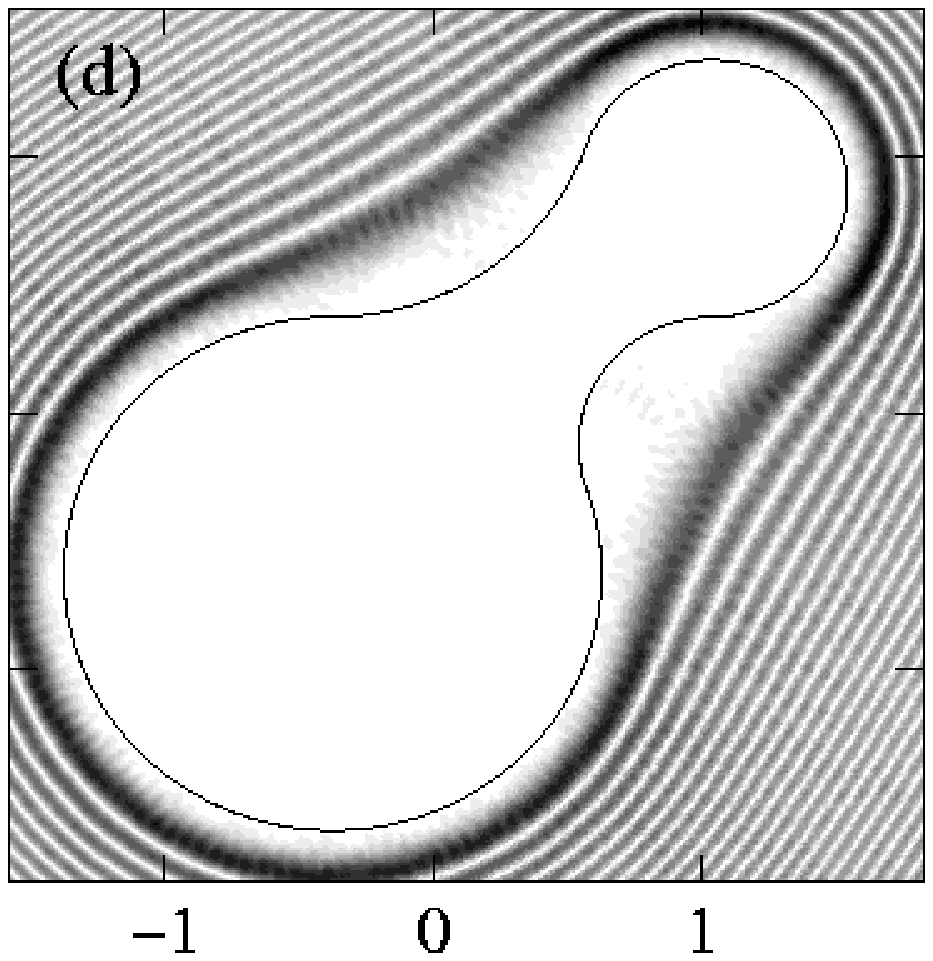} 
    \figurecaption{%
      Interior and exterior wave functions of the skittle shape around
      the one-thousandth interior eigenstate, at fixed
      cyclotron radius $\rho=1.2$. The
      plotted shade is proportional to $|\psi|$ and the thin line
      indicates the boundary $\Gamma$. 
      (a) A typical interior wave function, $\nu\simeq32.9880$. 
      (b) A bouncing ball mode, $\nu\simeq33.1203$.
      (c) A typical exterior edge state, $\nu\simeq32.8474$.
      (d) A typical exterior bulk state, $\nu\simeq32.50025$.
[figure quality reduced]
      }%
    \label{fig:skittle}%
  \end{center}%
\end{figure}
\begin{figure}[tbp]%
  \begin{center}%
    \includegraphics[scale=0.70] 
{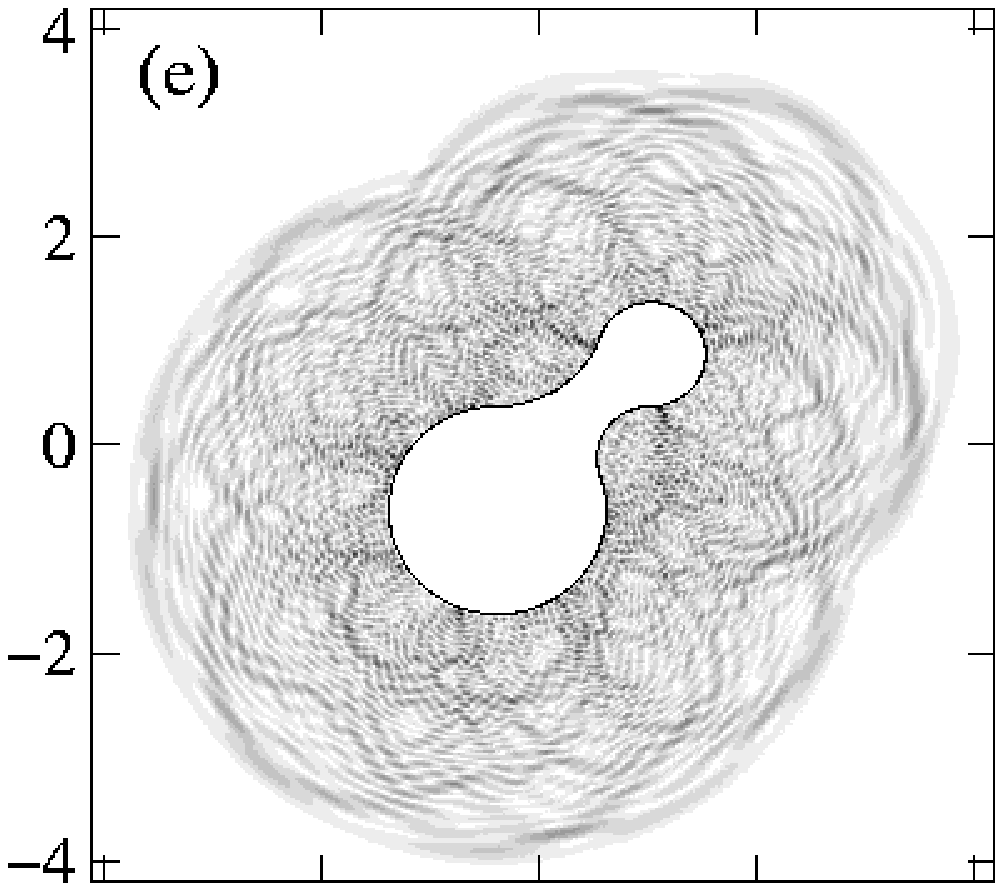}
    \includegraphics[scale=0.70] 
{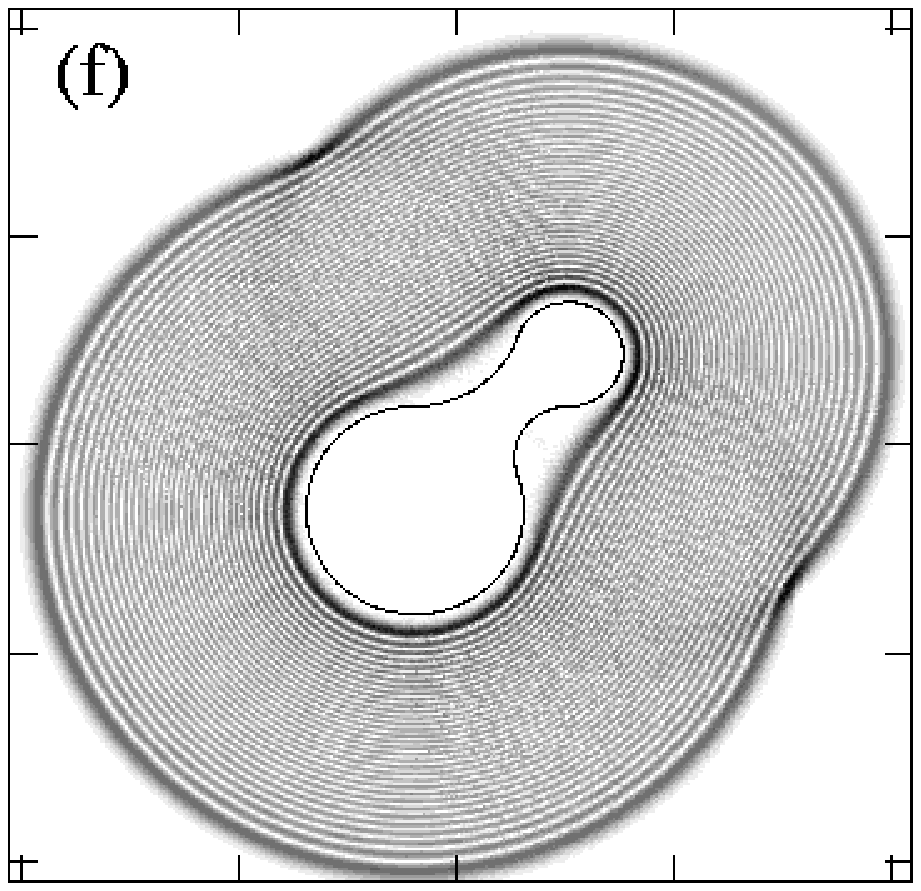}
    \\[1mm]
    \includegraphics[scale=0.70] 
{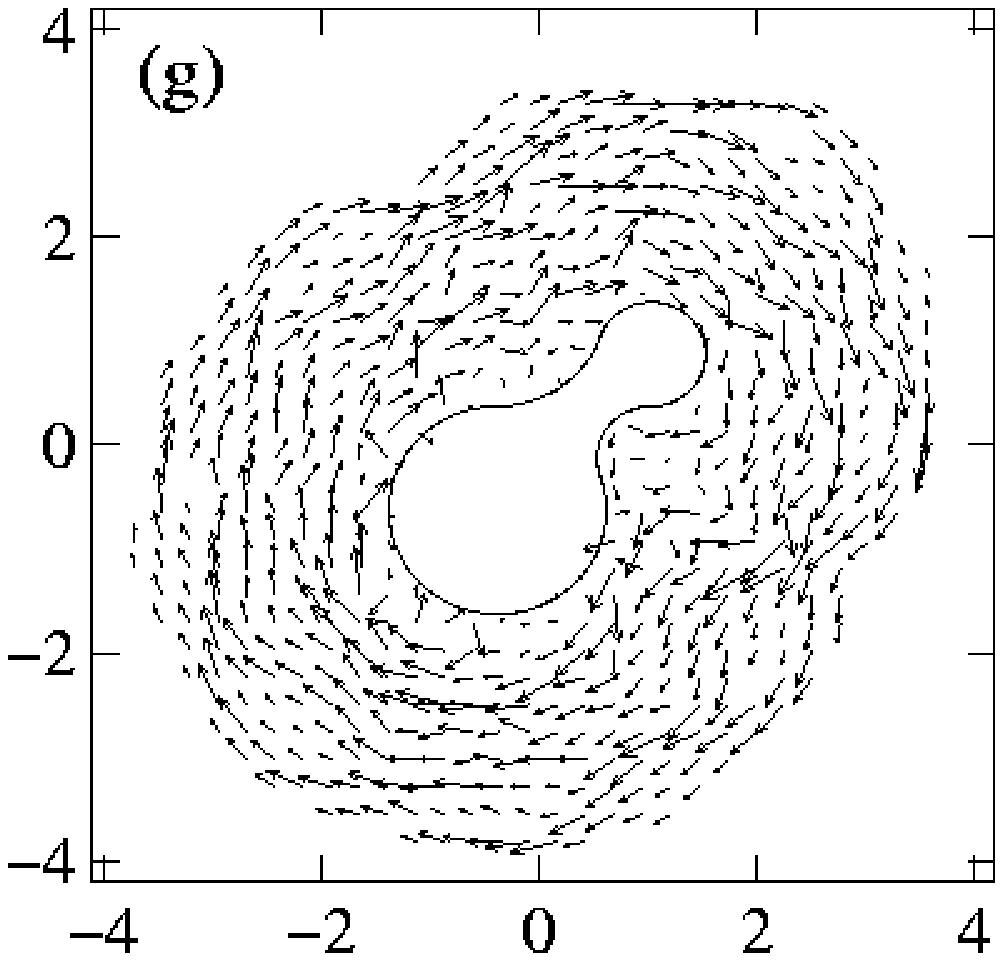} 
    \includegraphics[scale=0.70] 
{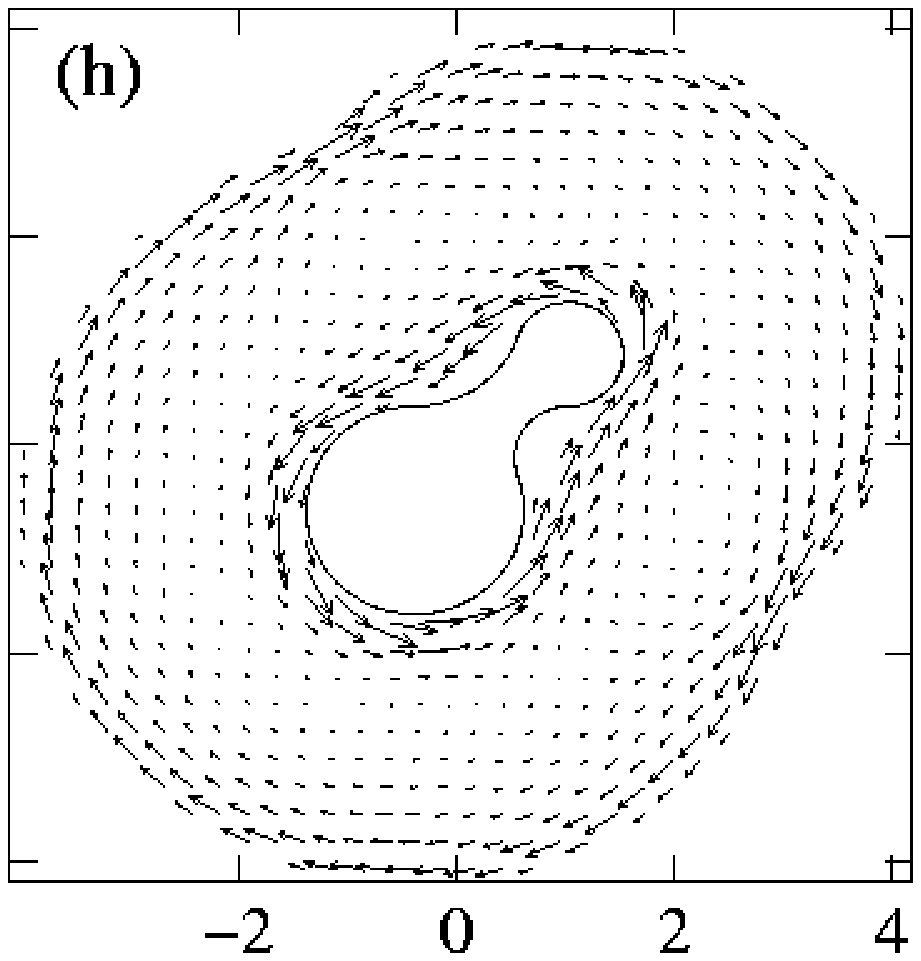} 
    \figurecaption{%
      (continued) 
      (e),(f) The same exterior edge and bulk states as in (c) and (d),
      respectively, on a larger scale. (g),(h) The current densities
      of the  edge and bulk states (e) and (f),  respectively. Here, 
      the length of the arrows is proportional to
      the magnitude of the current density (measured at the at the
      tails of the arrows). [figure quality reduced]
      }%
  \end{center}%
\end{figure}

Figure \ref{fig:skittle}(a) shows the density plot of a typical
interior wave function around the one-thousandth eigenstate.  As
expected for a classically chaotic system it spreads  throughout
the whole domain and  has the features of a random wave. 
Occasionally, one also encounters so called
\emph{bouncing-ball} modes. These states are localized on a
manifold of marginally stable periodic orbits (which have zero measure
in phase space).  
In the skittle a prominent manifold consists of  orbits with
period 2 bouncing in the larger circular part of the billiard.  The
wave function of a corresponding bouncing-ball mode is given in
Fig.~\ref{fig:skittle}(b).

We turn to the eigenstates of the exterior billiard.  The wave
function of a typical example may be found in
Fig.~\ref{fig:skittle}(c).  It belongs to an energy close to the one
of Fig.~\ref{fig:skittle}(a) and is displayed on the same scale.
Again, one observes the typical features of a chaotic wave function.
When viewed on a larger scale, cf Fig.~\ref{fig:skittle}(e), we find
that this state seems bound to the billiard and vanishes rapidly after
a distance smaller than a cyclotron diameter. In addition, circular
structures are faintly visible in the probability distribution with a
radius given by the classical cyclotron radius $\rho=1.2$.  This state
is clearly an edge state corresponding classically to a skipping
motion around the billiard. This is also evident from
Fig.~\ref{fig:skittle}(g) which displays the distribution of the
current density of the state.

Figure~\ref{fig:skittle}(d) shows a quite different exterior state.
Its energy is still in the same range as that of
Fig.~\ref{fig:skittle}(c) but now close to a Landau level.  One
observes that, unlike the edge state, the wave function shows no
appreciable amplitude close to the boundary.  Moreover, it displays a
rather regular structure consisting of rings of maximal probability
density which encircle the billiard.  This is seen clearly on a larger
scale, cf Fig.~\ref{fig:skittle}(f).  The band running around the
billiard has a width of the cyclotron diameter, $2\rho$, and in
general consists of $N+1$ rings if the energy is close to $\nu=N+\oh$
(here $N=32$).  This band moves outwards and gets more circular as one
goes to bulk energies which are increasingly close to the Landau level
(this way the bulk states in the sequence sweep over the whole plane).
Clearly, we are dealing with a bulk state. Its wave function corresponds to a
superposition of unperturbed cyclotron orbits which are placed around
the billiard.  This view is  supported again  by the distribution of the
current density, cf Fig.~\ref{fig:skittle}(h).  Near to the
boundary it displays an opposite orientation compared to at a distance of
$2\rho$. This renders the the net current around the billiard
exponentially small --- unlike the edge state  Fig.~\ref{fig:skittle}(h)
which displays a large finite current.

In order to see this separation into edge and bulk states more clearly
we turn to even more semiclassical energies 
and a
symmetric shape of the boundary.
\begin{figure}[tbp]%
  \begin{center}%
    \includegraphics[scale=0.45] 
{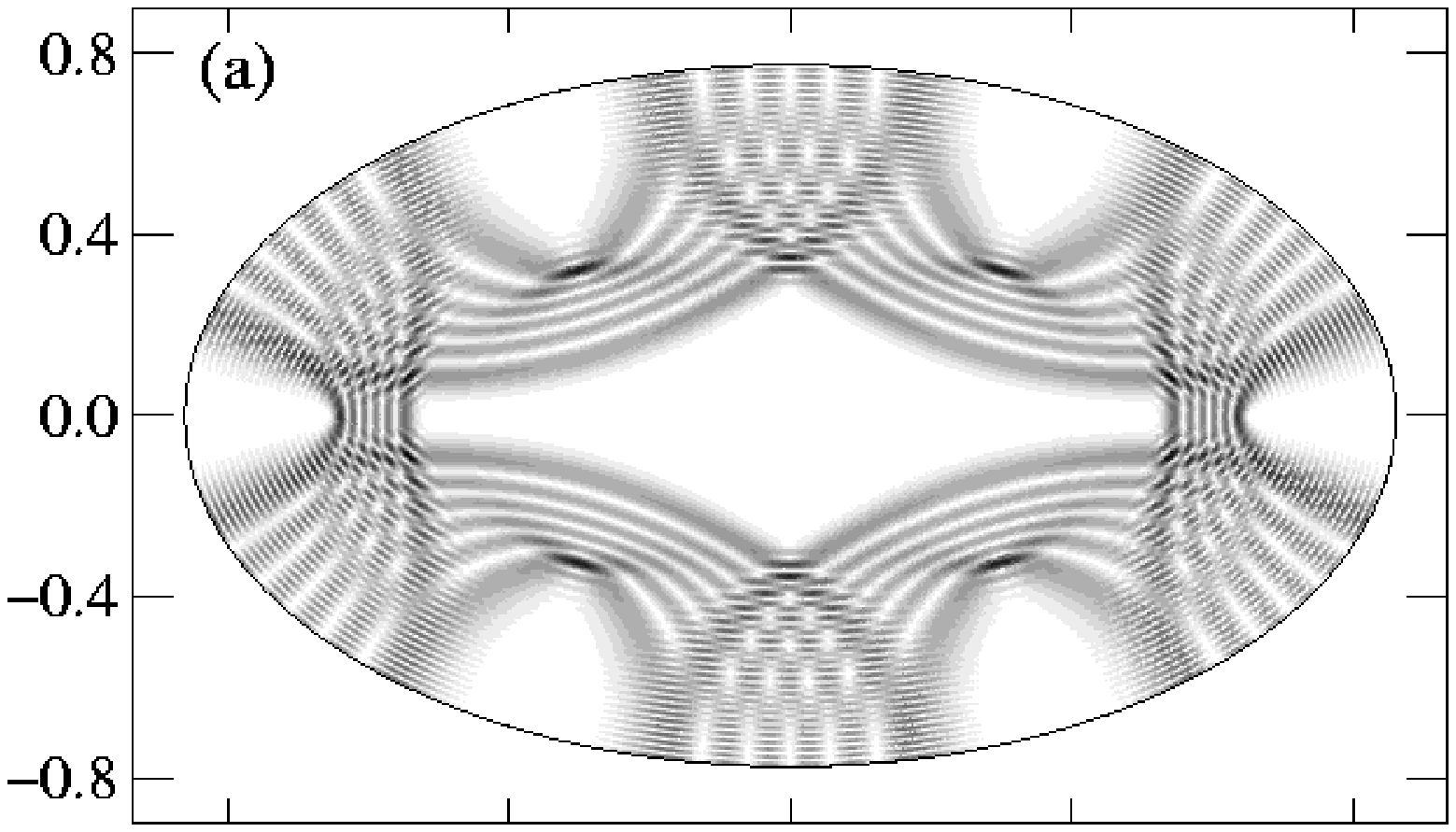}
    \includegraphics[scale=0.45] 
{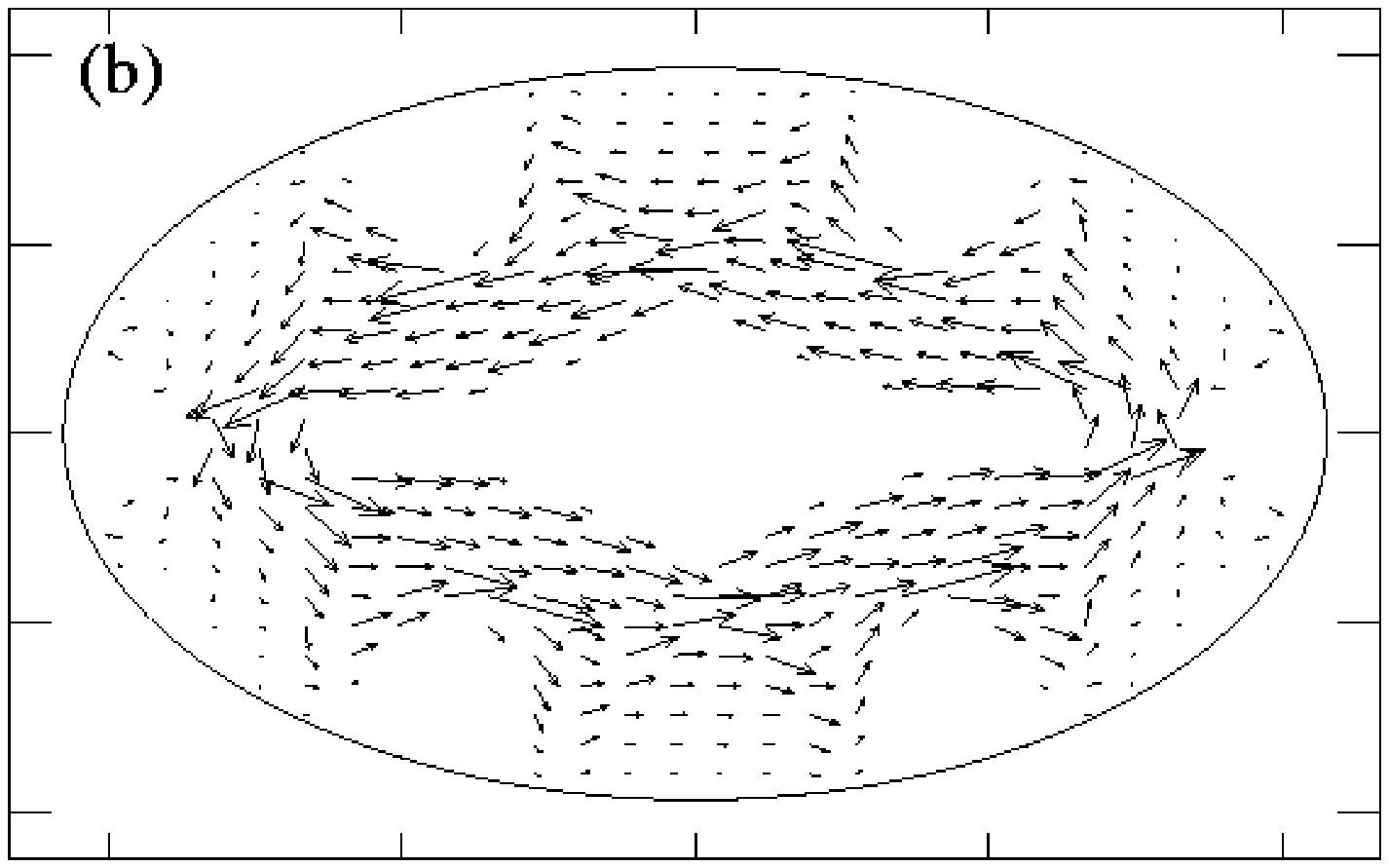}
    \\[1mm]
    \includegraphics[scale=0.45] 
{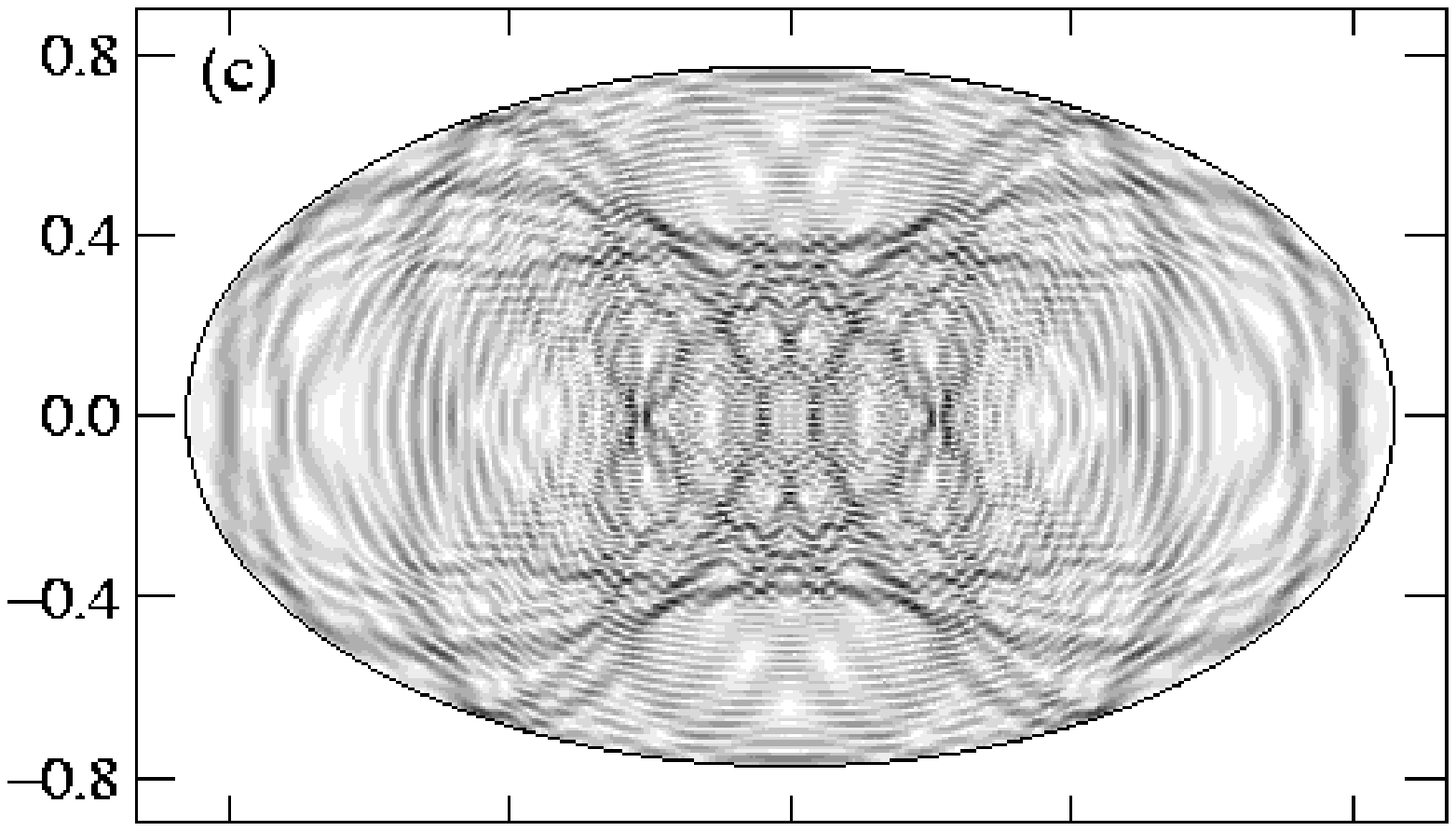}
    \includegraphics[scale=0.45] 
{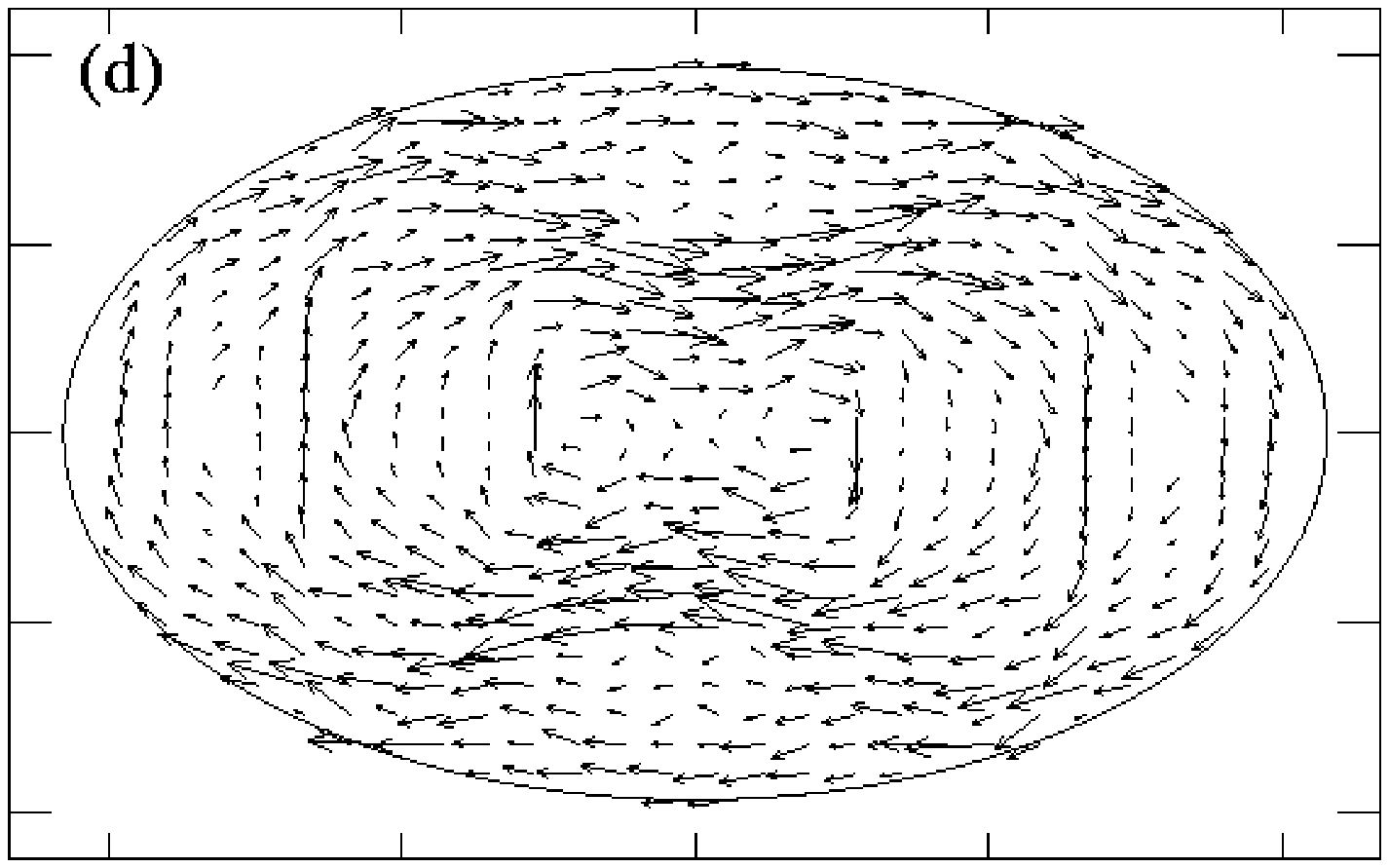}
    \\[1mm]
    \includegraphics[scale=0.45] 
{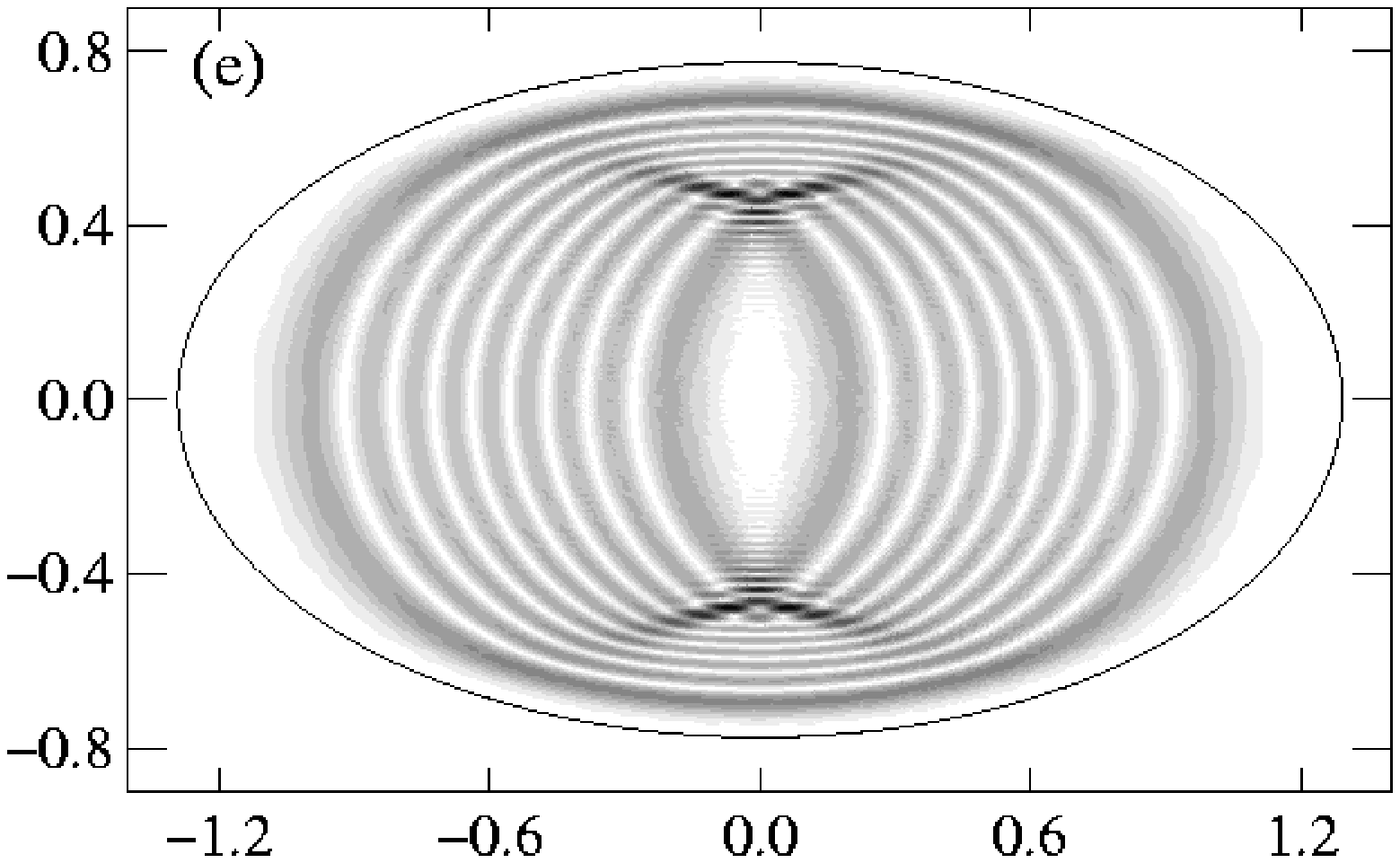}
    \includegraphics[scale=0.45] 
{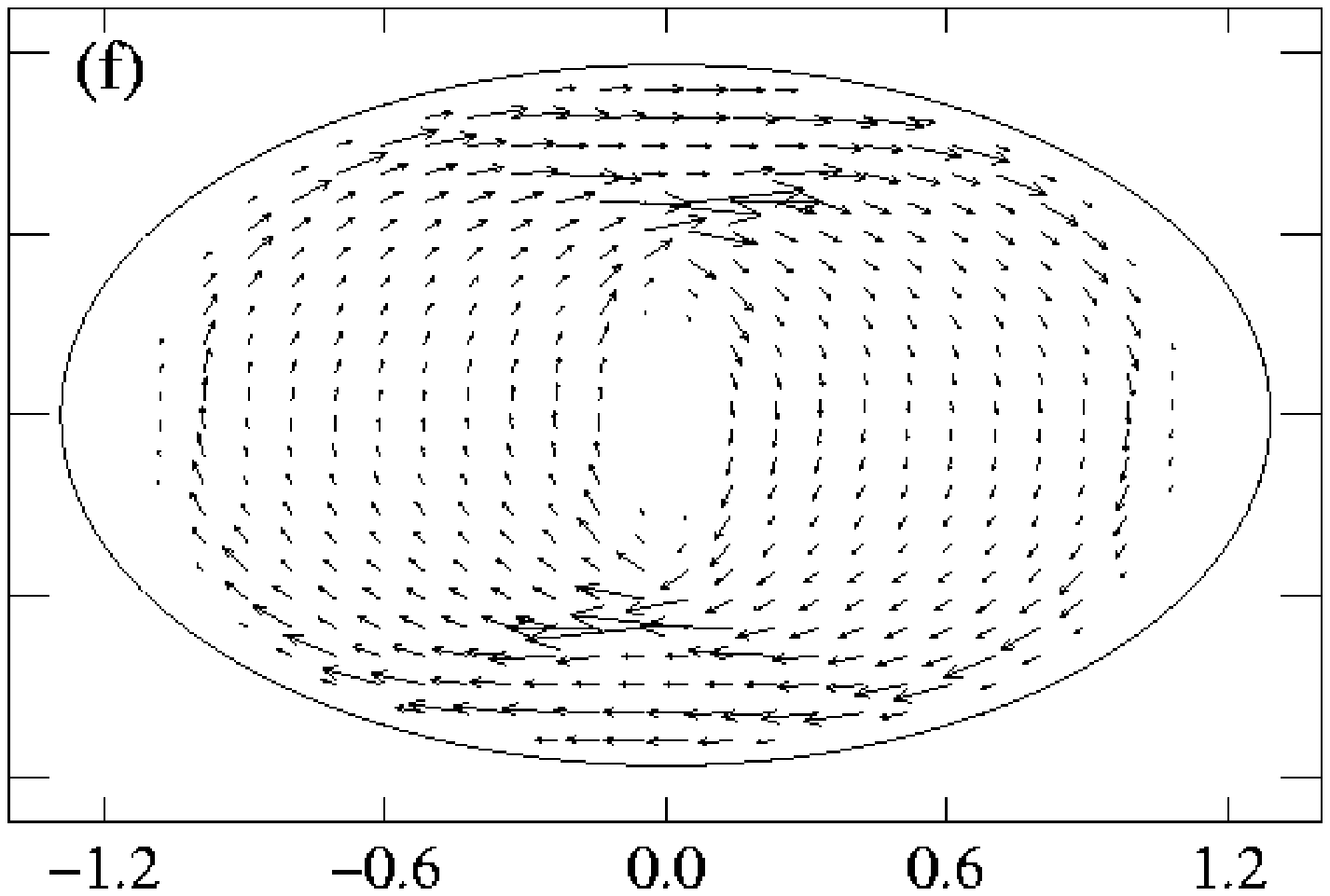}
    \figurecaption{%
      Probability densities (left) and current density distributions 
      (right) in an
      elliptic domain at $\rho=0.6$, around the ten-thousandth
      interior eigenstate. 
      (a,b) An edge state localized along a stable periodic orbit,  
      $\nu\simeq60.0602$.
      (c,d) A state which covers the whole domain
      $\nu\simeq60.1664$. Note that the gross direction of the current
      changed compared to (b).
      (e,f) An interior bulk state, $\nu\simeq60.50031$. 
      The wave function is exponentially small
      close to the boundary and represents a
      superposition of cyclotron motion. [figure quality reduced]
    }%
  \label{fig:ellipseint}%
  \end{center}%
\end{figure}

\begin{figure}[tbp]%
  \begin{center}%
    \includegraphics[scale=0.53] 
{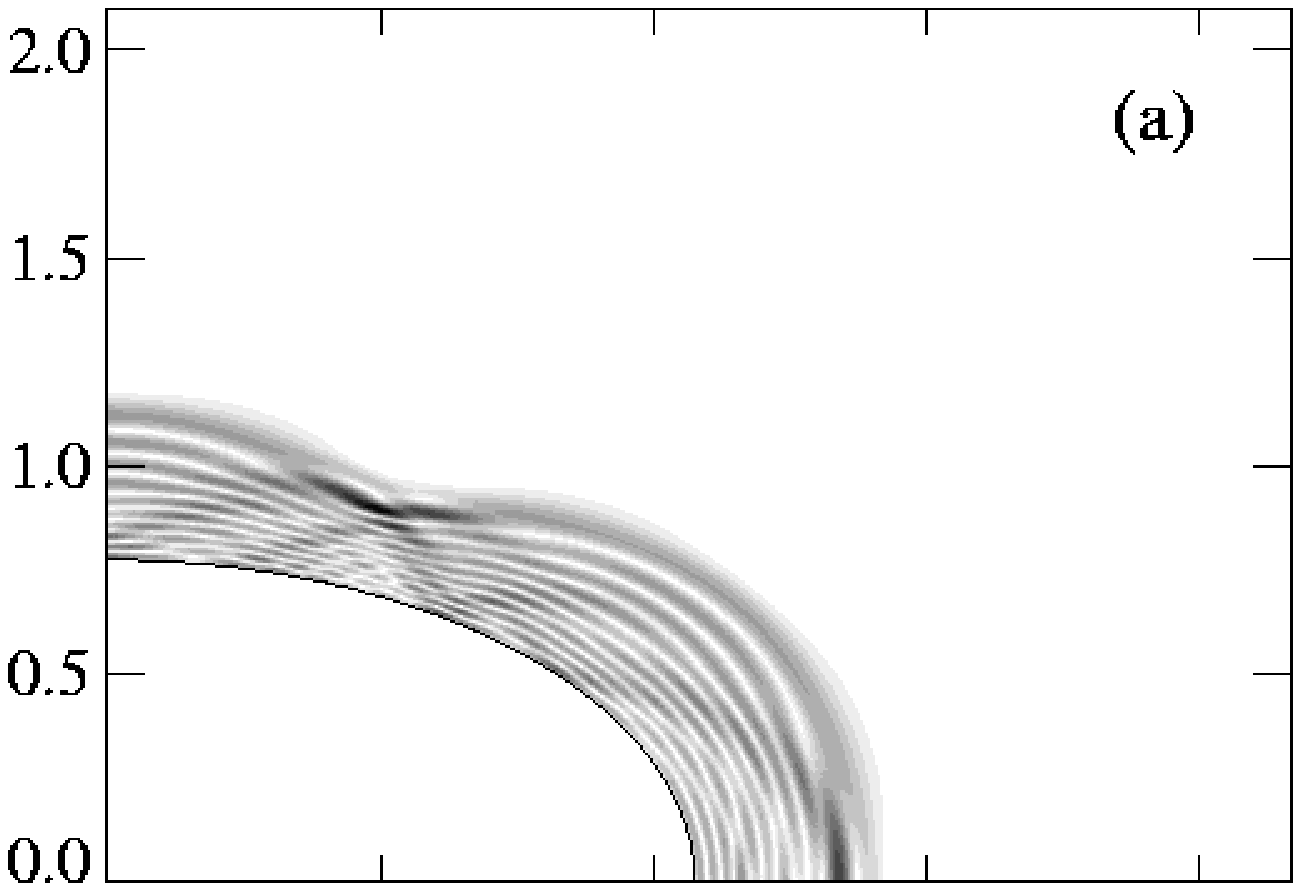}
    \includegraphics[scale=0.53] 
{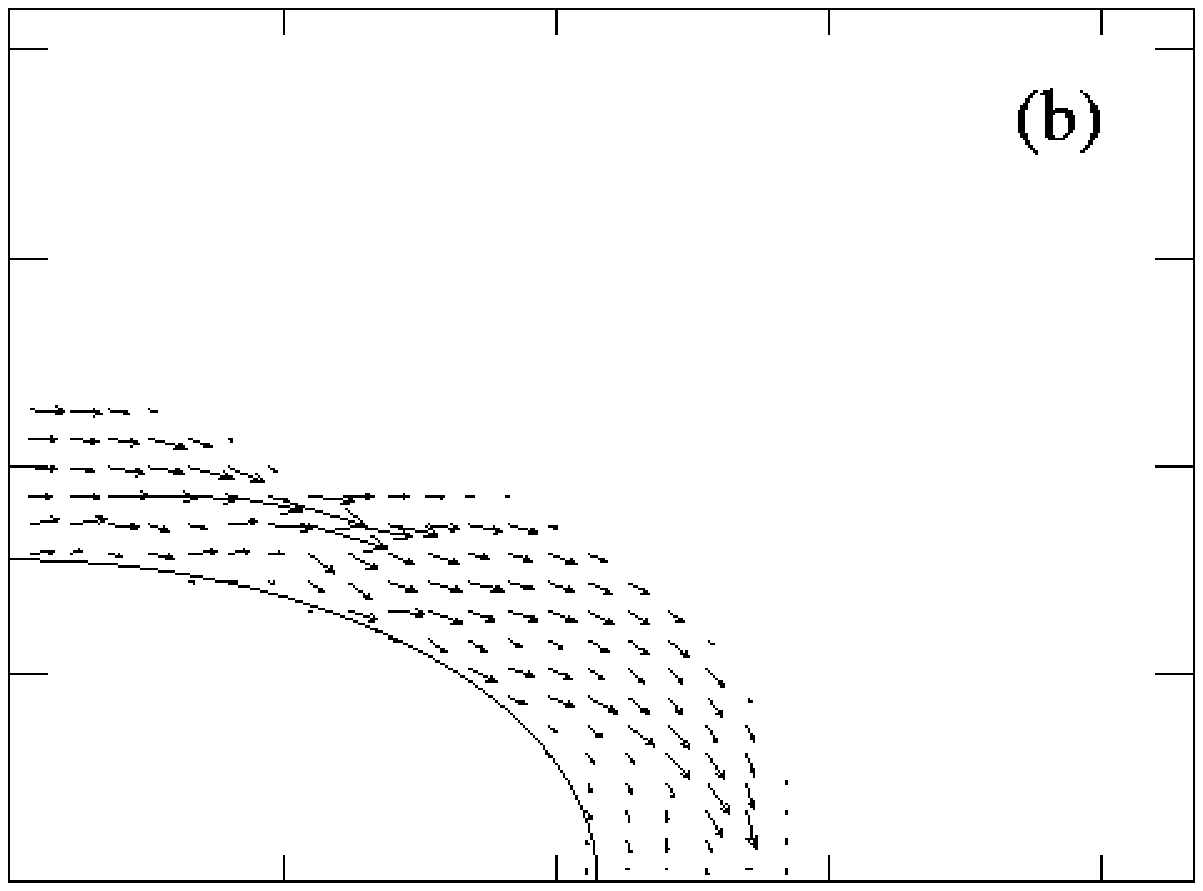}
    \\[1mm]
    \includegraphics[scale=0.53] 
{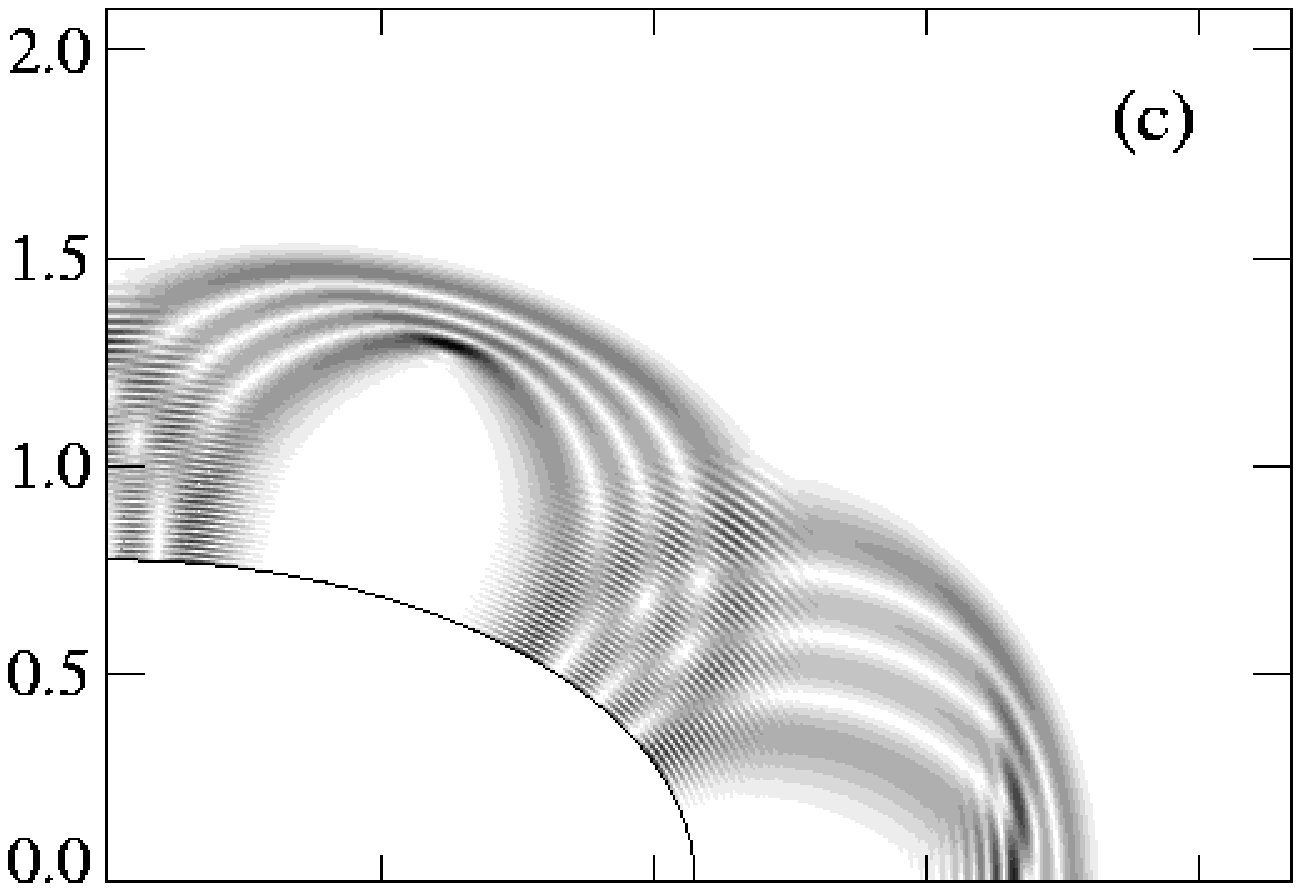}
    \includegraphics[scale=0.53] 
{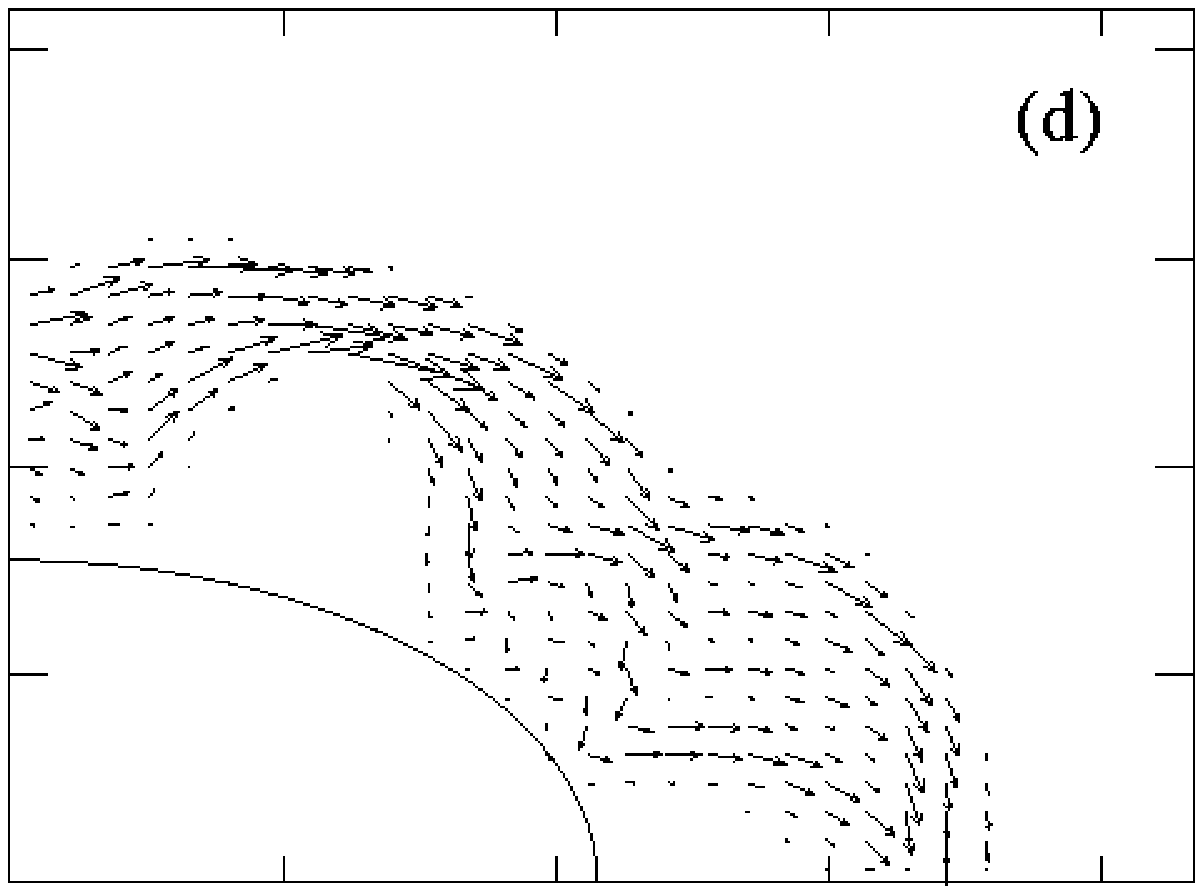}
    \\[1mm]
    \includegraphics[scale=0.53] 
{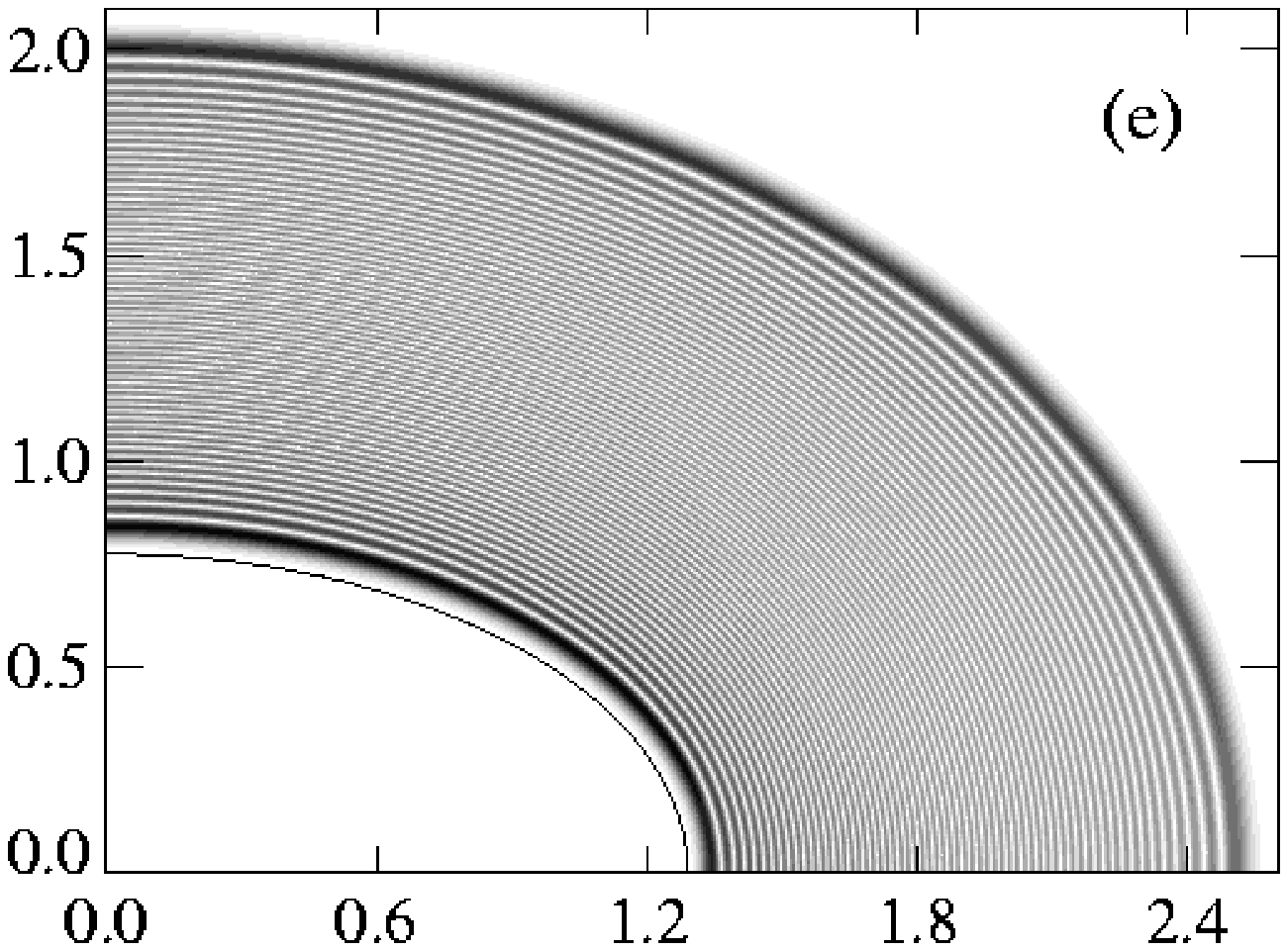}
    \includegraphics[scale=0.53] 
{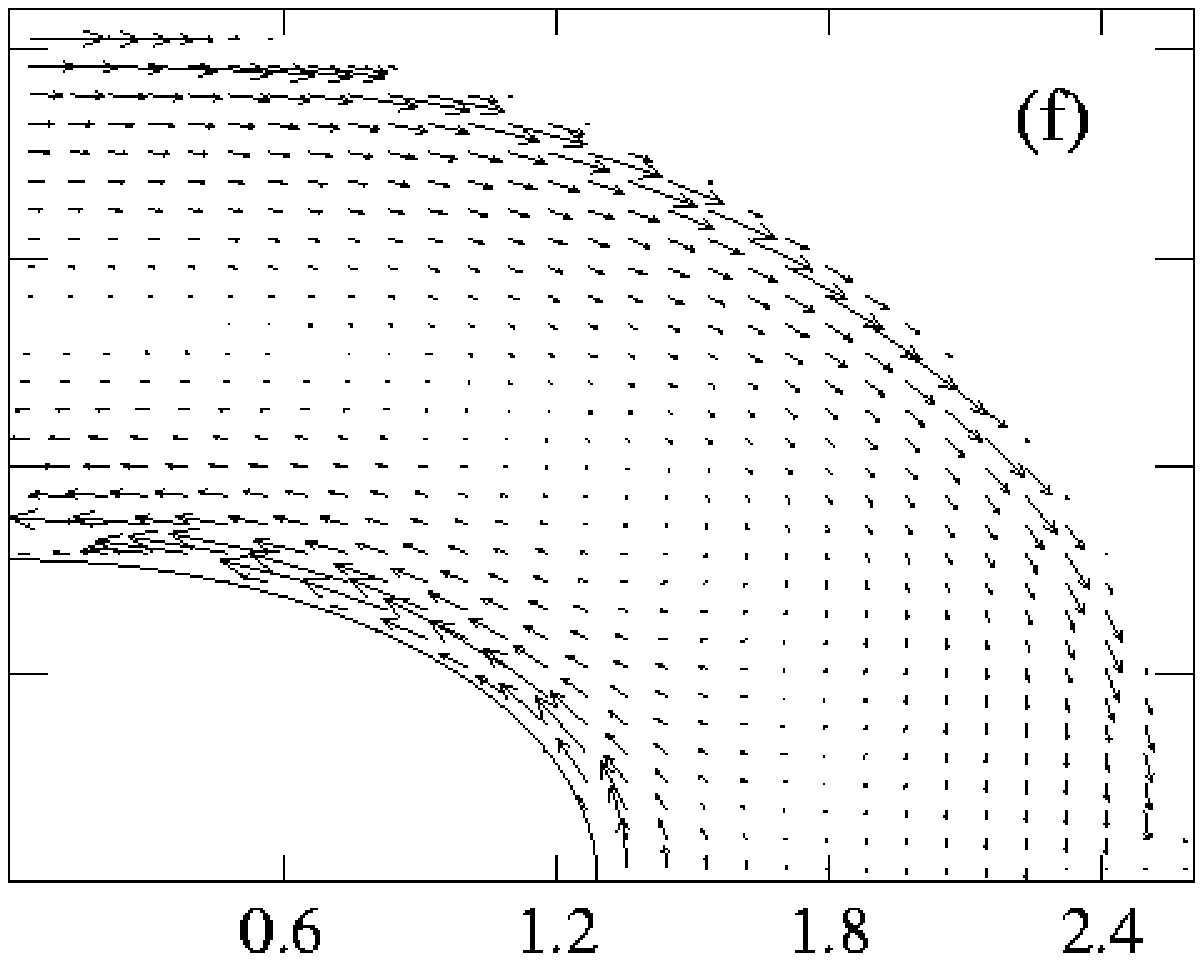}
    \figurecaption{%
      Exterior probability densities (left) and current density distributions 
     (right) at  $\rho=0.6$ and at similar energies as in
      Fig.~\ref{fig:ellipseint}. For each state only the first quarter
      of the picture is shown. 
     (a,b) A typical edge state, $\nu\simeq60.2087$.
     (c,d) An edge state localized along a stable periodic orbit, 
           $\nu\simeq60.2220$.
     (e,f) A typical bulk state, $\nu\simeq60.50016$. [Possible Moire
     patterns in (e) are an effect of low resolution printing.] [figure quality reduced]
      }%
    \label{fig:ellipseext}%
  \end{center}%
\end{figure}

\subsubsection*{The ellipse}

Next, we choose an elliptic boundary (of eccentricity $0.8$ and area
$\pi$) and even more semiclassical energies. The cyclotron radius is
taken to be $\rho=0.6$ which is small enough for complete cyclotron
orbits to fit into the interior domain of the billiard.  The classical
dynamics of the skipping motion is mixed chaotic in this case
\cite{RB85}, see Fig.~\ref{fig:pport} for a phase space portrait.
Going to the extreme semiclassical limit -- the ten-thousandth
interior eigenstate -- we expect the wave functions to mimic the
structures of the underlying classical phase space.
Indeed, Figure \ref{fig:ellipseint}(a) displays a wave function which
is localized along a stable interior periodic orbit.  This orbit has period 36
traveling six times around the billiard, with six reflections each
time, before it repeats itself.  In contrast, the state in
Fig.~\ref{fig:ellipseint}(c) is localized on a (mixed) chaotic part of
phase space. We note that the gross circulation of the current
density, Fig.~\ref{fig:ellipseint}(d), is opposite to that of the
first state,  Fig.~\ref{fig:ellipseint}(b).
Since $\rho$ is small enough for closed cyclotron orbits to fit into
the ellipse we find bulk states also in the interior, see
Fig.~\ref{fig:ellipseint}(e) for an example.  Again, it almost
vanishes at the billiard boundary and may be viewed semiclassically as
due to a superposition of closed cyclotron orbits.  This view is
supported by the distribution of the current density displayed in
Fig.~\ref{fig:ellipseint}(f).

Similar states are also found in the exterior, as displayed in Figure
\ref{fig:ellipseext}.  
To show more details we give only the
righthand-top quarter of the figure (the others follow by symmetry).
The first edge state, Fig.~\ref{fig:ellipseext}(a), corresponds to a
classical motion with creeps along the boundary.  It is the analogue
of a whispering-gallery mode.  Figure~\ref{fig:ellipseext}(c) displays
an exterior edge state which extends much further into the plane. Like
Fig.~\ref{fig:ellipseint}(a) it is clearly localized on a stable
skipping periodic orbit.  The bulk state Fig.~\ref{fig:ellipseext}(e)
is close to $\nu=60+\oh$. It consists of 61 concentric rings of
increased amplitude and shows no appreciable net current around the
billiard.

We emphasize that all the interior and exterior wave functions shown
above are calculated throughout the entire displayed area.  They turn
out to be \emph{numerically} zero in the complementary domains (ie,
either the exterior or the interior) as expected from theory.

\subsection{General boundary conditions}
\begin{figure}[p]%
  \begin{center}%
    \includegraphics[width=0.9\linewidth] 
    {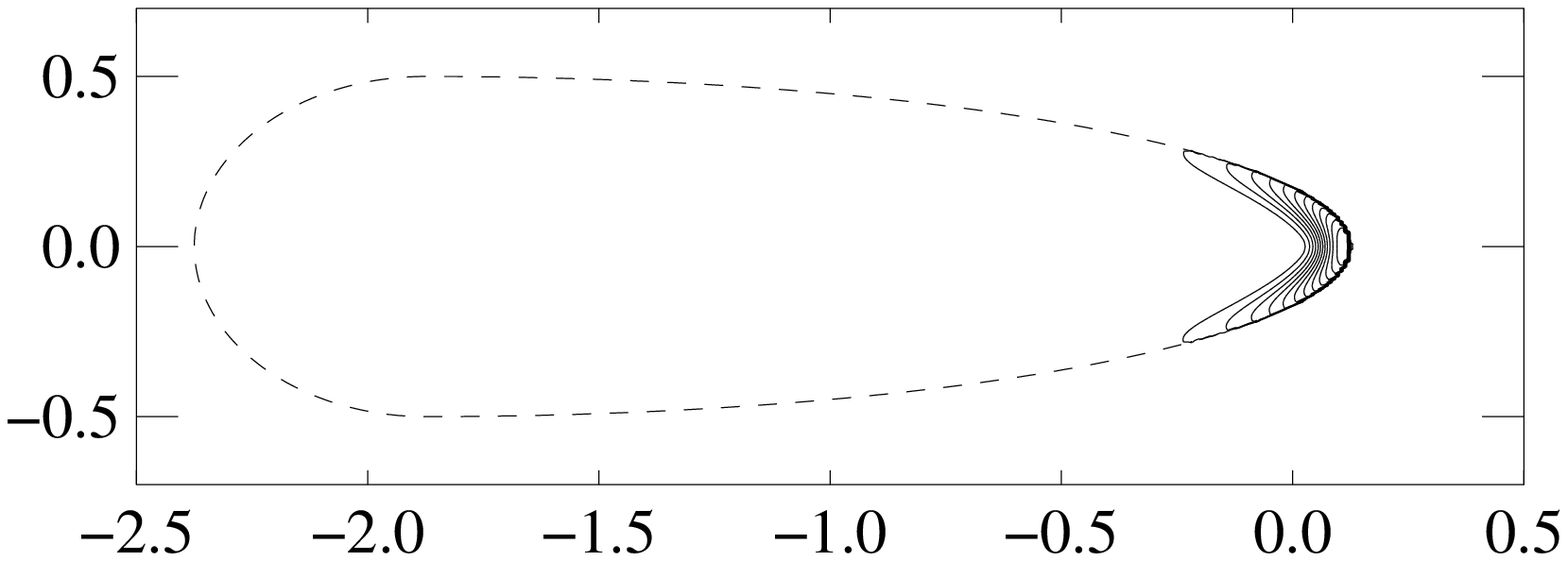}
    \figurecaption{%
      Contour plot of the ground state wave function (absolute value
      on a linear scale)
      for \emph{Neumann} boundary conditions and strong field,
      $b=0.05$, $\nu\simeq 0.2763$. The wave function is
      \emph{localized} at the boundary point of maximum curvature, as
      predicted by a recent theorem \cite{HM01}.  Here the billiard
      domain is given by the union of a half-circle and a half-ellipse
      (dashed line).}
\label{fig:helffer1}%
\end{center}%
\end{figure}
\begin{figure}[p]%
  \begin{center}%
    \psfrag{s}{$s$}
    \includegraphics[width=0.8\linewidth] 
    {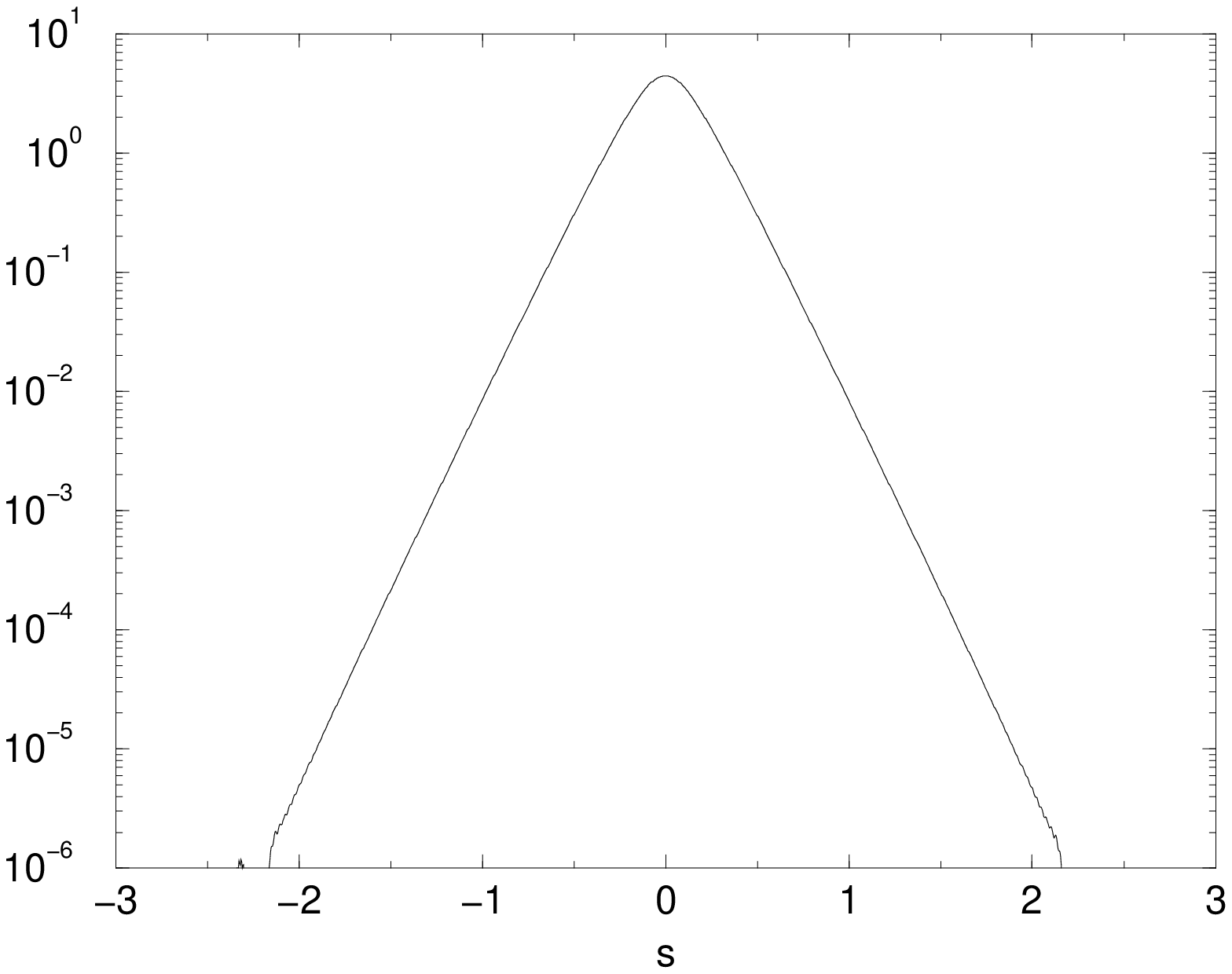}
    \figurecaption{%
      Boundary values of the ground state wave function of
      Fig.~\ref{fig:helffer1}. As predicted in \cite{HM01} it is
      localized  \emph{exponentially} at the point $s=0$ of maximum
      curvature. }
\label{fig:helffer2}%
\end{center}%
\end{figure}

\subsubsection*{The Neumann ground state }

So far, we only considered Dirichlet boundary conditions. They are the
natural choice from a physical point of view if one considers the
billiard boundary as due to an infinite wall potential.  On the other
hand, the Neumann boundary conditions, $\Lambda^{-1}=0$, are
frequently employed in spectral theory \cite{Helffer94}. They have the
advantage that the ground state energy lies below the first Landau
level, which facilitates its mathematical analysis.
Here, we are able to 
observe the manifestation of a recent theorem of spectral theory
\cite{HM01}. It states that the Neumann ground state of a magnetic
billiard is \emph{exponentially localized} around the boundary point
of maximum curvature.  In order to deal with a unique boundary point
of maximum curvature we choose the union of a half-circle and a
half-ellipse (with half-axes $\check{a}=2, \check{b}=0.5$) as
billiard boundary.  Choosing a magnetic length of $b=0.05$ (which
corresponds to a very strong field), we find the ground state energy
$\nu=0.2763$.  The Figures \ref{fig:helffer1} and \ref{fig:helffer2} display
the ground state wave function in the billiard and on the boundary,
respectively.  Indeed, one observes an exponential localization over
six orders of magnitude.

\begin{figure}[t]%
  \begin{center}%
    \includegraphics[width=\linewidth] 
    {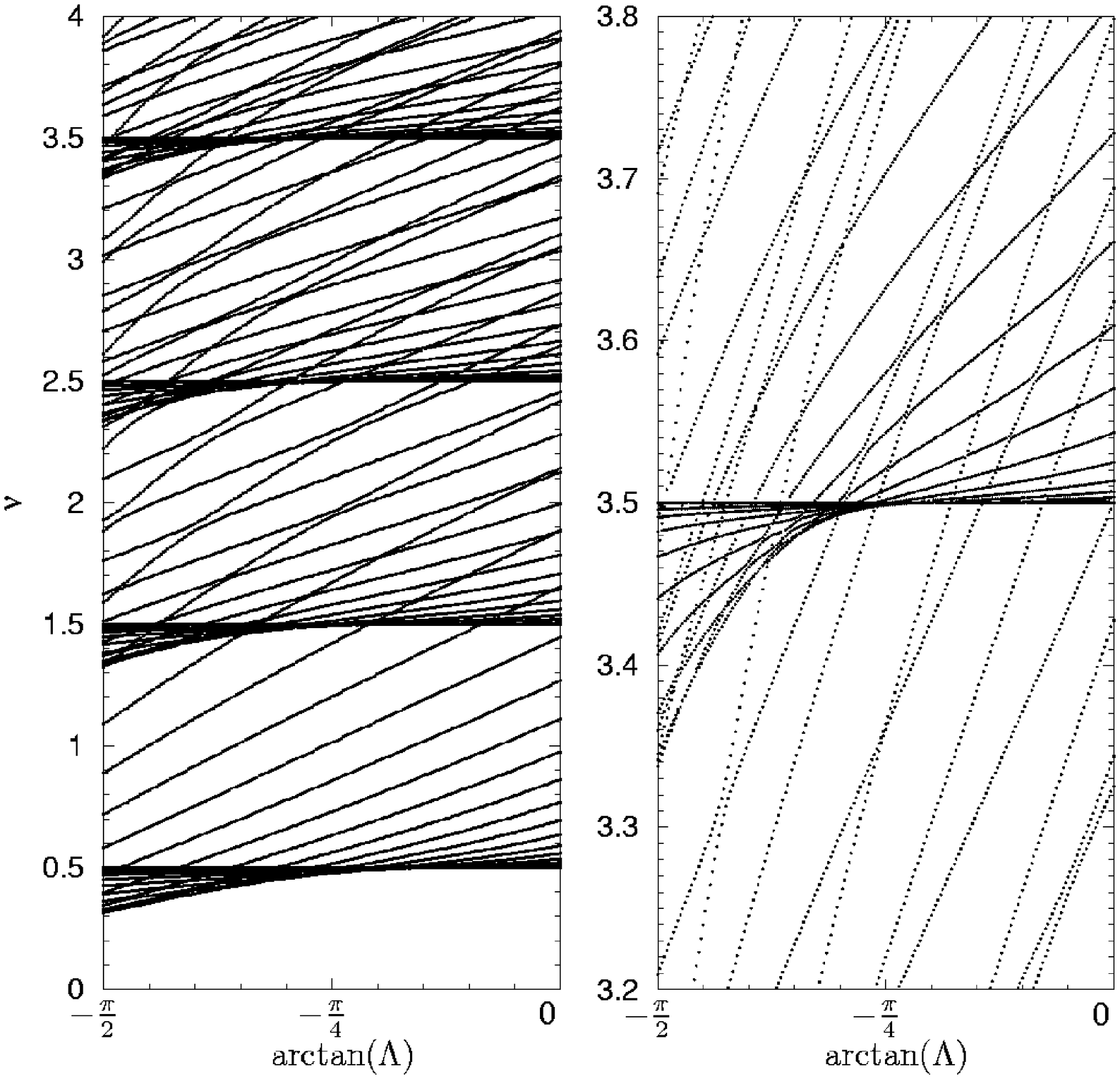}
    \figurecaption{%
      The parametric dependence of the \emph{exterior} spectrum on the
      boundary condition (for the asymmetric stadium, ie
      $\Len= 5.39724$, at fixed $b=0.25$).  The parameter $\Lambda$
      interpolates between Neumann ($\arctan\Lambda=-\piot$) and
      Dirichlet ($\arctan\Lambda=0$) boundary conditions.  The right
      graph shows details around the fourth Landau level. 
    [figure quality reduced]  }
\label{fig:lambdadyn}%
\end{center}%
\end{figure}

\subsubsection*{Parametric dependence on the mixing parameter }

As a last point, we show the parametric dependence of a spectrum on
the type of boundary conditions. Figure \ref{fig:lambdadyn} presents
the \emph{exterior} spectrum of the asymmetric stadium
as a function of the scaled mixing parameter $\Lambda\in (-\infty,0]$,
cf \eref{eq:Lambdadef}. It is chosen negative to ensure that the
transformation from Neumann ($\Lambda=-\infty$) to Dirichlet
($\Lambda=0$) boundary conditions is continuous. For positive
$\Lambda$ this would not be the case, which is a restriction similar
to the one for the field free case \cite{SPSUS95}.  ( We use the the
$\arctan$ function in Fig.~\ref{fig:lambdadyn} to transform the
infinite range of $\Lambda$ into a bounded interval.)

One observes that all the energy levels  increase monotonically
as $\Lambda$ is increased.  The energies clustering around the Landau
levels $\nu=N+\tfrac{1}{2}, N\in\Nnull$ belong to bulk states.  One
observes that they are lifted from the Landau levels to higher
energies at Dirichlet boundary conditions, whereas in the Neumann case
they are always shifted to smaller energies.  A semiclassical theory
which describes the exponential approach of the bulk states to the
Landau levels and their behavior as a function of $\Lambda$ is given
in Appendix \ref{app:uniform} and Section \ref{sec:bulkasymp}.  We
shall come back to Fig.~\ref{fig:lambdadyn} not only there, but also
in Chapter \ref{chap:edge}, when we define the edge state density.

\section{Semiclassical Quantization}%
\label{chap:trace}
 
In Chapter \ref{chap:bim} the boundary integral equations were found
to yield an efficient method for obtaining the exact quantum spectrum
of magnetic billiards.  It will be shown in the sequel that the same
equations are as important for the semiclassical quantization: They
serve as the starting point for the derivation of the semiclassical
trace formulas.

\myparagraph{Periodic orbit formulas for magnetic billiards}

The celebrated trace formulas of Gutzwiller
\cite{Gutzwiller71,Gutzwiller67} and Berry \& Tabor \cite{BT76,BT77}
allow the semiclassical quantization of systems  in terms of their classical
motion.  They were derived assuming a continuous Hamiltonian flow.
The corresponding formulas for field free billiards are known to
exhibit additional phase factors which account for the billiard
boundary conditions.  In order to
show that the same holds for magnetic billiards we shall explain {how}
the corresponding trace formulas are obtained from the exact boundary
integral formalism.  To our knowledge no such derivation has been
published for magnetic billiards to date.  The natural approach is to
follow the lines of Balian and Bloch's treatment of field-free
billiards \cite{BB72,HS92}, in analogy to the surface-of-section
method \cite{Bogomolny92} and the scattering approach \cite{DS92} for
non-magnetic systems.  Those attempts {failed} so far for magnetic
billiards due to the appearance of an abundance of unphysical
``ghost'' orbits which could not be handled.  To resolve this problem
we take advantage of the analysis
performed in Chapter \ref{chap:bim}. There it was found that the
boundary integral equations include spurious solutions which belong
to a particular complementary problem.
We will show that the semiclassical spectral determinant can be
factorized, accordingly, into an interior and exterior part. Each of
them leads to a trace formula incorporating only the physical periodic
orbits in the appropriate domain.

Like in the field-free case
\cite{BB72,HS92,Bogomolny92,DS92,GP95a,THS97,HST99} the semiclassical
quantization will be based on the double layer boundary integral
equation.  Apart from the spurious solutions, the main complication
arising at finite magnetic field is the inherently asymmetric form of
the respective integral kernel.
Unlike the case of field-free billiards \cite{Smilansky95}, the latter
is not simply related to the semiclassically unitary map operator
derived from the generating function.

In Section~\ref{sec:sbops} we deduce the semiclassical approximants
to the boundary integral operators of Chapter \ref{chap:bim}. After
that, in Sect.~\ref{sec:pop}, special map operators are introduced
in order to transform the spectral function of the double-layer
boundary integral equation. As a result, the number counting function
is given in terms of the traces of powers of the map operators.
The traces are evaluated semiclassically in Section~\ref{sec:trhyp}
assuming \emph{hyperbolic} 
skipping motion. We show why
only classically allowed periodic orbits contribute and how
their stability properties enter. The section concludes with the trace
formula for the density of states and the magnetization density.
In Section~\ref{sec:trdisk} the traces are evaluated assuming
\emph{integrable} dynamics. As a result we obtain the explicit
periodic orbit formula for the spectral density of states in the
magnetic disk billiard.  Section ~\ref{sec:disksep} gives the
corresponding WKB solution.

\subsection{The semiclassical  boundary integral  operators}
\label{sec:sbops}

In Section \ref{sec:bops} the boundary integral operators were defined
in terms of the free Green function and derivatives thereof.  To
obtain the semiclassical approximations of the operators one simply
replaces the Green function by its approximant. The latter is the leading order
asymptotic expression  in the semiclassically small
parameter $\nu^{-1}$, which was derived in Sect.~\ref{sec:Gsc}.  To
remain at a consistent level of approximation, the derivatives
appearing in the single-layer Neumann and the double-layer operators
\eref{eq:Opdef2} -- \eref{eq:Opdef4} are to be evaluated to the same
leading order.  This means in practice, that only the phase of the
Green function \eref{eq:Gsctsl} must be differentiated.
Accordingly, in the remainder of this report all equalities involving
semiclassical quantities are understood to be semiclassical in the
sense that they hold to leading order in $\nu^{-1}$.

In order to obtain expressions which have a semiclassically intuitive
and useful form it will be important to use the representation
\eref{eq:Gsctsl} 
which contains the actions of the short and long arcs
separately.
We found the geometric parts of the corresponding scaled actions
\eref{eq:deftsl} to be given by
\begin{align}
  \label{eq:tdef} 
  \ga_{\rm S}(\rvec;\rvec_0)
  &= \frac{1}{\pi}  
  \bigg(\arcsin\left(\frac{|\rvec-\rvec_0|}{2\rho}\right)
    +\frac{|\rvec-\rvec_0|}{2\rho}
    \sqrt{1-\left(\frac{\rvec-\rvec_0}{2\rho}\right)^2}
    - \frac{\rvec\times\rvec_0}{2\rho^2}
  \bigg)
  \intertext{and}
  \label{eq:trel}
  \ga_{\rm L}(\rvec;\rvec_0)
  &= 
  1- \ga_{\rm S}(\rvec_0;\rvec)
  \PO
\end{align}
As a first step, we note their gradients with respect to the initial and
the final points.
\begin{align}
  \label{eq:grad1}
  \grad_{\!\rvec_0}\, \ga_\SL(\rvec;\rvec_0)
  &=
  \frac{1}{\pi\rho} \left(
    \mp\frac{\rmrn}{|\rmrn|}
    \sqrt{1-\left(\frac{\rvec-\rvec_0}{2\rho}  \right)^2}
    -\frac{1}{2\rho}
    {-y\choose x}
    \right)
    \\
  \label{eq:grad2}
  \grad_{\!\rvec}\, \ga_\SL(\rvec;\rvec_0)
  &=
  \frac{1}{\pi\rho} \left(
    \pm\frac{\rmrn}{|\rmrn|}
    \sqrt{1-\left(\frac{\rvec-\rvec_0}{2\rho}  \right)^2}
    +\frac{1}{2\rho}
    {-y_0\choose x_0}
    \right)
\end{align}
Here, the upper and lower signs of the first summands stand for the
short arc and long arc contribution, respectively.
It will be  useful to state the distance
between the initial and the final point in terms of the positive angle
\begin{gather}
  \label{eq:alphadef}
  \alpha(\rvec;\rvec_0) \defas 
  \arcsin\left(\frac{|\rvec-\rvec_0|}{2\rho}\right)
  \PO
\end{gather}
In addition, the direction of the normal vectors at the initial and the final
points are measured by their (signed) angles with respect to the
distance vector connecting the two points.
\begin{align}
  \label{eq:betadef}
  \beta(\rvec;\rvec_0) \defas \Angle(\nvec;\rvec-\rvec_0)
  \qq
  \beta^{0}(\rvec;\rvec_0) \defas \Angle(\nvec_0;\rvec-\rvec_0)
\end{align}
\begin{figure}[tbp]%
  \begin{center}%
    \includegraphics[width=\linewidth] {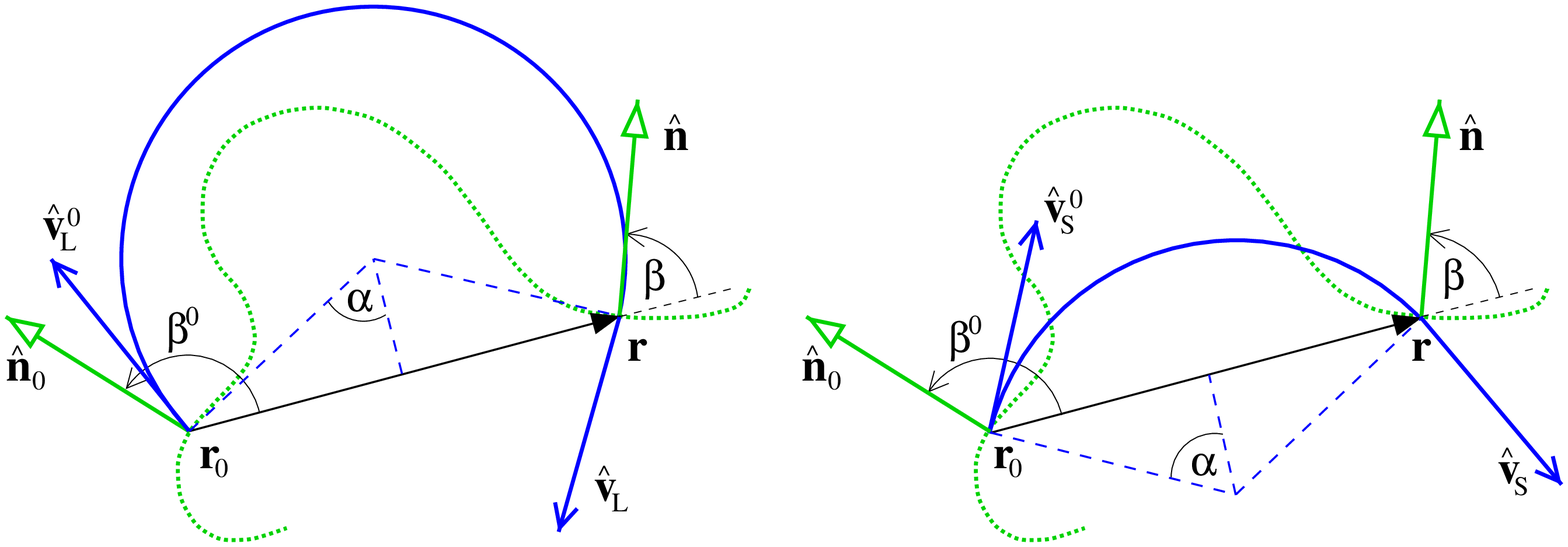}%
    \figurecaption{The angles $\alpha$, $\beta^0$, and $\beta$ are
      defined with respect the vector $\rvec-\rvec_0$ connecting the
      initial and the final point. They measure its length and the
      relative direction of the boundary normals, respectively. These
      quantities do not depend on the type of the arc (left: long,
      right: short), unlike the relative direction of the initial and
      the final velocities $\vvech^0$ and $\vvech$. The latter may be
      expressed in terms of $\alpha$, $\beta^0$, and $\beta$, cf
      \eref{eq:vn}. (The dotted line indicates the billiard boundary.)}%
    \label{fig:alphabeta}%
  \end{center}%
\end{figure}%
Now consider the classical arcs connecting the initial and the final
points.  They define the direction of the classical velocities at the
points of reflection and incidence (see Figure
\ref{fig:alphabeta} for  a sketch of the situation). 
The normal components are given by
\begin{alignat}{2}
  \label{eq:vn}
  \vvech_{\rm S}\,\nvec
  &=\cos(\beta+\alpha)
  &\qq
  \vvech^0_{\rm S}\,\nvec_0
  &=\cos(\beta^0-\alpha)
  \nnn
  \vvech_{\rm L}\,\nvec
  &=-\cos(\beta-\alpha)
  &\qq
  \vvech^0_{\rm L}\,\nvec_0
  &=-\cos(\beta^0+\alpha)
\end{alignat}
for short and long arcs, respectively.
Here, the velocity at the \emph{initial} point of the arc is denoted with a
\emph{zero} superscript, and the hats indicate that the velocity vectors are
normalized.

\myparagraph{The semiclassical Dirichlet operators}

We proceed to calculate the semiclassical approximation
to the kernel of the double layer Dirichlet operator \eref{eq:Opdef3}
by inserting \eref{eq:Gsctsl}.
For the short arc term one has to evaluate the gauge
invariant derivative

\begin{align}
  \label{eq:dnnbS}
  \dnnb[2\pi\rmi\nu \ga_{\rm S}-\chit_0]+\rmi\At_{n_0}
  &=
  2\rmi\sqrt{\nu}\left(
  -\frac{(\rmrn)\,\nvec_0}{|\rmrn|}\cos(\alpha)
  -\frac{(\rmrn)\times\nvec_0}{2\rho}\right)
  \nnn
  &=  -2\rmi\sqrt{\nu}\, \cos(\beta^0-\alpha)   =  
  -2\rmi\sqrt{\nu}\, (\vvech^0_{\rm S}\,\nvec_0)
  \CO
\end{align}
where we used eqs \eref{eq:grad1} and \eref{eq:vn}.  Apart from the
sign, it is given by the normal component of the classical velocity
after reflection since $2\sqrt{\nu}$ is the magnitude of the scaled
velocity. Note that $(\vvech^0_{\rm S}\,\nvec_0)$ is a non-symmetric
function of the initial and the final point and depends on the energy
through $\alpha$.
For the long arc term one obtains  the analogous expression
\begin{align}
  \label{eq:dnnbL}
  \dnnb[2\pi\rmi\nu \ga_{\rm L}-\chit_0]+\rmi\At_{n_0}
  &=  +2\rmi\sqrt{\nu}\, \cos(\beta^0+\alpha)   =  
  -2\rmi\sqrt{\nu}\, (\vvech^0_{\rm L}\,\nvec_0)
  \PO
\end{align}
It follows that the semiclassical approximation of the double-layer
Dirichlet kernel \eref{eq:QexpdlD} can be stated in a particularly
simple form,
\begin{multline}
  \label{eq:bopsc2}
  {\rm q}^{\rm D(sc)}_{\rm dl}(\rvec;\rvec_0)\defas
  \dnnb \Gsc_\nu +\rmi \At_{n_0}\Gsc_\nu 
  =   \frac{1}{2(1+\rme^{2\pi\rmi\nu})} \,
  \frac{1}{(2\pi\rmi)^\oh} \,\rme^{\rmi(\chit-\chit_0)}
  \\
  \times
  \left\{
    \frac{-\vvech^0_{\rm S}\,\nvec_0}{(\sin\alpha\cos\alpha)^\oh}
    \,\rme^{2\pi\rmi\nu \ga_{\rm S}}
    +
    \rme^{-\rmi\piot}
    \,\frac{-\vvech^0_{\rm L}\,\nvec_0}{(\sin\alpha\cos\alpha)^\oh}
    \,\rme^{2\pi\rmi\nu \ga_{\rm L}}
  \right\}
\PO
\end{multline}
It will be an important ingredient in the derivation of the trace
formulas.\footnote{%
  The semiclassical operators derived in Sect. \ref{sec:sbops} allow
  the computation of spectra within a ``semiquantum'' approximation by means
  of the boundary integral method of Chapter \ref{chap:bim}.  One
  merely  replaces the exact kernels \eref{eq:QexpslD} --
  \eref{eq:QexpdlN} by their approximants \eref{eq:Gsccos},
  \eref{eq:bopsc2}, \eref{eq:bopsc3}, \eref{eq:bopsc4} and calculates
  the respective determinants numerically without further
  approximation.  As an advantage of this scheme it applies
  irrespective of the type of classical motion (including mixed
  chaotic dynamics). However, it may \emph{not} be regarded as a
  proper semiclassical quantization since the degree of approximation
  is not consistent throughout the calculation. Also 
  the Fredholm determinant must be evaluated to leading order in $\nu$, see
   Sections \ref{sec:trhyp} and \ref{sec:trdisk}.
  }%
For completeness we note that the semiclassical single-layer
Dirichlet kernel
is simply given by the semiclassical Green function itself, $ {\rm
  q}^{\rm D(sc)}_{\rm sl}(\rvec;\rvec_0)\defas \Gsc_\nu$, as an
immediate consequence of \eref{eq:Opdef}.

\myparagraph{The semiclassical Neumann operators}

The kernels of the single- and double layer Neumann operators, eqs
\eref{eq:Opdef2} and \eref{eq:Opdef4}, involve gauge
invariant gradients with respect to the final point of the Green function. 
One finds
\begin{align}
  \label{eq:dnb}
  \dnb[2\pi\rmi\nu \ga_{\rm S}+\chit]-\rmi\At_{n}
  &=  +2\rmi\sqrt{\nu}\, \cos(\beta+\alpha)   =  
  +2\rmi\sqrt{\nu}\, (\vvech_{\rm S}\,\nvec)
  \CO
  \\
  \dnb[2\pi\rmi\nu \ga_{\rm L}+\chit]-\rmi\At_{n}
  &=  -2\rmi\sqrt{\nu}\, \cos(\beta-\alpha)   =  
  +2\rmi\sqrt{\nu}\, (\vvech_{\rm L}\,\nvec)
  \CO
\end{align}
similar to eqs \eref{eq:dnnbS} and \eref{eq:dnnbL}.  This way
the semiclassical single-layer Neumann kernel \eref{eq:QexpslN}
assumes the form 
\begin{multline}
  \label{eq:bopsc3}
  {\rm q}^{\rm N(sc)}_{\rm sl}(\rvec;\rvec_0)\defas
  \dnb \Gsc_\nu -\rmi \At_{n} \Gsc_\nu = 
  \frac{1}{2(1+\rme^{2\pi\rmi\nu})} \,
  \frac{1}{(2\pi\rmi)^\oh} \,\rme^{\rmi(\chit-\chit_0)}
  \\
  \times
  \left\{
    \frac{+\vvech_{\rm S}\,\nvec}{(\sin\alpha\cos\alpha)^\oh}
    \,\rme^{2\pi\rmi\nu \ga_{\rm S}}
    +
    \rme^{-\rmi\piot}
    \,\frac{+\vvech_{\rm L}\,\nvec}{(\sin\alpha\cos\alpha)^\oh}
    \,\rme^{2\pi\rmi\nu \ga_{\rm L}}
  \right\}
\end{multline}
It is worth noting how the mutual adjointness of the operators
\eref{eq:QexpslN} and \eref{eq:QexpdlD} shows up in the semiclassical
case. By permuting $\rvec$ and $\rvec_0$ the prefactors of the 
short and long arc terms change their roles,
\begin{align}
\label{eq:vnadj}
  (\vvech_{\rm S}\,\nvec)
  \equiv
  \vvech_{\rm S}(\rvec;\rvec_0)\,\nvec(\rvec)
  =
  \vvech^0_{\rm L}(\rvec_0;\rvec)\,\nvec(\rvec_0)
  \equiv
  (\vvech^0_{\rm L}\,\nvec_0)^\dagger
  \CO
\end{align}
and likewise $ (\vvech_{\rm L}\,\nvec)=(\vvech^0_{\rm
  S}\,\nvec_0)^\dagger$.  As for the phases, it is the factor
$(1+\rme^{2\pi\rmi\nu})^{-1}$ whose conjugation provides the term
$\rme^{2\pi\rmi\nu}$  needed in conjunction with the relation
\eref{eq:trel} to prove the mutual adjointness.

The kernel of the semiclassical double-layer Neumann operator follows
from applying the gauge invariant derivative \eref{eq:dnb} to the
single-layer Dirichlet expression \eref{eq:bopsc2}, cf
\eref{eq:Opdef4}. One obtains
\begin{multline}
  \label{eq:bopsc4}
  \begin{aligned}
  \,&{\rm q}^{\rm N(sc)}_{\rm dl}(\rvec;\rvec_0)\defas\,
  (\dnnb+\rmi \At_{n_0})   (\dnb \Gsc_\nu-\rmi \At_{n} \Gsc_\nu)\,
  \\
  \,&=\,
  \frac{1}{2(1+\rme^{2\pi\rmi\nu})} \,
  \frac{2\rmi\sqrt{\nu}}{(2\pi\rmi)^\oh}  \,\rme^{\rmi(\chit-\chit_0)}
  \end{aligned}
  \\
  \times
  \left\{
    \frac{-(\vvech^0_{\rm S}\,\nvec_0)(\vvech_{\rm S}\,\nvec)}
    {(\sin\alpha\cos\alpha)^\oh}
    \,\rme^{2\pi\rmi\nu \ga_{\rm S}}
    +
    \rme^{-\rmi\piot}
    \,\frac{-(\vvech^0_{\rm L}\,\nvec_0)(\vvech_{\rm L}\,\nvec)}
    {(\sin\alpha\cos\alpha)^\oh}
    \,\rme^{2\pi\rmi\nu \ga_{\rm L}}
  \right\}
  \PO
\end{multline}
Like the exact kernel \eref{eq:QexpdlN} this semiclassical version is
self-adjoint. This follows again from the observation that the two
summands simply change roles when the adjoint operator is formed.

\subsection{From  boundary to  map operators}
\label{sec:pop}

Let us now consider the semiclassical double-layer equation for
Dirichlet boundary conditions in more detail. As known from Chapter
\ref{chap:bim}  the corresponding Fredholm determinant
\eref{eq:detdl}
is a spectral function.  Its roots yield the Dirichlet spectrum of the
domain considered, conjoint with the Neumann spectrum of the
complementary domain.
    
The semiclassical spectral function is obtained by substituting
\eref{eq:bopsc2} into \eref{eq:detdl} with $\lambda=0$:
\begin{align}
  \label{eq:scdet}
\xi^{\rm (sc)}(\nu) \defas&
    \det\left[\oh\pm {\sf Q}_{\rm dl}^{\rm D(sc)}\right]
\end{align}
Here, the upper or lower sign indicates whether the originally
considered domain is of the interior or the exterior type (like in
Chapter \ref{chap:bim}).

\myparagraph{The map operators}

By defining the operator
\begin{gather}
  \label{eq:defP}
  {\sf P}
  \defas   2 (1+\rme^{2\pi\rmi\nu})\, {\sf Q^{\rm D\,(sc)}_{\rm dl}}
  \PO
\end{gather}
we factorize the determinant \ref{eq:scdet} into two parts,
\begin{gather}
  \label{eq:scdeta}
\xi^{\rm (sc)}(\nu)=
  \det\left[ \frac{1}{2(1+\rme^{2\pi\rmi\nu})} \right] \, \det\left[
    1+\rme^{2\pi\rmi\nu}\pm {\sf P}
  \right]
\PO
\end{gather}
This reflects the partitioning of the underlying classical phase space
into cyclotron orbits, which are detached from the boundary, and
skipping trajectories, see Sect.  \ref{sec:restr}.  The first
determinant in \eref{eq:scdeta} does not depend on the boundary.  Its
operator is diagonal and {singular} at the energies $\nu_N=N+\oh$,
$N\in\Nnull$, of the Landau levels.  Apparently, it represents the
semiclassical contribution of the bulk states to the spectrum
and its divergence at the Landau energies is due to the infinite
number of degenerate bulk states found in the exterior.
(The exponentially small lifting of the degeneracy observed in the
exact spectrum is not seen here since the semiclassical Green function
\eref{eq:Gsccos} does not describe tunneling effects.)

As will become clear in the following, the second factor in
\eref{eq:scdeta} yields the contribution of the \emph{skipping}
trajectories to the spectral function. It is described by the
\emph{map} operator $ {\sf P}$ defined in \eref{eq:defP}.  We will see
that it can be related to the classical billiard map
\eref{eq:Birmap} describing the motion of skipping trajectories.
The map operator consists of a short arc and a long arc
term and it is advantageous to split it accordingly,
\begin{align}
  \label{eq:Psplit}
  {\sf P} = {\sf P}_{\rm S}+{\sf P}_{\rm L}  
  \CO
\end{align}
with the corresponding integral kernels  given by 
\begin{align}
  \label{eq:psk}
  {\rm p}_{\rm S}(s,s_0)
  &\defas
  \frac{1}{(2\pi\rmi)^\oh}\,
  \frac{-\vvech^0_{\rm S}\,\nvec_0}{(\sin(\alpha)\cos(\alpha))^\oh}\,
  \,
  \rme^{2\pi\rmi\nu\ga_{\rm S}}
  \,
  \rme^{\rmi\chit-\rmi\chit_0}
\intertext{and}
  \label{eq:plk}
  {\rm p}_{\rm L}(s,s_0)
  &\defas
  \frac{1}{(2\pi\rmi)^\oh}\,
  \frac{-\vvech^0_{\rm L}\,\nvec_0}{(\sin(\alpha)\cos(\alpha))^\oh}\,
  \,
  \rme^{-\rmi\piot}\,
  \rme^{2\pi\rmi\nu\ga_{\rm L}}
  \,
  \rme^{\rmi\chit-\rmi\chit_0}
  \CO
\end{align}
see \eref{eq:bopsc2}, with $\rvec=\rvec(s), \rvec_0=\rvec(s_0)$. 

We note that the operator {\sf P} differs from the standard map operator
defined in terms of the generating function $\GF$ of the classical map
\cite{Smilansky95},
\begin{align}
  {\sf S} = 
  -
  \frac{1}{(2\pi\rmi)^\oh}
  \left(
    \frac{\partial^2\GF}{\partial s\partial s_0}
  \right)^\oh
  \rme^{\rmi\GF}
  \CO
\end{align}
which is the analogue of Bogomolny's transfer operator ${\sf T}$
\cite{Bogomolny92}.  The main difference is not the fact that ${\sf
  P}$ consists of two distinct parts.  Rather, it is the inherent
asymmetry in the coordinates $s,s_0$ which introduces difficulties not
encountered in the field-free treatment: The prefactors cannot in
general be stated as mixed derivatives of the relevant phase.
Moreover, the parts ${\sf P}_{\rm\!S/L}$ are not semiclassically
unitary (but satisfy equation \eref{eq:Prel}).

From now on we focus on the non-singular factor in
\eref{eq:scdet}. It gives the spectral function
\begin{align}
\label{eq:skipdet}
\xi^{\rm (sc)}_{\rm skip}(\nu) =&
  \det\left[
    1+\rme^{2\pi\rmi\nu}\pm {\sf P}
  \right]
\end{align}
of the states which correspond to skipping motion.
In the standard procedure to obtain a trace formula one would now
compute the imaginary part of the logarithm of \eref{eq:skipdet}.
Making use of the identity $\log\det=\tr\log$ \cite{MM92} 
one would like to evaluate the trace of powers of the
operator \eref{eq:Psplit} in stationary phase approximation.
However, unlike the case of field-free billiards the corresponding
saddle point condition selects classical periodic orbits in the
interior \emph{and} in the exterior. This 
is not surprising given the fact that the double-layer equation
includes solutions of the complementary domain.  To make matters
worse, an abundance of saddle-point configurations arises which do not
have a physical meaning at all.  In order to avoid these severe
difficulties it is vital to write the spectral function
\eref{eq:skipdet} as a product such that each factor yields the
spectrum in either the interior or the exterior domain.

\myparagraph{Factorizing the spectral function}
To facilitate the factorization of the determinant \eref{eq:skipdet}
we split the short and long arc operators once more,
${\sf P}_{\rm S}={\sf P}_{\rm S}^{\rm int}-  {\sf P}_{\rm S}^{\rm  ext}$
and
${\sf P}_{\rm L}={\sf P}_{\rm L}^{\rm int}-  {\sf P}_{\rm L}^{\rm  ext}$.
Ultimately, the parts labeled by ``int''and ``ext'' should exclusively
account for the motion in the interior and in the exterior,
respectively.  To that end, the splitting is defined by the signs of
the prefactors of the integral kernels which are functions of the
initial and the final points.
\begin{alignat}{2}
  \label{eq:pkern}
  {\rm p}_{\rm S}^{\rm int}(s,s_0) 
  &\defas
  \Theta(-\nvec_0\vvech^0_{\rm S})\, {\rm p}_{\rm S}(s,s_0) 
  &\qq                                              
  {\rm p}_{\rm L}^{\rm int}(s,s_0)                  
  &\defas                                           
  \Theta(-\nvec_0\vvech^0_{\rm L})\, {\rm p}_{\rm L}(s,s_0) 
  \nnn
  {\rm p}_{\rm S}^{\rm ext}(s,s_0)                  
  &\defas                                           
  -\Theta(\nvec_0\vvech^0_{\rm S})\, {\rm p}_{\rm S}(s,s_0) 
  &\qq                                              
  {\rm p}_{\rm L}^{\rm ext}(s,s_0)                  
  &\defas                                           
  -\Theta(\nvec_0\vvech^0_{\rm L})\, {\rm p}_{\rm L}(s,s_0) 
  \nnn
\end{alignat}
(The minus sign in front of the exterior kernels is introduced for
convenience.)  According to these definitions the
``interior'' part of the operators vanishes whenever the initial and
the final points have positions such that the corresponding classical
arc points into the exterior domain, and vice versa.
This crucial property is embodied in the operator  equations
\begin{align}
  \label{eq:Prel}
   {\sf P}_{\rm S}^{\rm int}\,  {\sf P}_{\rm L}^{\rm ext}
   + {\sf P}_{\rm L}^{\rm int}\,  {\sf P}_{\rm S}^{\rm ext}
   &= - \,\rme^{2\pi\rmi\nu}\, {\sf id}
\intertext{and}
  \label{eq:Prel2}
   {\sf P}_{\rm S}^{\rm int}\,  {\sf P}_{\rm S}^{\rm ext}
   + {\sf P}_{\rm L}^{\rm int}\,  {\sf P}_{\rm L}^{\rm ext}
   &=0
   \CO
\end{align}
derived in Appendix \ref{app:prod}.  
With their help it
follows immediately that the determinant in eq \eref{eq:skipdet}
factorizes into an interior and an exterior part,
\begin{align}
  \label{eq:fac}
\xi^{\rm (sc)}_{\rm skip}(\nu) &=  
  \det\left[
    1+\rme^{2\pi\rmi\nu}\pm 
    \big(
  {\sf P}_{\rm S}^{\rm int}
  -  {\sf P}_{\rm S}^{\rm ext}
  +{\sf P}_{\rm L}^{\rm int}
  -  {\sf P}_{\rm L}^{\rm ext}
    \big)
  \right]
  \nnn
  &= 
  \det\left[
    \Big(
    1\pm \big({\sf P}_{\rm S}^{\rm int}+{\sf P}_{\rm L}^{\rm int}\big)
    \Big)
    \Big(
    1\mp \big({\sf P}_{\rm S}^{\rm ext}+{\sf P}_{\rm L}^{\rm ext}\big)
    \Big)
  \right]
  \nnn
  &= 
  \det\left[
    1\pm {\sf P}^{\rm int}
  \right]
  \,
  \det\left[
    1\mp{\sf P}^{\rm ext}
  \right]
\PO
\end{align}
Here we merely replaced the term $\rme^{2\pi\rmi\nu}$ 
by the operators \eref{eq:Prel} and included \eref{eq:Prel2}.  In the
last equality we introduced the interior and exterior map operators
${\sf P}^{\rm int}\defas{\sf P}_{\rm S}^{\rm int}+{\sf P}_{\rm L}^{\rm
  int}$ and ${\sf P}^{\rm ext}\defas{\sf P}_{\rm S}^{\rm ext}+{\sf
  P}_{\rm L}^{\rm ext}$, respectively.\footnote{
  We remind the reader that the signs $\pm$ in \eref{eq:fac} choose
  the domain of interest and stem from the boundary integral
  equations. The labels ``${\rm int/ext}$'' refer only to the type of
  the classical arcs which are included in the definition of the
  corresponding operators in \eref{eq:pkern}.}

The factorization of the spectral function \eref{eq:fac} 
into an interior and exterior part is in accordance with the
observation that the double layer equation provides the spectra of
both the interior and the exterior problems, see the discussion in
Sect.~\ref{ssec:spurious}.  In the following section we shall show
that using the upper sign, ie, starting originally with the interior
problem, one gets the trace formula for the interior Dirichlet
spectrum from the first factor and the exterior Neumann spectrum from
the second one.  In the same way, if we use the lower sign we get the
trace formula for the exterior Dirichlet   spectrum from the second  factor
and the interior Neumann spectrum from the first.

We finally note that the operators ${\sf P}^{\rm int}$ and ${\sf
  P}^{\rm ext}$ are \emph{semiclassically} unitary, as shown in
Appendix \ref{app:prod}.  
Moreover, the interior and exterior map operators
obey the relation
\begin{align}
\label{eq:durel}
  {\sf P}^{\rm int}{\sf P}^{\rm ext}=-\rme^{2\pi\rmi\nu} {\sf id}
\CO
\end{align}
which follows from \eref{eq:Prel} and \eref{eq:Prel2}.  
It has an intuitive form: Propagating a boundary state first in
the exterior and then in the interior
one arrives again at the \emph{same} state, 
augmented by the
global phase $2\pi\nu$ of
a complete cyclotron orbit (plus the Maslov correction $\pi$). It is
the semiclassical manifestation of the classical interior-exterior
duality.

\subsection{Trace formula for hyperbolic billiards}
\label{sec:trhyp}

To obtain the fluctuating part of the number counting function of the
skipping spectrum one has to take the imaginary part of the logarithm
of the spectral function \eref{eq:fac}.  We start by computing the
contribution of the first factor in \eref{eq:fac},
\begin{align}
\label{eq:firstNskipint}
\N^{\rm skip(int)}_{\rm osc}(\nu)=& 
  -  \frac{1}{\pi}\Im\log\det
  \left[
    1\pm {\sf P}^{\rm int}
  \right]
  =  \frac{1}{\pi}\Im
  \sum_{n=1}^{\infty}
  \frac{(\mp)^n}{n} \,
  \tr\!\left[
    \big({\sf P}^{\rm int}\big)^n
    \right]
\end{align}
To obtain the periodic orbit formula we can now follow the lines of
the derivation of the trace formula for
field free billiards
\cite{BB72,Bogomolny92,HS92,Boasman94,GP95a,THS97,SPS97,Sieber98,HST99}.
The trace in \eref{eq:firstNskipint} amounts to an $n$-di\-men\-sio\-nal
integral of the form
\begin{align}
  \label{eq:trace1}
  &\tr\!
  \left[
    \big({\sf P}^{\rm int}\big)^n
  \right]
    =\frac{1}{(2\pi\rmi)^{n/2}}    
    \int    
    \frac{\rmd s_1 \ldots \rmd s_n}{b^n}\,
    \exp\Big({\rmi\sum_{j=1}^n(\Chit(s_{j+1})-\Chit(s_{j}))}\Big)
    \nnn
    &\begin{aligned}
      \times  
      \prod_{j=1}^n
      \Bigg[  &
      \frac{
        -(\vvech^0_{\rm S}\nvec_0)_j\,
        \Theta\big(-(\vvech^0_{\rm S}\nvec_0)_j\big)
        }
      {\big(\sin(\alpha_j)\cos(\alpha_j)\big)^\oh}
      \exp\Big({2\pi\rmi\nu\ga_{\rm S}(s_{j+1};s_{j})}\Big)
      \nnn
      &+
      \frac{
        -(\vvech^0_{\rm L}\nvec_0)_j\,
        \Theta\big(-(\vvech^0_{\rm L}\nvec_0)_j\big)
        }
      {\big(\sin(\alpha_j)\cos(\alpha_j)\big)^\oh}
      \exp\Big({2\pi\rmi\nu\ga_{\rm L}(s_{j+1};s_{j})-\rmi\piot}\Big)
      \Bigg]
      \PO
     \end{aligned}
\\
\end{align}
Here, the abbreviation $(\vvech^0\nvec_0)_j\defas
\vvech^0(\rvec(s_{j+1});\rvec(s_{j}))\,\nvec(\rvec(s_{j}))$ is used,
together with \eref{eq:ajdef} and the convention $s_0\equiv
s_n$.
Note that the gauge dependent factor
(involving the $\Chit(s_{j})$)
vanishes identically as a consequence of the cyclic permutability of
the integration variables.  This renders the trace \eref{eq:trace1} a
gauge invariant quantity.
It is now  evaluated to leading semiclassical order using the
stationary phase approximation \eref{eq:stphaseNd}.

\subsubsection{The saddle point conditions}
\label{sec:sp}

For each of the $2^n$ integrands in \eref{eq:trace1} the condition of
a stationary phase leads to $n$ saddle point equations
\begin{gather}
  \label{eq:spcond}
  \frac{\rmd}{\rmd s_j} \Big[ 2\pi\nu \ga_{\SoL_j}(s_j;s_{j-1})
  + 2\pi\nu \ga_{\SoL_{j+1}}(s_{j+1};s_{j})
  \Big]
  \stackrel{!}{=}0
  \CO
  \q j\in \{1,\ldots,n\}
  \PO
\end{gather}
Here, the indices
 $ \SoL_j \in \{ {\rm S}, {\rm L} \}$
account for the $2^n$ different sequences of short and long arc
operators under the trace. We shall treat all these equations
simultaneously by noting for any solution $\underline{s}$ of
\eref{eq:spcond} not only the configuration of saddle points but also
the corresponding sequence of types of arcs,
$\underline{s}=((s_1,\SoL_1),..,(s_n,\SoL_n))$.

\begin{figure}[t]%
  \begin{center}%
    \includegraphics[width=0.62\linewidth] {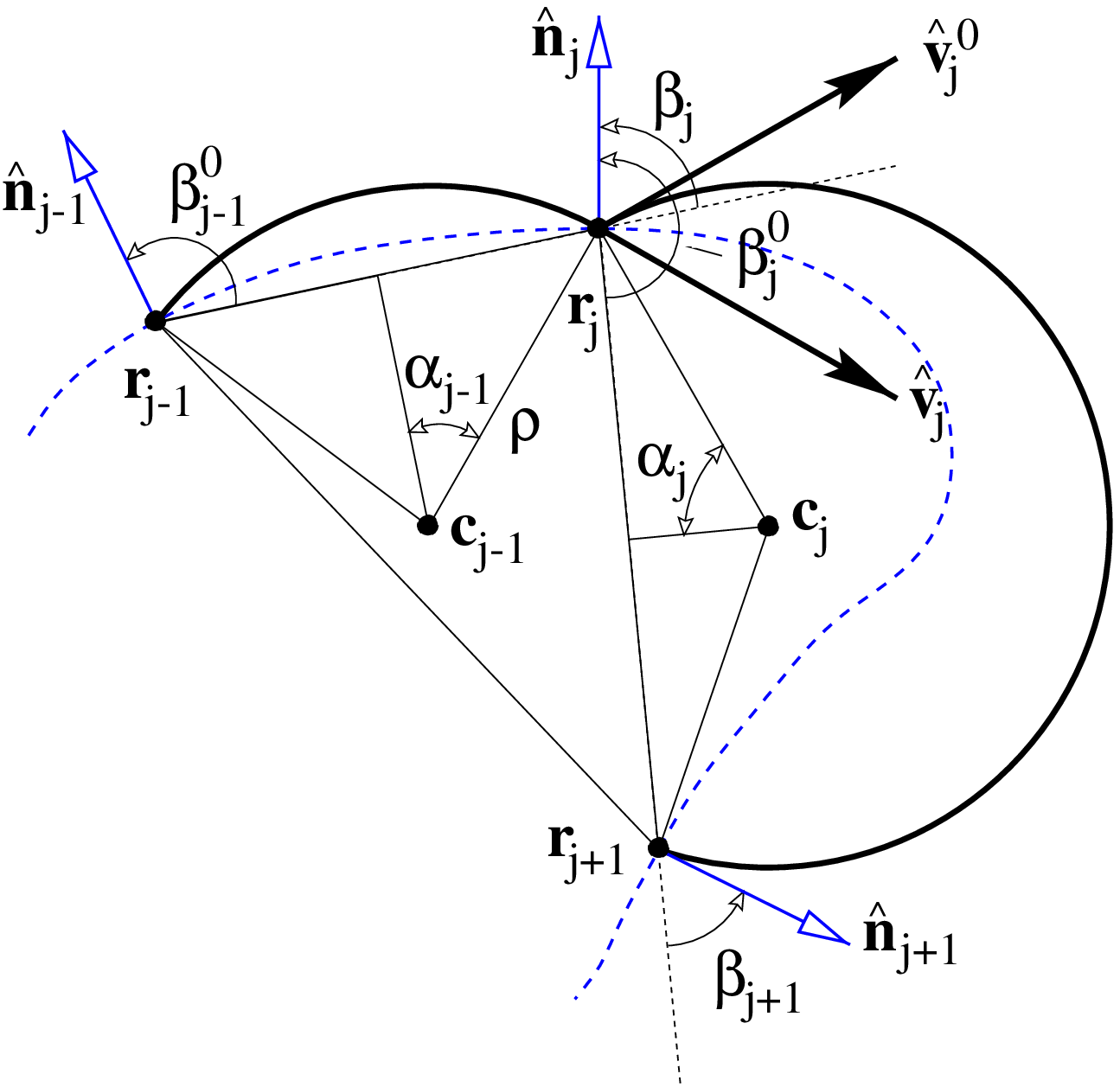}
    \figurecaption{The angles entering the $j$th saddle point
      condition. (The dashed line indicates the boundary.)}%
    \label{fig:saddle}%
  \end{center}%
\end{figure}

\begin{table}[t]
  \begin{center}
  \fbox{
   \begin{minipage}{.7\linewidth}
     \small
     \begin{alignat*}{2}
        \nvec_j\,\vvech^{0}_{{\rm S}\,j}
        &=
        +\cos(\beta^{0}_j-\alpha_j)
        &\qq
        \tvec_j\,\vvech^{0}_{{\rm S}\,j}
        &=
        -\sin(\beta^{0}_j-\alpha_j)
        \nnn
        \nvec_j\,\vvech^{0}_{{\rm L}\,j}
        &=
        -\cos(\beta^{0}_j+\alpha_j)
        &\qq
        \tvec_j\,\vvech^{0}_{{\rm L}\,j}
        &=
        +\sin(\beta^{0}_j+\alpha_j)
        \nnn
        \nvec_{j}\,\vvech_{{\rm S}\,j}
        &=
        +\cos(\beta_{j}+\alpha_{j-1})
        &\qq
        \tvec_{j}\,\vvech_{{\rm S}\,j}
        &=
        -\sin(\beta_{j}+\alpha_{j-1})
        \nnn
        \nvec_{j}\,\vvech_{{\rm L}\,j}
        &=
        -\cos(\beta_{j}-\alpha_{j-1})
        &\qq
        \tvec_{j}\,\vvech_{{\rm L}\,j}
        &=
        +\sin(\beta_{j}-\alpha_{j-1})
        \\
      \end{alignat*}
    \end{minipage}
  }
\end{center}
\vspace*{0.5\baselineskip}
\figurecaption{
  Components of the incident and reflected  velocities. 
   For the geometrical interpretation see Fig.~\ref{fig:saddle}. 
  }
\label{tab:vnt}
\end{table}  

In order to obtain a geometric interpretation of the saddle point
conditions we fix the positions $ \rvec_j\defas\rvec(s_j)$ and
extend the definition of the angles
\eref{eq:alphadef} and \eref{eq:betadef} to a sequence of $n$ points.
\begin{gather}
  \label{eq:ajdef}
  \alpha_j \defas \arcsin\left(\frac{|\rvec_{j+1}-\rvec_j|}{2\rho}\right)
\intertext{and}
\label{eq:bjdef}
  \beta^{0}_j \defas \Angle(\nvec_j;\rvec_{j+1}-\rvec_{j})
  \qq
  \beta_{j+1} \defas \Angle(\nvec_{j+1};\rvec_{j+1}-\rvec_j)
  \PO
\end{gather}
This definition implies 
\begin{alignat}{2}
  \cos(\beta_j)
  &=
  \frac{(\rvec_j-\rvec_{j-1})\,\nvec_j}{|\rvec_j-\rvec_{j-1}|}
  &\qq
  \sin(\beta_{j})
  &=
  \frac{(\rvec_j-\rvec_{j-1})\times\nvec_{j}}{|\rvec_j-\rvec_{j-1}|}
  \nnn
  \cos(\beta_j^{0})
  &=
  \frac{(\rvec_{j+1}-\rvec_{j})\,\nvec_j}{|\rvec_{j+1}-\rvec_{j}|}
  &\qq
  \sin(\beta_{j}^{0})
  &=
  \frac{(\rvec_{j+1}-\rvec_{j})\times\nvec_{j}}{|\rvec_{j+1}-\rvec_{j}|}
  \CO
\end{alignat}
and
\begin{gather}
  \cos(\beta_{j+1}-\beta_{j}^{0})
  =\nvec_j\,\nvec_{j+1}
  \qq
  \sin(\beta_{j+1}-\beta_{j}^{0})
  =\nvec_j\times\nvec_{j+1}
  \PO
\end{gather}
Again, $\alpha_j$ determines the angles of the incident and the
reflected velocity vectors with respect to the direction given by
$\rvec_{j+1}-\rvec_j$.  
It follows that the normal and tangential
components of the velocity are  given by
the expressions in Table \ref{tab:vnt}.
They allow to state the derivative of the action with respect to the
arc length $s$ along the boundary
in a particularly convenient form:
\begin{align}
  \frac{\rmd}{\rmd s_j}
  \ga_{\rm S}(s_j;s_{j-1})
  &=\frac{1}{\pi}
  \left(
    2\sqrt{1-\left(\frac{\rvec_j-\rvec_{j-1}}{2\rho}\right)^2}\,
    \frac{(\rvec_j-\rvec_{j-1})\tvec_j}{|\rvec_j-\rvec_{j-1}|2\rho}
    -\frac{\tvec_j\times\rvec_{j-1}}{2\rho^2}
  \right)
  \nnn
  &=\frac{1}{\pi\rho}
  \left(
     -\cos(\alpha_{j-1})\sin(\beta_j)+\frac{\rvec_{j-1}\,\nvec_j}{2\rho}
  \right)
\end{align}
Similarly, one finds
\begin{align}
  \label{eq:dsjpo}
    \frac{\rmd}{\rmd s_j}
  \ga_{\rm S}(s_{j+1};s_{j})
  &=\frac{1}{\pi\rho}
  \left(
     +\cos(\alpha_{j})\sin(\beta^0_j)-\frac{\rvec_{j+1}\,\nvec_j}{2\rho}
  \right)
  \\
  \tag{\ref{eq:dsjpo}a}
  \frac{\rmd}{\rmd s_j}
  \ga_{\rm L}(s_j;s_{j-1})
  &=\frac{1}{\pi\rho}
  \left(
    +\cos(\alpha_{j-1})\sin(\beta_j)+\frac{\rvec_{j-1}\,\nvec_j}{2\rho}
  \right)
  \\
  \tag{\ref{eq:dsjpo}b}
  \frac{\rmd}{\rmd s_j}
  \ga_{\rm L}(s_{j+1};s_{j})
  &=\frac{1}{\pi\rho}
  \left(
     -\cos(\alpha_{j})\sin(\beta^0_j)-\frac{\rvec_{j+1}\,\nvec_j}{2\rho}
  \right)
  \PO
\end{align}
As a result, an explicit expression for the $j$th saddle point
condition is obtained in terms of the vectors $\rvec_{j-1}, \rvec_{j},
\rvec_{j+1}$, and $\nvec_j$.
Naturally, the condition depends on the type of 
the two operators involved.
\begin{align}
  \label{eq:spcond2}
  \frac{(\rvec_{j+1}-\rvec_{j-1})\,\nvec_j}{2\rho}
  =
  \begin{cases}
    -\sin(\beta_j)\cos(\alpha_{j-1})+\sin(\beta^0_j)\cos(\alpha_j)
    &\text{if $(\SoL_{j-1},\SoL_{j})=({\rm S},{\rm S})$}
    \\
    -\sin(\beta_j)\cos(\alpha_{j-1})-\sin(\beta^0_j)\cos(\alpha_j)
    &\text{if $(\SoL_{j-1},\SoL_{j})=({\rm S},{\rm L})$}
    \\
    +\sin(\beta_j)\cos(\alpha_{j-1})+\sin(\beta^0_j)\cos(\alpha_j)
    &\text{if $(\SoL_{j-1},\SoL_{j})=({\rm L},{\rm S})$}
    \\
    +\sin(\beta_j)\cos(\alpha_{j-1})-\sin(\beta^0_j)\cos(\alpha_j)
    &\text{if $(\SoL_{j-1},\SoL_{j})=({\rm L},{\rm L})$.}
  \end{cases}
\end{align}
The left hand side of this equation can be written in terms of the
angles appearing on the right side after adding and subtracting the
expression $(\rvec_j\,\nvec_j)/(2\rho)$.
\begin{gather}
  \frac{(\rvec_{j+1}-\rvec_{j})\,\nvec_j}{2\rho}
  +
  \frac{(\rvec_{j}-\rvec_{j-1})\,\nvec_j}{2\rho}
  =\cos(\beta^0_j)\sin(\alpha_j)+\cos(\beta_j)\sin(\alpha_{j-1})
\end{gather}
Combining the last two equations,
the saddle point condition assumes a form, 
\begin{alignat}{2}
  \label{eq:spcond3}
  \sin(\beta_j+\alpha_{j-1}) &= \sin(\beta^0_j-\alpha_j)
  &\qq&
  \text{if  $(\SoL_{j-1},\SoL_{j})=({\rm S},{\rm S})$}
  \nnn
  \sin(\beta_j+\alpha_{j-1}) &= -\sin(\beta^0_j+\alpha_j)
  &\qq&
  \text{if  $(\SoL_{j-1},\SoL_{j})=({\rm S},{\rm L})$}
  \nnn
  \sin(\beta_j-\alpha_{j-1}) &= -\sin(\beta^0_j-\alpha_j)
  &\qq&
  \text{if  $(\SoL_{j-1},\SoL_{j})=({\rm L},{\rm S})$}
  \nnn
  \sin(\beta_j-\alpha_{j-1}) &= \sin(\beta^0_j+\alpha_j)
  &\qq&
  \text{if  $(\SoL_{j-1},\SoL_{j})=({\rm L},{\rm L})$}
  \CO
\end{alignat}
which should be compared to the expressions in Table \ref{tab:vnt} for the
components of the classical velocities. One observes that the
equations  \eref{eq:spcond3} simply amount to the condition
\begin{gather}
  \label{eq:spcond4}
  \tvec_j\,\vvech_{\SoL_{j}\,j}=  \tvec_j\,\vvech^0_{\SoL_j\,j}
  \CO
\end{gather}
for $j=1,\ldots\, n$, and any $\SoL_j\in\{\rm S,L\}$: The tangential
component of the classical velocities which correspond to the saddle
point configuration $\underline{s}$ are {continuous} in the point of
reflection.
Since the modulus of the velocity is a constant of the motion, the
trajectory is either continuous in this point or the normal component
changes its sign. In the first case the trajectory penetrates the
boundary which we call an unphysical solution. In the second case,
the trajectory corresponding to the saddle point configuration obeys
the law of \emph{specular reflection} in $\rvec_j$.

\begin{figure}[tb]%
  \begin{center}%
    \includegraphics[width=0.8\linewidth] {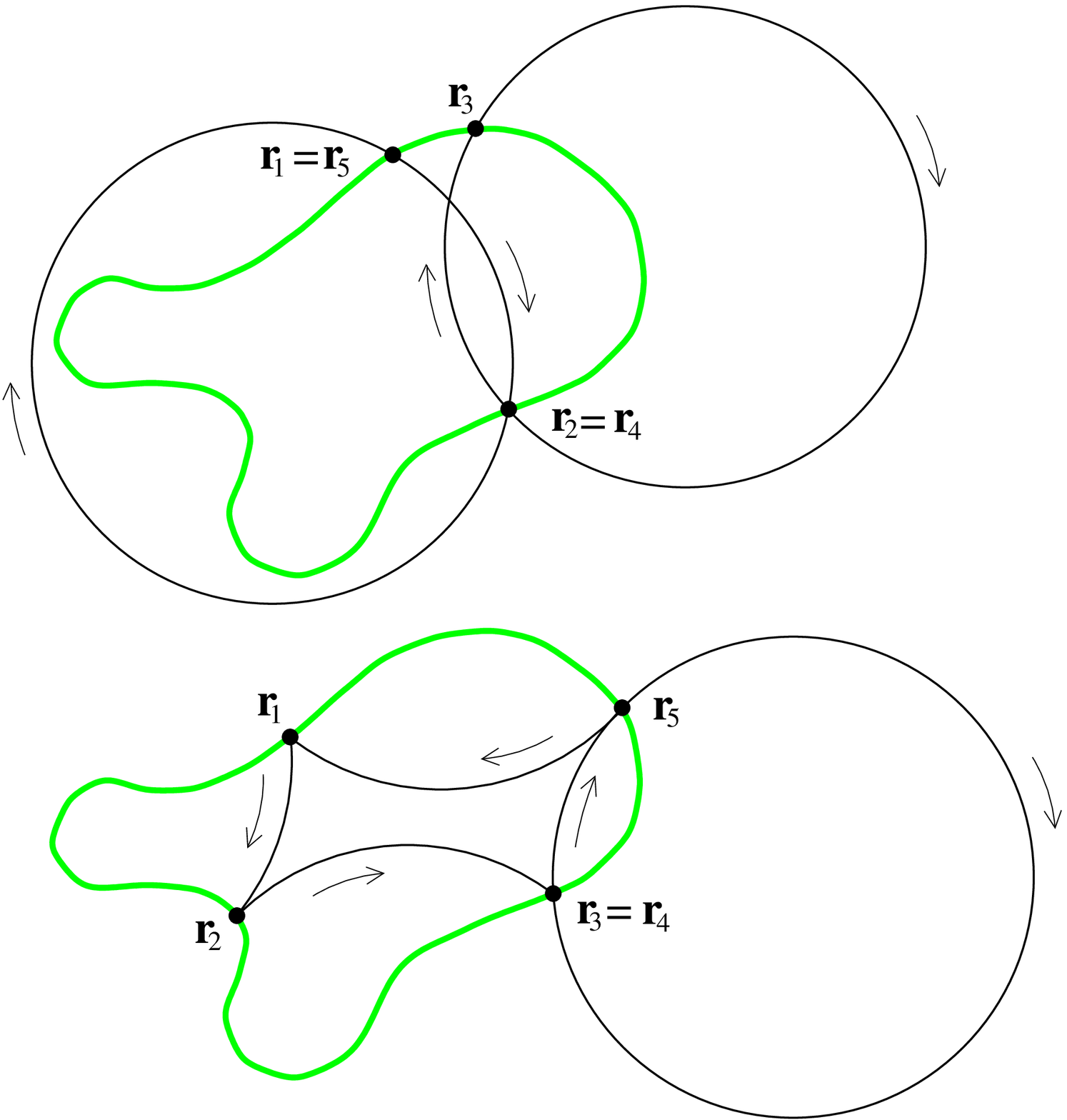}
    \figurecaption{%
      Typical saddle point configurations appearing in the
      semiclassical evaluation of the trace \eref{eq:trace1}. Both
      correspond to unphysical trajectories. The upper configuration
      has no relation to an orbit of the classical problem, while the
      lower one corresponds to a physical trajectory which is dressed
      by an additional cyclotron loop.
      }%
    \label{fig:ghost}%
  \end{center}%
\end{figure}

From the fact that \eref{eq:spcond4} must be satisfied simultaneously
at the $n$ points $\rvec_j$ it follows that any saddle point
configuration corresponds to a closed, periodic orbit. However, by no
means is this orbit necessarily a physically allowed classical
trajectory.  
Figure \ref{fig:ghost} sketches the two different types of saddle
point configurations which appear in magnetic billiards. Here we
choose $n=5$, ie, the saddle points correspond to periodic orbits of
period $5$.  Clearly, both of them are unphysical trajectories. The
one on the top features a specular reflection at $\rvec_2$. 
Then the boundary is penetrated at $\rvec_3$ giving rise to a full
cyclotron loop.  After one more reflection (this time from the
exterior) at $\rvec_4=\rvec_2$ the trajectory arrives at its
initial point.  It performs one more cyclotron orbit without even
displaying a boundary point at $\rvec_2$. This saddle point is a
legitimate solution of \eref{eq:spcond4} belonging to a dense and
two-dimensional set of stationary points
(since the boundary points $\rvec_1$ and $\rvec_2$ may be
shifted independently without changing the picture.)
It has clearly no relation to a physical periodic orbit.
The saddle point shown on the bottom part of Fig.~\ref{fig:ghost}, on
the other hand, does exhibit the boundary points of a physical
periodic orbit (with period 4).  Nonetheless, the depicted trajectory is
unphysical since it leaves the interior domain, performing a
cyclotron loop between the third and forth boundary points. 
Obviously, there is an infinite number of these unphysical saddle points
attached to any proper, physical periodic orbit. They merely dress the
original orbit with additional cyclotron loops.
It might be expected that these unphysical contributions can be
re-summed, leaving behind only the contributions of physical periodic
orbits of the interior and exterior problem.  This is a difficult
task, due to its combinatorial nature in conjunction with a number of
ambiguities.  A saddle point configuration may, for example,
incorporate an interior and and exterior periodic orbit at the same
time, leaving the question undetermined whether to assign the
contribution to the interior or to the exterior problem.  These
problems are resolved immediately by the splitting of the operator
\eref{eq:pkern} into interior and exterior types.  Here, it is the
Heaviside functions introduced by the splitting which guarantee that
only those saddle points contribute for which the corresponding
classical trajectory is directed into the correct domain at
\emph{each} point of reflection.  As a consequence, the unphysical
solutions discussed above are \emph{erased} from the sum.  The
remaining saddle points will be denoted by $\ipon$ and $\epon$,
\label{pag:podef} respectively.  They correspond to the periodic
orbits of period $n$ found in the classical interior and exterior
billiard problem.

\begin{figure}[tb]%
  \begin{center}%
  \includegraphics[width=\linewidth] {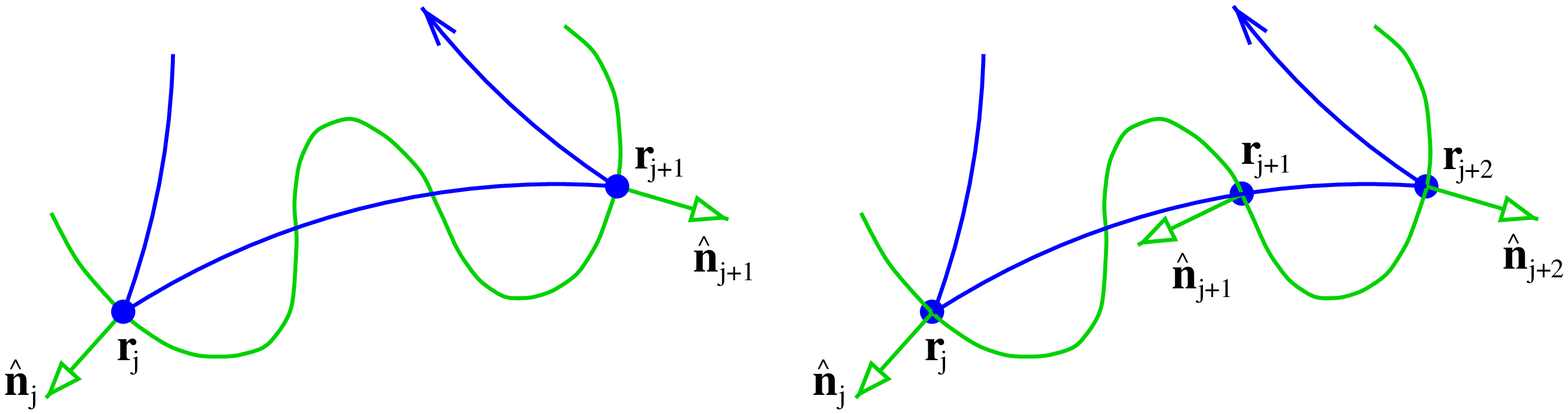}
  \figurecaption{Sketch of an interior ghost orbit, see text. 
 (The normals are pointing outwards.)}  
  \label{fig:ghost2}
\end{center}%
\end{figure}

Strictly speaking, the set of saddle points $\pon$ which are directed
into the correct domain at each point of reflection still includes the
so-called ghost orbits.  These are periodic orbits which leave (and
necessarily re-enter) the proper domain without exhibiting a component
of the saddle point (ie, a point of reflection) when leaving it.  The
left side of Fig.~\ref{fig:ghost2} shows the situation.  Like in the
case of non-magnetic billiards \cite{BB72} these saddle points
finally do not contribute to the sum over the traces. This is because
for any ghost orbit of period $n$ one finds another of period $n+1$,
with an additional boundary point at the position of re-entrance
(right side of Fig.~\ref{fig:ghost2}).  These two contributions differ
by a factor $(-1)$ due to the additional boundary point and therefore
cancel.

\subsubsection{The prefactors}

The next step is to transform the prefactors in the trace integral
\eref{eq:trace1}.  Eventually they should combine
with the determinant of the matrix of action derivatives introduced by
the stationary phase approximation \eref{eq:stphaseNd}.  The resulting
expression should then be given in terms of the physical properties of
the attributed periodic orbit $\pon$.

We start with the evaluation of the mixed derivatives of the actions
in terms of the angles \eref{eq:ajdef}, \eref{eq:bjdef} characterizing
the $j$th part of the trajectory.
If the  arc is of the short type one obtains the formula
\begin{align}
  \label{eq:d2ts}
  \frac{\rmd^2}{\rmd s_j \rmd s_{j+1}}\, 
  \ga_{\rm S}(s_{j+1},s_j)
  &=
  \frac{1}{\pi\rho}  \frac{\rmd}{\rmd s_{j+1}}
  \left(
    \cos(\alpha_{j})\sin(\beta^0_j)-\frac{\rvec_{j+1}\,\nvec_j}{2\rho}
  \right)
  \nnn
  &=
  \frac{1}{2\pi\rho^2}
  \Big(
  \tan(\alpha_j)
    \sin(\beta_{j+1})\sin(\beta^0_j)
  \nnn
   & \phantom{=\frac{1}{2\pi\rho^2}\Big(} -
   \cot(\alpha_j)
    \cos(\beta_{j+1})\cos(\beta^0_j)
    +\sin(\beta_{j+1}-\beta^0_j)
  \Big)
  \nnn
  &=
  \frac{-1}{2\pi\rho^2}\,
  \frac{(\vvech^0_{{\rm S}\,j}\nvec_{j})(\vvech_{{\rm S}\,j+1}\nvec_{j+1})}
  {\sin(\alpha_j)\cos(\alpha_j)}
  \PO
\end{align}
Here, the expressions given in \eref{eq:dsjpo} and Tab.~\ref{tab:vnt}
were employed, as well as 
\begin{align}
  \frac{\rmd}{\rmd s_{j+1}} \cos(\alpha_j)
  &= -\sin(\alpha_j)\,  \frac{\rmd}{\rmd s_{j+1}} 
  \arcsin\left(\frac{|\rvec_{j+1}-\rvec_{j}|}{2\rho}\right)
  \\
  &=\frac{-1}{2\rho}\,\frac{\sin(\alpha_j)}{\cos(\alpha_j)}\,
  \frac{(\rvec_{j+1}-\rvec_{j})\,\tvec_{j+1}}{|\rvec_{j+1}-\rvec_{j}|}
  =\frac{1}{2\rho}\,{\tan(\alpha_j)}\sin(\beta_{j+1})
  \nonumber
\intertext{and}
  \frac{\rmd}{\rmd s_{j+1}} \sin(\beta^0_j)
  &= \frac{\rmd}{\rmd s_{j+1}} 
  \left[
    \frac{(\rvec_{j+1}-\rvec_{j})\times\nvec_j}{|\rvec_{j+1}-\rvec_{j}|}
  \right]
  \nnn
  &= \frac{\tvec_{j+1}\times\nvec_j}{|\rvec_{j+1}-\rvec_{j}|}
  -\frac{(\rvec_{j+1}-\rvec_{j})\times\nvec_j}{|\rvec_{j+1}-\rvec_{j}|}
  \,\frac{(\rvec_{j+1}-\rvec_{j})\,\tvec_{j+1}}{(\rvec_{j+1}-\rvec_{j})^2}
  \nnn
  &=\frac{-1}{2\rho}\,\frac{\cos(\beta_{j+1})\cos(\beta^0_j)}{\sin(\alpha_j)}
  \PO
\end{align}
Hence, the mixed derivative \eref{eq:d2ts} is essentially determined
by the normal components of the velocities at the initial and the
final point of the corresponding arc.  Note that this expression is
manifestly \emph{positive} if the arc is part of a physical
trajectory meaning that it lies either in the interior or in the
exterior at both points (see also Fig.~\ref{fig:saddle}).

If the $j$th part of the action corresponds to a long arc one obtains
in a similar fashion
\begin{align}
  \label{eq:d2tl}
  \frac{\rmd^2}{\rmd s_j \rmd s_{j+1}}\, \ga_{\rm L}(s_{j+1},s_j)
  &=
  \frac{1}{2\pi\rho^2}\,
  \frac{\cos(\beta^0_j+\alpha_j)\cos(\beta_{j+1}-\alpha_j)}
  {\sin(\alpha_j)\cos(\alpha_j)}
  \nnn
  &=
  \frac{1}{2\pi\rho^2}\,
  \frac{(\vvech^0_{{\rm L}\,j}\nvec_{j})(\vvech_{{\rm L}\,j+1}\nvec_{j+1})}
  {\sin(\alpha_j)\cos(\alpha_j)}
  \PO
\end{align}
The form of this formula is analogous to \eref{eq:d2ts}, except for
the difference in sign.
It follows that equation \eref{eq:d2tl} is manifestly \emph{negative}
if the angles $\alpha_j, \beta^0_j,$ and $\beta_{j+1}$ describe a segment
of a physical trajectory --- again due to the change in the
orientation of the velocity vector.

The mixed derivatives \eref{eq:d2ts} and \eref{eq:d2tl} allow the 
transformation of the product under the trace integral \eref{eq:trace1} into a
symmetrized expression.  For a given saddle point $\pon$ we denote the
geometric part of the total action by
\begin{align}
  \label{eq:Gadef}
  \Ga(\pon)&\defas \sum_{j=1}^n \ga_{\SoL_j}(s_{j+1},s_j)
  \CO
\end{align}
cf \eref{eq:tdef}, \eref{eq:trel}.
In addition, the number of \emph{long} arcs appearing in $\pon$ will be
called $\ell_\po$.  
The product under the trace integral \eref{eq:trace1} can now be
stated in terms of $\Ga(\pon)$ and $\ell_\po$. For the interior
operators, it assumes the form
\begin{align}
  \label{eq:pre}
  &
  \rme^{2\pi\rmi\nu \Ga(\ipon)}
  \,
  \rme^{-\rmi\piot \ell_\po}\,
  \prod_{j=1}^n
    \frac{-(\vvech^0_{\SoL_j}\nvec_0)_j}
    {\big(\sin(\alpha_j)\cos(\alpha_j)\big)^\oh}
  \\\tag{\ref{eq:pre}a}\label{eq:prea}
  =\,  &
    \frac{ 
      \prod_{j=1}^n
      \left(  
        (\vvech^0_{\{\SoL_j\,j\}}\nvec_j) 
        (\vvech^0_{\{\SoL_{j+1}\,j+1\}}\nvec_{j+1})
      \right)^\oh
      }
    {
      \prod_{j=1}^n
      \left(  
        \sin(\alpha_j)\cos(\alpha_j)
      \right)^\oh
      }
  \,
  \rme^{-\rmi\piot \ell_\po}\,
  \rme^{2\pi\rmi\nu \Ga(\ipon)}
  \\\tag{\ref{eq:pre}b}\label{eq:preb} 
  =\,
  &
  \frac{ 
    \prod_{j=1}^n
    \left(  
      -
      (\vvech^0_{\{\SoL_j\,j\}}\nvec_j) 
      (\vvech_{\{\SoL_{j+1}\,j+1\}}\nvec_{j+1})
    \right)^\oh
    }
  {
    \prod_{j=1}^n
    \left(  
      \sin(\alpha_j)\cos(\alpha_j)
    \right)^\oh
    }
  \,
  \rme^{-\rmi\piot \ell_\po}\,
  \rme^{2\pi\rmi\nu \Ga(\ipon)}
  \\\tag{\ref{eq:pre}c}\label{eq:prec}
  =\,
  &
  \prod_{j=1}^n
  \left(
    2\pi\rho^2
    \frac{\rmd^2  \ga_{\SoL_j}(s_{j+1},s_j)}
    {\rmd s_j \rmd s_{j+1}}
  \right)^\oh
  \,
  \rme^{\rmi\piot (\ell_\po-\ell_\po)}\,
  \,
  \rme^{2\pi\rmi\nu \Ga(\ipon)}
  \\\tag{\ref{eq:pre}d}\label{eq:pred}
  =\,
  &
  (2\pi)^\frac{n}{2}
  \prod_{j=1}^n
  \left|
    \rho^2
    \frac{\rmd^2  \Ga(\ipon)}{\rmd s_j \rmd s_{j+1}}
  \right|^\oh
  \,
  \rme^{-\rmi\piot \ell_\po}
  \,
  \rme^{2\pi\rmi\nu \Ga(\ipon)}
  \PO
\end{align}
Here we used several times the fact that the saddle point
configurations described by $\ipon$ correspond to physical, interior
periodic orbit with $n$ reflections.  First, we noted the positivity
of the factor $-(\vvech^0_{\SoL_j}\nvec_0)_j$ to write it as a product
of square roots (and shifted one index by one). Second, the reflection
condition
\begin{align}
      (\vvech^0_{\{\SoL_{j}\,j\}}\nvec_{j})
      =
      -(\vvech_{\{\SoL_{j}\,j\}}\nvec_{j})
\end{align}
was employed to get from \eref{eq:prea} to \eref{eq:preb}. As a
result, the prefactors are symmetric in $s_j$ and $s_{j+1}$, which
allows stating them in terms of the mixed derivatives of the classical
action.  Finally, given the sign of each factor in \eref{eq:pred} the
sign of the product can be taken out.  It is $(-)^{\ell_\po}$, due to
the $\ell_\po$ long arcs in $\pon$.

Upon evaluating the trace of the exterior operators one is led to
the \emph{same} expression \eref{eq:pred}, with $\ipon$ replaced by
$\epon$. This is because the additional sign in the definition
\eref{eq:pkern} of the exterior operators cancels the change in the
orientation of the normals relative to the velocity vectors.

\begin{table}[tbp]
  \begin{center}
    \small
    \fbox{
    \begin{tabular}{r@{\quad}p{0.75\linewidth}c}
      $\rho$ &  cyclotron radius ($\rho>0$) & \eref{eq:rhobdef}
      \\
      $\alpha$  & relative distance of the initial and the final point
      ($0\le\alpha\le\piot$) &\eref{eq:alphadef}
      \\
      $\beta^0$ & relative direction of normal at the initial point 
      ($0\le\beta^0<2\pi$) & \eref{eq:betadef}
      \\
      $\beta$ & relative direction of normal at final point 
      ($0\le\beta<2\pi$) & \eref{eq:betadef}
      \\
      $\ga_{\rm S}$, $\ga_{\rm L}$  & geometric part of the
      action of a short (long) arc
      & \eref{eq:tdef} 
      \\
      $\po$, $\pon$ & physical periodic orbit (with $n$ reflections) 
      & page \pageref{pag:podef}
      \\
      $\Ga(\po)$  & geometric part of the
      action of the periodic orbit $\po$
      & \eref{eq:Gadef}
      \\
      $n_\po$ ($r_\po$)& number of reflections (repetitions) in
      $\po$ 
      & page \pageref{pag:nrdef}
      \\
      $\mu_\po$& Maslov index (number of conjugate points in $\po$) 
      & page \pageref{pag:mudef}
      \\
      ${\rm M}(\po)$& stability matrix of $\po$ 
      & \eref{eq:Mrel}
    \end{tabular}
    }
    \vspace*{0.5\baselineskip}
  \end{center}
  \figurecaption{Important geometric quantities}
  \label{tab:geo}
\end{table}

\subsubsection{Performing the  trace}
\label{sec:trace}

Now, with the prefactors written as mixed derivatives of
the action in \eref{eq:pred} we can follow the standard procedure to
derive the semiclassical trace formula.
We  apply the stationary phase approximation
to the $n$-dimensional trace integral.  At first, the contributing
saddle points are assumed to be isolated.  This amounts to the
assumption that the corresponding classical billiard dynamics is
{hyperbolic} \cite{Gutzwiller90}.  The case of an integrable system is
treated afterwards.

Combining eqs \eref{eq:trace1} and \eref{eq:pre}, together with
\eref{eq:stphaseNd}, yields
\begin{align}
  \label{eq:trace2}
  \tr\!\left[
    \big({\sf P}^{\rm int}\big)^n
  \right]
    &=\,
    \sum_{\po\in\big\{\ipons\big\}} \frac{n}{r_\po} 
    \,\frac{1}{\rho^n}\,
    \frac{  
      \prod_{j=1}^n
      \left|
        \rho^2
        \frac{\partial^2  \Ga(\po)}{\partial s_j \partial s_{j+1}}
      \right|^\oh
      }
    {
      \left|
        \det
          \left(      
            \frac{\partial^2 \Ga(\po)}{\partial s_k\partial s_{l}}
          \right)_{k,l}
        \right|^\oh
      }
    \,  \rme^{2\pi\rmi\nu \Ga(\po)}
    \, \rme^{-\rmi\piot (\ell_\po+\nu_\po)}
    \nnn
    &=\,
    \sum_{\po\in\big\{\ipons\big\}} \frac{n}{r_\po} 
    \,
    \frac{1}{|\tr({\rm M}(\po)-2|^\oh}
    \,  \rme^{2\pi\rmi\nu \Ga(\po)}
    \, 
    \rme^{-\rmi\piot\mu_\po}
    \PO
\end{align}
The factor $n/r_\po$  appears because the sum is taken over all
$n$-periodic orbits of the interior billiard rather than over all
contributing saddle points.  Each $n$-periodic orbit (with repetition
number $r_\po$ \label{pag:nrdef}) corresponds to $n/r_\po$ distinct
saddle points $\underline{s}$, which are related by a cyclic shift of
their components.  For the last equality in \eref{eq:trace2} we 
used once more the fact that $\ipon$ is a classical periodic orbit of a
billiard problem.  This implies a general relation between the
derivatives of the generating function of the billiard map
 $\Ga(\pon)$ and the stability matrix
${\rm M}(\pon)$ \cite{HST99},
\begin{align}
  \label{eq:Mrel}
  \det
  \left[
  \left(      \frac{\partial^2 \Ga(\pon)}{\partial s_k\partial s_{l}}
  \right)_{k,l}
  \right]
  =
  (-)^n \big[\tr{\rm M}(\pon)-2\big]
  \, 
  \prod_{j=1}^n  \frac{\partial^2 \Ga(\pon)}{\partial s_j\partial s_{j-1}}
  \PO
\end{align}
Its  modulus was taken to derive \eref{eq:trace2}.
The integer $\mu_\po\defas\ell_\po+\nu_\po$ \label{pag:mudef} denotes
the total number of conjugate points. Here, $\nu_\po$ is given by the
number of negative eigenvalues of the determinant in the dominator.
It counts those conjugate points along the trajectory, which are due
to the focusing and defocusing effect of the boundary.  The remaining,
trivial conjugate points, which show up at each long arc (after an
angle of $\pi$), are taken into account by $\ell_\po$.

For later reference, let us mention that the dual partner orbit
of $\pon$, denoted as $\dpon$, has
\begin{align}
  \label{eq:murel}
  \mu_\dpo=2n-\mu_\po
\end{align}
conjugate points:  As discussed in Section
\ref{sec:duality} the dual orbits consists of the arcs complementary
to those of $\pon$ and has opposite orientation. From \eref{eq:trel}
we find $\Ga(\dpon)=n-\Ga(\pon)$ and it follows that
$\nu_\dpo=n-\nu_\po$ since every element of the matrix of second
derivatives in \eref{eq:trace2} is multiplied by $(-1)$. By
definition we have $\ell_\dpo=n-\ell_\po$  leading to 
\eref{eq:murel}. Note also that the stabilities of dual periodic
orbits are equal, $ \tr{\rm M}(\dpo) = \tr{\rm M}(\po)$, which follows
from equation \eref{eq:Mrel}.

\myparagraph{The trace formula for the spectral counting function}

Inserting the expression for the trace \eref{eq:trace2} into
\eref{eq:firstNskipint} we obtain
the fluctuating number counting function attributed to the interior
map operator.
\begin{align}
  \label{eq:Trskipint}
  \N_{\rm osc}^{\rm skip(int)}
  &\defas  \frac{1}{\pi}\Im
  \sum_{n=1}^{\infty}
  \frac{(\mp)^n}{n} \,
  \tr\!\left[
    \big({\sf P}^{\rm int}\big)^n
  \right]
  \\
  &=
  \frac{1}{\pi}\Im
  \sum_{n=1}^\infty
  \sum_{\po\in\big\{\ipons\big\}}
  \frac{(\mp)^n}{r_\po}
  \frac{1}{
    \big|\tr{\rm M}(\po)-2\big|^\oh
    }
  \,  \rme^{2\pi\rmi\nu \Ga(\po)-\rmi\piot\mu_\po}
  \nnn
  \label{eq:Nskipint}
  \tag{\ref{eq:Trskipint}a}
  &=
  \frac{1}{\pi}
  \sum_{\po\in\{\ipo\}}
  \frac{(\mp)^{n_\po}}{
    r_\po\,\big|\tr{\rm M}(\po)-2\big|^\oh
    }
  \,  \sin
  \left({2\pi\nu \Ga(\po)
      -\piot\mu_\po}
  \right)
\end{align}
It is naturally associated with the \emph{interior} problem, since the
sum  includes all periodic orbits $\big\{\ipo\big\}$ of the
interior billiard problem (with $n_\po$ the number of reflections).
If we are originally interested in the Dirichlet spectrum of the
interior billiard we have to choose the upper sign in
\eref{eq:Trskipint}.  In this case each reflection is associated with
an additional phase  shift of $\pi$.  The lower sign is to be
taken if the spectral problem was originally formulated for the
exterior spectrum. In this case \eref{eq:Trskipint} provides the
spurious interior Neumann spectrum which is included by the double
layer equation.  It differs from the Dirichlet spectrum merely by the
fact that there is no phase shift associated with the reflections at
the billiard boundary.

The second factor of the spectral function \eref{eq:fac} yields a
number counting function which includes the trace over powers of the
{exterior} operators. In complete analogy to the treatment above one
obtains a periodic orbit sum like equation \eref{eq:trace2}. As the
only difference, the sum is over all the periodic orbits $\epo$ of the
\emph{exterior} classical billiard map,
\begin{align}
  \label{eq:Trskipext}
  \N_{\rm osc}^{\rm skip(ext)}
  &\defas  \frac{1}{\pi}\Im
  \sum_{n=1}^{\infty}
  \frac{(\pm)^n}{n} \,
  \tr\!\left[
    \big({\sf P}^{\rm ext}\big)^n
  \right]
  \\
  \label{eq:Nskipext}
  \tag{\ref{eq:Trskipext}a}  
  &=
  \frac{1}{\pi}
  \sum_{\po\in\{\epo\}}
  \frac{(\pm)^{n_\po}}{
    r_\po\,\big|\tr{\rm M}(\po)-2\big|^\oh
    }
  \,    \sin
  \left(
    2\pi\nu \Ga(\po)
    -\piot\mu_\po
  \right)
 \PO
\end{align}
Like above, an additional phase shift of $\pi$ is associated with each
reflection if the original double layer equation was formulated for
the same domain as the orbits are taken from (upper sign in
\eref{eq:Nskipext}). Again there is no shift if the periodic orbit sum
represents the spurious solutions of the double layer equation which
belong to the complementary domain (lower sign).  The fact that the
trace formulas for Dirichlet and Neumann boundary conditions differ
only by a phase is also known from the theory of non-magnetic quantum
billiards \cite{SPSUS95}.

We conclude that for either the interior or the exterior Dirichlet
problem the fluctuating number function is given by
\begin{align}
  \label{eq:Nskiposc}
  \N_{\rm osc}^{\rm skip}(\nu)
  =
  \frac{1}{\pi}
  \sum_{\po}
  \frac{1}{
    r_\po\,\big|\tr{\rm M}(\po)-2\big|^\oh
    }
  \,    \sin
  \left(
    2\pi\nu \Ga(\po)
    -\pi      n_\po
    -\piot\mu_\po
  \right)
  \CO
\end{align}
where the sum is over all periodic orbits in the respective domain.

This final result in complete agreement with the standard trace
formulas. One could have used them without the preceeding derivation.
However, the exclusion of the non-physical trajectories would remain an
act of faith. The derivation above provides a sound basis for the
intuitively sound results.

\subsubsection{Geometric interpretation}
 
At this point a brief discussion of the geometric meaning of a 
trajectory's scaled action is in order.  We start with the observation
that the actions of short and long arcs are given by
\emph{identical} expressions once the parameter
\begin{align}
  \label{eq:sigmadef}
  \sigma_j 
  \defas
  \frac{(\rvec_{j+1}-\cvec_j)\times(\rvec_{j+1}-\rvec_{j})}
  {\rho\,|\rvec_{j+1}-\rvec_{j}|}
  =
  \begin{cases}
    -\cos(\alpha_j) & \text{if ``short'' arc}
    \\
    +\cos(\alpha_j) & \text{if ``long'' arc}
  \end{cases}
\end{align}
is introduced to describe the $j$-th arc.  Unlike the angle $\alpha_j$
\eref{eq:ajdef}, it is not
just a function of $\rvec_j$ and $\rvec_{j+1}$ but it contains
information on the \emph{type} of the arc through its sign:
$\sigma_j$ is negative for short arcs and positive for long ones. 
The geometric parts of the actions of short and long arcs, \eref{eq:tdef} and
\eref{eq:trel}, now assume the {common} form
\begin{align}
  \label{eq:ga}
  \ga(\rvec_{j+1};\rvec_j)=
  \frac{1}{\pi}
  \left(\frac{\pi}{2}+\arcsin(\sigma_j)-\sigma_j\sqrt{1-\sigma_j^2}
  -\frac{\rvec_{j+1}\times\rvec_j}{2\rho^2}\right)
\CO
\end{align}
which is a remarkable simplification.\footnote{%
  The derivation of the trace formulas would have been considerably
  more complicated, had we introduced this parameterization earlier.  }
It allows to show immediately that a periodic orbit $\po$ (of period
$n$) exhibits a geometric action \eref{eq:Gadef}
\begin{align}
  \label{eq:Ga2}
  \Ga(\pon)&=
  \sum_{j=1}^n \ga(\rvec_{j+1},\rvec_j)
 = \frac{\rho\, \Len_\po \pm \Area_\po}{\rho^2\pi} 
\CO
\end{align}
which is given by the length of the trajectory,
\begin{align}
 \label{eq:lenpodef}
  \Len_\po
  &\defas 
  \rho \sum_{j=1}^n  \left({\pi}+2\arcsin(\sigma_j)\right)
  \\
  \tag{\ref{eq:lenpodef}a}
  \label{eq:lenpodefa}
  &=
  \rho\, \frac{\rmd}{\rmd\nu} \big[2\pi\nu\Ga(\pon)\big]
\CO
\end{align}
 and the enclosed area,
\begin{align}
 \label{eq:areapodef}
  \Area_\po
   &\defas
   \Area^{\rm poly}_\po \mp 
   \sum_{j=1}^n \left(\frac{\pi}{2}+\arcsin(\sigma_j)
   +\sigma_j\sqrt{1-\sigma_j^2}
   \right) \rho^2 
\PO
\end{align}
Here, $\Area^{\rm poly}_\po$ is the area of the polygon defined by the
points of reflection $\{\rvec_j\}$ and each of the summands in
\eref{eq:areapodef} is equal to the area enclosed by the $j$-th arc
and the cord connecting its initial and final points, cf
\eref{eq:areaint}. (Overlapping parts of the enclosed area are counted
according to their multiplicity.)

Equation \eref{eq:lenpodefa} follows bearing in mind that $\rho$ and
$\sigma_i$ are functions of $\nu$, cf \eref{eq:defnu}.  It illustrates
the fact that the excursion time of a trajectory is given by the
derivative of its action with respect to energy.  Using the proper
scaled energy $\widetilde{E}=2\nu$ (cf the discussion of
\eref{eq:defnu}) we obtain the scaled time of flight $\tau_\po$ of
the periodic orbit,
\begin{align}
\label{eq:deftau}
  \tau_\po = \frac{\rmd}{\rmd(2\nu)} \Big[2\pi\nu\Ga(\pon)\Big]
  =  \sum_{j=1}^n  \left({\piot}+\arcsin(\sigma_j)\right)
\PO
\end{align}

\myparagraph{Density of skipping states}

The formula for the fluctuating part of the density of skipping states
\eref{eq:dskip} follows by taking the derivative of the number
counting function \eref{eq:Nskiposc} with respect to $\nu$,
\begin{align}
  \label{eq:dskiposc}
  d_{\rm osc}^{\rm skip}(\nu)
  =
  \frac{2}{\pi}
  \sum_{\po}
  \frac{\tau_\po}{
    r_\po\,\big|\tr{\rm M}(\po)-2\big|^\oh
    }
  \,    \cos
  \left(
    2\pi\nu \Ga(\po)
    -\pi      n_\po
    -\piot\mu_\po
  \right)
  \PO
\end{align}
It must be emphasized, however, that the applicability of this
expression is rather restricted so far. It is valid only for the
interior billiard and  only if the entire phase space consists of skipping
trajectories (ie, for weak fields only).
In all other cases any attempt to include
the cyclotron contributions ``by hand'' yields unsatisfactory results
\cite{BB97a,BB97c}.

\myparagraph{Magnetization density}

Another derivative of the action occurs in the definition of the
scaled magnetization density \eref{eq:magdef} which was discussed in
Sect.~\ref{sec:mag}. We find that it is determined by the area
$\Area_\po$ enclosed by the trajectory \eref{eq:areapodef},
\begin{align}
 \label{eq:magdiff}
   \left(
     -b^2\frac{\rmd}{\rmd b^2}
     - \nu \frac{\rmd}{\rmd \nu}
   \right)
\Big[2\pi\nu\Ga(\po) \Big]
=  \pm \frac{2}{b^2} \Area_\po
\PO
\end{align}
The semiclassical expression for the fluctuating part of the scaled
magnetization density is obtained by applying the derivatives in
\eref{eq:magdiff} to the trace formula for $\N_{\rm osc}$, cf
\eref{eq:magdef}.  Assuming that all periodic orbits are isolated and
of the skipping type we find
\begin{align}
\label{eq:mmosc}
  \mmt^{\rm osc}(\nu)
  &=
  \pm
  {2}
  \sum_{\po}
  \frac{{\Area_\po}/({b^2\pi})}{
    r_\po\,\big|\tr{\rm M}(\po)-2\big|^\oh
    }
  \,    \cos
  \left(
    2\pi\nu \Ga(\po)
    -\pi      n_\po
    -\piot\mu_\po
  \right)
  \PO
\end{align}
Hence, compared to the density of skipping states \eref{eq:dskiposc}
each periodic orbit contribution to the scaled magnetization density
includes   the enclosed area in units of $b^2\pi$, ie, the magnetic
moment of the classical orbit rather than the scaled time of flight.
Again, the expression \eref{eq:mmosc} is only applicable for the
interior problem at weak fields.
The corresponding, less intuitive semiclassical expression for the
\emph{conventional} magnetization at weak fields may be found in
\cite{PAKGEC94}.

\subsection{Trace formula for the integrable case}
\label{sec:trdisk}

In the previous section the classical billiard map was assumed to be
hyperbolic. We now shift to the other extreme, the disk billiard,
which exhibits integrable motion.

\subsubsection{The disk billiard}
 
The periodic orbit formula for the density of states in the interior of
the magnetic disk was derived recently by Blaschke \etal \cite{BB97a}.
These authors used the trace formula by Creagh and Littlejohn
\cite{CL91} to account for the continuous circular symmetry of the
disk. 

In the following, we derive the trace formula starting from the
boundary integral equation. This demonstrates how the integrable case
is treated in the framework of the boundary map operators and yields
an explicit formula in a straightforward manner.
Moreover, the exterior case is easily included in
our treatment.

Many results of the last section still apply. In particular,
the factorization of the spectral function \eref{eq:fac} does not
depend on the type of motion, hence, we can start directly with the
equations \eref{eq:Trskipint} and \eref{eq:Trskipext} for the interior
and the exterior counting functions.
However, the trace of powers of the map operators cannot be evaluated
like in the hyperbolic case since the periodic orbits are not
isolated but appear in continuous families \cite{BT77}.

The classical motion is governed by one parameter, the ratio
\begin{align}
  \label{eq:Gamdef}
  \Gam\defas \frac{R}{\rho}
\end{align}
between the radius of the disk $R$ and the cyclotron radius.  
For \emph{weak} fields, $\Gam<1$, any two points on the boundary can be
connected in the interior only by short arcs and in the exterior
(only) by long ones.  The field is \emph{strong}, $\Gam>1$, if
complete cyclotron orbits fit into the interior.  The skipping motion
then displays both types of arcs in the interior and the exterior, and two
points on the boundary are  no longer necessarily connected by an
arc.

It is advantageous to use the polar angles $\phi=s/R$.
To be definite, we shall choose the angles always such, that adjacent
points differ at most by $\pi$.
Simple geometry tells that the positive angle $\alpha$, as defined in
\eref{eq:alphadef},\footnote{For the sake of clarity we use $\alpha$
  \eref{eq:alphadef} rather than $\sigma$ \eref{eq:sigmadef} in this
  section.} obeys
\begin{align}
\label{eq:sadisk}
  \sin(\alpha)=\Gam\sin\!\left(\frac{|\phi-\phi_0|}{2}\right).
\end{align}
Moreover, we note the relation
\begin{align}
  \oh\Gam^2\, |\sin(\phi-\phi_0)|
  \,\gtrless \,
  \sin(\alpha)\cos(\alpha)
  \q\text{for}\q
  \Gam \,\gtrless\, 1
\end{align}
which is needed in proving almost all the equations below.
Finally, geometry tells that 
the normal components of the reflected velocities 
are given by
\begin{align}
  \label{eq:vndisk}
  -\vvech^0_\SL\,\nvec_0
  =
  \frac{1}{\Gam}
  \left(
    \oh\Gam^2\sin(\phi-\phi_0)\pm\sin(\alpha)\cos(\alpha)
  \right)
  \CO
\end{align}
for the short arc and long arc, respectively.  They allow  stating
 the prefactors of the map operators \eref{eq:psk} and
\eref{eq:plk} explicitely in terms of the  angle increment $\phi-\phi_0$.

\subsubsection{Operators for the integrable map}

Upon choosing the symmetric gauge, $\Chi=0$, one finds that  the
actions of short and long arcs are merely functions of the difference
of the initial and the final coordinate,
\begin{align}
  \ga_{\rm S}(\phi-\phi_0)
  \defas
  \ga_{\rm S}(R\phi;R\phi_0)
  =\frac{1}{\pi}
  \left(
    \alpha+\sin(\alpha)\cos(\alpha)+\tfrac{1}{2}\Gam^2\sin(\phi-\phi_0)
  \right)
\CO
\end{align}
and likewise $   \ga_{\rm L}(\phi-\phi_0) \defas  \ga_{\rm L}(R\phi;R\phi_0) =
1-  \ga_{\rm S}(\phi_0-\phi)$.
This reflects the integrability of the classical motion. 

For the special case of the disk billiard the map operators
\eref{eq:pkern} can be related directly to the magnetic generalization
of the ${\sf T}$ operator \cite{Bogomolny92}.
Following \cite{Bogomolny92} we define two
operators, ${\sf T}_{\rm S}$ and ${\sf T}_{\rm L}$,
entirely in terms of the actions of a short and long arc,
$\SoL\in\{{\rm S,L}\}$, respectively, with kernels 
\begin{align}
  {\rm t}_{\rm \SoL}(\phi;\phi_0)
  &\defas
  \frac{1}{(2\pi\rmi)^\oh}\,
  \left(
    \frac{\rmd^2 (2\pi\nu\ga_{\rm \SoL})}
    {\rmd\phi\,\rmd \phi_0}(\phi-\phi_0)
  \right)^\oh
  \,\rme^{2\pi\rmi\nu\ga_{\rm \SoL}}
  \PO
\end{align}
Evaluating the mixed second derivatives of the actions, one finds that
they may be stated in a form
\begin{align}
  \frac{\rmd^2 \ga_{\rm S}}{\rmd\phi\,\rmd \phi_0}(\phi-\phi_0)
  &=\,\phantom{-}
  \frac{1}{2\pi}
  \,
  \frac{(\sin(\alpha)\cos(\alpha)+\oh\Gam^2\sin(\phi-\phi_0))^2}
  {\sin(\alpha)\cos(\alpha)}
\\
  \frac{\rmd^2 \ga_{\rm L}}{\rmd\phi\,\rmd \phi_0}(\phi-\phi_0)
  &=\,
  - \frac{1}{2\pi}
  \,
  \frac{(\sin(\alpha)\cos(\alpha)-\oh\Gam^2\sin(\phi-\phi_0))^2}
  {\sin(\alpha)\cos(\alpha)}
\end{align}
which permits the direct comparison with equation \eref{eq:vndisk}.  
It follows that the operators $ {\sf P}_{\rm S}^{\rm int}$ and ${\sf
  P}_{\rm S}^{\rm ext}$ (cf eq \eref{eq:pkern}) are given essentially
in terms of ${\sf T}_{\rm S}$:
\begin{align}
  \label{eq:pdisk1}
  {\rm p}_{\rm S}^{\rm int}(R\phi,R\phi_0)
  &=
  {\rm t}_{\rm S}\left(\phi;\phi_0\right)
  \,
  \frac{b}{R}\,
  \begin{cases}
    \Theta(\phi-\phi_0)&\text{if $\Gam>1$}
    \\
    1 &\text{if $\Gam<1$}
  \end{cases}
\\
  \label{eq:pdisk2}
  {\rm p}_{\rm S}^{\rm ext}(R\phi,R\phi_0)
  &=
  {\rm t}_{\rm S}\left(\phi;\phi_0\right)
  \,
  \frac{b}{R}\,
  \begin{cases}
    \Theta(\phi_0-\phi)&\text{if $\Gam>1$}
    \\
    0 &\text{if $\Gam<1$}
  \end{cases}
\end{align}
They vanish whenever there is no classically allowed
trajectory connecting the initial and the final point in the considered domain.
Similarly, the operators ${\sf P}_{\rm L}^{\rm int}$ and ${\sf P}_{\rm
  L}^{\rm ext}$ are given as restrictions of ${\sf T}_{\rm L}$.
\begin{align}
  \label{eq:pdisk3}
  {\rm p}_{\rm L}^{\rm int}(R\phi,R\phi_0)
  &=
  - {\rm t}_{\rm L}\left(\phi;\phi_0\right)
  \,
  \frac{b}{R}\,
  \begin{cases}
    \Theta(\phi-\phi_0)&\text{if $\Gam>1$}
    \\
    0 &\text{if $\Gam<1$}
  \end{cases}
\\
  \label{eq:pdisk4}
  {\rm p}_{\rm L}^{\rm ext}(R\phi,R\phi_0)
  &=
  - {\rm t}_{\rm L}\left(\phi;\phi_0\right)
  \,
  \frac{b}{R}\,
  \begin{cases}
    \Theta(\phi_0-\phi)&\text{if $\Gam>1$}
    \\
    1 &\text{if $\Gam<1$}
  \end{cases}
\end{align}
Here we assume $|\phi-\phi_0|\leq\pi$ (as throughout this section).

\subsubsection{The explicit trace formula}

To obtain a semiclassical expression for the number counting function
we start by calculating the kernel of the $\n$th power
$({\sf P}^{\rm int})^\n$
at coinciding initial and final point $s_0$.  It is given by a
$(\n-1)$-dimensional integral,
\begin{align}
  \label{eq:pintn}
  \left({\rm p}^{\rm int}\right)^\n(s_0,s_0)
  &=\int
  \prod_{j=1}^\n
  \left[
    \big({\rm p}_{\rm S}^{\rm int}+{\rm p}_{\rm L}^{\rm int}\big)
    (s_j,s_{j-1})
  \right]
  \frac{\rmd s_1 \ldots \rmd s_{\n-1}}{b^{\n-1}}
\CO
\end{align}
with \emph{fixed} $s_\n \equiv s_0$.  This integral may be evaluated by
the stationary phase method.  For the same reason as above (Sect.
\ref{sec:sp}) only the  saddle points contribute which
correspond to a physically allowed trajectory.  However, they are now
required to start and end at the point $s_0$.  Each saddle point is
characterized by the constant angular increment
$\Delta\phi$, the $j$th component given by
\begin{align}
  \phi_j=\phi_0+j\,\Delta\phi \CO\qq j=0,\ldots,\n-1
\PO
\end{align}
For given $\n$ there is a finite number of possible increments 
\begin{align}
\label{eq:Mint}
\Delta\phi\in
  \PSet_{\rm int}^\n
  =
  \begin{cases}
    +2\pi\frac{\m}{\n}; \q \m=1,2,\ldots \m_{\rm max}
    &\text{if $\Gam>1$}
    \\
    \pm 2\pi\frac{\m}{\n}; \q \m=1,2,\ldots \m_{\rm max}
    &\text{if $\Gam<1$.}
  \end{cases}
\end{align}
Here, the second index $\m$ has the meaning of a winding
number.\footnote{%
  We use capital letters for the indices $\n,\m$ in this section to
  avoid confusion with the radial and angular momentum quantum
  numbers, see \eref{eq:xidisksc}.
  } It gives the number of times the trajectory encircles the origin.
The maximum value is given by
\begin{align}
  \label{eq:Mmax}
  \m_{\rm max}=  
  \begin{cases}
       [\arcsin(1/\Gam)\,\n/\pi] &\text{if $\Gam>1$}
       \\
       \q\q[\n/2]  &\text{if $\Gam<1$,}
  \end{cases}
\end{align}
where $[\cdot]$ indicates the integer part.
The stationary phase approximation \eref{eq:stphaseNd} brings about a
($\n-1$)-dimensional matrix of second derivatives. Its determinant is
easily calculated since the difference between adjacent angles is
constant:
\begin{align}
  \label{eq:diskdet}
    \det
    \left(      
      \frac{\partial^2 \sum\ga(\phi_{j+1}-\phi_j)}
      {\partial \phi_k\partial \phi_l}
    \right)_{k,l=1\ldots \n-1}
    \!\!=&
    \,\left(\ga''(\Delta\phi)\right)^{\n-1}
    \,
    \det
    \begin{pmatrix}
      \textstyle
      2&-1&&0
      \\
      \textstyle
      -1&\ddots&\ddots&
      \\
      \textstyle
      &\ddots&\ddots&-1
      \\
      \textstyle
      0&&-1&2
    \end{pmatrix}
    \nnn
    = &
    \,\n\,    \left(\ga''(\Delta\phi)\right)^{\n-1}
\end{align}
The number of negative eigenvalues is $\nu_{\sf A}=0$ or $\nu_{\sf
  A}=\n-1$, respectively, for positive or negative sign of $
\ga''(\Delta\phi)$ (ie, for long or short arcs).

Taking the square-root of \eref{eq:diskdet} cancels all but one of the
prefactors in the integrand of eq \eref{eq:pintn}.  Altogether, 
the
kernel
$ \left({\rm p}_{\rm S}^{\rm int}+{\rm p}_{\rm L}^{\rm
    int}\right)^\n(s_0,s_0) $ is given  by
\begin{align}
  \label{eq:diskn}
  \frac{1}{(2\pi\rmi)^\oh}\,
  \frac{1}{\sqrt{\n}}
  \frac{b}{R}
  \sum_{\Delta\phi\in\PSet_{\rm int}^\n}
    \bigg\{&
    \left|\frac{\rmd^2(2\pi\nu\ga_{\rm S}(\Delta\phi))}{\rmd\phi^2}\right|^\oh
    \rme^{\n 2\pi\rmi\nu\ga_{\rm S}(\Delta\phi)-\rmi\piot(\n-1)}
    \\
    +&
    \left|\frac{\rmd^2(2\pi\nu\ga_{\rm L}(\Delta\phi))}{\rmd\phi^2}\right|^\oh
    \rme^{\n 2\pi\rmi\nu\ga_{\rm L}(\Delta\phi)-\rmi\piot \n}
    \,\Theta(\Gam-1)
    \bigg\}
    \nn
    \PO
\end{align}
It is a sum over all \emph{families} of interior periodic orbits
where each family is represented by the orbit starting at $s_0$.

The $n$th power of the \emph{exterior} operators, $({\sf P}_{\rm
  S}^{\rm ext}+{\sf P}_{\rm L}^{\rm ext})^\n$, assumes the same form
except for the Heaviside function which appears in the short arc term
of the sum.  Naturally, the summation is now over the exterior
periodic orbit families, the respective increments given by the set
\begin{align}
\label{eq:Mext}
  \PSet_{\rm ext}^\n
  =
  \begin{cases}
    -2\pi\frac{\m}{\n}; \q \m=1,2,\ldots \m_{\rm max}
    &\text{if $\Gam>1$}
    \\
    \pm 2\pi\frac{\m}{\n}; \q \m=1,2,\ldots \m_{\rm max}
    &\text{if $\Gam<1$\PO}
  \end{cases}
\end{align}
As the last step in forming the trace $\tr\{({\sf P}^{\rm
  int})^\n\}$ we have to integrate $s_0$.
Since the expression \eref{eq:diskn} does not depend on $s_0$ this
simply adds the factor $2\pi R/b$.

It follows that the fluctuating number function due to the skipping
orbits \eref{eq:Trskipint} assumes the form
\begin{align}
  \label{eq:Ndiskint}
  \N_{\rm osc}^{\rm skip(int)}&
  =
  \left(\frac{2\nu}{\pi}\right)^\oh
  \sum_{\n=2}^\infty\,
  \frac{1}{\n^{3/2}}
  \sum_{\Delta\phi\in\PSet_{\rm int}^\n}
  \bigg\{
  \\
  &\begin{aligned}[t]
    & \frac{\oh\Gam^2\sin(\Delta\phi)+\sin(\alpha)\cos(\alpha)}
    {(\sin(\alpha)\cos(\alpha))^\oh}
    \sin\left(2\pi\nu \n\,\ga_{\rm S}(\Delta\phi)+\n\frac{\pi}{2}+\piof\right)
    \\
    +&
    \frac{\oh\Gam^2\sin(\Delta\phi)-\sin(\alpha)\cos(\alpha)}
    {(\sin(\alpha)\cos(\alpha))^\oh}
    \sin\left(2\pi\nu \n\,\ga_{\rm L}(\Delta\phi)+\n\frac{\pi}{2}-\piof\right)
    \Theta(\Gam-1)
    \bigg\},
  \end{aligned}
  \nn
\end{align}
with $\alpha\equiv\arcsin(\Gam\sin(|\Delta\phi|/2))$.
Analogously, the periodic orbit sum for the exterior problem is given
by
\begin{align}
  \label{eq:Ndiskext}
  \N_{\rm osc}^{\rm skip(ext)}
  &=
  \left(\frac{2\nu}{\pi}\right)^\oh
  \sum_{\n=2}^\infty\,
  \frac{1}{\n^{3/2}}
  \sum_{\Delta\phi\in\PSet_{\rm ext}^\n} \bigg\{
  \\
  &
  \begin{aligned}[t]
    -&
    \frac{\sin(\alpha)\cos(\alpha)+\oh\Gam^2\sin(\Delta\phi)}
    {(\sin(\alpha)\cos(\alpha))^\oh}
    \sin\left(2\pi\nu \n\,\ga_{\rm S}(\Delta\phi)+\n\frac{\pi}{2}+\piof\right)
    \Theta(\Gam-1)
    \\
    +&
    \frac{\sin(\alpha)\cos(\alpha)-\oh\Gam^2\sin(\Delta\phi)}
    {(\sin(\alpha)\cos(\alpha))^\oh}
    \sin\left(2\pi\nu \n\,\ga_{\rm L}(\Delta\phi)+\n\frac{\pi}{2}-\piof\right)
    \bigg\}.
  \end{aligned}
  \nn
\end{align}

\myparagraph{The conventional density of states}

The semiclassical expression for the density of states is obtained by
taking the derivative of the number function with respect to the
energy. In order to compare with the result of Blaschke and Brack, which
is in units of the conventional energy $E$, we have to take the
derivative
\begin{align}
  \label{eq:ddE}
  \frac{\rmd}{\rmd E} =
  \frac{1}{E}
  \left(
  \nu  \frac{\rmd}{\rmd \nu}
  -\oh\Gam  \frac{\rmd}{\rmd \Gam} 
  \right)
  \PO
\end{align}
Applying this differential to \eref{eq:Ndiskint} 
yields the fluctuating part of the density for the interior problem
\begin{align}
  \label{eq:dskipint}
    &d_{\rm osc}^{\rm\, skip(int)}(E) = 
    \frac{1}{E}
    \frac{(2\nu)^{\frac{3}{2}}}{\pi^\oh}
    \sum_{\n=2}^\infty
    \sum_{\Delta\phi\in\PSet_{\rm int}^\n}
    \frac{1}{\sqrt{\n}}
  \nnn
  &\times
  \begin{aligned}[t]
    \bigg\{&
    \alpha\;
    \frac{\oh\Gam^2\sin(\Delta\phi)+\sin(\alpha)\cos(\alpha)}
    {(\sin(\alpha)\cos(\alpha))^\oh}
    \cos\left(2\pi\nu \n\,\ga_{\rm S}(\Delta\phi)+\n\frac{\pi}{2}+\piof\right)
    \\
    +&
    {(\pi-\alpha)}   \,
    \frac{\oh\Gam^2\sin(\Delta\phi)-\sin(\alpha)\cos(\alpha)}
    {(\sin(\alpha)\cos(\alpha))^\oh}
    \cos\left(2\pi\nu \n\,\ga_{\rm L}(\Delta\phi)+\n\frac{\pi}{2}-\piof\right)
     \Theta(\Gam-1) 
    \bigg\}
    \PO
  \end{aligned}
  \nnn
\end{align}
This periodic orbit formula is identical to the result in \cite{BB97a}.
It approximates the quantum spectrum of the interior magnetic disk
only for weak fields $\Gam<1$, when all trajectories are of the
skipping type.  For strong fields, $\Gam>1$, complete cyclotron orbits
occur in the interior.  One might wish to include the latter ``by hand''
into the periodic orbit sum.
However, it was shown in \cite{BB97a} that energies close to the
Landau levels cannot be reproduced this way.  Rather than trying to
refine the semiclassical approximation, we shall define a new spectral
density of edge states below which will resolve the problem of the bulk
contributions.

\subsection{The separable case}
\label{sec:disksep}

We proceed to quantize the disk billiard for a second time
-- now using the separability of the quantum problem in a specific
gauge.  This way closed expressions for the spectral functions may be
obtained, which yield explicit formulas for important quantities, such
as the magnetization.  Moreover,
by formulating the relation
to the periodic orbit formula derived in the preceeding section
we can examine the effect of general boundary conditions on the trace
formula.

\subsubsection{The disk billiard revisited}
\label{ssec:diskseprev}

The magnetic disk turns into a separable problem if we choose the
symmetric gauge, $\Chi=0$, and place the center of the disk at the
origin. In this case the canonical angular momentum $L$ is conserved
and the eigenstates are characterized by the quantum number
\begin{align}
\label{eq:defmdisk}
m=\frac{L}{\hbar}=\frac{c^2-\rho^2}{b^2}
\PO
\end{align}
In the second part of \eref{eq:defmdisk} we state the scaled
angular momentum in terms of the radial distance $c$ of the center of
motion, cf
\eref{eq:Aconvsym}.
Along with the cyclotron radius $\rho$ the latter determines whether
the classical motion is of the skipping type. This is the case for
$R-\rho<c<R+\rho$. Hence, a quantum state (of energy $\nu$)
corresponds to classically skipping motion if its angular momentum
quantum number $m$ is bounded from above and below by
\begin{align}
   \label{eq:minmax}
   m_{\rm max} &= \Rt^2+2\sqrt{\nu}\Rt \intertext{and}
   \tag{\ref{eq:minmax}a}
   \label{eq:minmaxa}
   m_{\rm min} &= \max\big(\Rt^2-2\sqrt{\nu}\Rt,-\nu\big) \CO
\end{align}
respectively. Here, the scaled radius $\Rt\defas R/b$ enters as the
only external parameter.

We start with the traditional
Bohr-Sommerfeld quantization method and proceed to discuss its
relation to the periodic orbit formula of Sect.  \ref{sec:trdisk}.
The exact quantization in terms of special functions
is discussed in Appendix \ref{app:diskexact}.

\subsubsection{Semiclassical  quantization}
\label{sssec:scdisk}

Using polar coordinates $(r,\vartheta)$, the ansatz
\begin{gather}
  \psi(r,\vartheta) = \frac{\phi(r/b)}{\sqrt{r/b}}\,\rme^{\rmi m
    \vartheta}
\end{gather}
transforms equation \eref{eq:Schreq2}
into the form of a one-dimensional Schr\"odinger equation for the
radial function $\phi(\rt)$.
\begin{gather}
  -\frac{1}{4} \phi''(\rt) +\left( \frac{1}{4} \frac{(\rt^2-m)^2-
      \frac{1}{4}}{\rt^2} -\nu \right) \phi(\rt) =0
\end{gather}
It may be solved to leading order in $b^2$ using the standard WKB
technique, see eg \cite{LaLi3,BM72}.

\myparagraph{The semiclassical wave function}

It follows that in the energetically allowed region the resulting
semiclassical wave function has the form
\begin{gather}
\label{eq:psidiscsc}
\psi^{\rm (sc)}(r,\vartheta)= \mathcal{N}_{\rm disk}\,
\frac{\cos\big(\Phi^{\rm int/ext}_{\rm
    disk}(\nu,m,\frac{r}{b})-\piof\big)}
{\big(4\nu\frac{r}{b}-\big((\frac{r}{b})^2-m\big)^2\big)^\frac{1}{4}}
\,\rme^{\rmi m \vartheta}
\PO
\end{gather}
Here, the phases $\Phi^{\rm int}_{\rm disk}$ and $\Phi^{\rm ext}_{\rm disk}$
are obtained by an integration starting at the interior and exterior
classical turning points of the radial motion, respectively.
\begin{align}
   \label{eq:Phiint}
  \Phi^{\rm int}_{\rm disk}(\nu,m,\rt) &= \oh
  \sqrt{4\nu\rt^2-(\rt^2-m)^2} -(\nu+\frac{m}{2}) \arctan\!  \left(
    \frac{2\nu+m-\rt^2}{\sqrt{4\nu\rt^2-(\rt^2-m)^2}} \right) \nnn
  &\phantom{=}- \frac{m}{2} \arctan\!  \left(
    \frac{(2\nu+m)\rt^2-m^2}{m\sqrt{4\nu\rt^2-(\rt^2-m)^2}} \right)
  +\piot\left(\nu+\frac{m-|m|}{2}\right)
\intertext{and} 
   \label{eq:Phiext}
   \Phi^{\rm ext}_{\rm disk}(\nu,m,\rt)
   &=\pi\left(\nu+\frac{m-|m|}{2}\right)- \Phi^{\rm int}_{\rm
     disk}(\nu,m,\rt)
\PO
\end{align}
As for the normalization factor $\mathcal{N}_{\rm disk}$,
we find \cite{LaLi3}
\begin{align}
  \label{eq:normalisation}
  (\mathcal{N}_{\rm disk})^{-2} &\defas \frac{\pi^2}{4} \mp
  \piot\arctan\!  \left(
    \frac{2\nu+m-\Rt^2}{\sqrt{4\nu\Rt^2-(\Rt^2-m)^2}} \right) \CO
\end{align}
where again the upper sign stands for the interior problem.

\myparagraph{A spectral function}

Allowing for general boundary conditions \eref{eq:bcond} at the disk
radius $r=R$, we obtain the quantization condition
\begin{align}
  \label{eq:scqc}
  &\pm \cot\big(\Phi^{\rm int/ext}_{\rm disk}(\nu,m,\Rt)-\piof\big) \nnn
  &= - \frac{
    (\pm\Lambda)\,\big({4\nu\Rt^2-(\Rt^2-m)^2}\big)^\frac{3}{2}}
  {2\sqrt{\nu}\Rt\big({4\nu\Rt^2-(\Rt^2-m)^2}\big)+
    (\pm\Lambda)\Rt^2\big(2\nu+m-\Rt^2\big)} 
  \PO
\end{align}
The boundary condition enters on the right side through the
dimensionless\footnote{%
  The dimensionless mixing parameter \eref{eq:Lambdadef} is introduced
  for convenience. Strictly, it is not an independent variable but
  should be replaced by $2\sqrt{\nu}\lambda/b$ everywhere (to avoid
  energy dependent boundary conditions). This distinction does not
  matter, ultimately, since we are only interested in the derivatives
  at $\Lambda=0$, see \eref{eq:dnudLdisksc}, \eref{eq:defweight}.}
mixing parameter $\Lambda$ \eref{eq:Lambdadef} which vanishes for
Dirichlet boundary conditions.  In order to transform the dependence
on the boundary condition into a \emph{phase shift} $\pshift$, we
define
\begin{align}
\label{eq:pshiftdisk}
  \pshift(\nu,m,\Rt)\defas\arctan\!  \left( \frac{(\pm\Lambda)\,
      \big({4\nu\Rt^2-(\Rt^2-m)^2}\big)^\frac{3}{2}}
    {2\sqrt{\nu}\Rt\big({4\nu\Rt^2-(\Rt^2-m)^2}\big)
      +(\pm\Lambda)\,\Rt^2\big(2\nu+m-\Rt^2\big)} \right)
\end{align}
The semiclassical quantization condition \eref{eq:scqc} is then
readily brought into a form,
\begin{align}
\label{eq:cossec}   \cos\Big(\Phi^{\rm int/ext}_{\rm disk}(\nu,m,\Rt)
  \mp\pshift(\nu,m,\Rt)-\piof\Big)=0
 \CO
\end{align}
which permits a spectral function $\xi$ to be written in terms of two
quantum numbers, the number of radial nodes $n$, and the angular
momentum $m$,
\begin{align}
\label{eq:xidisksc}
\xi_{\rm disk}^{\rm (sc)} \Big(\nu;n,m,\Lambda,\frac{R}{b}\Big)
\defas \Phi^{\rm int/ext}_{\rm disk}\Big(\nu,m,\frac{R}{b}\Big) \mp
\pshift\Big(\nu,m,\frac{R}{b}\Big) -\Big(n+\frac{3}{4}\Big)\pi \CO
\end{align}
with $n\in\Nnull, m_{\rm min}\le m \le m_{\rm max}$
\eref{eq:minmax}.  Its zero in $\nu$ yields the semiclassical energy
of a state with given radial and angular quantum numbers $n$ and $m$.
Although the energies are defined implicitly by \eref{eq:xidisksc},
the spectral function yields explicit formulas for the infinitesimal
change of the energies as an external parameter is varied.
For the derivative of the energy with respect to the boundary
mixing parameter at Dirichlet boundary conditions ($\Lambda=0$) we
obtain
\begin{align}
\label{eq:dnudLdisksc}
  \left.\frac{\rmd \nu}{\rmd \Lambda}\right|_{\Lambda=0} = -
  \,\frac{\ds\;\frac{\rmd}{\rmd\Lambda}\xi_{\rm disk}^{\rm (sc)}\;}
  {\ds\frac{\rmd}{\rmd\nu}\xi_{\rm disk}^{\rm (sc)}} &=
  \frac{\ds\sqrt{1-\frac{(\Rt^2-m)^2}{4\nu\Rt^2}}} {\ds
    \piot\mp\arctan\!\left(
      \frac{2\nu+m-\Rt^2}{\sqrt{4\nu\Rt^2-(\Rt^2-m)^2}} \right)} \PO
\end{align}
This short formula is further compressed below and needed soon.

\myparagraph{The magnetic moment}

It was shown in Section~\ref{sec:mag} that the scaled magnetic moment
of a quantum state in the magnetic billiard is 
determined by the derivative of its energy with respect to the
magnetic length,
see  \eref{eq:mag}.  From the semiclassical spectral function
\eref{eq:xidisksc} we find
\begin{align}
\label{eq:dnudbdisksc}
  b^2\,\frac{\rmd \nu}{\rmd b^2} = -\, \frac{\ds b^2\,\frac{\rmd}{\rmd
      b^2}\xi_{\rm disk}^{\rm (sc)}\;}
  {\ds\frac{\rmd}{\rmd\nu}\xi_{\rm disk}^{\rm (sc)}} &= \pm\oh\,
  \frac{\ds \sqrt{4\nu\Rt^2-(\Rt^2-m)^2}} {\ds \piot\mp\arctan\!\left(
      \frac{2\nu+m-\Rt^2}{\sqrt{4\nu\Rt^2-(\Rt^2-m)^2}} \right)} \PO
\end{align}
Alternatively, the expectation value may be calculated directly, using
the semiclassical wave function \eref{eq:psidiscsc}. We obtain indeed,
after lengthy transformations,
\begin{align}
  &\oh\bra{\psi^{\rm (sc)}}(\rvect\times\vvect)_{\rm
    sym}\ket{\psi^{\rm (sc)}} = \int\!  \Im\Big[ \psi^*_{\rm
    (sc)}(\partial_\vartheta-\rmi\,\rt^2)\psi^{\rm (sc)}\Big]
  \,\rmd\vartheta\, \rt\, \rmd\rt \nnn
  &= \pm\oh\, \frac{\ds \sqrt{4\nu\Rt^2-(\Rt^2-m)^2}} {\ds
    \piot\mp\arctan\!\left(
      \frac{2\nu+m-\Rt^2}{\sqrt{4\nu\Rt^2-(\Rt^2-m)^2}} \right)}
  \,-\,\nu \CO
\end{align}
in agreement with \eref{eq:mag}. In the above radial integration
(which is limited by the disk radius and the interior or exterior
turning point, respectively), the strongly fluctuating $\cos^2$-term
was replaced by its mean.
The fact that the exact relation \eref{eq:mag}
is reproduced shows that this approximation is consistent with the
semiclassical one.

\myparagraph{The bulk states}

States with angular momenta beyond the bounds given by
\eref{eq:minmax} are not included in the spectral function
\eref{eq:xidisksc}. Classically, they correspond to cyclotron motion.
The semiclassical energies of these bulk states are determined by the
condition that the two wave functions \eref{eq:psidiscsc} defined from
the interior and exterior turning points must match.  They are given
by the Landau energies $\nu=n+\oh$, and the wave functions are readily
shown to exhibit a magnetic moment of $-\nu$.

However, the exact quantization, which is 
given in Appendix
\ref{app:diskexact}, does not distinguish between edge and
bulk states and the  bulk energies
exhibit  deviations from the Landau energies.  An asymptotic
treatment of these exponential corrections to the bulk energy is given
in Sect.  \ref{sec:bulkasymp}.

\subsubsection{Relation to the periodic orbit formula}
\label{sec:rpo}

In the preceeding section \ref{sssec:scdisk}
the semiclassical quantization was carried out according to the
 traditional Bohr-Sommerfeld rule for separable systems.  It is based
 on the \emph{quantizing tori}, ie, those invariant manifolds in phase
 space whose \emph{scaled} actions are integers.  This should be
 contrasted to the periodic orbit formula for the magnetic disk
 derived in Sect. \ref{sec:trdisk}. The latter is a sum over the
\emph{rational tori}, whose \emph{classical} frequencies are
commensurate \cite{BT77}.
In order to sketch how the trace formula is connected to the
Bohr-Sommerfeld quantization we follow the work of Berry and Tabor
\cite{BT76} who derived the trace formula for general integrable
systems.  In particular, this permits us to show how the trace formula
is modified if one allows for general boundary conditions
\eref{eq:bcond}.

The semiclassical spectrum is given by the energies $\nu(n,m)$, which
are implicitly defined as the roots of $\xi_{\rm disk}^{\rm (sc)}$
\eref{eq:xidisksc}.  We may write the spectral density as a sum over
the two quantum numbers,
\begin{align}
  d(\nu_0) 
  &= \sum_{n,m} \delta\big(\nu_0-\nu(n,m)\big) 
  \nnn
  &= 
  \overline{d}(\nu_0)+ {\sum_{N,M=-\infty}^\infty}  \int
  \rme^{2\pi\rmi(Nn+Mm)} \delta\big(\nu(n,m)- \nu_0\big) \,\rmd n\,
  \rmd m 
  \nnn 
\label{eq:ddd}
  &= 
  \overline{d}(\nu_0)+ \sum_{N,M=-\infty}^\infty\int
  \Big|\frac{\rmd n}{\rmd \nu}\Big|\,
  \rme^{2\pi\rmi(Nn(\nu_0,m)+Mm)}\, \rmd m
\CO
\end{align}
where the Poisson summation formula (eg \cite{Titchmarsch48}) was
employed to transform the sum into an integral. (Boundary corrections
which are to higher order in $\nu$ are neglected.) The sum excludes
the term with $N=M=0$, which yields the mean density $\overline{d}$.
Upon integrating $n$ the $\delta$-function selects the real valued
``number'' of radial nodes which is known explicitely from above.
\begin{align}
  n(\nu_0,m)=\frac{1}{\pi} \Big( \Phi^{\rm int/ext}_{\rm
    disk}(\nu_0,m,\Rt) \mp \pshift(\nu,m,\Rt) -\frac{3\pi}{4} \Big)
\end{align}
We evaluate the remaining integral in \eref{eq:ddd} by the
stationary phase approximation.  The phase shift $\pshift$ should be
neglected in the saddle point condition
\begin{align}
  \label{eq:spcdisk}
  -2\frac{\rmd}{\rmd m}  \big[\Phi^{\rm int/ext}_{\rm disk}\mp \pshift\big]
  \,\stackrel{!}{=} \,
   2\pi \frac{M}{N}
  \equiv\Delta\phi
\end{align}
since $\Phi$ is of order $\nu$ (while $\pshift$ is of order $1$). A
detailed calculation shows, that the angles $\Delta\phi$ selected by
\eref{eq:spcdisk} are indeed given by the sets $\PSet_{\rm
  int/ext}^N$ defined in \eref{eq:Mint} and \eref{eq:Mext} (modulo
$2\pi$).  It is now convenient to characterize the corresponding
skipping trajectories by the signed sine of the angle of incidence
\begin{align}
\label{eq:epsilondef}
  \epsilon
  \defas 
  \frac{c^2-R^2-\rho^2}{2R\rho}
  =
  \nvec\times\vvech
\CO
\end{align}
such that the former quantum number $m$ is  given by the real value
\begin{align}
  m=\Rt^2+ 2\sqrt{\nu}\Rt\,\epsilon
\PO
\end{align}
One finds, after a lengthy calculation, that
\begin{align}
  2 \Phi^{\rm int/ext}_{\rm disk} + \frac{M}{N} 2\pi m
  &=
 \pi\nu+2\nu\arcsin(\sigma)-2\nu\sigma\sqrt{1-\sigma^2}+\Rt^2\sin(\Delta\phi)
\nnn
 &= 2\pi\nu\, \ga
\end{align}
with $\sigma=\mp(\epsilon R+\rho)/c$ defined in \eref{eq:sigmadef},
and ``$\ga$'' the geometric action \eref{eq:ga} of one arc.
Transforming the summation in \eref{eq:ddd} to positive $N$ we obtain,
observing \eref{eq:stphase1d},
\begin{align}
\label{eq:dskipdisk2}
  \dosc^{\rm skip}(\nu_0) =&\, \frac{2}{\sqrt{\pi}}
  \!\!\!\sum_{\begin{smallmatrix}N\in\Natural,M\in\Z: \\
      \Delta\phi=2\pi\frac{M}{N}\in\PSet^N_{\rm int/ext}
\end{smallmatrix}}
\!\!\!\frac{1}{N^\oh}\,
\frac{
  \frac{\rmd}{\rmd\nu}\Phi^{\rm int/ext}_{\rm disk}
  } 
{
  \left| \frac{\rmd^2}{\rmd m^2}\Phi^{\rm int/ext}_{\rm disk} \right|^\oh
}
\\
&\times\cos\Big(
  N 2\pi\nu a + N \piot 
  +\piof\sgn\big(N\partial_m^2\Phi^{\rm int/ext}_{\rm
    disk}\big) 
 \mp 2 N \pshift 
\Big)
\PO
\nn
\end{align}
We note the derivatives 
\begin{align}
  \frac{\rmd^2}{\rmd m^2} \Phi^{\rm int/ext}_{\rm disk}
  &= -\frac{1}{2\Rt\tilde{c}} 
  \,\frac{\sigma}{\sqrt{1-\epsilon^2}}
\intertext{and}
  \frac{\rmd}{\rmd \nu} \Phi^{\rm int/ext}_{\rm disk}
  &= 
  \piot+\arcsin(\sigma)
  = \oh\, \frac{\rmd}{\rmd \nu}  \big(2\pi\nu a\big)
\PO
\end{align}
The last equality permits to integrate the spectral density
immediately. It yields the oscillatory part of the number counting
function,
\begin{align}
\label{eq:Noscdisk2}
  \Nosc^{\rm skip}(\nu_0) =&\, \Big(\frac{2\nu}{\pi}\Big)^\oh
\sum_{N=2}^\infty\sum_{\Delta\phi\in\PSet^N}
\frac{1}{N^\frac{3}{2}}\,
\frac{\left(\frac{R c}{\rho^2}\, 
\sqrt{1-\epsilon^2}\sqrt{1-\sigma^2}\right)^\oh}
{\left| \sigma\sqrt{1-\sigma^2} \right|^\oh}
\\
&\times\sin\Big(
  2\pi\nu a N +  \piot N
  +\piof\sgn(\sigma)
 \mp 2 N \pshift 
\Big)
\CO
\nn
\end{align}
which may be compared to the trace formulas \eref{eq:Ndiskint} and
\eref{eq:Ndiskext} obtained from the boundary integral equations.  The
agreement of the prefactors follows after a tedious discrimination of
the various cases (interior/exterior, short/long arcs, and
$R\gtrless\rho$.)
As the only difference compared to the Dirichlet result of Section
\ref{sec:trdisk} we observe the non-vanishing phase factor $\mp
2N\pshift$ for finite $\Lambda$.

\subsubsection*{The effect of general boundary conditions}

This result suggests that, compared to Dirichlet boundary conditions, the only
effect of a finite mixing parameter is the appearance of an additional
\emph{phase shift} at every point of reflection,
\begin{align}
\label{eq:ps}
  \mp 2\pshift
  &=-2 \arctan\!\left( \Lambda\, \frac{\sqrt{1-\epsilon^2}} {\ds
      1\pm\Lambda\,\frac{1}{4\nu}\,
      \frac{\rho/R+\epsilon}{1-\epsilon^2}
      } \right)
\\[1ex]
\tag{\ref{eq:ps}a}
\label{eq:pssl}
  &=-2 \arctan\!\left( \Lambda\, \sqrt{1-\epsilon^2} \right)
  + \ \Or(\Lambda^2)\CO
  \qq\text{as $\Lambda\to 0$.}
\end{align}
Here, we stated \eref{eq:pshiftdisk} in terms of the geometry of the
periodic orbit \eref{eq:epsilondef} and of $\nu$.
One might be tempted to ``generalize'' the result \eref{eq:ps} to
arbitrarily shaped billiards, by replacing the disk radius $R$ by the
radius of curvature at the point of reflection.
However, the phase shift at a point of zero curvature (which is given
in appendix \ref{app:line}) is not reproduced correctly this way. Only
the limiting expression \eref{eq:pssl} for small $\Lambda$ matches
with its zero curvature analogue.  The latter is determined merely by
the (unsigned) angle of incidence
with respect to the normal at
the point of reflection,
\begin{align}
  \label{eq:theta}
  \sqrt{1-\epsilon^2}=
   |\nvec\,\vvech| 
\PO
\end{align}
Its form coincides with the non-magnetic result \cite{SPSUS95}.  This
generality suggests that at small $\Lambda$ any billiard exhibits the
additional phase \eref{eq:pssl} at the points of reflection.  All what
will be needed below is this dependence to first order in $\Lambda$.
It shows up in the derivative \eref{eq:dnudLdisksc} which we can now
rewrite in terms of the geometric quantities $\sigma$ and $\epsilon$,
see \eref{eq:sigmadef} and \eref{eq:epsilondef}, describing the length
of the arc and the angle of incidence, respectively.  This way the
dependence of the energy on $\Lambda$ assumes a particularly simple
form,
\begin{align}
\label{eq:dnudLdisksc2}
  \left.\frac{\rmd \nu}{\rmd \Lambda}\right|_{\Lambda=0} 
  &= 
  \frac{\sqrt{1-\epsilon^2}}{\piot+\arcsin(\sigma)}
  \PO
\end{align}
\section{A spectral measure for edge states}
\label{chap:edge}

In the following, two different quantitative definitions for 
edge states are introduced.
We discuss their relation, the asymptotically smooth form of the
corresponding spectral densities and their semiclassical
interpretation.

\subsection{Bulk states and edge states}

In Chapter \ref{chap:boundary} we discussed the existence of two types
of states in the spectra of magnetic billiards. The wave functions of
a few typical representatives were given in Chapter \ref{chap:numres},
in Figs. \ref{fig:skittle} -- \ref{fig:ellipseext}.  Observing these
images one might think that it is an easy task to separate the
spectrum into two disjunct sets, edge states and bulk states,
respectively --- similar to the classical trajectories which are
either skipping or cyclotron orbits.  However, in general there is no
way to perform such a strict partitioning.  Rather, a general wave
function may share some features of both, edge states and bulk states,
to a certain degree and there is a gradual transition taking place
between the characteristics of the two types of states.

On the other hand, there is a clear need for an objective way to
separate the edge from the bulk contributions in the spectrum.  Bulk
states are very uninteresting. They do not contribute to transport and
tend to accumulate in the vicinity of Landau levels.  Moreover, their
number often dominates the spectrum.  In the exterior a mean density
of states cannot be defined (as a derivative of a mean counting
function) due to the infinite number of bulk states showing up in the
vicinity of each Landau level.  As a consequence, the oscillatory part
of the spectral density cannot be extracted --- which seems to impede any
statistical or semiclassical analysis of the exterior problem.  Also
in the interior the accumulation of bulk states severely complicates
the analysis of the spectrum in terms of the classical skipping
motion.

To the best of our knowledge, no general and objective criterion for
what constitutes an edge state has been proposed so far.\footnote{In
  \cite{AANS98} Akkermans \etal\ propose special ``chiral'' boundary
  conditions. In the case of a disk billiard (and only there) this
  leads to a gap in the level diagram which may be interpreted as a
  separation into edge and bulk states. A disadvantage of this
  approach is that the obtained spectrum is unrelated to the standard
  Dirichlet spectrum.}  Clearly, any reasonable definition must take
into account the fact that there exist transitional states between
pure edge and pure bulk states \cite{HS01a,HS02a}.
We illustrate this gradual transition by referring to Figure
\ref{fig:lambdadyn} on page \pageref{fig:lambdadyn}. It displays an
{exterior} spectrum as a function of the boundary mixing parameter
$\Lambda$ \eref{eq:Lambdadef}.  In the level diagram one observes that
the infinitely many states which accumulate near the Landau levels are
hardly affected by changes of the boundary condition.  These are
\emph{bulk states}.  The extreme insensitivity of their energies with
respect to $\Lambda$ is explained by the fact that bulk wave functions
are not localized at the the boundary.  They approach it with an
exponentially damped tail giving rise to exponentially small energy
shifts. (This will be discussed quantitatively in
Sect.~\ref{sec:bulkasymp}.)  Other states depend strongly on $\Lambda$
because they are localized at the boundary.  They are naturally
associated with \emph{edge states}.
The fact that states may have a transitional nature can now be seen in
the right part of Figure \ref{fig:lambdadyn}. One observes a sequence
of bulk states which originate from the Landau level and gradually
turn into edge states with a strong dependence on the boundary, ie, a
large slope.  Obviously, it would be inappropriate to split this
sequence at an arbitrary point into two distinct parts.

In order to quantify the notion of edge states we propose to attribute
positive, real valued number $w_n >0$ to each eigenstate $\psi_n$
which gives a measure for the degree to which the state has the
character of an edge state.  This way a \emph{density of edge states}
can be defined which applies in the interior as well as in the
exterior and which consistently accounts for the gradual transition
from edge to bulk.  As compared to the standard density \eref{eq:ddef}
each $\delta$-contribution is \emph{weighted} individually in our
definition:
\begin{align}
  \label{eq:dedgedef}
  \dedge(\nu)
  &:=  
  \sum_{n=1}^{\infty}
  w_n\,
  \delta(\nu-\nu_n)
\PO
\end{align}
Hence, the sum still extends over \emph{all} states in the spectrum,
but for a proper choice of the \emph{quantum weights} $w_n$ the bulk
states are effectively suppressed by their small values.  There are a few
requirements which are naturally imposed on the definition of the quantum
weights $w_n$:
\begin{enumerate}
\renewcommand{\labelenumi}{(\roman{enumi})}
\item The mean density of edge states $\dedgesm$ must be well defined
  in the exterior.
\item The interior and the exterior mean densities should be equal to
  leading order.
\item As a sequence of bulk states approaches a Landau level their
  weights must decrease  at least exponentially.
\item Last not least: The weights should admit a semiclassical
  interpretation which complies with the intuitive notion of edge
  states.
\end{enumerate}
To make the last requirement more specific
consider the semiclassical periodic orbit formula for the oscillatory
part of the density of edge states. It should be a sum over the
interior or  exterior skipping periodic orbits which differs from the
expression \eref{eq:dskiposc} for the standard density at most by a
\emph{classical weight} $w_\po$ attributed to each periodic orbit
contribution $\po$.
\begin{equation}
  \label{eq:dedgeosc}
  d_{\rm edge}^{\rm osc}(\nu) =
  \frac{2}{\pi}
  \sum_\po 
  \frac{w_\po \, \tau_\po }
  {r_\po\,\big|\tr{\rm M}(\po)-2\big|^\oh}
  \cos\Big(
    2\pi\nu \Ga(\po)
    -\pi      n_\po
    -\piot\mu_\po
    \Big)
\PO
\end{equation}
Similar to the quantum weights $w_n$ which must consistently fade out
the bulk contributions, the classical weights $w_\po$ should vanish
gradually for periodic orbits which are increasingly close to being
detached from the boundary.\footnote{Equation \eref{eq:dedgeosc} is
  stated for completely chaotic classical dynamics. In the case of
  integrable dynamics the corresponding periodic orbit sum is
  modified analogously by the same classical weight.}

In the following sections we introduce two different definitions of
the quantum weights $w_n$ satisfying the abovementioned requirements.
The first one, which is very convenient from a mathematical point of
view, is discussed in Sect.~\ref{sec:edgelambda}.  It 
has the property that it renders the leading term of the mean edge
density $\dedgesm$ proportional to the circumference $\Len$ of the
billiard rather than the area $\Area$.

The second definition, given in Sect.~\ref{sec:edgemag}, is 
the most natural choice from a physical point of view.  It has the
property that the interior weights assume unit value for large
cyclotron radii, ie, it approaches the standard density if bulk states
cannot exist in the interior. Consequently, the interior mean edge
state density equals the standard mean density for this second definition.  
The relation between the two different definitions of the quantum
weight is also discussed below.
\subsection{A spectral density based on the boundary conditions}
\label{sec:edgelambda}
Figure~\ref{fig:lambdadyn} suggests that
the \emph{slope} in the level diagram provides a quantitative
criterion for the degree to which a state is of the edge type.
We therefore propose to
weight each Dirichlet eigenstate
$\ket{\psi_n}$ by the derivative of its energy $\nu_n$ with
respect to the boundary mixing parameter at Dirichlet boundary
conditions $\Lambda=0$,
\begin{gather}
  \label{eq:defweight}
  w_n \defas
    \frac {\rmd\nu_n}{\rmd \Lambda}
  \Big|_{\Lambda=0}
  \equiv\,
  \frac{b}{2\sqrt{\nu}}\,
  \frac {\rmd\nu_n}{\rmd \lambda}
  \Big|_{\lambda=0}
\CO
\end{gather}
which is positive valued.
It will be shown in the sequel that this definition complies with the
requirements for a definition of edge states stated above.  In
particular it
admits a semiclassically meaningful interpretation, as discussed in
Sect.~\ref{sec:dedgesc}.  

For the mean density we obtain the simple expression (cf
Sect.~\ref{ssec:mecf})
\begin{gather}
  \label{eq:dedgesm}
  \overline{d}_{\rm edge}(\nu) = \frac{\Len}{2\pi b}\, \nu^{\oh} \mp \oh
  \CO
\end{gather}
where the upper sign stands for the interior problem.  The
leading order term is proportional to the {circumference} $\Len$
of the billiard.  
As argued below, the second order term may be related to the mean
curvature of the billiard boundary (which is positive from the
interior and negative from the exterior).

Before discussing the various asymptotic properties which come along
with the definition \eref{eq:defweight} we present a few examples of edge
spectra of interior and exterior billiards.  They provide a first
indication that the quantum weights succeed to sort out the bulk
contributions consistently.  A more quantitative check of this
assertion will then be given in  Chapter \ref{chap:stat} where we
perform a statistical analysis of the edge spectra.

\begin{figure}[tp]%
  \begin{center}%
    \psfrag{nu}{$\nu_n$}
    \psfrag{wn-int}{\hspace*{-1em}interior $w_n$}
    \psfrag{wn-ext}{\hspace*{-1em}exterior $w_n$}
    \includegraphics[width=\linewidth] 
    {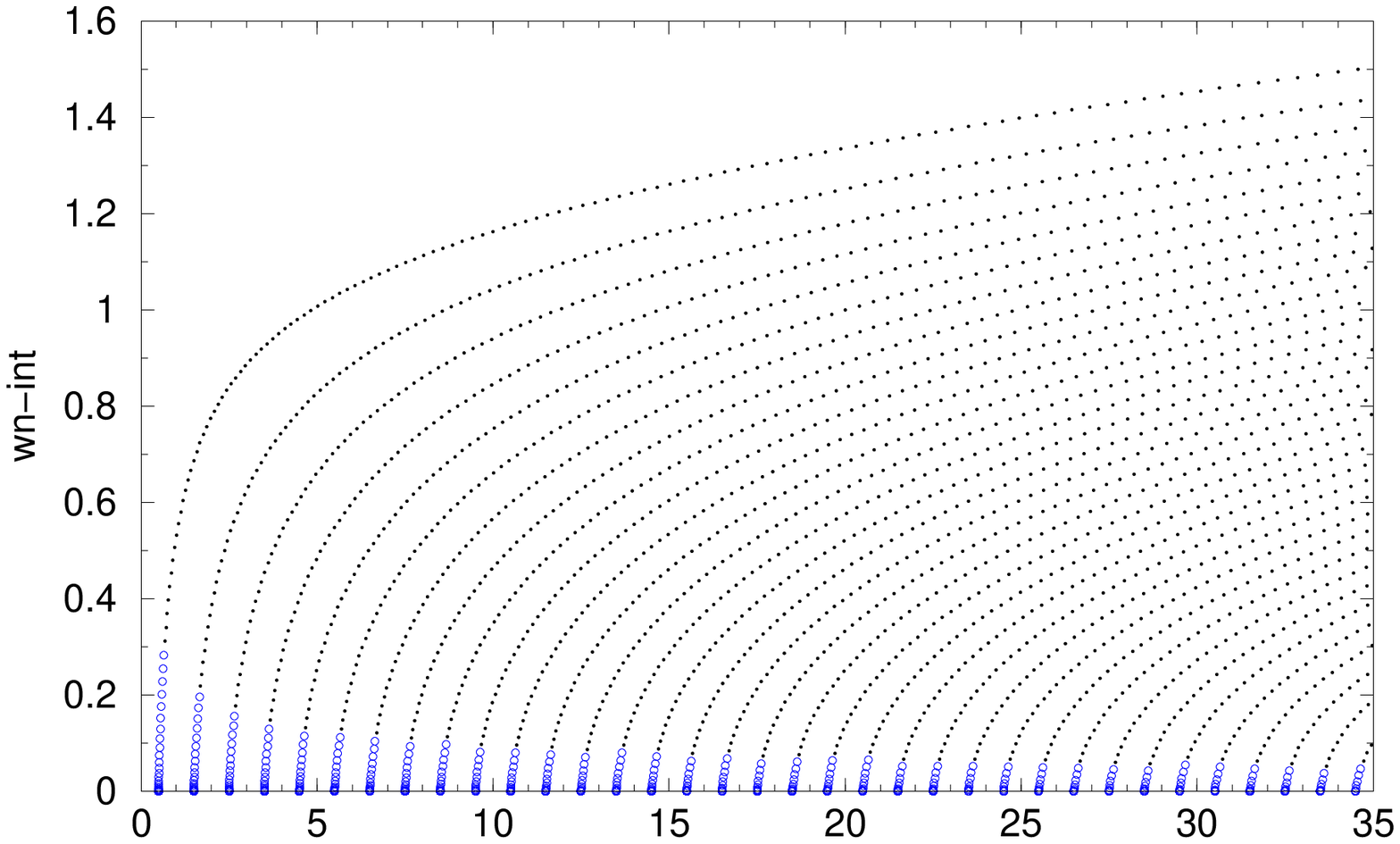}
    \includegraphics[width=\linewidth] 
    {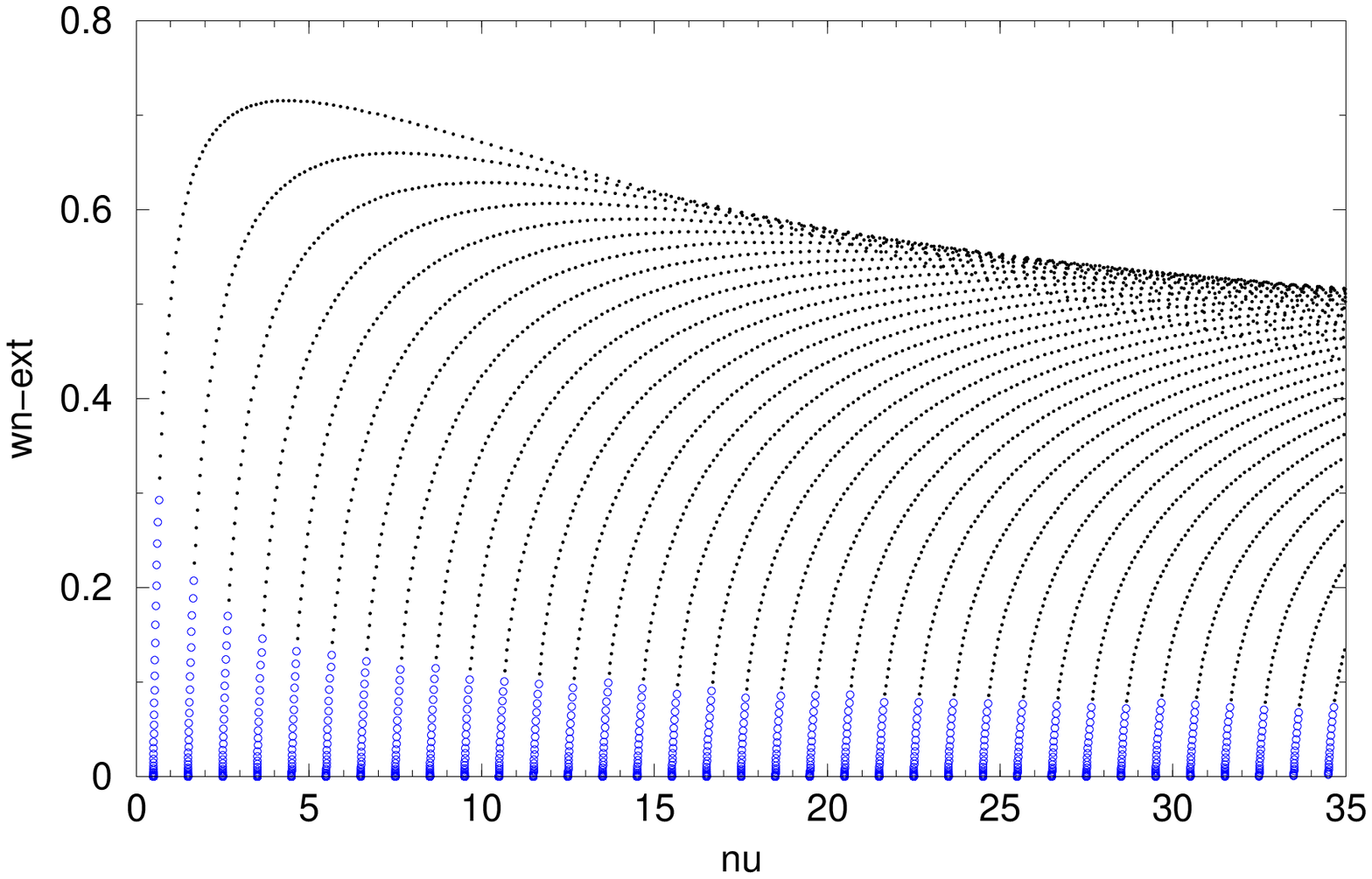}
    \figurecaption{%
      Weighted spectra of the interior (top) and the exterior (bottom)
      magnetic disk (area $\Area=\pi, b=0.1$).  Each dot (open and
      filled) corresponds to an eigenstate $\ket{\psi_n}$ with the
      energy $\nu_n$ given by the abscissa. The ordinate indicates the
      attributed quantum weight $w_n$ defined in eq
      \eref{eq:defweight}.  It serves to distinguish edge states (with
      large $w_n$) from bulk states.  The latter accumulate at the
      Landau levels $\nu=N+\oh, N\in\Natural$, and are characterized
      by vanishingly small weights $w_n$.
      A sequence of transitional states emanates from each Landau
      level and connects with the edge states.  As an alternative
      criterion, the angular momentum quantum number permits to decide
      whether the state corresponds classically to skipping motion
      (full dots) or cyclotron motion (open dots), see text.
}
\label{fig:wsdiskintext}%
\end{center}%
\end{figure}

\begin{figure}[p]%
  \begin{center}%
    \psfrag{nun}{$\nu_n$}
    \psfrag{wn-int}{\hspace*{-1em}interior $w_n$}
    \psfrag{wn-ext}{\hspace*{-1em}exterior $w_n$}
    \includegraphics[width=\linewidth] 
    {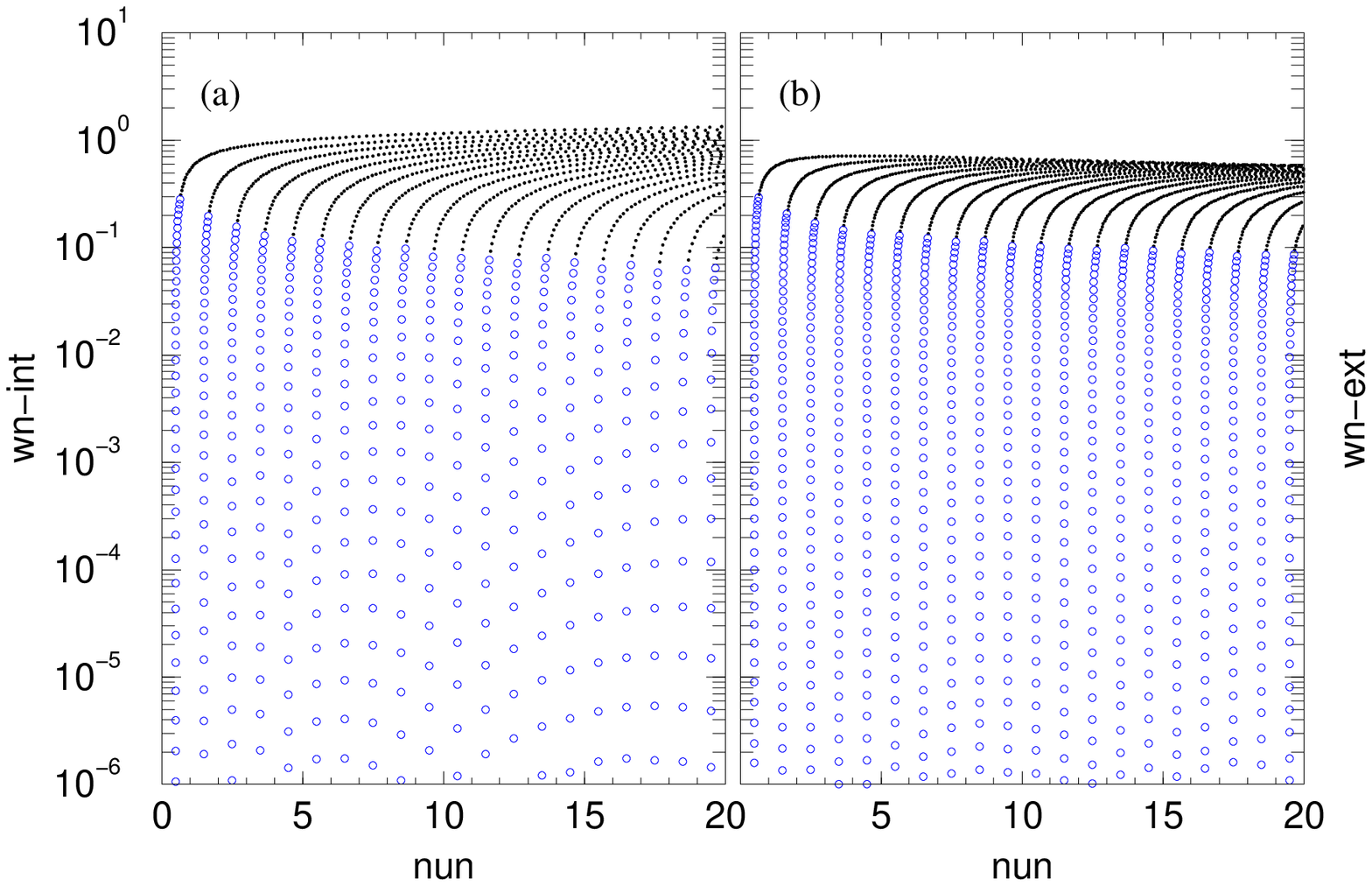}
    \figurecaption{%
      Weighted spectra of (a) the interior and (b) the exterior
      magnetic disk in a semilogarithmic plot (same data as in
      Fig.~\ref{fig:wsdiskintext}). The logarithmic scale highlights the
      states which correspond classically to cyclotron motion (open
      dots). One observes that the the respective weights decrease
      exponentially.  }
\label{fig:wsdiskbothlog} 
\end{center}%
\end{figure}

An edge spectrum $\{(\nu_n,w_n)\}$ consists of the energies $\nu_n$
and the attributed weights $w_n$.  Figure \ref{fig:wsdiskintext} gives
an example of an \emph{interior} (top) and an \emph{exterior} (bottom)
{edge spectrum} at a strong magnetic field.
The spectra belong to a disk billiard of unit radius and were obtained from
eqs \eref{eq:xidisk} and \eref{eq:wdisk}.
In these plots each
point belongs to one eigenstate and indicates the weights versus the
energy.  One observes how the weights  segregate  edge
states with large $w_n$ from the bulk states. The latter
accumulate at the Landau levels $\nu=N+\oh, N\in\Nnull,$ with vanishingly
small weights.  
They are highlighted in Fig.~\ref{fig:wsdiskbothlog} which shows the
same data as Fig.~\ref{fig:wsdiskintext} on a logarithmic scale.
A sequence of bulk states can be seen emanating from each Landau
level and gradually turning into edge states.

Since the disk billiard has a second quantum number we can compare our
characterization of edge states based on weights with a classical
criterion.  As discussed in Sect.~\ref{ssec:diskseprev} a state
corresponds classically to skipping motion if the angular momentum
quantum number lies within the bounds given by \eref{eq:minmax}.
In the Figs.~\ref{fig:wsdiskintext} and \ref{fig:wsdiskbothlog} we
indicate those states with constants of the motion which belong to a
skipping trajectory by a full dot. The others are given by a large
open dot.  One observes that the effective separation produced by the
weights complies with the classical criterion. At the same time it
seems more appropriate to formulate the separation in terms of a
continuous quantity.  This is the more so since a second quantum
number does not exist for shapes other than the disk.

\begin{figure}[p]%
\begin{center}%
    \psfrag{nun}{$\nu_n$}
    \psfrag{wn}{\hspace*{-1em} exterior $w_n$}
    \includegraphics[width=\linewidth] 
    {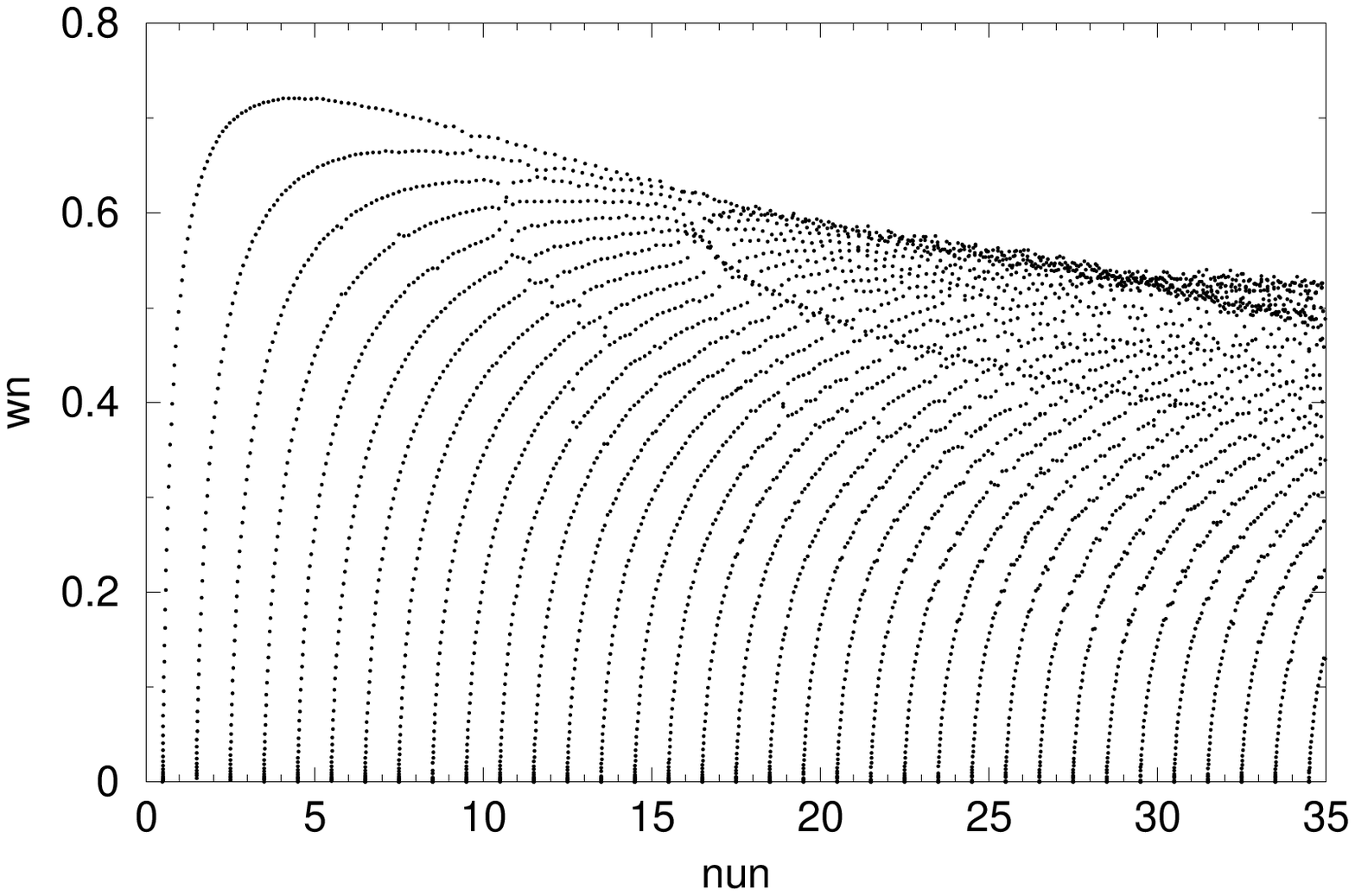}
    \figurecaption{%
      Weighted spectrum of the \emph{exterior} ellipse billiard (with
      eccentricity 0.8, area $\Area=\pi$, magnetic length $b=0.1$).
      It should be compared to the exterior disk,
      Fig.~\ref{fig:wsdiskintext} (bottom).  While the bulk states are
      very similar, one observes that the edge weights no longer lie
      on smooth curves but tend to cluster. These structures can be
      related to the classical (mixed chaotic) phase space.
     }
\label{fig:wspecELext}
\end{center}%
\end{figure}

Figure \ref{fig:wspecELext} shows the {exterior} edge spectrum of an
\emph{ellipse} billiard (which is not integrable in the magnetic
field).  We took the same area $\Area$ and magnetic length $b$ as for
the disk in Fig.~\ref{fig:wsdiskintext}.  Comparing the ellipse
spectrum to the disk one observes that they resemble in their gross
features. In particular, the bulk states behave very similarly.
However, for the ellipse there are additional structures showing up in
the distribution of the weights of edge states. These can be related
to features of the classical (mixed chaotic) phase space, as will be
shown below.

\subsubsection{Edge state counting functions}

Upon integrating the density one obtains the \emph{edge state counting
  function}
\begin{equation}
  \label{eq:Nedgedef}
  \Nedge(\nu)
  :=
  \int^\nu_0\! d_{\rm edge}(\nu') \,\rmd\nu'
  =
  \sum_{n=1}^{\infty}
  w_n\,  \Theta(\nu-\nu_n)
\CO
\end{equation}
which is a \emph{weighted} staircase.  It jumps by $w_n$ at the
corresponding spectral point $\nu_n$. Again, the sum formally includes
the bulk states.  We expect their contribution to be effectively
eliminated by the rapid decay of the weights such that the edge state
counting function
should bear no marks of the Landau levels. According to
\eref{eq:dedgesm} its smooth part
is given by
\begin{equation}
  \label{eq:Nedgesm}
  \Nedgesm(\nu) = \frac{2}{3} \frac{\Len}{2\pi b} \nu^\frac{3}{2}
  \mp \oh \nu
  \q+\Or(1)
\PO
\end{equation}
Note that the leading order exhibits the same functional dependence as
the phase space estimate of the skipping states for the periodic
straight line problem \eref{eq:Nskipsm}.
The only difference is an additional prefactor of
$\oh$.

\begin{figure}[tb]%
  \begin{center}%
    \includegraphics[width=0.9\linewidth] 
    {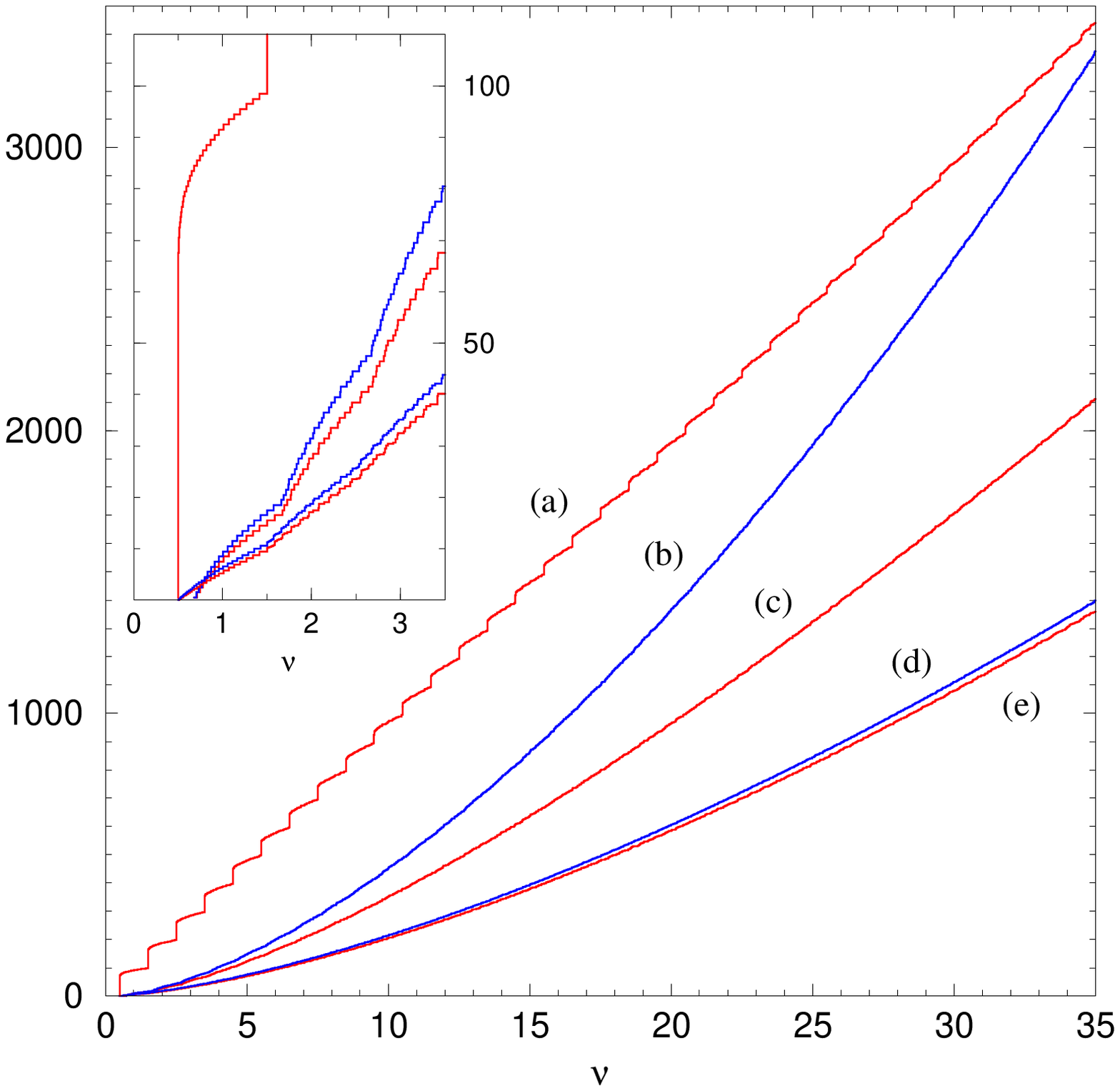}
    \figurecaption{%
      Spectral counting functions for the magnetic disk ($R/b=10$).
      (a) Total number of interior states. (b,c) Number of (b)
      exterior and (c) interior skipping states according to the
      angular momentum criterion. (d,e) Weighted number of edge states
      for the (d) exterior and (e) interior problem.  When averaged
      the curves are well reproduced by the smooth counting functions,
      \eref{eq:Nsmooth}, \eref{eq:Nskipsmde}, \eref{eq:Nskipsmdi}, and
      \eref{eq:Nedgesm}, respectively (not shown).  The inset gives
      the counting functions for the first four Landau levels. The
      small kinks seen in $\N_{\rm skip}$ and $\N_{\rm edge}$ are
      damped away at higher energies.}
\label{fig:Ndisk}%
\end{center}%
\end{figure}

In Figure \ref{fig:Ndisk} we compare various spectral counting
functions obtained from the magnetic disk spectra given in
Fig.~\ref{fig:wsdiskintext}. Curve (a) shows the \emph{total} number
of states in the interior. It exhibits distinct steps at the energies
of the Landau levels. In the exterior a total counting function does
not exist but the angular momentum criterion \eref{eq:minmax} enables
counting the exterior states of the \emph{skipping} type, see curve
(b).  The corresponding number of interior skipping states is
indicated by curve (c). As one expects these two counting functions
hardly exhibit steps at the Landau levels but they show a different
functional dependence (given by eqs \eref{eq:Nskipsmde},
\eref{eq:Nskipsmdi}).  In contrast, the \emph{weighted} exterior and
interior edge state counting functions, curves (d) and (e),
respectively, display the same mean values to leading order. Their
average is reproduced by eq \eref{eq:Nedgesm} and no marks of the
Landau levels are visible.

\begin{figure}[tb]%
  \begin{center}%
    \psfrag{y}{\hspace*{-1em}$\Nedge^{\rm osc}(\nu)$}
    \psfrag{xn}{$\nu$}
    \includegraphics[width=0.9\linewidth] 
    {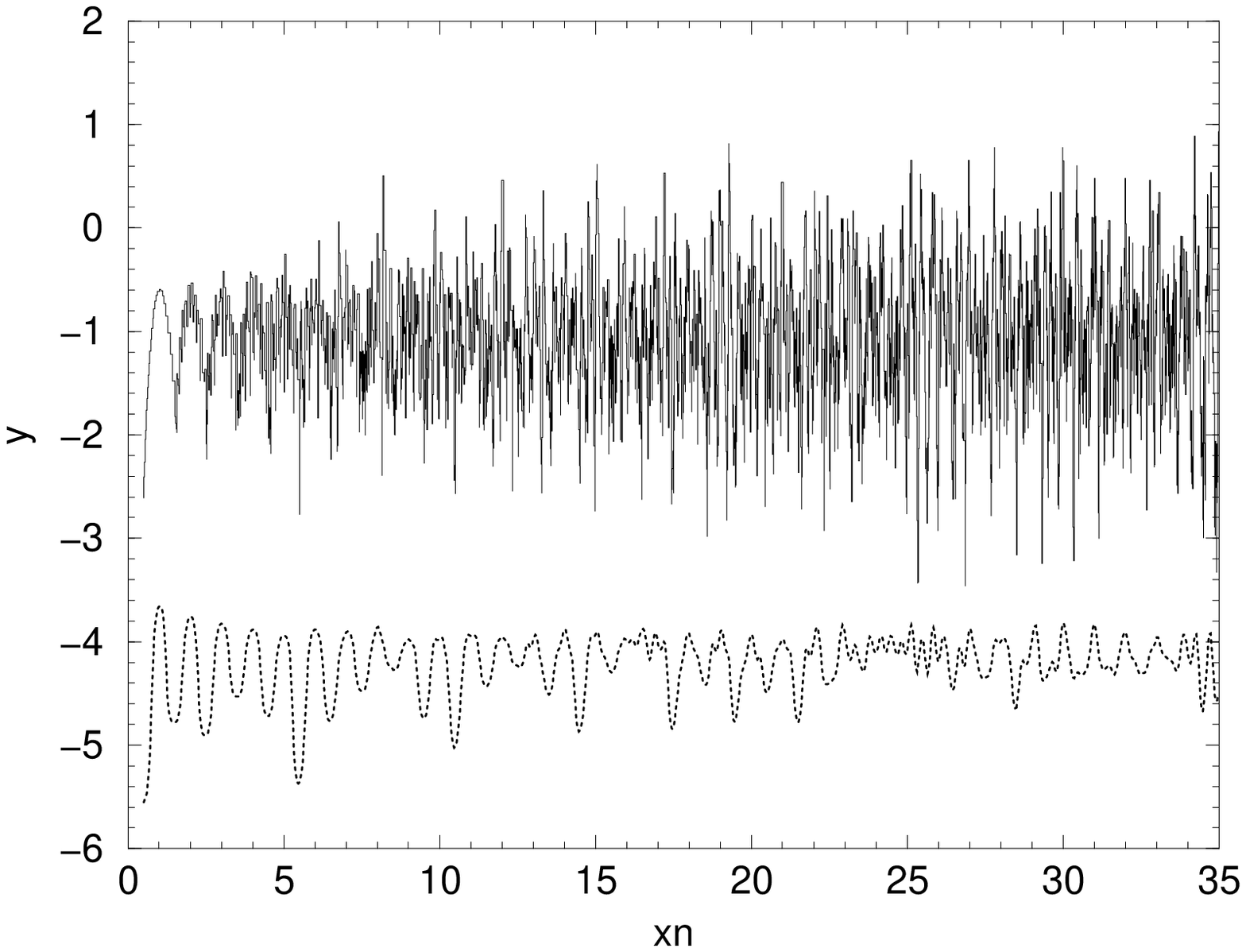}
    \figurecaption{%
      Solid line: Oscillatory part \eref{eq:defNosc} of the edge state
      counting function for the \emph{exterior} magnetic disk
      ($R/b=10$).
      (The unspecified constant part of $\Nedgesm$ was not
      subtracted.)  Dotted line: The oscillatory part convoluted by a
      Gaussian of width $\sigma=0.1$ minus an offset of $3.0$.   }
\label{fig:NedgeDi}%
\end{center}%
\end{figure}

To examine more closely the suppression of the bulk states we plot the
\emph{fluctuating part} of the edge state counting function,
\begin{equation}
  \label{eq:defNosc}
  \N^{\rm osc}_{\rm edge}(\nu) =
  \Nedge(\nu) - \Nsm_{\rm edge}(\nu)
\PO
\end{equation}
Figure \ref{fig:NedgeDi} shows this quantity as obtained from the
exterior spectrum.  One observes that the fluctuating function hardly
exhibits a signature of the Landau levels.  The dotted line depicts $
\N^{\rm osc}_{\rm edge}(\nu)$ after convolution with a narrow Gaussian
which smoothes out the fluctuations.  Its oscillations are due to the
remnant contributions of the bulk states which are slowly damped out
for higher energies.  The fluctuating part of the exterior
\emph{ellipse} spectrum shown in Fig.~\ref{fig:wspecELext} looks very
similar \cite{Hornberger01}.

\subsubsection{The semiclassical density of edge states}
\label{sec:dedgesc}

The standard spectral density \eref{eq:ddef} is given by the
derivative of the number counting function $\N(\nu)$ with respect to
energy. The edge state density \eref{eq:dedgedef} may also be
formally defined as a derivative, now with respect to the boundary
mixing parameter $\Lambda$, at Dirichlet boundary conditions
($\Lambda=0$),
\begin{equation}
  \label{eq:dedgedef2}
  \dedge(\nu)
  = -
  \left .
    \frac {\rmd \N(\nu)}{\rmd \Lambda}
  \right|_{\Lambda=0}
  \equiv\,
  -
  \frac{b}{2\sqrt{\nu}}\,
  \left .
  \frac {\rmd \N(\nu)}{\rmd  \lambda}
  \right|_{\lambda=0}
\PO
\end{equation}
Hence, the periodic orbit expression for the oscillatory part of the edge
state density can be deduced immediately once we have the semiclassical
formula for $\Nosc$ at hand.
For the time being, we restrict ourselves to hyperbolic systems.
Combining the results of the previous chapter
(eqs \eref{eq:Nskiposc}, \eref{eq:pssl}, and \eref{eq:theta}), the
number of states based on the skipping part of phase space is given by
\begin{align}
  \label{eq:Nosc}
  \N_{\rm osc}^{\rm skip}(\nu;\Lambda) =
  \frac{1}{\pi}
  \sum_{\po}&
  \frac{(-)^{n_\po}}{
    r_\po\,\big|\tr{\rm M}(\po)-2\big|^\oh
    }
  \,    
  \\
  \times&
  \sin
  \Big(
    2\pi\nu \Ga(\po)
    -\piot\mu_\po
    -2 \Lambda     \sum_{j=1}^{n_\po}|\nvec_j\vvech_j| 
    \Big)
    \ +\Or(\Lambda^2)
    \nn
\PO
\end{align}
Compared to \eref{eq:Nskiposc} the leading order dependence on
$\Lambda$ is included, as discussed in Sect.~\ref{sec:rpo}. (See
Tables \ref{tab:vnt} and \ref{tab:geo} for the definition of the
various quantities in \eref{eq:Nosc}.)  Since the semiclassical bulk
states do not depend on the boundary condition their contribution
vanishes when taking the derivative.  Using \eref{eq:dedgedef2} one
obtains the semiclassical trace formula for the edge state density at
Dirichlet boundary conditions, 
\begin{equation}
  \label{eq:dedgeosc2}
  d_{\rm edge}^{\rm osc}(\nu) =
  \frac{2}{\pi}
  \sum_\po
  \frac{\sum_{j=1}^{n_\po}|\nvec_j\vvech_j| }
  {r_\po\,\big|\tr{\rm M}(\po)-2\big|^\oh}
  \cos\Big(
    2\pi\nu \Ga(\po)
    -\pi      n_\po
    -\piot\mu_\po
    \Big)
\PO
\end{equation}
This expression should be compared to that of the unweighted density
of states \eref{eq:dskiposc} which was obtained by taking the
derivative with respect to the energy $\nu$.  It exhibits the scaled
time of flight $\tau_\po$  \eref{eq:deftau} in the numerator.
Hence, the periodic orbit sum \eref{eq:dedgeosc2} differs from the
standard spectral density only by the prefactors
\begin{gather}
  \label{eq:defwp}
  w_\po \defas
  \frac{\sum_{j=1}^{n_\po} |\nvec_j\vvech_j| }  { \tau_\po}
  \PO
\end{gather}
They attribute an individual \emph{classical weight} to each skipping
periodic orbit $\po$.  The classical weights are given by the time
averaged value for the normal component of the velocity
$|\nvec\,\vvech|$ at the points of reflection. They vanish for
cyclotron orbits.
Similar to the quantum weights, the $w_\po$ lead to a gradual
transition from edge to bulk contributions.  It is easy to see that in
the limit of a ``grazing'' trajectory of increasingly many short arcs
variations in the curvature of the boundary may be neglected and the
classical weights $w_\po$ approach a constant value.  In the opposite
case of an orbit which is almost detached from the boundary the
weights vanish since the normal components of the velocities approach
zero at a finite time of flight in the denominator of \eref{eq:defwp}.

\begin{figure}
  \begin{center}%
\includegraphics[width=0.9\linewidth]{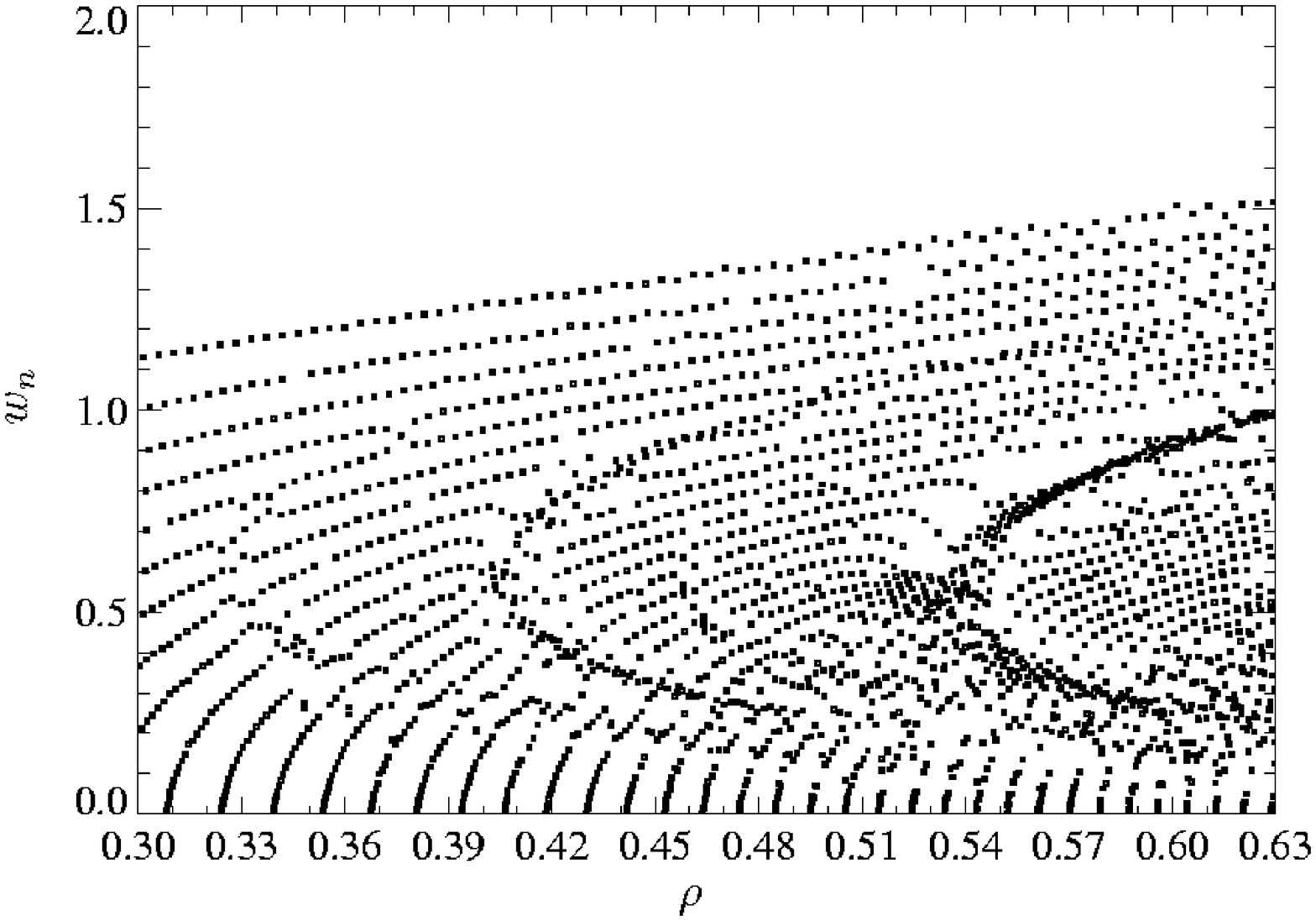}
\includegraphics[width=0.9\linewidth]{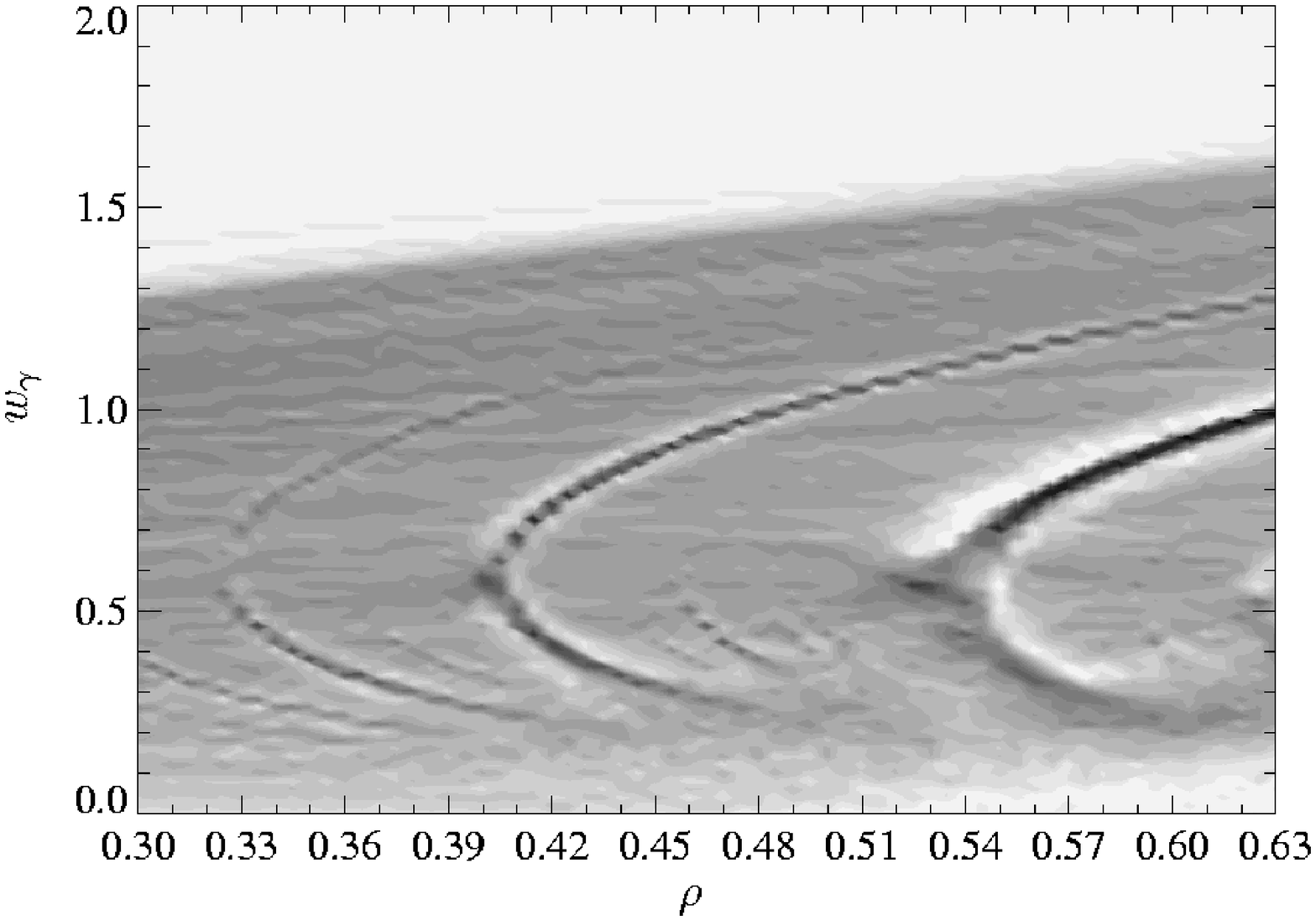}
  \figurecaption{%
    Weighted edge spectrum \eref{eq:dedgedef} (top) and phase space
    distribution of the classical weights \eref{eq:defwp} (bottom) for
    the interior ellipse.  To ease comparison, also the quantum
    spectrum (calculated at constant $b=0.1$, as in Fig.
    \ref{fig:wspecELext}) is given in terms of the classical cyclotron
    radius ( $\rho=b\times\sqrt{\nu}$.)  One observes that the quantum
    weights tend to mimic the structures in the distribution of
    classical weights (which are due to stable islands in phase
    space, cf Fig.~\ref{fig:pport}). [figure quality reduced] }
\label{fig:qmclswt}
\end{center}
\end{figure}

It is instructive to compare the distributions of quantum and
classical weights.  A direct comparison is not possible since the
classical and the quantum weights are associated with different
objects, eigenvalues and periodic orbits, respectively.  In
Fig.~\ref{fig:qmclswt} we compare the distribution of classical
weights to the corresponding weighted quantum spectrum.  The data were
obtained for the interior elliptic billiard, and are given in both
cases as a function of the classical cyclotron radius $\rho$. The
shade in the distribution of classical weights gives the probability
for obtaining a certain weight if the trajectories are chosen randomly
with respect to the invariant measure.  It was approximated
numerically by the histogram over a finite number of trajectories
taken uniformly from phase space.\footnote{In the numerical
  calculation we could use general (non-periodic) trajectories to
  approximate the periodic orbits which are dense in phase space.}
Remarkably, one observes that the characteristic features of both
distributions {coincide}.  This shows that the quantum weights may be
considered the expectation values of an observable which has a
classical limit, ie, they measure a classical property. This holds in
spite of the fact that the $w_n$ are defined in terms of the boundary
condition, which has no classical analogue.

The bifurcating structures seen in Fig.~\ref{fig:qmclswt} are due to
stable periodic orbits.  At the bifurcation points periodic orbits
$\pon$ with a fixed number of reflections $n_\po$ exhibit the smallest
possible cyclotron radius ($n_\po=6$ in the case of the rightmost
structure).  As the cyclotron radius increases, the orbits turn into
pairs with either longer or shorter arcs. (Some of the corresponding
islands of stability in phase may be identified in 
Fig.~\ref{fig:pport}, left column.)

\subsection{Asymptotic properties of edge and bulk states}
\label{sec:asymp}

We proceed to briefly discuss the leading order behavior of the bulk
energies and the smooth part of the edge counting function.  
Both estimates are obtained in the semiclassical limit such that it is
legitimate to substitute the boundary by the straight line with
periodic boundary conditions discussed in Appendix~\ref{app:line}.
The finite mean curvature and variations of the boundary curvature are
expected to appear only as higher order corrections, see the
discussion in Sect.~\ref{sec:acf}.

\subsubsection{Bulk state energies and weights}
\label{sec:bulkasymp}

\begin{figure}[tb]%
  \begin{center}%
    \psfrag{deltanum}{$\Delta\nu_m$}
    \psfrag{wm}{$w_m$}
    \includegraphics[width=0.8\linewidth] 
    {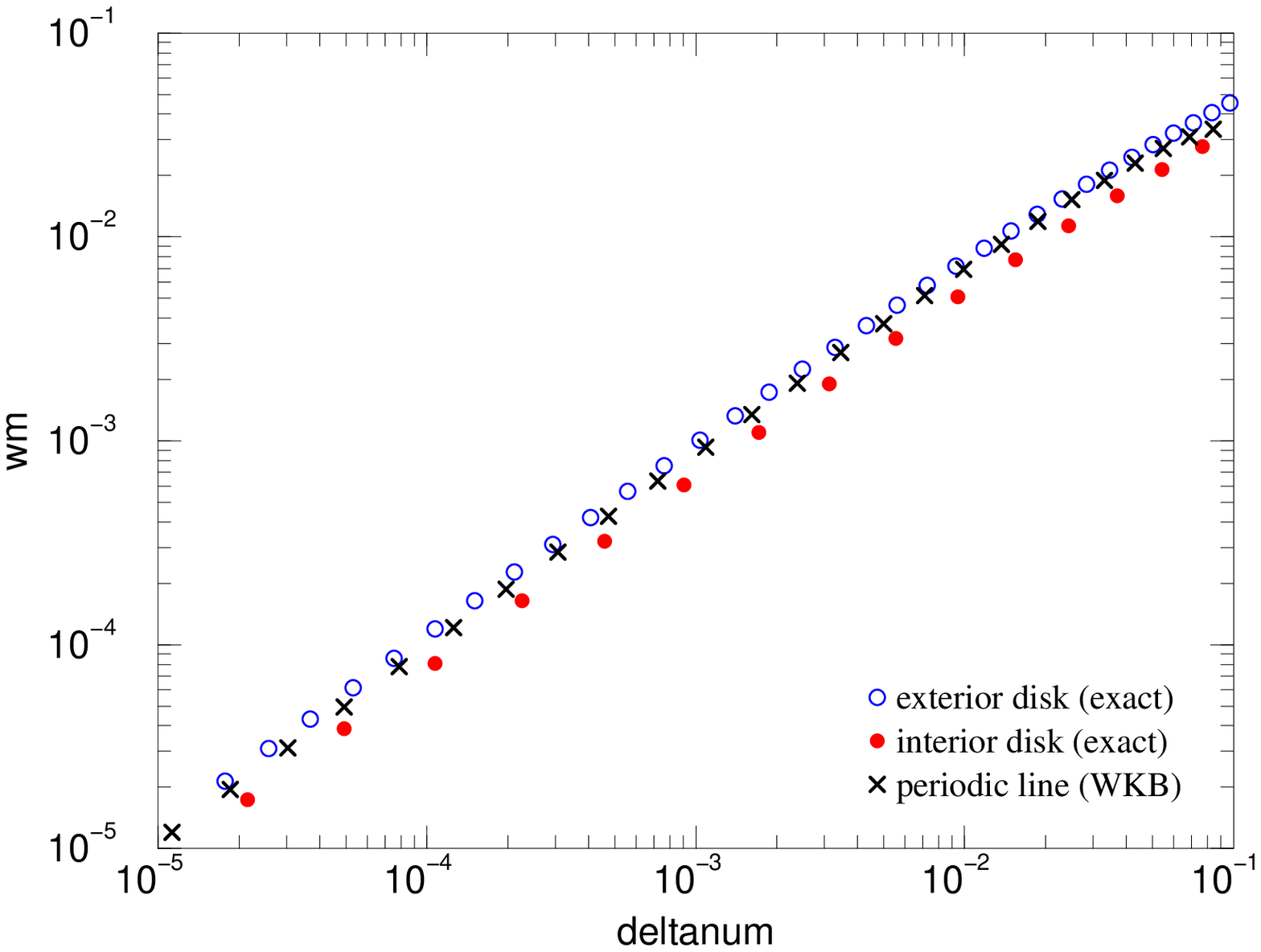}
    \figurecaption{%
      Bulk state energy shifts $\Delta\nu_n$, and weights $w_n$, for
      the magnetic disk ($R/b=15.0111$) at the 50th Landau level (on a
      double-logarithmic scale).  The interior ($\fullcircle$)
      and exterior ($\opencircle$) bulks states are approximated by
      the periodic line ($\boldsymbol{\times}$), cf
      \eref{eq:bulkshiftwkb} and \eref{eq:weightswkb}.  }
    \label{fig:LLdisk}
  \end{center}%
\end{figure}

The energy shift of a bulk state for general boundary condition
is derived in Appendix \ref{app:line} by a uniform approximation, see
\eref{eq:bulkshiftuni}.  An asymptotic expansion, which amounts to the
WKB approximation in the energetically forbidden region, yields the
expression
\begin{align}
  \label{eq:bulkshiftwkb}
  \Delta\nu_m
  &\simeq \frac{1}{2\pi}
  \exp\Big(-2\nu\big(\qm\sqrt{\qm^2-1}-\log\big(\qm+\sqrt{\qm^2-1}\big)
  \big)\Big)
  \\
  \tag{\ref{eq:bulkshiftwkb}a}
  \label{eq:bulkshiftwkba}
  &\sim  
  \frac{1}{2\pi}\,
  \bigg(\frac{2\pi b m}{\Len({N+\oh})^\oh}\bigg)^{2N+1}  \!\!
   \exp\Big({N+\oh-2\Big(\frac{\pi b m}{\Len}\Big)^2}\Big)
   \q\text{as $m\to\infty$\CO}
\end{align}
with $N\in\Natural$ the Landau level and $\qm$ the quantized distance of the
cyclotron center to the boundary, as defined in \eref{eq:qdef}.  We
observe that the bulk state energies approach the Landau levels
exponentially fast (indeed, like a Gaussian) as the integer $m$
increases, ie, as the distance of the cyclotron motion from the
boundary grows.

The weights of bulk states follow likewise, by taking the derivative
of equation \eref{eq:bulkshiftuni}. Essentially, they decay as fast as
the shifts of the bulk energies:
\begin{align}
  \label{eq:weightswkb}
  w_m \simeq 2 \,
  \bigg(   \frac{\pi^2 b^2 m^2}{\Len^2(N+\oh)}-1    \bigg)^\oh
  \Delta\nu_m
\end{align}
In Figure \ref{fig:LLdisk} we show exact bulk state energies and
weights in a double-logarithmic representation. They belong to the
interior and the exterior disk at $R/b=15.0111$ and to the 50th Landau
level.  The crosses indicate the zero curvature estimates according to eqs
\eref{eq:bulkshiftwkb} and \eref{eq:weightswkb}. One observes that the
asymptotic weights and the spacing between the asymptotic energies
match approximately the exact values, and lie between  those of the
finite curvature case. 
The approximation is improved when both $m$ and $N$ increase.

\subsubsection{The mean edge counting function}
\label{ssec:mecf}

As a second application, the periodic straight line problem allows the
straightforward derivation of the leading term of the mean edge state
counting function \eref{eq:Nedgesm}.  We simply identify the
transverse quantum number $n$ as a partial counting function for
states with fixed quantum number $m$. An explicit formula for $n$,
which includes the dependence on $\Lambda$ follows from $\smash{
  \xi_{\rm line}^{\rm (sc)}}=0$ \eref{eq:xilinesc}.  The sum
\begin{align}
  \N_{\rm skip}(\nu,\Lambda)
  = 
  \sum_{m}
n(m,\Lambda)
\end{align}
yields the total number of states corresponding to skipping motion.
Taking the derivative with respect to $\Lambda$ and  replacing
the summation by an integral we obtain the leading order of the smooth
edge state density,
\begin{align}
\label{eq:dedgesm2}
\dedgesm(\nu) &= 
  \left .
-    \frac{\rmd \Nsm_{\rm skip}(\nu;\Lambda)}{\rmd \Lambda}
  \right|_{\Lambda=0}
  =\frac{1}{\pi} \int_{-\sqrt{\nu}}^{\sqrt{\nu}} 
  \Big({1- \frac{{\tilde{c}_y}^2}{\nu}}\Big)^\oh \,
  \frac{\Len\, \rmd \tilde{c}_y}{\pi b}  \ +\Or(1)
\nnn
&= \frac{\Len}{2\pi b}\sqrt{\nu} \ +\Or(1)
\PO
\end{align}
The same result can be derived from the magnetic disk problem
discussed in Sect.~\ref{sssec:scdisk}. 
It follows from  \eref{eq:cossec}
that the semiclassical spectrum is obtained by requiring that the phase 
\begin{align}
  \phi^{\rm int/ext}_m(\nu,\Rt,\Lambda)=
  \Phi^{\rm int/ext}_{\rm disk}(\nu,m,\Rt)
  \mp\pshift(\nu,m,\Rt)-\piof
\end{align}
is an integer multiple of $\pi$. Hence, the smooth counting function
of the skipping states is obtained by
\begin{align}
  \Nsm^{\rm int/ext}_{\rm skip}(\nu,\Rt,\Lambda)=\frac{1}{\pi}
  \sum_m  \phi^{\rm int/ext}_m(\nu,\Rt,\Lambda)
\CO
\end{align}
where the sum  can be replaced by an integral over the
interval $|m-\Rt^2|<2\nu^\oh \Rt$. Using the
\eref{eq:pshiftdisk} one gets
\begin{align}
\dedgesm(\nu) &= 
  \left .- 
  \frac{\rmd\Nsm^{\rm int/ext}_{\rm skip}}{\rmd \Lambda}\right|_{\Lambda=0}
  \simeq
  \nu^\oh \Rt = \frac{\Len}{2\pi b}\nu^\oh
\end{align}
which reproduces \eref{eq:dedgesm2}.
This is further evidence that the periodic line problem yields the
leading order terms consistently.  The second order term in
\eref{eq:dedgesm} is not obtained this way.  It will be deduced in the
next section by relating the quantum weights of the disk to the
magnetic moments of the states.

\subsection{Edge magnetization as a spectral measure}
\label{sec:edgemag}

The preceeding section showed that our first definition
\eref{eq:defweight} of the quantum weights yields an efficient and
mathematically natural way to separate edge from bulk. However, the
physical interpretation of the mixed boundary conditions is not
immediate \cite{SPSUS95}. We therefore propose an alternative
definition of the weights which is physically more accessible.
It is obtained from the expectation value of the magnetic moment of
the state.

In Section \ref{sec:mag} we introduced the edge magnetization
\eref{eq:Medge} of an interior billiard.  It gives the scaled \emph{excess}
magnetization that is induced by the presence of the billiard
boundary and was defined as
\begin{align}
  \label{eq:Mag2}
  \Magt_{\rm edge}  (\nu) &=  
  \sum_{n=1}^\infty   b^2\frac{\rmd \nu_n}{\rmd b^2}\, \Theta(\nu-\nu_n)
\PO
\end{align}
Like the edge state counting function \eref{eq:Nedgedef} this is a
weighted staircase.  The size of the steps are now given by a
derivative with respect to the magnetic length rather than $\Lambda$.
Since the Landau levels do not depend on $b$ the bulk states
contribute merely to a negligible degree to \eref{eq:Mag2}, as
discussed in Sect.~\ref{sec:mag}.  Hence, it is reasonable to extend
the definition \eref{eq:Mag2}
 of the edge magnetization 
to the exterior problem.  

The exterior edge current shows an orientation which is opposite to
the interior (see Sect.~\ref{sec:wavefunctions}). Therefore, one expects
$\Magt_{\rm edge} $
to turn negative in the exterior -- which is indeed found.  Moreover,
the mean is \emph{finite} in both cases and given by
\begin{align}
\label{eq:Magsmboth}
   \overline{\Mag}_{\rm edge}  (\nu) &= 
   \pm \oh\,\frac{\Area}{b^2 \pi}\nu^2 
   - \frac{1}{3}\, \frac{\Len}{2\pi b} \nu^\frac{3}{2}
\PO
\end{align}
The interior case (upper sign) follows from eq \eref{eq:mmedgesm}
while the exterior one (lower sign) is suggested by symmetry and
confirmed empirically.  Like in the case of the edge counting function
\eref{eq:Nedgedef} the moduli of the mean interior and exterior edge
magnetizations are equal to leading order.
This suggests to use the
edge magnetization density
\begin{align}
\label{eq:mmedge2} 
    \widetilde{\mm}_{\rm edge}(\nu)
    =  \frac{\rmd }{\rmd \nu}\,   \Mag_{\rm edge}  (\nu)
    =   \sum_{n=1}^\infty b^2\, \frac{\rmd \nu_n}{\rmd b^2}\, \delta(\nu-\nu_n)
\CO
\end{align}
which was introduced in
Sect.~\ref{sec:mag}, to define a physically motivated spectral measure
for the edge states:
\begin{align}
\label{eq:dmagdef} 
    \dedge^\magn(\nu) 
\defas \pm\frac{1}{\nu} \, \widetilde{\mm}_{\rm edge}(\nu)
    =   \sum_{n=1}^\infty w_n^\magn\, \delta(\nu-\nu_n)
\end{align}
The index ${\scriptstyle(\mathcal{M})}$ is used to distinguish this
magnetization based density of edge states from the former definition
\eref{eq:defweight}.  The weights are now given by
\begin{align}
\label{eq:wnmagdef} 
 w_n^\magn
\defas \pm \frac{b^2}{\nu} \frac{\rmd \nu_n}{\rmd b^2}
=   
\pm
 \frac{
   \bra{\psi_n}{\frac{\rvect\times\vvect}{2}}\ket{\psi_n}
    +\nu
   }{\nu}
\PO
\end{align}
Again, these positive quantities may be obtained as derivatives of the
energies with respect to an external parameter (which is the magnetic
length $b$ in the present case). At the same time they are expressed in
terms of the (symmetrized) expectation values of the scaled magnetic
moment \eref{eq:mag}.  The weights vanish as the Landau levels are
approached since for interior and exterior bulk states the scaled
magnetic moment approaches the diamagnetic value
$\bra{\psi_n}{\frac{\rvect\times\vvect}{2}}\ket{\psi_n}\to-\nu$ from
above and below, respectively.  As for the smooth edge state density,
the expression is obtained immediately from \eref{eq:Magsmboth}:
\begin{align}
\label{eq:dmagsm} 
    \dedgesm^\magn(\nu) = 
    \pm \frac{1}{\nu}  \frac{\rmd \overline{\Mag}_{\rm edge}}{\rmd\nu}
    =   \frac{\Area}{b^2\pi} \mp \oh\frac{\Len}{2\pi b}\nu^{-\oh}
\end{align}
It \emph{coincides} with the  standard mean density for the interior
case.  Moreover, the interior weights approach unity as the magnetic
field $B$ is decreased and bulk states no longer exist in the
interior. This is seen immediately if we write the 
weight \eref{eq:wnmagdef} in terms of  conventional units
\begin{align}
\label{eq:wnmagconv} 
 w_n^\magn
=   
\pm
  \left(
     \frac{q\,B}{2 E}\bra{\psi_n}\rvec\times\vvec\ket{\psi_n}
    +1
  \right)
\PO
\end{align}
In the limit $B\to 0$ the interior weights assume unit value and the
interior edge state density turns into the standard density. This
limit does not make sense in the exterior since in this case the
spectrum ceases to be discrete; at finite field the exterior weights
remain always  non-negative.

The expression \eref{eq:wnmagconv} shows   that the weighted spectrum
may be obtained immediately by measuring the magnetic moments of the
states.  
The scaled magnetization density \ref{eq:magdef} follows from the
conventional magnetization density \eref{eq:macconv} by a simple
multiplication with the magnetic field:
\begin{align}
  \mmt(\nu,b^2)=B\, \mm\big(E=\frac{2\hbar^2}{\mass
    b^2},B=\frac{2\hbar}{q b^2}\big)
\end{align}
This relation to an experimentally measurable quantity is a clear
advantage of the present definition of the edge state density.  It
is achieved at a price --- the leading order of the mean edge density
is now determined by the area of the billiard rather than by its
circumference, see \eref{eq:dmagsm}. It indicates that with this
measure the quasi one-dimensional character of the edge states is not
accounted for to the same degree as by the former definition of
$\dedge$. However,
it does an equally good job in consistently suppressing the bulk
contributions.  Moreover, in the case of a disk  both spectral
densities, $\dedge$ and $ \dedge^\magn$,  are \emph{identical} up to a
factor.
This is seen from  \eref{eq:wmagrel} which leads to
the equation
\begin{align}
\label{eq:wnrelat}
  w^\magn_n =  w_n\, \frac{R}{b}\,\nu^{-\oh}
\PO
\end{align}
This relation is as surprising as fortuitous and does not hold for
general billiard shapes.  Nonetheless, it allows to deduce the second,
constant term of the mean density \eref{eq:dedgesm} by comparison with
the smooth edge state density \eref{eq:dmagsm}.  Being the next order
after the circumference term it is determined by the mean curvature
which is equal for all simply connected boundaries, $\int_\Gamma
\kappa(s) \rmd s=$ $\pm2\pi$ (according to the Gauss-Bonnet theorem).

\begin{figure}
  \begin{center}%
  \includegraphics[width=0.9\linewidth,clip]
{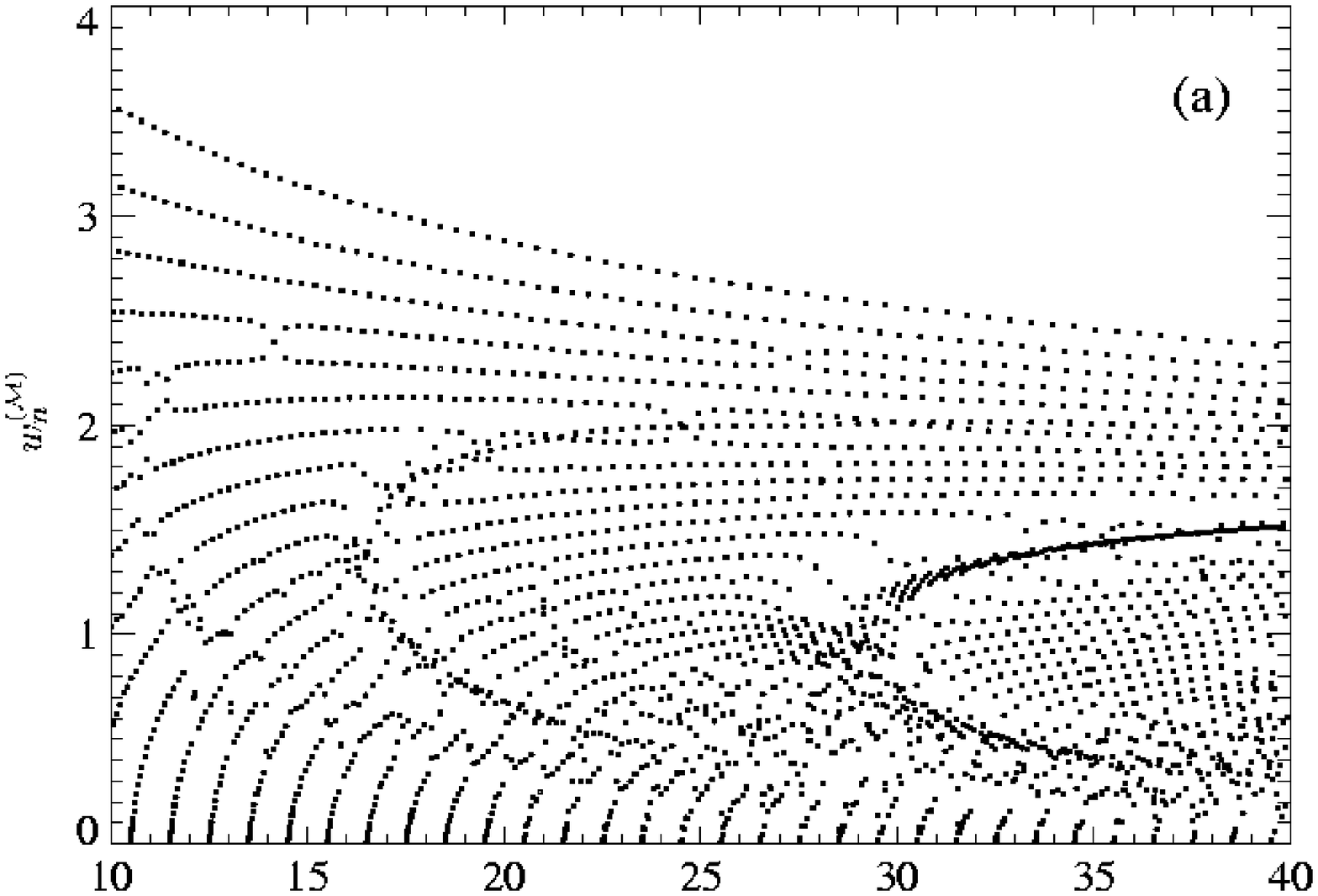}
  \includegraphics[width=0.9\linewidth,clip]
{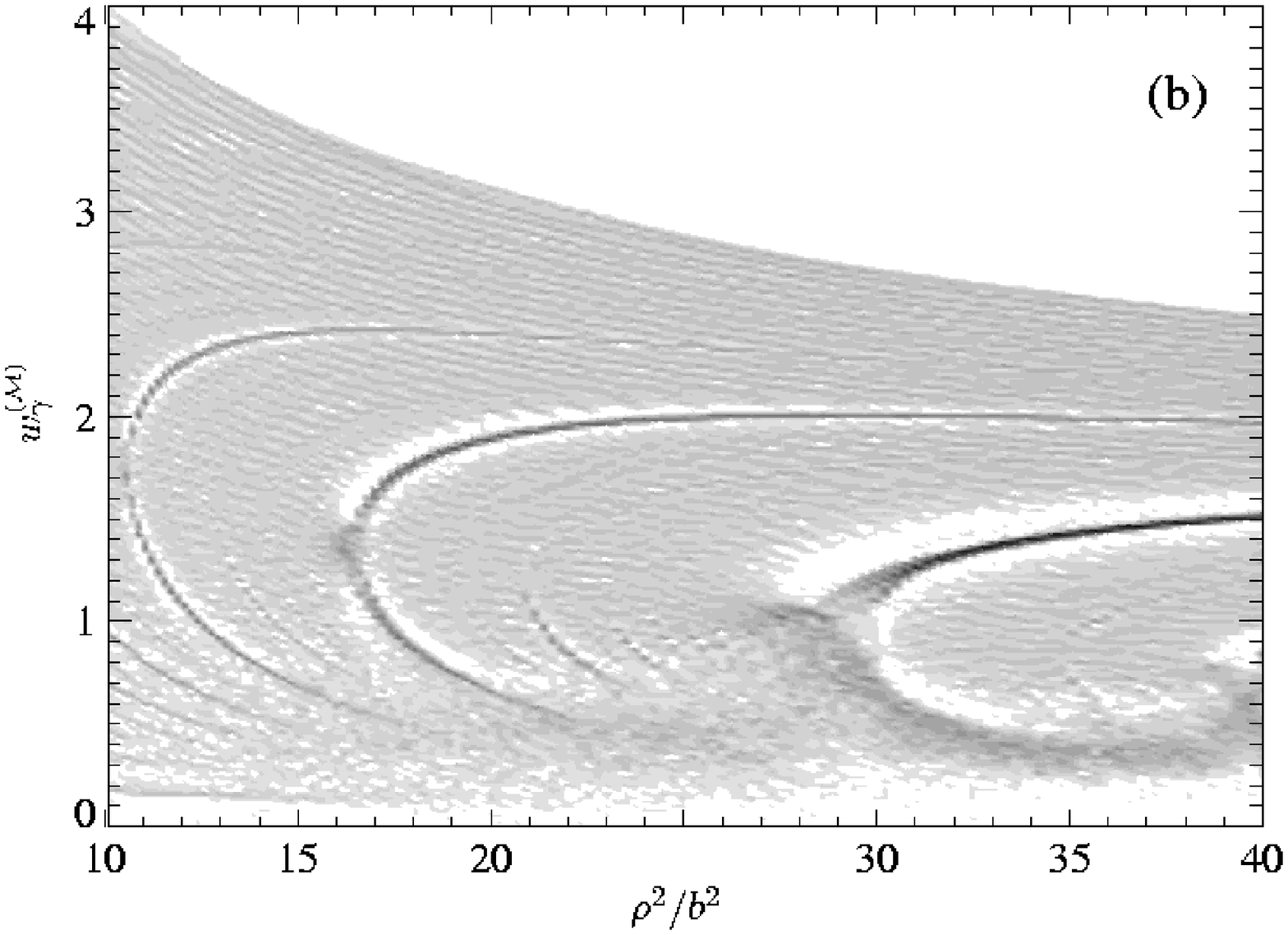}
\end{center}
  \figurecaption{%
    (a) Edge state spectrum using the edge magnetization weights
    \eref{eq:wnmagdef} compared to (b) the phase space distribution of
    the classical weights \eref{eq:wpomagdef} for the interior
    ellipse.  [figure quality reduced]
  }
\label{fig:qmclmag}
\end{figure}

It follows from \eref{eq:wnrelat} that for the disk the magnetization
based edge spectra differ from the spectra shown in
Fig.~\ref{fig:wsdiskintext} merely by a geometric transformation and
there is no need to reproduce them here. As for the ellipse, Figure
\ref{fig:qmclmag} shows the weights as obtained from the edge
magnetization.  Like in Fig.~\ref{fig:qmclswt} the structures in the
distribution are reproduced by the probability density of the
corresponding classical weights.

\myparagraph{Semiclassical edge magnetization}

The semiclassical periodic orbit formula for the complete
magnetization density is given in \eref{eq:mmosc}. Likewise, one finds
that $\pm \mmt^{\rm osc}_{\rm edge}/\nu$ has the form of the trace
formula for the standard density with each periodic orbit now weighted
individually by
\begin{align}
  \label{eq:wpomagdef}
  w_\po^\magn &\defas
  \mp\,\frac{\ds b^2\frac{\rmd}{\rmd b^2}\big[2\pi\nu\Ga(\po)\big]}
  {\ds \nu \frac{\rmd}{\rmd \nu}\big[2\pi\nu\Ga(\po)\big]}
  = \pm\,
  \frac{\ds \sum\nolimits_{j=1}^{n_\po} 
    \Big(
    -\sigma_j\sqrt{1-\sigma_j^2}-\frac{\rvec_{j+1}\times\rvec_{j}}{2\rho^2}
    \Big)
    }{\ds
    \sum\nolimits_{j=1}^{n_\po} 
    \Big(
    \piot+\arcsin(\sigma_j)
    \Big)
    }
  \nonumber\\[3ex]
  &= \frac{ 2 \Area_\po\pm\rho\Len_\po}{\rho\Len_\po}
\CO
\end{align}
see \eref{eq:lenpodefa} and \eref{eq:magdiff}.  Like in the case
of $w_\po$, the classical weights $w^\magn_\po$ vanish as trajectories
get almost detached from the boundary (since the numerator approaches
$2\Area$, while $\Len_\po\to\infty$).

Equipped with well-defined spectral densities for edge states and the
corresponding trace formulas we can now proceed with a statistical
and semiclassical study of edge state spectra.  We shall not only
consider the statistics within  a given edge state spectrum (in Chapter
\ref{chap:stat}) but look also for  cross correlations between different,
classically related spectra (in Chapter \ref{chap:cross}).

\section{Properties of edge state spectra}
\label{chap:stat}

We apply the weighted spectral densities discussed in the previous
chapter to analyze two aspects of the interior and the exterior edge
spectra. First we perform a statistical analysis comparing the
results to the predictions of random matrix theory. Second, we check
the validity of the semiclassical trace formula taking the disk
billiard as an example.

\subsection{Universal auto-correlations}
\label{sec:auto}

One of the central goals in the field of quantum chaos is to
understand how the statistical properties of the quantum spectrum
reflect the nature of the underlying classical dynamics
\cite{LesHouches91}. We extend these studies to magnetic billiards by
making use of the spectral measure of edge states introduced in the
previous chapter. It was constructed to focus on the non-trivial part
of phase space which is determined by the billiard boundary map
\eref{eq:Birmap}.

As a first point we  check whether the edge spectra of both interior
\emph{and} exterior magnetic billiards display the universal
characteristics of random matrix theory (RMT) if the corresponding
skipping motion is hyperbolic.  Our quantity of choice to characterize
the spectrum statistically is the spectral form factor $K(\tau)$.  It
is sensitive to correlations of the eigenenergies beyond the mean
level spacing \cite{Bohigas91}.  The standard form factor was already
used in Chapter \ref{chap:numres} to study the two-point correlations
in the unweighted spectra of interior billiards.  For edge spectra
$K(\tau)$ is readily defined in terms of the 2-point autocorrelation
function of the edge density,
\begin{equation}
  R_{\nu_0}(\nu)=\int
  \dedge^{\rm osc}\Big(\nu'+\frac{\nu}{2}\Big)
  \,
  \dedge^{\rm osc}\Big(\nu'-\frac{\nu}{2}\Big)
  \,
  g_1(\nu'-\nu_0)
  \,
  \rmd\nu'
  \PO
\end{equation}
Here, we included a normalized Gaussian window function $g_1$
to pick up a spectral interval centered at $\nu_0$.

Before comparing to RMT it is advantageous to remove the trivially
system-dependent properties of the spectrum by ``unfolding'' it
\cite{Bohigas91}.  This is a transformation of the spectrum
which renders it dimensionless and of unit mean density.  Dealing with a
\emph{weighted} spectrum the unfolding procedure must transform both
the energies and the weights.  The natural choice involves the smooth
edge state counting function $\Nsm_{\rm edge}$ and the average weight
$\langle w^2\rangle / \langle w \rangle$ in the spectral interval
considered:
\begin{align}
  \label{eq:unfoldnuandw}
  \unfolded{\nu}_n 
  \defas 
  \frac{ \langle w \rangle}{ \langle w^2 \rangle}
  \,
  \Nsm_{\rm edge}(\nu_n)
\qq\text{and}\qq
  \unfolded{w}_n 
  \defas
  \frac{ \langle w \rangle}{ \langle w^2 \rangle}
  \,
  w_n
\PO
\end{align}
Here,  the first and second moments of the weights,
\begin{align}
  \langle w\rangle &= \sum_{n=1}^{\infty} w_n
  \;
  g(\Nsm_{\rm edge}(\nu_n)
  -\unfolded{\nu}_0)
\intertext{and}
  \langle w^2\rangle &= \sum_{n=1}^{\infty} w_n^2
  \;
  g(\Nsm_{\rm edge}(\nu_n)
  -\unfolded{\nu}_0)
\CO
\end{align}
are taken locally in the spectrum in terms of the window function $g$
(a normalized Gau{ss}ian of width $\sigma_g$.)  As a result of this
unfolding, both the weights and the weighted density have unit mean.

Since we are dealing with a discrete spectrum, the form factor must be
averaged to be well-defined. The standard procedure is to take the
spectral average over non-overlapping parts of the spectrum,
\begin{align}
  \label{eq:formfactordef}
  K(\tau)=
   \left\langle
     \int
     \e^{2\pi\rmi\unfolded{\nu}\tau}
     R_{\unfolded{\nu}_0}(\unfolded{\nu})\,
     g_2(\unfolded{\nu})\,
     \rmd\unfolded{\nu}
  \right\rangle_{\unfolded{\nu}_0}
\end{align}
as indicated by the triangular brackets.  According to the spectral
ergodicity hypothesis \cite{BFFMPW81} this should be equivalent to an
ensemble average for hyperbolic systems.

\begin{figure}[tp]%
\begin{center}
  \includegraphics[width=\linewidth]
  {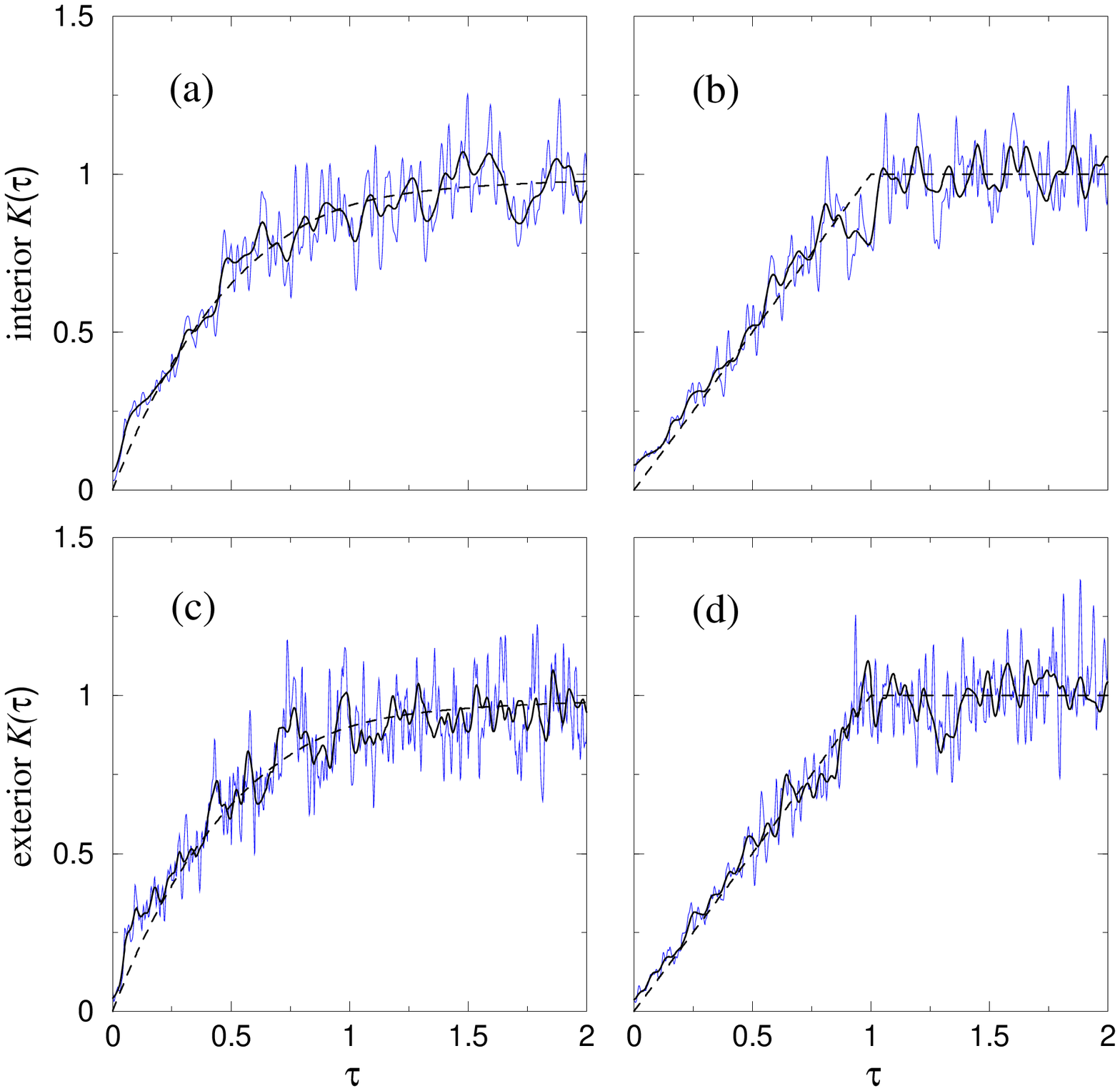}
\figurecaption{%
  Form factors \eref{eq:formfactorqm} of the interior (a,b) and
  exterior (c,d) edge state spectra for the asymmetric stadium (a,c)
  and skittle (b,d) billiard at $\rho=1.2$. The shapes are defined in
  Fig.~\ref{fig:shapes}.  The functions follow the RMT predictions of
  the GOE and GUE ensembles \cite{Bohigas91}, respectively (dashed
  lines). The heavy lines correspond to stronger spectral averaging
  than the thin lines ($\sigma_g=10$ and $\sigma_g=3$, respectively.)
  }
\label{fig:edgeformfactor}
\end{center}
\end{figure}

If we choose the widths of the Gaussians $g_1$ and $g_2$ as
$\sigma_g/\sqrt{2}$ and $\sigma_g\sqrt{2}$, respectively, the Fourier
transform of the autocorrelation function leads directly to the power
spectrum.  The form factor is then given by the \emph{weighted} sum
\begin{align}
  \label{eq:formfactorqm}
  K(\tau)=
  \left\langle
    \frac{2\sqrt{2\pi}\sigma}{ \langle w^2\rangle }
    \left| \sum_{n=1}^{\infty}
      \unfolded{w}_n\, 
      \rme^{2\pi\rmi ( \unfolded{\nu}_n- \unfolded{\nu}_0)\tau}
      g(\unfolded{\nu}_n-\unfolded{\nu}_0)
      -\hat{g}(\tau)
    \right|^2
  \right\rangle_{\unfolded{\nu}_0} \ ,
\end{align}
where the Fourier transform of  $g$ is denoted by $\hat{g}$.

The previous discussion holds for both definitions,
\eref{eq:defweight} and \eref{eq:wnmagdef}, of the spectral density of
edge states. However, it is necessary to keep the type of the
underlying classical motion unchanged during the spectral averaging.
This is conveniently done by taking the spectrum in the semiclassical
rather than the conventional direction, see Sects.~\ref{sec:scaling}
and \ref{sec:qspec}.  In this case the quantum weights $w_n$ are
simply obtained by taking the derivatives with respect to $\Lambda$ at
fixed $\rho$, see Appendix \ref{app:rho}.  In the present section
we use only this first definition of a spectral density of edge states
since it is not possible to define magnetization-based weights for the
semiclassical direction.

\begin{figure}[t]%
\begin{center}
  \psfrag{x}{$w$}
  \psfrag{y}{$p(w)$}
  \includegraphics[width=0.8\linewidth]
  {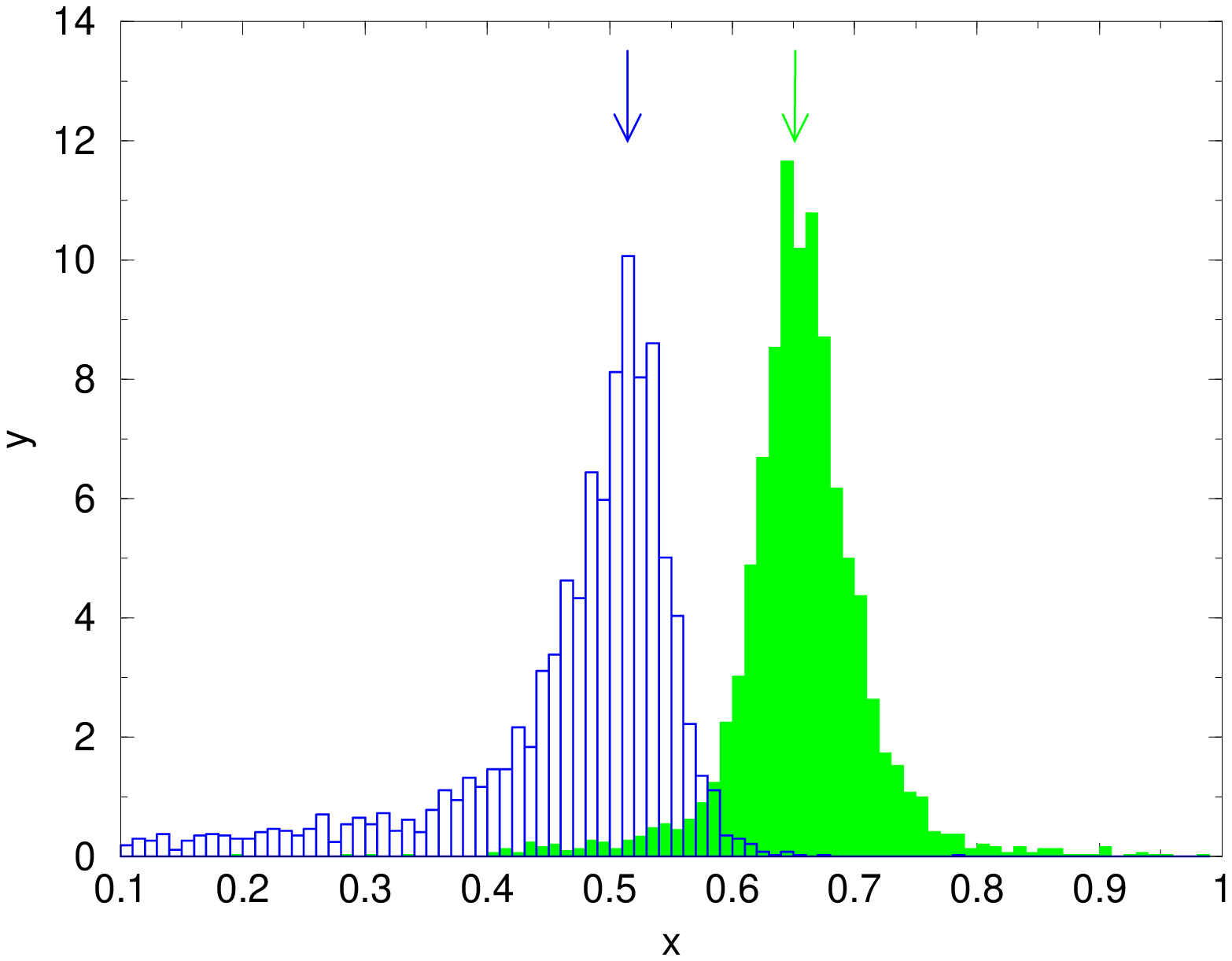}
  \figurecaption{%
    Distribution of the quantum weights $w_n>0.1$ of the interior
    (shaded) and exterior (transparent) skittle spectrum at
    $\rho=1.2$. 
    The histograms show peaks whose positions are well reproduced by the
    phase space estimates \eref{eq:wavint} and \eref{eq:wavext}
    (indicated by the arrows).  Unlike the interior case, the exterior
    distribution shows a tail due to the transitional states which ranges
    to the small weights. (For normalization (bulk) states with
    weights smaller than $0.1$ had to be disregarded.)
    }
  \label{fig:histoweights}
\end{center}
\end{figure}

Figure \ref{fig:edgeformfactor}  shows the form factors for the
interior (top) and exterior (bottom) edge state spectra for the
asymmetric stadium (left) and skittle (right) billiard, respectively.
The spectra were obtained in the semiclassical direction, at fixed
$\rho=1.2$, ie, for the same situation as in
Fig.~\ref{fig:formfactor}. The weights were obtained by numerical
differentiation with respect to $\Lambda$.
We observe that the interior form factors follow the RMT prediction of
the Gaussian Orthogonal and Gaussian Unitary Ensembles, respectively,
as expected from the specific symmetry properties of the Hamiltonians.
In the interior case this is not surprising.  To ensure essential
hyperbolicity of the classical motion the value of $\rho$ had to be
chosen large such that the interior phase space consists only of
skipping trajectories which cover it ergodicly.  As a consequence, one
expects that all \emph{interior} states are edge states to an equal
degree.  Indeed, the interior weights are distributed narrowly around
a mean value $\overline{w}$, given by the ratio of weighted and
unweighted mean densities,
\begin{align}
  \label{eq:wavint}
  \overline{w}=\frac{\dedgesm^{(\rho)}(\nu)}
  {\overline{d}_{\rm tot}^{(\rho)}(\nu)}
  = \frac{\Len\rho}{4\Area}
\CO
\end{align}
as can be observed from the shaded histogram in 
Fig.~\ref{fig:histoweights}.  The weights do not provide   additional
information in this case, which explains why $K(\tau)$ reproduces the
RMT prediction, like in the unweighted case.

In contrast, the standard form factor -- like any other standard
statistical function -- does not even exist for the exterior spectrum,
which is dominated by infinitely many bulk states.  Nonetheless, we
find that the exterior spectrum closely obeys the predictions of
random matrix theory (bottom row of Fig.~\ref{fig:formfactor}) if
viewed in an appropriate way, ie, by means of the edge state density.
This way, a crucial test for the consistency of the spectral measure
of edge states is passed.  The quantum weights succeed to filter out
\emph{selectively} the relevant edge states, which in turn exhibit the
universal characteristics expected for chaotic motion.

The distribution of the exterior weights is given by the transparent
histogram in Figure \ref{fig:histoweights}. 
One observes that the
distribution of large weights is peaked like in the interior case.
Again, the peak position is well described by the ratio of weighted
and unweighted densities,
\begin{align}
  \label{eq:wavext}
  \overline{w}=
  \frac{\dedgesm^{(\rho)}(\nu)}
  {\overline{d}_{\rm skip}^{(\rho)}(\nu)}
  \simeq 
  \frac{\Len\rho}{2\Area_{\rm skip}^{\rm ext}(\rho)}
  \CO
\end{align}
with the mean unweighted density now given by the phase space estimate
\eref{eq:dskip} of skipping states. Unlike the interior case, the
distribution has a tail of transitional states which ranges to the
infinitely many bulk states with small weights.

\begin{figure}[t]%
\begin{center}
  \psfrag{t}{$t$}
  \psfrag{y1}{\hspace*{-5.1em}
    $-|\widehat{d}^{\rm (\rho, sc)}_{\rm edge}\!(t)|
    \q|\widehat{d}^{\rm (\rho)^{osc}}_{\rm edge}\!(t)|$}
  \psfrag{y2}{\hspace*{-5.1em}
    $-|\widehat{d}^{\rm (\rho, sc)}_{\rm edge}\!(t)| 
    \q|\widehat{d}^{\rm (\rho)^{osc}}_{\rm edge}\!(t)|$}
  \includegraphics[width=\linewidth]
  {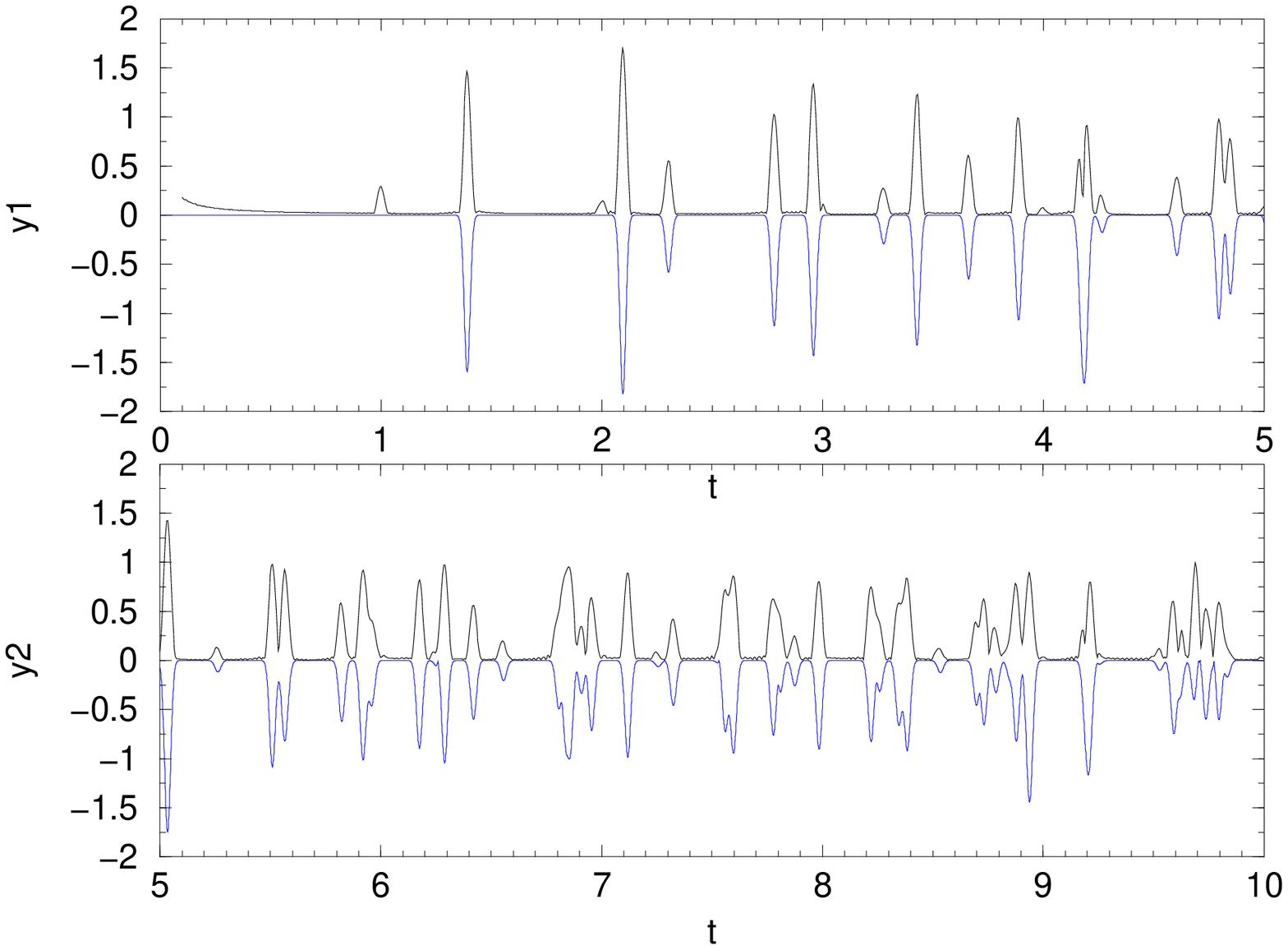}
  \figurecaption{%
    Action spectrum of the \emph{exterior} disk at $\rho=2 R$.  The
    positive values give the Fourier transform \eref{eq:dftdef} of the
    exterior edge density (absolute values).  The positions of the
    peaks are well reproduced by the trace formula \eref{eq:dedgecirc}
    (negative values) -- except for the small peaks at integer $t$
    which are remnants of the bulk states. The peak heights match well
    in most cases; they are expected to fit better if a spectral
    interval larger than $\nu\in[0;48]$ is used.  }
  \label{fig:actionsext}
\end{center}
\end{figure}

\begin{figure}[pt]%
\begin{center}
  \psfrag{y1}{\hspace*{-4.1em}\small
    $-|\widehat{d}^{\rm (\rho, sc)}_{\rm edge}\!(t)|
    \q|\widehat{d}^{\rm (\rho)^{osc}}_{\rm edge}\!(t)|$}
  \psfrag{y2}{\hspace*{-4.1em}\small
    $-|\widehat{d}^{\rm (\rho, sc)}_{\rm edge}\!(t)| 
    \q|\widehat{d}^{\rm (\rho)^{osc}}_{\rm edge}\!(t)|$}
  \psfrag{y3}{\hspace*{-4.1em}\small
    $-|\widehat{d}^{\rm (\rho, sc)}_{\rm edge}\!(t)| 
    \q|\widehat{d}^{\rm (\rho)^{osc}}_{\rm edge}\!(t)|$}
  \psfrag{t}{$t$}
  \includegraphics[width=\linewidth]
  {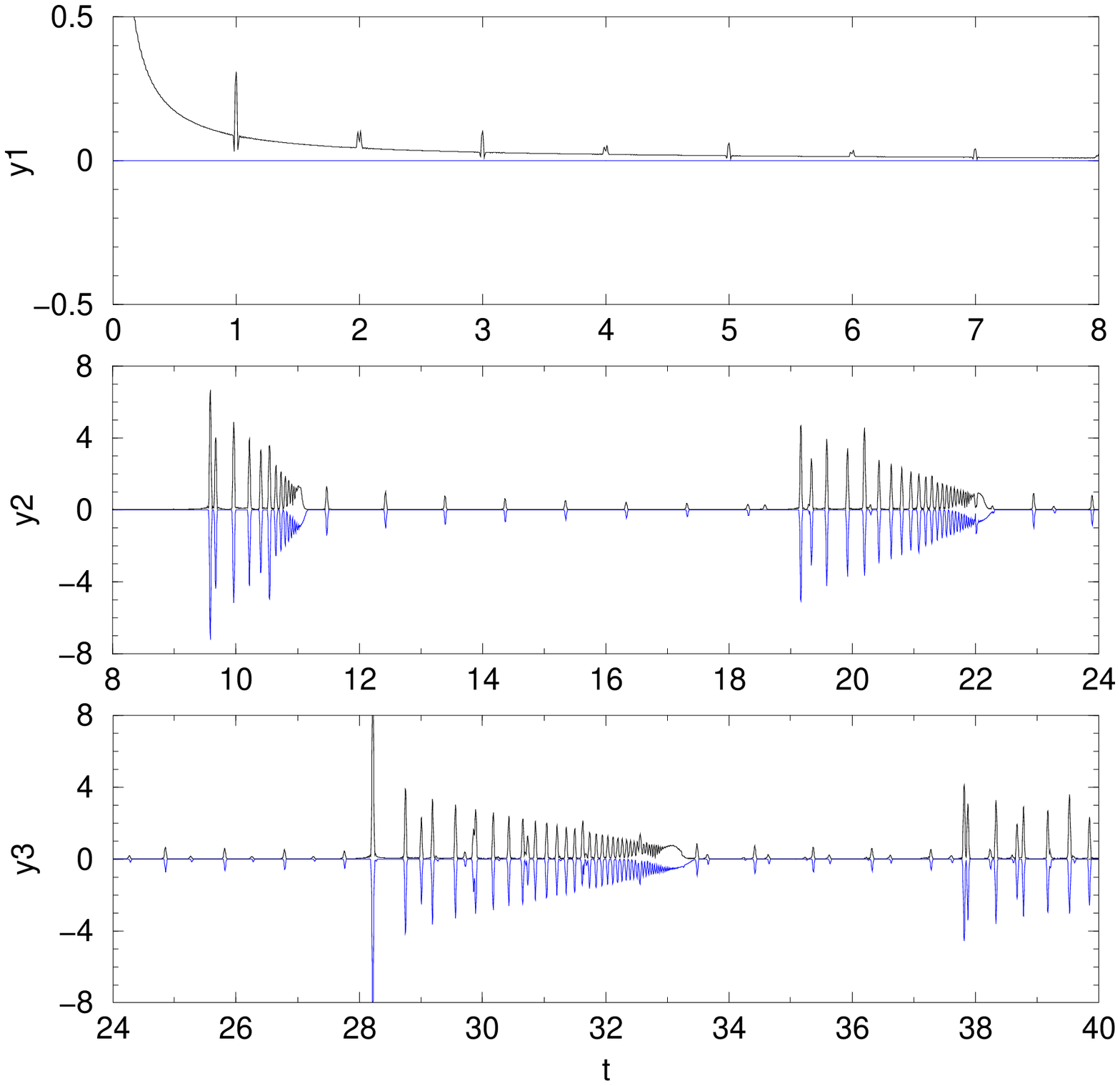}
  \figurecaption{%
    Action spectrum of the \emph{interior} disk at a cyclotron radius
    $\rho=0.4\times R$ small enough to enable bulk states.  The
    Fourier transform \eref{eq:dftdef} of the interior edge density
    (positive values, $\nu\in[0;60]$) is well reproduced by the trace
    formula (negative values).  Note that in the top part, which shows
    the remnant peaks of bulk contributions, the y-axis has a different
    scale.}
  \label{fig:actionsint}
\end{center}
\end{figure}

\subsection{The action spectrum}
\label{sec:acspec}

We turn from the statistical analysis of edge spectra to their
semiclassical description.  Here, the main purpose is to show that the
trace formula for the edge state density -- which rates each periodic
orbit with a classical weight -- succeeds in approximating the exact
edge spectrum.

We choose the disk billiard for which an explicit periodic orbit
formula is readily obtained from equation \eref{eq:Noscdisk2}. For the
exterior case and $\Gam=R/\rho<1$ we find, see \eref{eq:dedgedef2},
\begin{align}
  \label{eq:dedgecirc}
    d^{\rm osc}_{\rm edge}(\nu) = &
  \left(\frac{2\nu}{\pi}\right)^\oh
  \sum_{\n=2}^\infty\,
  \frac{2}{\n^{1/2}}
  \sum_{\Delta\phi\in\PSet_{\rm ext}^\n} 
  \Big|\sin\Big(\alpha-\frac{\Delta\phi}{2}\Big)\Big|
  \\
 & \times
  \frac{\sin(\alpha)\cos(\alpha)-\oh\Gam^2\sin(\Delta\phi)}
    {(\sin(\alpha)\cos(\alpha))^\oh}
    \cos\left(2\pi\nu \n\,\ga_{\rm L}(\Delta\phi)+\n\frac{\pi}{2}-\piof\right)
\CO
\nn
\end{align}
with $\alpha$ defined by equation \eref{eq:sadisk}. (The term
$|\sin(\alpha-\frac{\Delta\phi}{2})|$ corresponds to the normal
component of the velocity, $|\nvec\,\vvech|$,
in \eref{eq:dedgeosc2}.)
Moreover, the exact quantum spectrum of the disk is calculated
relatively easily in terms of the roots of special functions, see
App.~\ref{app:diskexact}.  We calculated spectral intervals large
enough so that the Fourier transformation of the spectral densities,
\begin{align}
  \label{eq:dftdef}
  \widehat{d}^{{\rm osc}}_{\rm edge}\!(t) =
  \int \rme^{2\pi\rmi\nu t}  d^{{\rm osc}}_{\rm edge}\!(\nu) 
  \,h(\nu-\nu_0)\, \rmd\nu
  \CO
\end{align}
resolves the classical actions $t$ of the underlying periodic orbits.
Here, the function $h$ is a suitable window centered on the midpoint
$\nu_0$ of the spectral interval.
This \emph{action spectrum} may be readily compared to the
semiclassical  prediction based on
\eref{eq:dedgecirc}.

Like in the previous section, it is  convenient to take the 
spectrum in the semiclassical direction, at constant $\rho$.
In Figure \ref{fig:actionsext} we show the action spectrum for the
\emph{exterior} disk at a cyclotron radius $\rho=2 R$ (positive
values). The corresponding prediction of the trace formula
\eref{eq:dedgecirc} is given by the negative values.  One observes
that the peak positions match very well with the predictions of
semiclassical theory. The only exception are the small peaks at
integer actions which are not reproduced semiclassically.  They
are remnants of the infinite number of bulk states.
The peak heights are well reproduced most of the time, except if two
peaks overlap too strongly.  These deviations are expected to fade
as a larger spectral interval is used and the the widths of
the peaks decrease.
This is also seen in Figure \ref{fig:actionsint} where we present the exact
and semiclassical action spectra of the \emph{interior} magnetic disk -- based
on a large spectral interval ($\nu\in[0;60]$ at $\rho=0.4\times R$). Here,
the cyclotron radius was chosen small enough for bulk
states to exist in the interior.
One observes again that the latter are very efficiently
suppressed in the action spectrum giving rise only to the small peaks
at integer values (shown in the top part of Fig.~\ref{fig:actionsint}). 
In the Fourier transform of the \emph{unweighted} density, in
contrast, the bulk states  obliterate the edge contributions such
that not a single action is resolved (not shown).  Blaschke and Brack
\cite{BB97a} analyzed semiclassically the spectrum for the unweighted
interior problem. The contribution of the bulk states was estimated
and added by hand resulting in an unsatisfactory agreement between the
semiclassical and the quantum spectra.  

In conclusion, we find that the the semiclassical trace formula
succeeds in reproducing the quantum edge state density. It does so by
weighting each periodic orbit contribution with a classical weight
which vanishes for cyclotron orbits. This removes the bulk
contributions analogous to -- and consistent with -- the quantum
weights of the edge state density.

\begin{figure}[t]%
  \begin{center}%
    \psfrag{sqrt(5)/2}{$\ds\oh\sqrt{5}$}
    \psfrag{1.0}{$1.0$}
    \includegraphics[width=0.9\linewidth] 
    {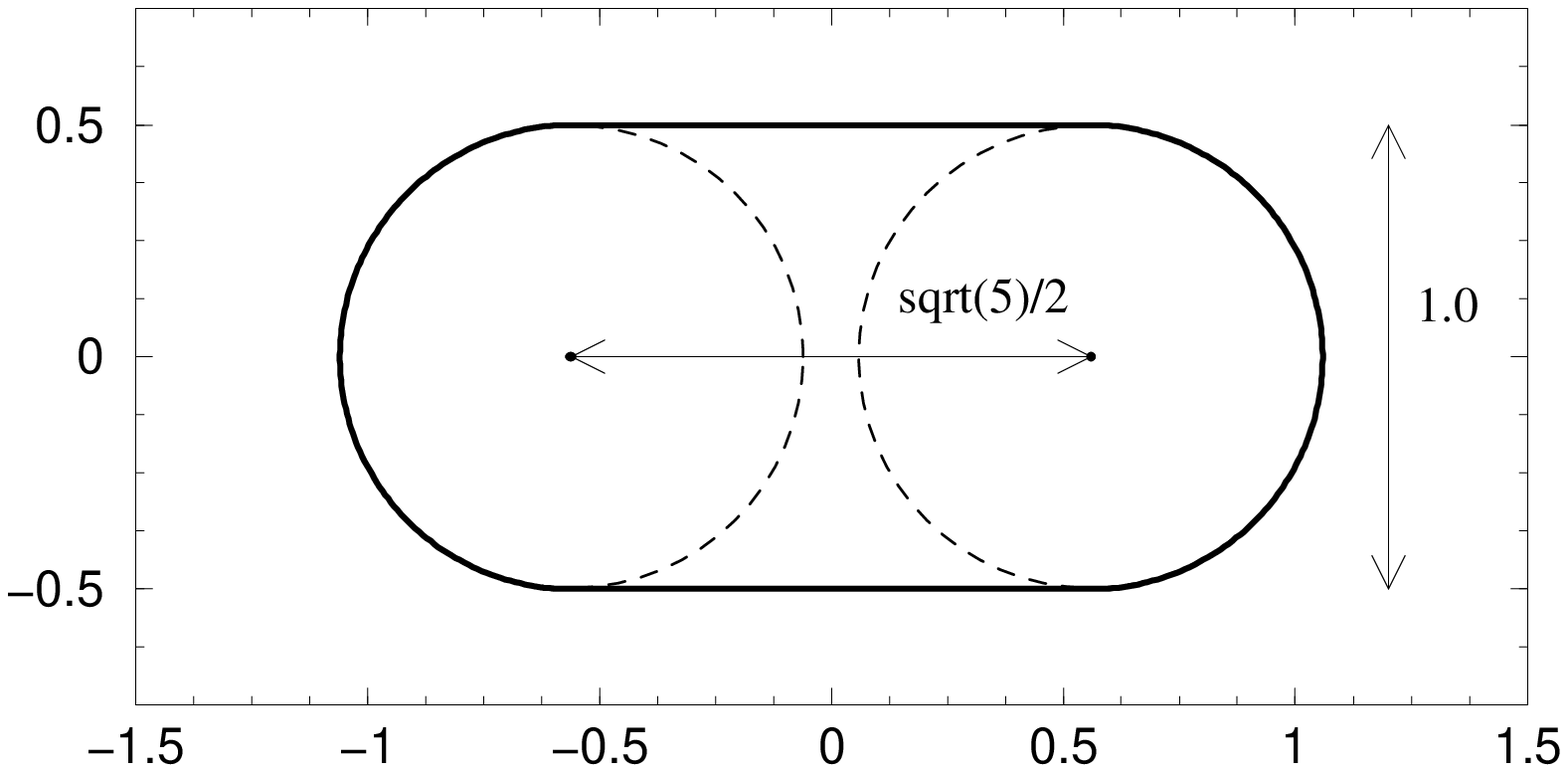}
    \figurecaption{%
      Definition of the Bunimovich stadium used in
      Sect.~\ref{sec:magstadium} and Chapter \ref{chap:cross}. 
      }
    \label{fig:s14shape}
  \end{center}%
\end{figure}

\subsection{Using the edge magnetization}
\label{sec:magstadium}

Finally, let us demonstrate that the edge state density may as well be
defined in terms of the magnetization as discussed in Section
\ref{sec:edgemag}.

Choosing the Bunimovich stadium billiard (defined in
Fig.~\ref{fig:s14shape}) we calculated the interior and exterior
magnetization spectrum in the high-energy direction, at $b=0.2$.  The
selected spectral interval $\nu\in[100;135]$ corresponds to large
cyclotron radii $\rho\in[2;2.32]$ giving rise to essentially
hyperbolic\footnote{The term ``essentially hyperbolic'' means that
  although there might be small integrable parts in phase space their
  combined area is much smaller than the uncertainty product
  $(b^2\pi)^2$.} classical motion.
Quantum mechanically, the problem exhibits one unitary and one
anti-unitary symmetry (rotation by $\pi$ and reflection at one axis,
respectively).  Hence, the spectrum decomposes into two symmetry
classes (a feature which will be used in the next chapter) while each
class should obey the characteristics of the Gaussian Orthogonal
Ensemble \cite{Bohigas91}.

\begin{figure}[tp]%
  \begin{center}%
    \psfrag{x}{$\nu_n$}
    \psfrag{yext}{\hspace*{-2.0em}exterior $w^\magn_n$}
    \psfrag{yint}{\hspace*{-2.0em}interior $w^\magn_n$}
    \includegraphics[width=0.9\linewidth] 
    {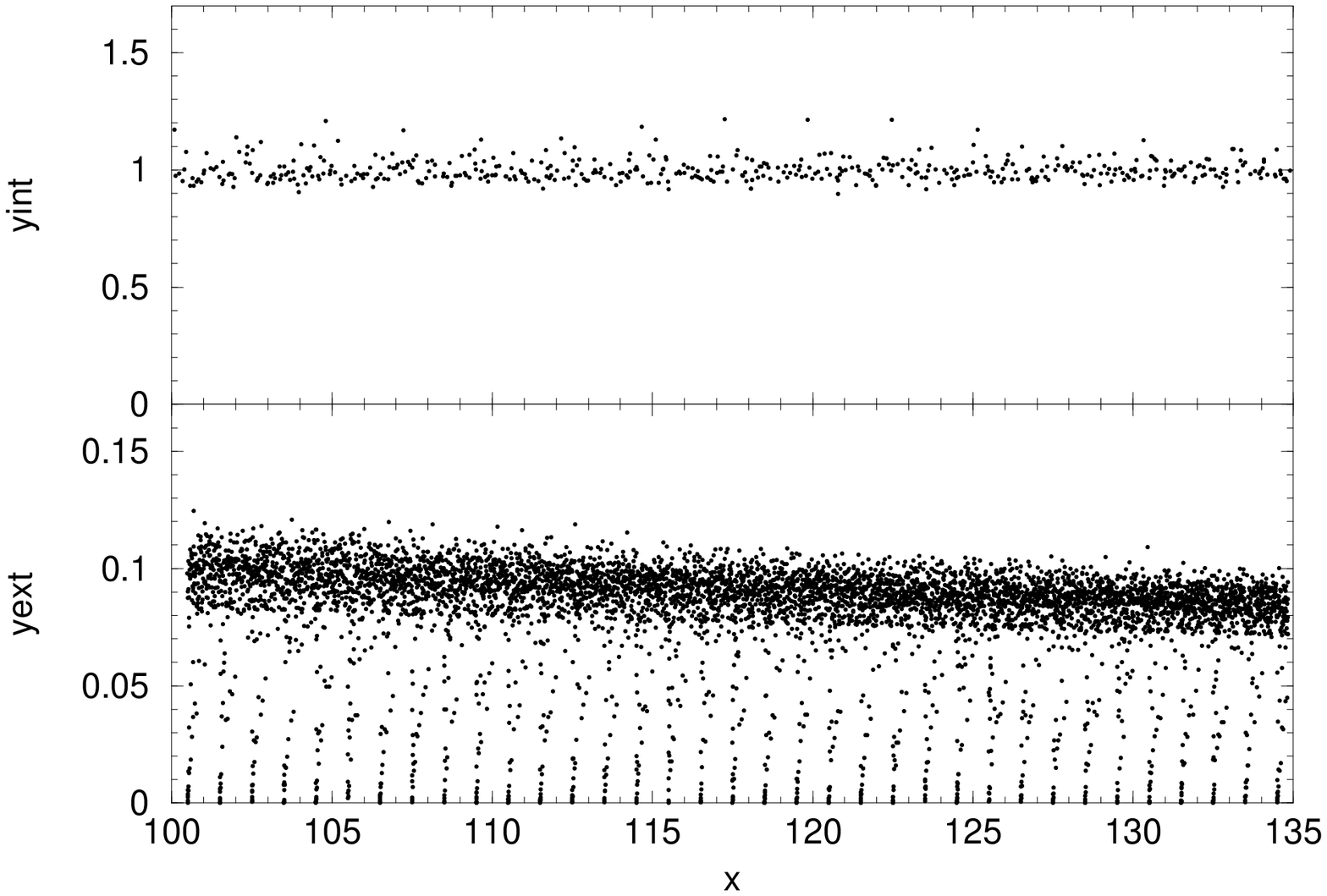}
    \figurecaption{%
      Weights as obtained from the edge magnetization for the stadium
      billiard at $b=0.2$ and high energy.  Note that the values of
      the weights differ in the (top) and exterior (bottom) by a
      factor of about ten.  Notwithstanding, the mean edge state
      densities are equal to leading order. The classical cyclotron
      radius which corresponds to this part of the spectrum is large,
      $\rho\in[2;2.32]$, giving rise to essentially hyperbolic
      classical motion.
 }
    \label{fig:s14magspec}
  \end{center}%
\end{figure}

\begin{figure}[tp]%
  \begin{center}%
    \includegraphics[width=0.66\linewidth] 
    {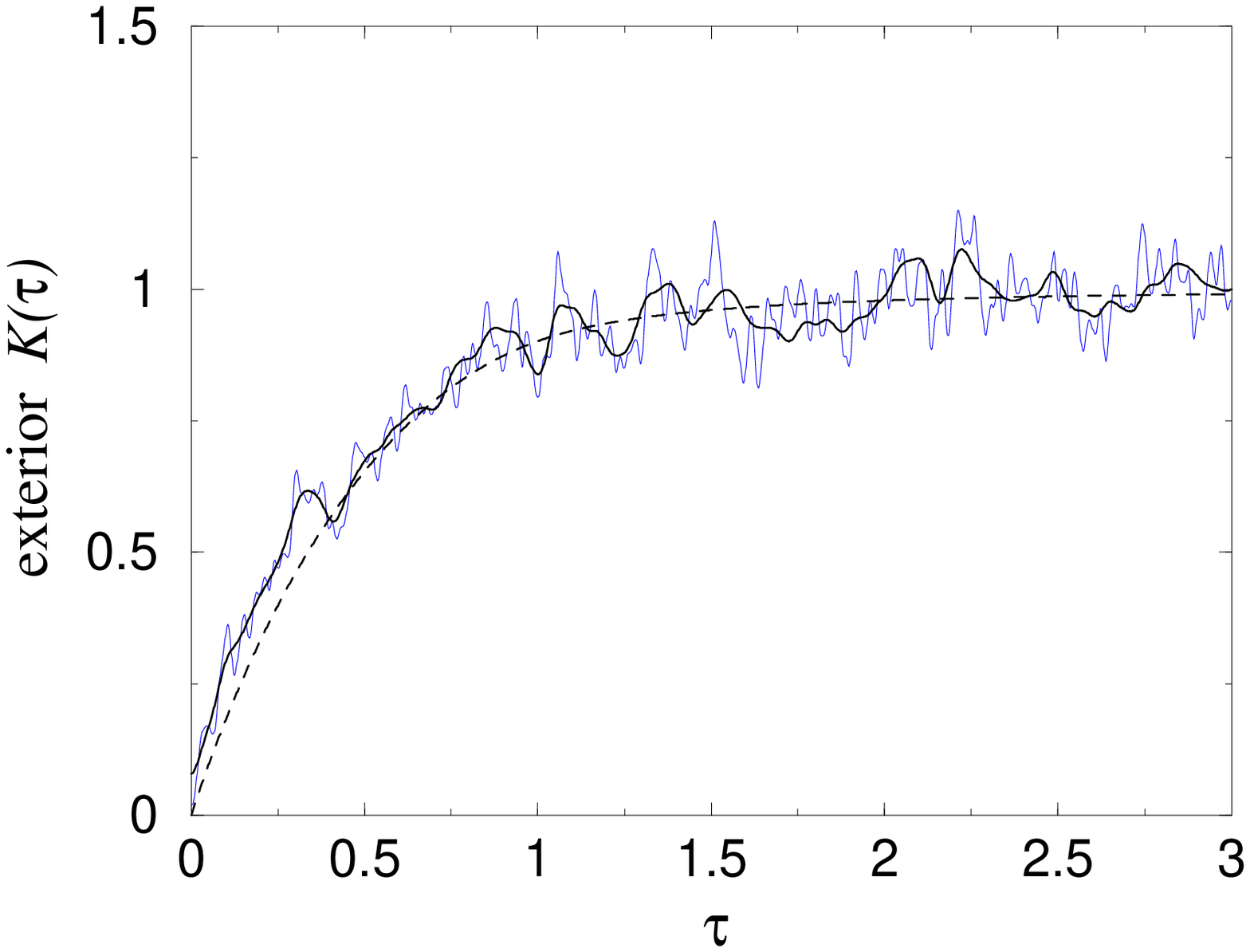}
    \figurecaption{%
      Form factor of the \emph{exterior} Bunimovich stadium
      (Fig.~\ref{fig:s14shape}) computed from the edge magnetization
      spectrum shown in Fig.~\ref{fig:s14magspec}, bottom part.  }
    \label{fig:mform}
  \end{center}%
\end{figure}

The weighted spectra are shown in Figure \ref{fig:s14magspec}.  Note
that the weights are very different in magnitude, although they lead
to the same average edge magnetization \eref{eq:Magsmboth}.  This
is explained by the different areas $\Area_{\rm skip}$ of the interior
and exterior skipping motion since the mean weight is asymptotically
determined by the ratio
\begin{align}
  \label{eq:meanu}
  \overline{w}^\magn = 
  \frac{\pm\overline{\mm}_{\rm edge}(\nu)}
  {\nu\,\overline{d}_{\rm skip}(\nu)}
  \simeq
 \frac{\Area}{\Area_{\rm skip}(\rho=\sqrt{\nu}b)}
\CO
\end{align}
see \eref{eq:dskip} and \eref{eq:mmedgesm}; (in the interior case
$\Area_{\rm skip}=\Area$). 
Similar to Fig.~\ref{fig:histoweights} 
the distributions
of the magnetization weights are localized
at $\overline{w}^\magn$ (not shown).

Figure \ref{fig:mform} presents the form factor
\eref{eq:formfactordef} of the exterior magnetization spectrum
restricting the energies to a single symmetry class. As one expects
the form factors follows the GOE prediction (dashed line).  This
indicates that the weights \eref{eq:wnmagdef} based on the
magnetization succeed to filter the bulk states consistently.  They
perform as well as the weights \eref{eq:defweight} based on the
boundary condition.

\section{Spectral cross correlations: The interior-exterior duality}
\label{chap:cross}

The previous chapter focused on the correlations within a given
interior or exterior edge spectrum and their relation to the
corresponding classical dynamics.  We now turn to a different type of
question, namely, whether one can relate the interior and the exterior
spectra belonging to the same billiard.

From a spectral theory point of view there is no apparent reason why the
spectra of the magnetic Laplacian defined on complementary domains
should have anything in common.  However, it was shown in Section
\ref{sec:cbilliard} that the classical periodic orbits of the interior
and the exterior problem are in general intimately related. They come
in dual pairs with equal stability and their actions adding up to an
integer multiple of the action of a cyclotron orbit.  Since the
semiclassical spectra are determined by the sets of periodic orbits
one expects that the correlation in the classical dynamics carries
over to the quantum case inducing a relation between the interior and
exterior spectra.

This observation being made it comes as a surprise that one does not
find any signature of a cross-correlation if the spectra are analyzed
with standard statistical methods, ie, the interior and exterior
spectra of a billiard seem to be statistically independent (not
shown).  Nonetheless, it will be found in the present chapter that
there exists indeed a strong cross-correlation between the two
spectra.  It can be observed only if a quantitative definition of edge
states, as developed above, is at hand.

\subsection{A semiclassical theory of spectral cross correlations}

In order to unravel the connection between interior and exterior edge
state energies a special cross-correlation function is needed.  It not
only involves the Dirichlet energies of the edge states but also
relies crucially on the information provided by their weights.

\myparagraph{The cross correlation function}

As the first step to obtain the appropriate correlator, we
 \emph{formally} extend  the definition of the edge state density to
finite boundary mixing parameters $\Lambda$.  
\begin{align}
  \label{eq:dedlam}
  d_{\rm edge}(\nu;\Lambda)
  \defas 
  - \frac{\rmd}{\rmd\Lambda} 
  \N(\nu;\Lambda)
\end{align}
The dependence of the spectral density on $\Lambda$ will be needed
only in the vicinity of the Dirichlet boundary condition, $\Lambda=0$,
\eref{eq:bcond}, where an expansion to first order in $\Lambda$ is
allowed. The spectral density \eref{eq:dedlam} can then be written
only in terms of the Dirichlet energies and Dirichlet weights,
\begin{align}
  \label{eq:degela}
  d_{\rm edge}(\nu;\Lambda)
  &=
  \sum_{n=1}^\infty
  \frac{\rmd \nu_n}{\rmd \Lambda}(\Lambda)\;
  \delta\big(\nu-\nu_n(\Lambda)\big)
  \nnn
  &\cong
  \sum_{n=1}^\infty
  \frac{\rmd \nu_n}{\rmd \Lambda}(0)\;
  \delta\Big(
    \nu
    -\nu_n(0)
    -\Lambda\, \frac{\rmd \nu_n}{\rmd \Lambda}(0)
  \Big)
  \nnn
  &=
  \sum_{n=1}^\infty
  \delta\Big(\frac{\nu-\nu_n}{w_n} -\Lambda\Big)
\CO
\end{align}
which follows from \eref{eq:defweight} and the properties of the
$\delta$-function.  The cross-correlation function is now defined as
an integral over energy and boundary parameter
\begin{equation}
  \label{eq:Cdef}
  C(\nu_0)=
  \int\!\!\!\int
  d^{\rm osc(int)}_{\rm edge}(\nu;\Lambda)  \,
  d^{\rm osc(ext)}_{\rm edge}(\nu;-\Lambda)  \,
  h(\Lambda)  \,
  g(\nu-\nu_0)  \,
  \rmd\Lambda   \,
  \rmd\nu
\CO
\end{equation}
with normalized Gaussian window functions $h$ and $g$.  Here, $h$
serves to restrict the integration over $\Lambda$ to the range where
the linear approximation in \eref{eq:degela} is valid and may have a
width of order one. The function $g$ is needed to regularize the pair
distribution $ d^{\rm osc(int)}_{\rm edge}(\nu) \, d^{\rm
  osc(ext)}_{\rm edge}(\nu)$.
It selects a narrow energy interval
centered around the energy $\nu_0$ and should have the width of a few
effective nearest neighbor spacings. 

Inserting expression \eref{eq:degela} the cross-correlation function
turns into a double sum over the interior and exterior edge spectrum,
\begin{equation}
  \label{eq:qmcorr}
  C(\nu_0)=
  \sum_{i,j=1}^\infty
  \frac{w_i w_j'}{w_i+w_j'}
  \,
  g\!\left(
    \frac{\ds  \frac{\nu_i-\nu_0}{w_i}-\frac{\nu_0 - \nu_j' }{w_j'}       }
    {\ds  \frac{1}{w_i}+\frac{1}{w_j'} }
  \right)
  h\!\left(
    \frac{\nu_i-\nu_j'}{w_i+w_j'}
  \right)
  - C_{\rm bg}
\CO
\end{equation}
where the primes label the exterior energies and
weights for the sake of brevity.

This pair correlation function is far from being standard since it
relies heavily on the weights attributed to the individual levels.
However, this function is the most natural choice to accentuate the
spectral cross-correlations which originate from the underlying
classical interior-exterior duality. This will be shown below.  The
important point to note is that due to the small width of $g$ only a
few pairs of interior and exterior spectral points will contribute
appreciably at a given $\nu_0$.  It is the pairs with equal
\emph{weighted distances} from the left and right, respectively, to
the reference energy $\nu_0$.  Here, the energy differences are scaled
individually by the reciprocal weight attached to each spectral point.
The function $h$, in contrast, limits the absolute energy distance.
Note also that the prefactor in \eref{eq:qmcorr} ensures that pairs
which include at least one bulk state do not contribute appreciably to
the sum.

The term $C_{\rm bg}$ in \eref{eq:qmcorr} subtracts the  background.
It is approximated by
\begin{equation}
  \label{eq:Cbg}
  C_{\rm bg}
  \cong
  \overline{d}_{\rm edge}(\nu_0)
  \left(
  \sum_{i=1}^\infty  h\Big(\frac{\nu_i-\nu_0}{w_i}\Big)
  +
  \sum_{j=1}^\infty  h\Big(\frac{\nu_j'-\nu_0}{w_j'}\Big)
  -\overline{d}_{\rm edge}(\nu_0)
  \right),
\end{equation}
if we neglect the width of $g$ and disregard the fact that the
interior and exterior mean edge densities differ in the higher order
terms.  We shall discuss the correlation function further after we
derive its main properties using the semiclassical approximation.

\myparagraph{The semiclassical correlator }

We turn now to the semiclassical evaluation of the correlation
function using the periodic orbit formula \eref{eq:dedgeosc} discussed
in Section \ref{sec:dedgesc}. It applies to completely chaotic
systems.  One obtains a double sum over the skipping interior and
exterior periodic orbits:
\begin{align}
  \label{eq:sccor1}
    C(\nu_0)=
    \int\!\! \rmd\nu\,& g(\nu-\nu_0)
    \,\,
    \frac{2}{\pi^2}
    \sum_{\po,\po'}
    \frac{    w_\po\,\tau_\po}{r_\po\, |\tr{\rm M}(\po)-2|^\oh}
    \,
    \frac{w_{\po'}\,\tau_{\po'}}{r_{\po'}\, |\tr{\rm M}(\po')-2|^\oh}
    \qq\qq
    \nnn
    \times
    \Bigg\{
    &\cos\left(
      2\pi\nu\big(\Ga(\po)+\Ga(\po')\big)
      -\pi(n_\po+n_{\po'})-\frac{\pi}{2}(\mu_\po+\mu_{\po'})
    \right)
    \nnn
    &\times\,
    \hat{h}\bigg(
      \frac{1}{\pi}\sum_{j=1}^{n_\po} |\nvec_j\vvech_j|
      -\frac{1}{\pi}\sum_{j=1}^{n_{\po'}} |\nvec_j'\vvech_j'|
    \bigg)
    \nnn
    +
    &\cos\left(
      2\pi\nu\big(\Ga(\po)-\Ga(\po')\big)
      -\pi(n_\po-n_{\po'})-\frac{\pi}{2}(\mu_\po-\mu_{\po'})
    \right)
    \nnn
    &\times\,
    \hat{h}\bigg(
      \frac{1}{\pi}\sum_{j=1}^{n_\po} |\nvec_j\vvech_j|
      +\frac{1}{\pi}\sum_{j=1}^{n_{\po'}} |\nvec_j'\vvech_j'|
    \bigg)
    \Bigg\}
\end{align}
Here, $\hat{h}$ is the Fourier transform of the window function $h$,
and the exterior quantities are again marked with a prime.  The width
of $\hat{h}$ is small compared to the sum over $ |\nvec_j\vvech_j|$
(which is of order $n_\po$).
As a result, the second term in the curly brackets of \eref{eq:sccor1}
is  suppressed. 
In the first term of equation \eref{eq:sccor1}, $\hat{h}$ reduces the
sum effectively to those pairs with approximately equal sums of angles
of incidence $\sum_j |\nvec_j\vvech_j|=$ $\sum_j |\nvec_j'\vvech_j'|$.
The dual pairs of periodic orbits discussed in Section
\ref{sec:duality}
have precisely this property.  Hence, the only systematic contribution
to the correlator will come from these pairs.  In
Sect.~\ref{sec:trace} we discussed the relations
between $\po$ and its dual partner orbit $\dpo$, which may be
summarized as
\begin{align}
  \begin{array}{rcl}
    \Ga(\po)+\Ga(\dpo) &=&  n_\po = n_\dpo \CO
    \\
    \tr{\rm M}(\dpo) &=& \tr{\rm M}(\po) \CO
    \\
    |{\nvec}_j{\vvech}_j|_{(\dpo)}
    &=&  |\nvec_{j-n_\po}\vvech_{j-n_\po}|_{(\po)} \CO
  \end{array}
  \qq\q
  \begin{array}{rcl}
    r_\dpo &=& r_\po \CO
    \\
    \mu_\dpo &=& 2 n_\po - \mu_\po \CO
    \\
    w_\dpo\, \tau_\dpo &=&  w_\po\, \tau_\po \PO
  \end{array}
\end{align}

If we retain  only the contributions of the dual pairs the cross-correlation
function simplifies to a single sum over
interior (or exterior) periodic orbits.  Assuming global classical
duality we obtain
\begin{equation}
  \label{eq:sccor2}
    C(\nu_0)
    =
    \frac{2}{\pi^2}   \sum_\po
    \frac{    w_\po^2 \,\tau_\po^2}{r_\po^2\, |\tr{\rm M}(\po)-2|}
    \cos(2\pi n_\po(\nu_0-\tfrac{1}{2}))
    \,\, \hat{g}(n_\po) \ .
\end{equation}
The restriction of the double sum \eref{eq:sccor1} to the dual pairs
is tantamount to the \emph{diagonal approximation} used in the semiclassical
evaluation of the autocor\-re\-la\-tion function \cite{Berry85}. 
In the present case, the actions of the chosen pairs of periodic orbits
complement each other to an integer $n_\po$, while in the usual diagonal approximation it
is the resonant terms, $ \Ga(\po)-\Ga(\po')=0$, which give the
dominant ``diagonal'' contribution.

In deriving \eref {eq:sccor2} the energy dependence of the amplitudes
of the trace formula could be neglected since the variation of the
energy was assumed to be small on the classical scale in
\eref{eq:Cdef}. If $\nu_0$ is taken large (i.e. we are in the
semiclassical regime of the spectrum) the classical quantities in
\eref{eq:sccor2} will hardly change as $\nu_0$ is varied.  By grouping
together the contributions from all the periodic orbits with the same
number of reflections $n_\po$ we obtain
\begin{align}
  \label{eq:sccor3}
  C(\nu_0)
    &=
    \sum_{n=n_{\rm min}}^\infty
    f(n) \, \hat{g}(n)\,
    \cos(2\pi n(\nu_0-\tfrac{1}{2}))
\CO
    \\
\intertext{with}\qq
  \label{eq:fn}
f(n) &=
    \frac{2}{\pi^2}   \sum_{\po:n_\po=n}
    \frac{    w_\po^2 \,\tau_\po^2}{r_\po^2\, |\tr{\rm M}(\po)-2|}\PO
\end{align}
Assuming ergodicity, the weighted sum over classical $n$-orbits
\eref{eq:fn} can be calculated as a phase space average. For large $n$
it takes on the universal value $f(n)=n/8$.  At the same time, the
number of reflections $n_\po$ is geometrically bounded from
bellow, $n\ge n_{\rm min}$, for a given cyclotron radius.
Hence,
\begin{align}
  f(n)=
  \begin{cases}
    0 & \text{for $n<n_{\rm min}$}
    \\
    \frac{1}{8} n & \text{as $n\gg n_{\rm min}$\ .}
  \end{cases}
\end{align}
Equation \eref{eq:sccor3} makes a clear prediction on the form of the
cross-correlation function.  Even if the classical dynamics changes
slowly as $\nu_0$ is varied the infinite sum \eref{eq:sccor3} will be
appreciable only at energies $\nu_0=N+\tfrac{1}{2}$,
$N\in\mathbb{N}_0$, where the cosine terms are stationary.  We
therefore expect the cross-correlation function to display pronounced,
equidistant peaks at large energies.  Their positions are expected to
coincide with the Landau levels (although they have nothing to do with
bulk states) and their appearance provides a direct quantum
manifestation of the existence of classically dual orbits.

If the billiard exhibits a discrete symmetry the semiclassical theory
suggests also a natural way to test that the predicted structures in
$C(\nu_0)$ are not artefacts (eg due to the bulks states).  In this
case the cross correlation between exterior and interior spectra with
\emph{different} symmetries is derived in a similar fashion as
\eref{eq:sccor3}.  However, now we have
\begin{align}
  \label{eq:fndifsym}
   f(n) =
    \frac{2}{\pi^2}   \sum_{\po:n_\po=n} (-)^{s_\po}
   \frac{    w_\po^2 \,\tau_\po^2}{r_\po^2\, |\tr{\rm M}(\po)-2|}
\CO
\end{align}
where $s_\po$ counts the number of times the periodic orbit $\po$
crosses the symmetry line \cite{Robbins89}. Since in the sum $s_\po$
will be even or odd with equal frequency
the terms cancel on average and no correlation signal is expected.

\myparagraph{Action cross correlations}

Next, we consider the Fourier transform of the cross-correlation
function \eref{eq:Cdef} which highlights its fluctuating part.
The semiclassical theory predicts a sequence of equidistant
$\delta$-spikes at integer values,
\begin{equation}
  \label{eq:corrft}
  D(t)=  \int\!C(\nu_0)\,\e^{-2\pi\rmi\nu_0 t} \rmd \nu_0
  = \frac{1}{2}   \sum_{n=n_{\rm min}}^\infty (-)^n
  f(n)\,\hat{g}(n)\,\delta(n-t) 
\PO
\end{equation}
They correspond to the sums of the actions of dual pairs, which
complement each other to integer values, starting from the minimal
number of reflections $n_{\rm min}$.

\myparagraph{Using the edge magnetization}

Let us turn to the question whether the correlation is also seen if
one uses the spectral density $\dedge^\magn$ based on the edge
magnetization \eref{eq:mmedge2} as the spectral measure.  It has the
advantage of being easier to measure both numerically and in
experiments since one does not have to change the boundary conditions.
The theoretical treatment is completely analogous to the above with
the correlation function now involving an integration over the the
magnetic length parameter $b$ rather than the boundary condition:
\begin{align}
  \label{eq:Cmagdef}
  C_{\rm mag}(\nu_0)&=
   \int\!\!\!\int
  d^{\rm osc\magn}_{\rm edge(int)}(\nu;b^2)  \,
  d^{\rm osc\magn}_{\rm edge(ext)}(\nu;b^2)  \;
  h\Big(\frac{b^2-b_0^2}{b_0^2}\Big)  \, 
  \frac{\rmd (b^2)}{b_0^2}\,
  \nnn
  &\phantom{=\int\!\!\!\int}
  \times 
  g(\nu-\nu_0)  \,\rmd\nu
\PO
\end{align} 
The linear expansion of the
dependence of the energies on $b^2$ yields a double sum
like eq \eref{eq:qmcorr} with the   $w_n$ 
replaced by the magnetic weights $w_n^\magn$ \eref{eq:wnmagdef}.
Semiclassically, the integration over $\delta b^2$ selects those pairs
of interior and exterior orbits which satisfy
\begin{align}
\label{eq:mpairrel}
  w^\magn_\po\, \tau_\po &= 
  w^\magn_{\po'} \, \tau_{\po'}
\CO
\intertext{ie,}
\tag{\ref{eq:mpairrel}a}
  (\Len_\po+\Len_\po')\rho &= 2(\Area_\po+\Area_\po')
\CO
\end{align}
which is again the \emph{dual} pairs, $\po'=\dpo$. (This may be seen
from eqs \eref{eq:wpomagdef} and \eref{eq:deftau} since the signs of
the $\sigma_j$ \eref{eq:sigmadef} and the order of the points of
reflection $\rvec_j$ are reversed as one goes from an orbit to its
dual. Geometrically, it is evident
that dual orbits with $N$
reflections satisfy $\Len_\po+\Len_\dpo= 2\pi\rho N$ and
$\Area_\po+\Area_\dpo=\rho^2\pi N$.)

These predictions for the cross-correlation function are not
restricted to purely chaotic dynamics, although the bouncing map was
assumed to be hyperbolic, so far.  For the (integrable) disk one
obtains a completely analogous result.  The function $f(n)$ is not
universal in this case, but the prediction remains that $C(\nu_0)$ is
peaked at the energies of the Landau levels.  
Below, in Sect.~\ref{sec:pairrel} it will be shown that semiclassical
correlations can be predicted even for mixed chaotic systems (without
resorting to periodic orbit theory).

\subsection{Numerical   evidence}

In this section we provide numerical evidence supporting the above
semiclassical predictions. We start by analyzing the edge spectra of
the ellipse billiard.  The underlying classical motion is not
completely chaotic but still we expect most of the predictions of the
semiclassical analysis to hold. We choose this example because we
accumulated the most extensive numerical data for this system.

\begin{figure}[p]%
  \begin{center}%
    \includegraphics[width=0.8\linewidth] 
    {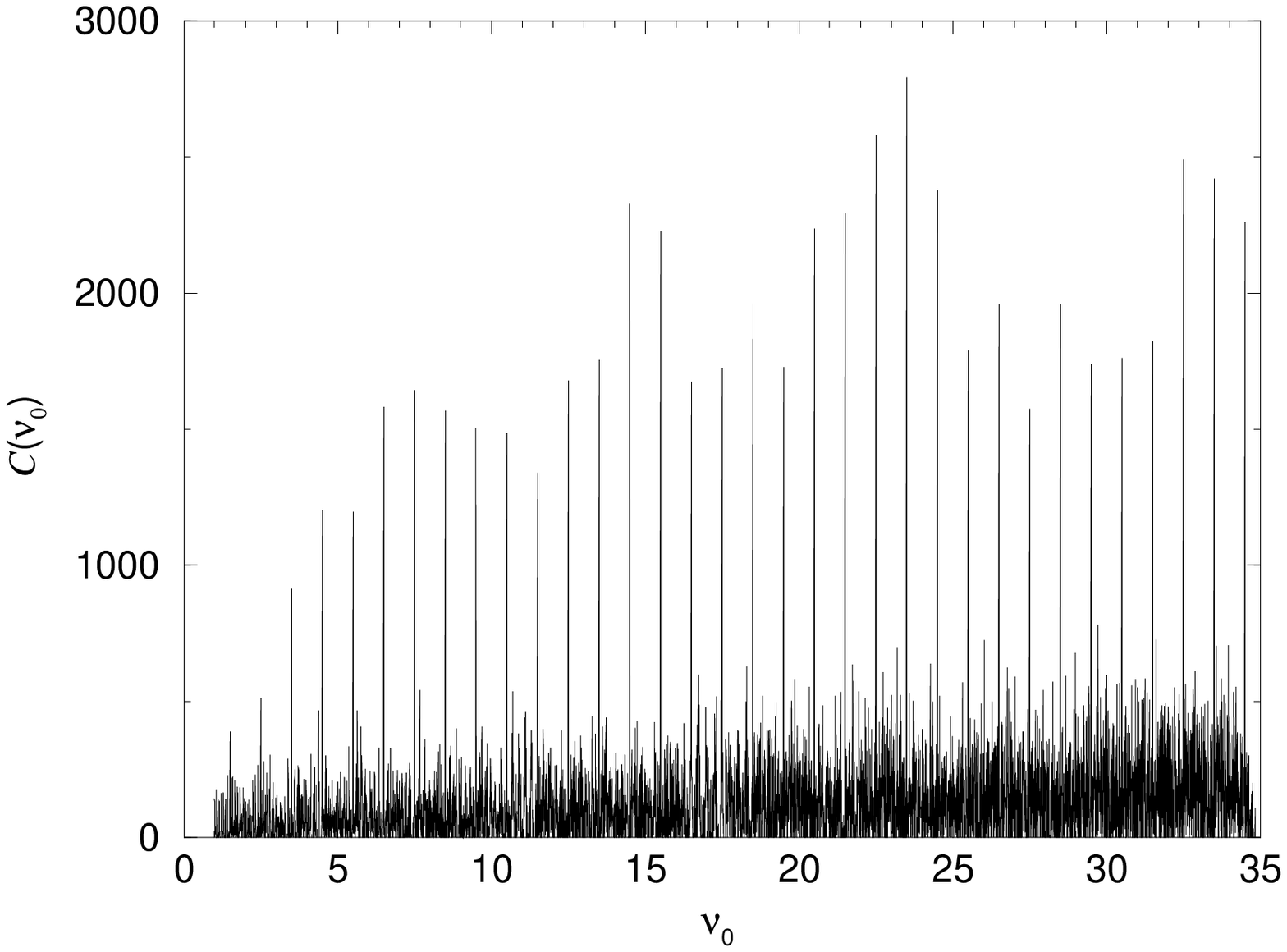}
    \figurecaption{%
      Cross-correlation function \eref{eq:qmcorr} for the elliptic
      billiard (eccentricity $\epsilon=0.8$, $b=0.1$,
      $\sigma_g=0.001$, $\sigma_h=1$, positive part). The pronounced
      peaks at $\nu_0=N+\tfrac{1}{2}$, $N\in\mathbb{N}_0$, indicate
      the existence of non-trivial correlations between interior and
      exterior edge states. (The figure remains unchanged if one
      removes all bulk states from the sum \eref{eq:qmcorr} by
     imposing a threshold on $w_n$; not shown.)  }
\label{fig:Cellipse}%
\end{center}%
\end{figure}
\begin{figure}[p]%
  \begin{center}%
    \includegraphics[width=0.8\linewidth] 
    {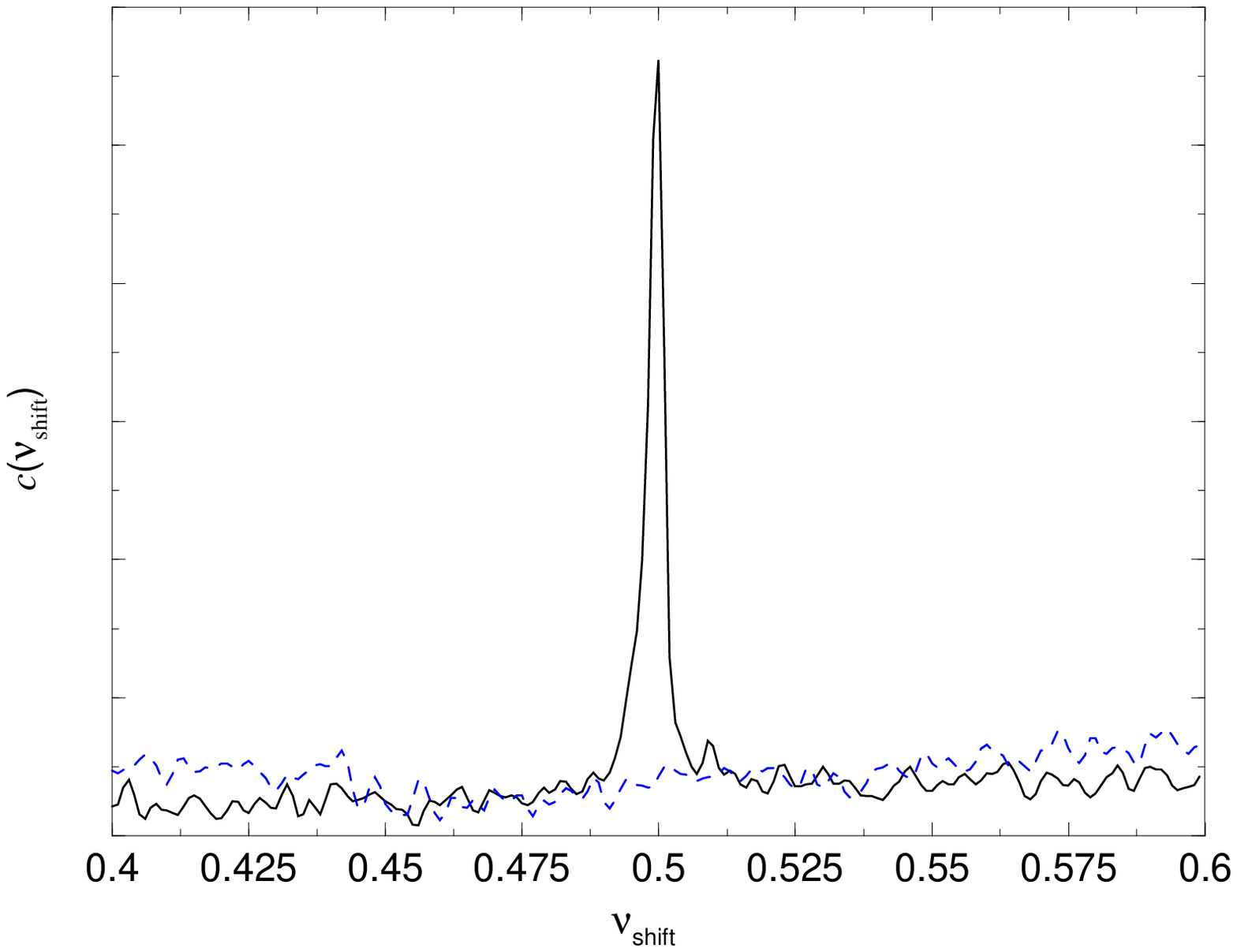}
    \figurecaption{%
      Cross-correlation function from Fig.~\ref{fig:Cellipse} summed
      over integer shifts of the argument, $c(\nu_{\rm shift}) =
      \sum_n C(n+\nu_{\rm shift})$, in order to focus on the positions
      of the peaks.  In the double sum \eref{eq:qmcorr} the energies
      were taken within the same symmetry class (solid line) and
      between different symmetry classes (dotted line.)}
\label{fig:modulo}%
\end{center}%
\end{figure}

In Figure \ref{fig:Cellipse} we show the cross-correlation function
\eref{eq:qmcorr} for the ellipse billiard at magnetic length $b=0.1$.
It was calculated from the edge spectra shown in Figures
\ref{fig:wspecELext} and \ref{fig:qmclswt}.  The corresponding
classical dynamics
exhibits a strict one-to-one correspondence between the interior and
the exterior classical dynamics up to $\nu=21.6$.  Beyond this energy,
when the cyclotron radius is greater than the minimum radius of
curvature, the classical duality still holds in a substantial part of
phase space.
One observes that $C(\nu_0)$ is strongly fluctuating and displays
pronounced, equidistant peaks at energies $\nu_0=N+\oh$, as predicted
by \eref{eq:sccor2}.  In Figure \ref{fig:modulo} we focus on these
dominant structures by plotting the cross-correlation function in
terms of $\nu_{\rm shift}=\nu_0({\rm mod}\ 1)$ around one half.  To
check that 
the resulting correlation signal
is not an artefact or due to the
accumulation of bulk states we make use of the fact that the spectra
of the ellipse decompose into two symmetry classes.  As shown above
\eref{eq:fndifsym} one expects $C(\nu_0)$ to be structureless if one
correlates edge spectra belonging to different symmetries.  This is
clearly supported by the numerical results shown as a dashed line in
Figure \ref{fig:modulo}.

\begin{figure}[t]%
  \begin{center}%
    \psfrag{t}{$t$}
    \psfrag{|D(t)|}{$|D(t)|$}
    \includegraphics[width=0.8\linewidth] 
    {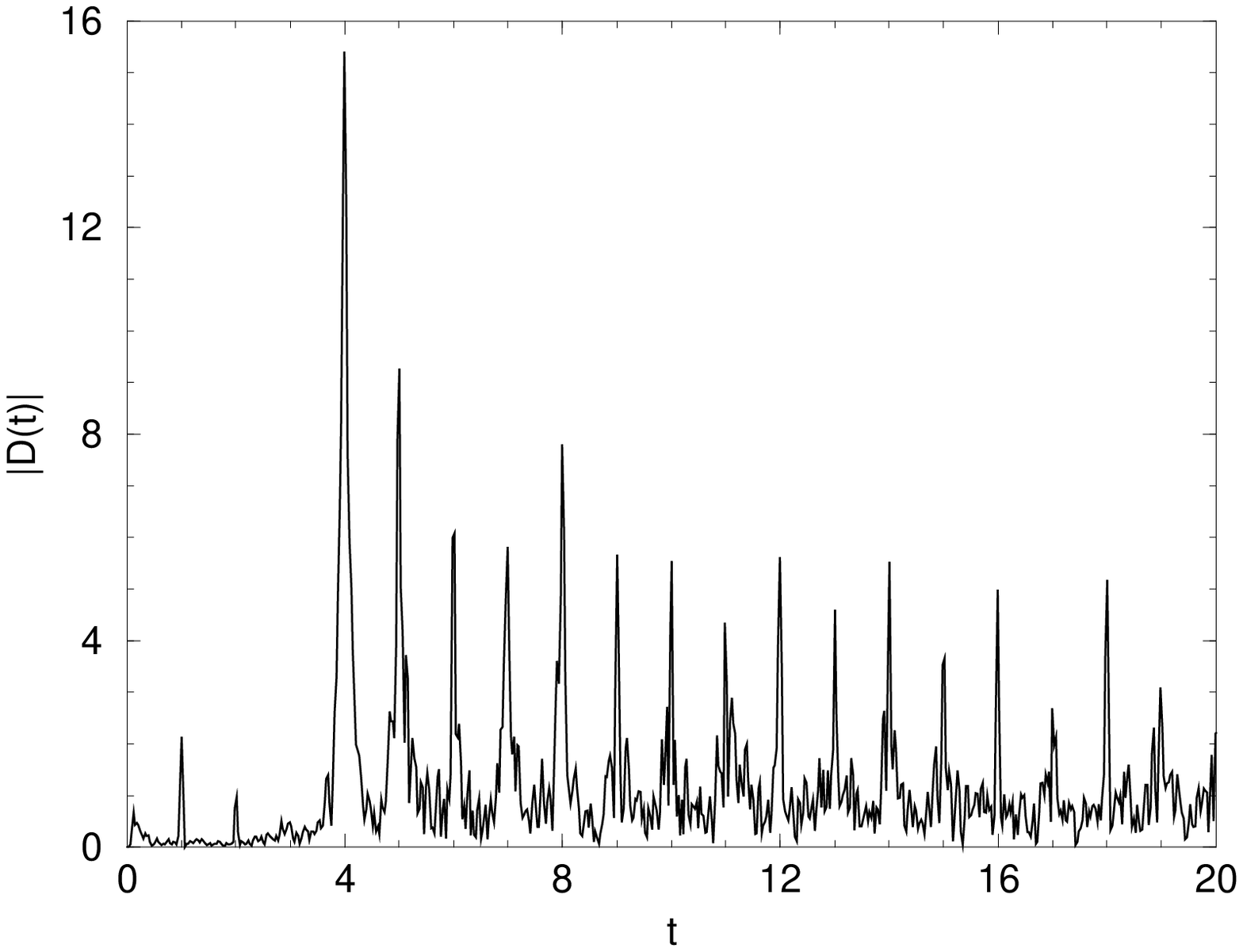}
    \figurecaption{%
      Fourier transformation $D(t)$ \eref{eq:corrft} of the
      cross-correlation function $C(\nu_0)$ given in
      Fig.~\ref{fig:Cellipse} (absolute value). The peaks at integer
      $t$ correspond to the combined actions of dual periodic orbits
      ($\sigma_g=5\times~10^{-4}$, $\sigma_h=4$.) 
       }
\label{fig:FTellipse}%
\end{center}%
\end{figure}

The Fourier transform \eref{eq:corrft} of the cross-correlation
function exposes the sums of actions of the contributing pairs of
periodic orbits.  The absolute value of $D(t)$, calculated for the
same spectrum as Fig.~\ref{fig:Cellipse}, is shown in Figure
\ref{fig:FTellipse}.  In this case, the periodic orbits of the
de-symmetrized ellipse have at least $n_{\rm min}=4$ reflections.  One
observes that $|D(t)|$ displays distinct spikes at integer values. The
real parts of the peaks have signs $(-)^n$, as expected from eq
\eref{eq:fn} (not shown).  As predicted by the semiclassical theory
the dominant peaks start at $n_{\rm min}=4$ which is a clear proof
for the classical origin of the edge state correlations.
The tiny peaks at $t=1,2,3$ vanish if one decreases the width of the
window function $g$ (which in turn deteriorates the statistical
significance of the result).  They are due to the remnant
contributions of the bulk states, and disappear if one removes
the bulk states from the correlator sum by setting a
threshold on the weights (not shown; the remaining peaks would not
change by this procedure).

\begin{figure}[tp]%
  \begin{center}%
    \psfrag{y}{$|D_{\rm mag}(t)|$}
    \psfrag{t}{$t$}
    \includegraphics[width=0.8\linewidth] 
    {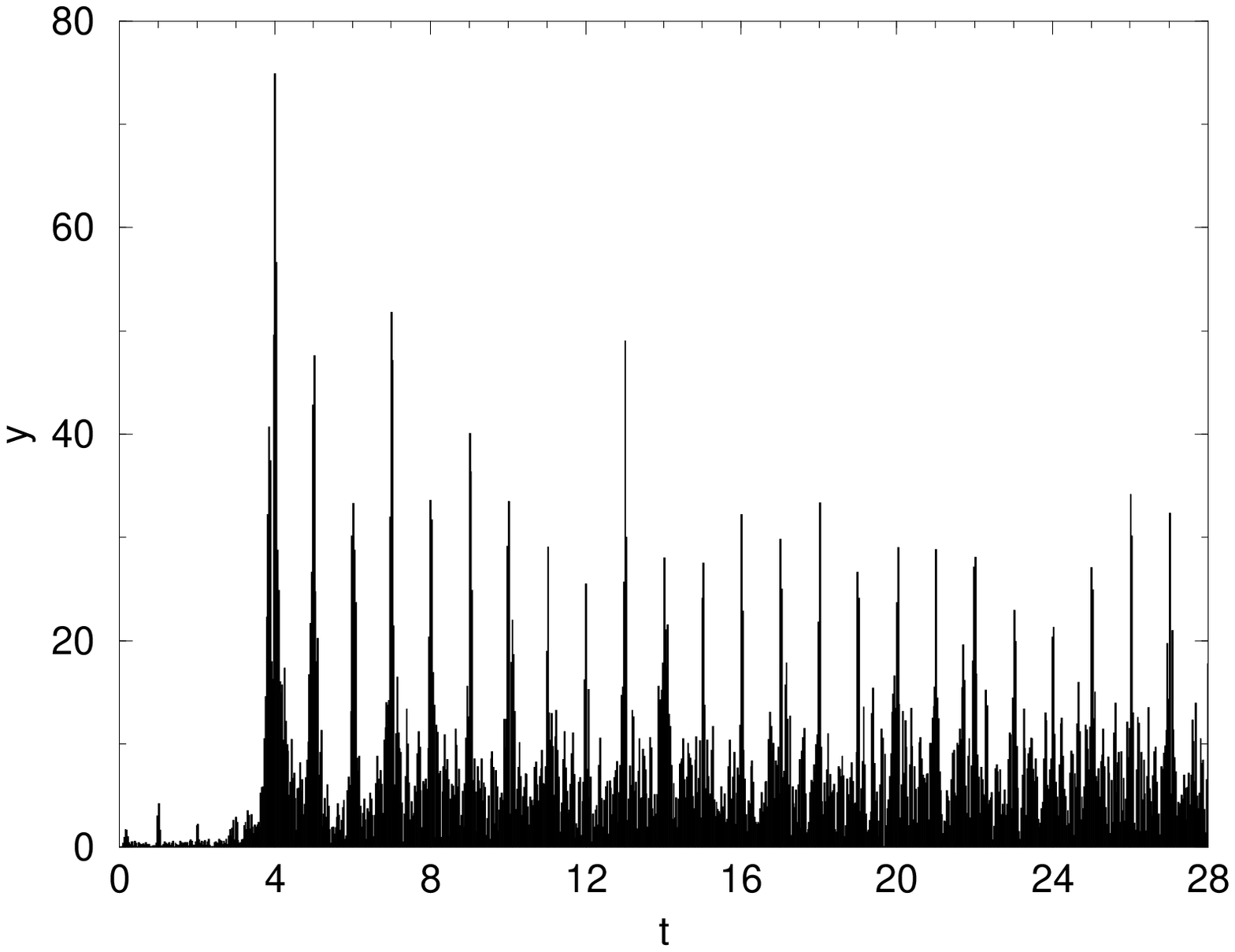}
    \figurecaption{%
      Fourier transform of the ellipse cross correlation function
      \eref{eq:Cmagdef} for the ellipse defined in terms of the edge
      magnetization density.  The graph should be compared to
      Fig.~\ref{fig:FTellipse}. 
      ($\sigma_g=5\times 10^{-4}$,
      $\sigma_h=0.5$) }
    \label{fig:mcorrellipseft}%
  \end{center}%
\end{figure}

We repeated the calculation of the cross-correlation function of the
ellipse spectrum now using $\dedge^\magn(\nu)$ as the spectral
measure. The resulting function exhibits peaks at the Landau energies
similar to Fig.~\ref{fig:Cellipse} (not shown). Its Fourier transform
is given in Figure \ref{fig:mcorrellipseft}.  Again, the peaks are
located at integer values starting at $t=4$.  This shows that the edge
magnetization density $\widetilde{\mm}_{\rm edge}$ succeeds to unravel
the cross correlations similar to the edge density $\dedge$ --- a
reassuring but not a surprising result.

\begin{figure}[t]%
  \begin{center}%
    \psfrag{x}{$\nu_{\rm shift}$}
    \psfrag{y}{$\hspace*{-5mm}c(\nu_{\rm shift})$}
    \includegraphics[width=0.8\linewidth] 
    {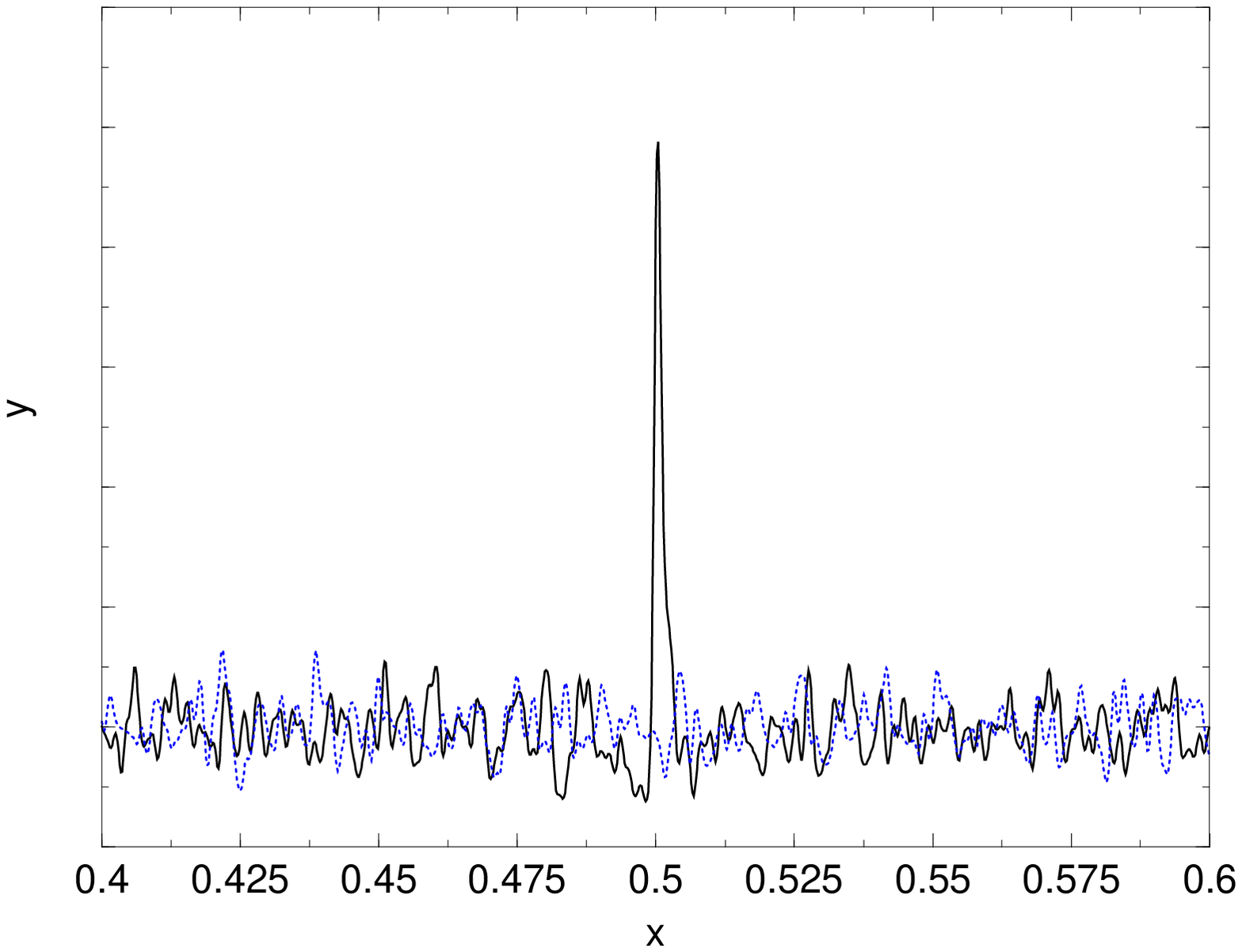}
    \figurecaption{%
      Cross-correlation summed over integer shifts of the argument
      like in Fig.~\ref{fig:modulo}. The data belongs to the stadium
      billiard in Fig.~\ref{fig:s14magspec} ($\nu=100-135$,
      $\sigma_g=5\times 10^{-4}$, $\sigma_h=0.2$, using the edge
      magnetization density \eref{eq:mmedge2}.) A clear
      cross-correlation exists between energies of the same symmetry
      class (full line), while there is no signal if the energies are
      taken from different symmetry classes (dashed line.)  }
\label{fig:stadiumshift}%
\end{center}%
\end{figure}

The ellipse spectrum considered so far exhibits generic, mixed chaotic
dynamics with relatively large integrable parts in phase space. As the
last point, we demonstrate that the correlations do exist also in a
system which is completely chaotic. We choose the spectrum of the
\emph{stadium billiard} defined in Fig.~\ref{fig:s14shape} and use the
edge magnetization to define the spectral density.
As discussed in Chapter \ref{chap:stat}, the spectral interval shown
in Fig.~\ref{fig:s14magspec} corresponds to cyclotron radii large
enough to ensure that the corresponding classical dynamics is
essentially hyperbolic.
Figure \ref{fig:stadiumshift} gives the corresponding
cross-correlation function. Like in Fig.~\ref{fig:modulo} the variable
is plotted modulo one in order to focus attention on the peaks. Again,
we observe a clear cross-correlation signal for pairs within the same
symmetry class (solid line) while the reference calculation from
different symmetry classes shows no peak (dashed line).

\subsection{The pair relation}
\label{sec:pairrel}

\begin{figure}[tb]%
  \begin{center}%
    \includegraphics[width=0.5\linewidth] 
    {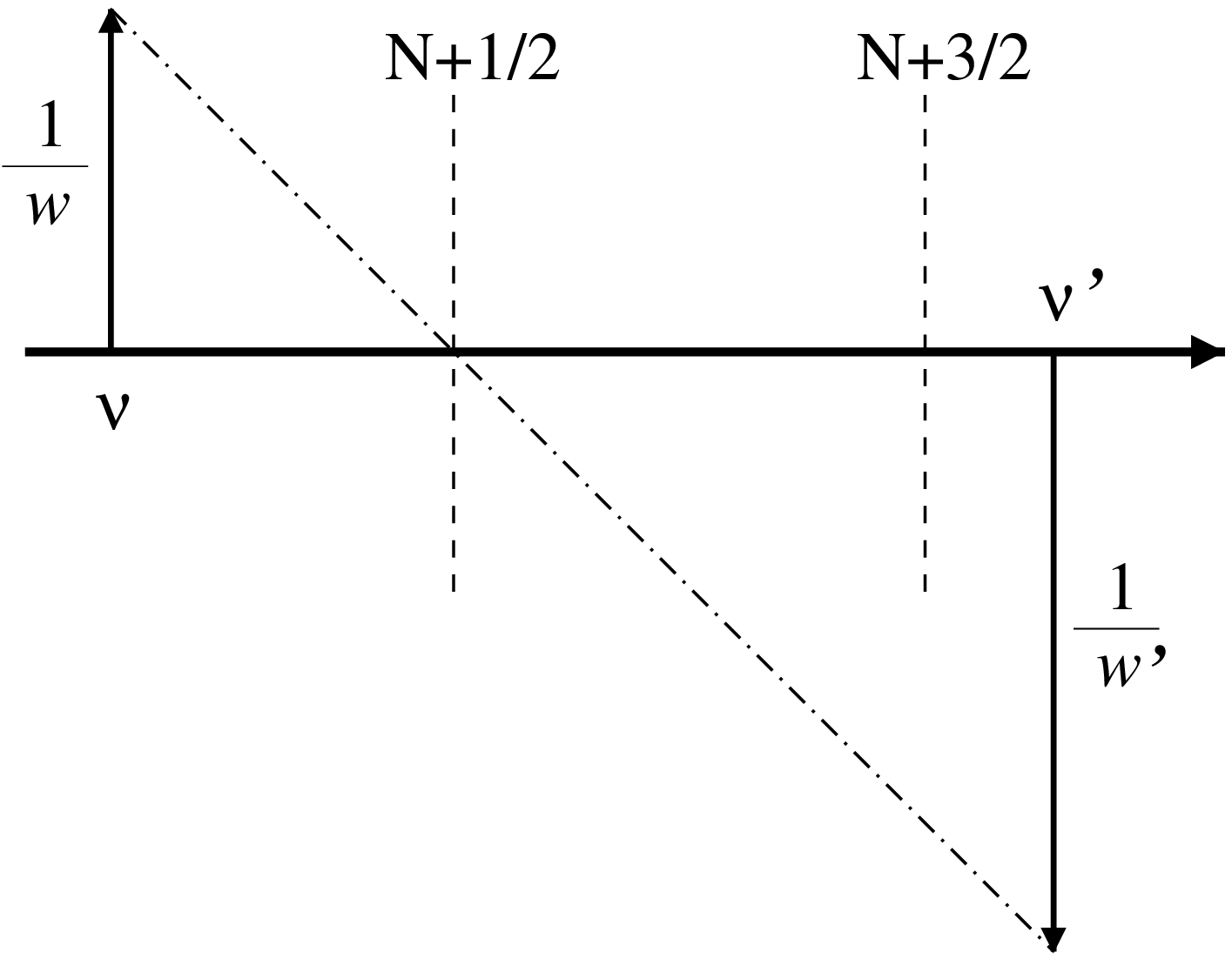}
    \figurecaption{%
      For every correlated pair of interior and exterior edge energies,
      $\nu$ and $\nu'$, there exists a Landau level $N+\oh$ such that
      the distances -- scaled individually by the reciprocal quantum
      weights, $w$ and $w'$ -- coincide.
      }
\label{fig:pairrelation}%
\end{center}%
\end{figure}

The peaks in $C(\nu_0)$ were attributed to the complementarity of the
actions of dual orbits.  Quantum mechanically their occurrence
\emph{implies} that there exists a  pairwise relation between
individual interior and exterior edge states.  This follows from the
discussion of the quantum correlator \eref{eq:qmcorr} above. We have
noted that pairs of edge energies contribute only if they have the
same {weighted} distance to the reference energy
from the left and right, respectively. Since the peaks appear at
$\nu_0=N+\tfrac{1}{2}$ the interior energies $\nu_i$ and exterior
energies $\nu_j'$ must appear in pairs which satisfy
\begin{equation}
  \label{eq:qmrel}
  \frac{\nu_i-(N-\tfrac{1}{2})}{w_i}
  \cong
  \frac{(N-\tfrac{1}{2})-\nu_{j}'}{w_{j}'}
\end{equation}
with integer $N$, see Fig.~\ref{fig:pairrelation}.  Although this is
not an exact relation, it will be the more precise the larger and the
closer the two energies are, since the semiclassical approximation
\eref{eq:dedgeosc} and the linearization \eref{eq:degela} then hold
the better.

It is clear from  \eref{eq:qmrel} that the information provided by
the ratio of the individual quantum weights plays a crucial role in
unraveling this pair correlation.  It explains why standard
correlation functions, which involve unweighted densities, do not show
any signal in general.  Moreover, the fact that the quantum weights
enter reciprocally in \eref{eq:qmrel} explains how a pairwise relation
between interior and exterior states can exist in spite of different
local unweighted densities.  It is consistent with the mean edge
densities \eref{eq:dedgesm} being equal in the interior
and exterior.
Using the weights \eref{eq:wnmagdef} which are based on the
magnetization we obtain the same formula. It implies that for
correlated pairs the individual ratios of the weights defined by
\eref{eq:defweight} and by \eref{eq:wnmagdef} are approximately equal.
This is indeed observed numerically.

For a given interior edge state it is of course not known, a priori,
which is the associated Landau level $N+\oh$ and the exterior weight.
Therefore, even in an asymptotic sense it is not possible to infer
an edge spectrum given the complementary one by just using the relation
\eref{eq:qmrel}.  However, having an interior and exterior edge
spectrum available, one can  decide whether they belong to the
same billiard.
In the spectra considered so far we could easily spot single pairs of
edge states by just using the relation \eref{eq:qmrel}. Examples are
given in Table \ref{tab:cpair}.

\begin{table}[tb]
  \begin{center}
    \fbox{
      \small
      \begin{tabular}{lcccccc}
        Pair in Fig.~\ref{fig:cpair} 
        & $\nu_i$  &  $w_i$  &  $\dfrac{\nu_i-\nu_0}{w_i}$ &
        $\dfrac{\nu_0-\nu'_j}{w'_j}$  &  $w'_j$  &  $\nu'_j$
        \\[2ex]
        top &
        $32.5367$ &
        $0.826$   &
        $0.0445$ &
        $0.0444$ &
        $0.506$ &
        $32.4775$
        \\
        middle &
        $33.5248$ &
        $0.489$ &
        $0.0507$ &
        $0.0501$ &
        $0.533$ &
        $33.4733$
        \\
        bottom & 
        $32.5082$ &
        $0.286$ &
        $0.0288$ &
        $0.0248$ &
        $0.508$ &
        $32.4874$
      \end{tabular}
      }
  \vspace*{\baselineskip}
  \end{center}
  \figurecaption{%
    Energies and weights of the correlated pairs 
    in Fig.~\ref{fig:cpair}, with the primes indicating exterior states.
    }
  \label{tab:cpair}
\end{table}

\subsubsection*{An alternative derivation of the pair relation}

An independent semiclassical derivation of the pair relation
\eref{eq:qmrel} can be obtained \emph{without} invoking periodic orbit
theory by inspecting the semiclassical map operators $\mathsf{P}^{\rm
  int}$ and $\mathsf{P}^{\rm ext}$ of the interior and the exterior,
see Sect.~\ref{sec:pop}.  We present here the derivation for the
magnetization based weights \eref{eq:wnmagdef}, the calculation for
the weights \eref{eq:defweight} is quite analogous.  If follows from
the discussion of \eref{eq:fac} that
the interior Dirichlet number counting function can be written in
terms of the eigenphases $\theta_\ell(\nu;b^2)$ of the unitary map operator
$\mathsf{P}^{\rm int}$,
\begin{align}
\label{eq:Nskipphase}
 \N_{\rm skip}^{\rm int}(\nu;b^2) = 
 \sum_\ell \Theta_{2\pi}\left(\theta_\ell(\nu;b^2)+\pi\right)
\end{align}
where $\Theta_{2\pi}(\theta)$ is the unit staircase function at
integer multiples of $2\pi$.  The interior edge state density
\eref{eq:dmagdef} is then given by
\begin{align}
\label{eq:dmagphase}
 \dedge^{\magn}(\nu;b^2) &= -\frac{b^2}{\nu}\frac{\rmd}{\rmd b^2} 
\N_{\rm skip}^{\rm int}(\nu;b^2)
\nnn
  &=-\frac{b^2}{\nu}
 \sum_\ell \frac{\rmd \theta_\ell}{\rmd b^2}(\nu;b^2)
 \, \delta_{2\pi}\left(\theta_\ell(\nu;b^2)+\pi\right)
\CO
\end{align}
with
$\delta_{2\pi}(\theta)=\frac{\rmd}{\rmd\theta}\Theta_{2\pi}(\theta)$
the $2\pi$-periodic $\delta$-function.  The unweighted spectral
density \eref{eq:ddef} is obtained in the same way by taking the
derivative of \eref{eq:Nskipphase} with respect to $\nu$.  Comparing
the two densities we conclude that the edge state weights are given by
\begin{align}
\label{eq:wtheta}
w^{\rm \magn} =\left. -  \frac{b^2}{\nu} \frac{\ds \frac{\rmd
      \theta_\ell}{\rmd b^2}(\nu;b^2)}{\ds \frac{\rmd \theta_\ell}{\rmd
      \nu}(\nu;b^2)} \right|_{\theta_\ell(\nu)=\pi \mod 2\pi}
\end{align}
The same equations hold for the exterior quantities (labeled by a
prime) with a plus sign in \eref{eq:dmagphase} and \eref{eq:wtheta}.
Now we make use of the duality relation \eref{eq:durel} between the
interior and the exterior map operators. It implies
\begin{align}
 \label{eq:thetarel}
  \theta_\ell(\nu;b^2)+\theta_\ell'(\nu;b^2) = 2\pi\left(\nu-\oh\right)
  +2\pi\widetilde{M}
\PO
\end{align}
with integer $\widetilde{M}$.  Take a pair of interior and exterior
energies $\nu$ and $\nu'$ which are determined by the $\ell$th
eigenphase, ie, $\theta_\ell(\nu)+\pi = 2\pi M$ and $\theta'_\ell(\nu')+\pi
= 2\pi M'$ with integer $M, M'$.  If we expand the eigenphases to
first order around the Landau level $\nu+0=M+M'-\widetilde{M}-\oh$
we obtain from \eref{eq:thetarel}
\begin{align}
 \label{eq:thetarel2}
 \frac{\Delta\nu}{\Delta\nu'} \cong
 -\frac{\tfrac{\rmd}{\rmd\nu}\theta'_\ell(\nu_0;b^2)}
  {\tfrac{\rmd}{\rmd\nu}\theta_\ell(\nu_0;b^2)}
  =
  \frac{
    \tfrac{b^2}{\nu_0}  
    \tfrac{\rmd}{\rmd b^2}\theta_\ell(\nu_0;b^2)
    / \tfrac{\rmd}{\rmd\nu}\theta_\ell(\nu_0;b^2)
  }{
    \tfrac{b^2}{\nu_0}  
    \tfrac{\rmd}{\rmd b^2}\theta'_\ell(\nu_0;b^2)
    / \tfrac{\rmd}{\rmd\nu}\theta'_\ell(\nu_0;b^2)
  }
 \cong - \frac{w^\magn}{{w^\magn}'}
\end{align}
with $\Delta\nu=\nu-\nu_0$ and $\Delta\nu'=\nu'-\nu_0$.  The first
equality holds if the distances to the Landau level $\Delta\nu$ and
$\Delta\nu'$ are sufficiently small. To the same degree of
approximation we can replace the Landau energy $\nu_0$ by the
eigenenergies $\nu$ and $\nu'$ and using \eref{eq:wtheta} we get the
pair relation \eref{eq:qmrel} in terms of the magnetic weights (last
equality).  The fact that it can be obtained without resorting to
periodic orbit theory shows that the pairwise cross-correlation is a
generic semiclassical feature of dual magnetic billiards and is not
related to the type of the classical motion.
We note that the duality relation \eref{eq:durel} also implies that in
the vicinity of a Landau level the operators $\mathsf{P}^{\rm int}$
and $\mathsf{P}^{\rm ext}$ are approximately inverse to each other.
Hence, for correlated pairs of eigenstates the normal derivatives at
the boundary are expected to be approximately equal (see the numerical
test below).

\begin{figure}[p]%
  \begin{center}%
    \includegraphics[bb=104 250 508 538,clip,keepaspectratio,
    width=0.68\linewidth] {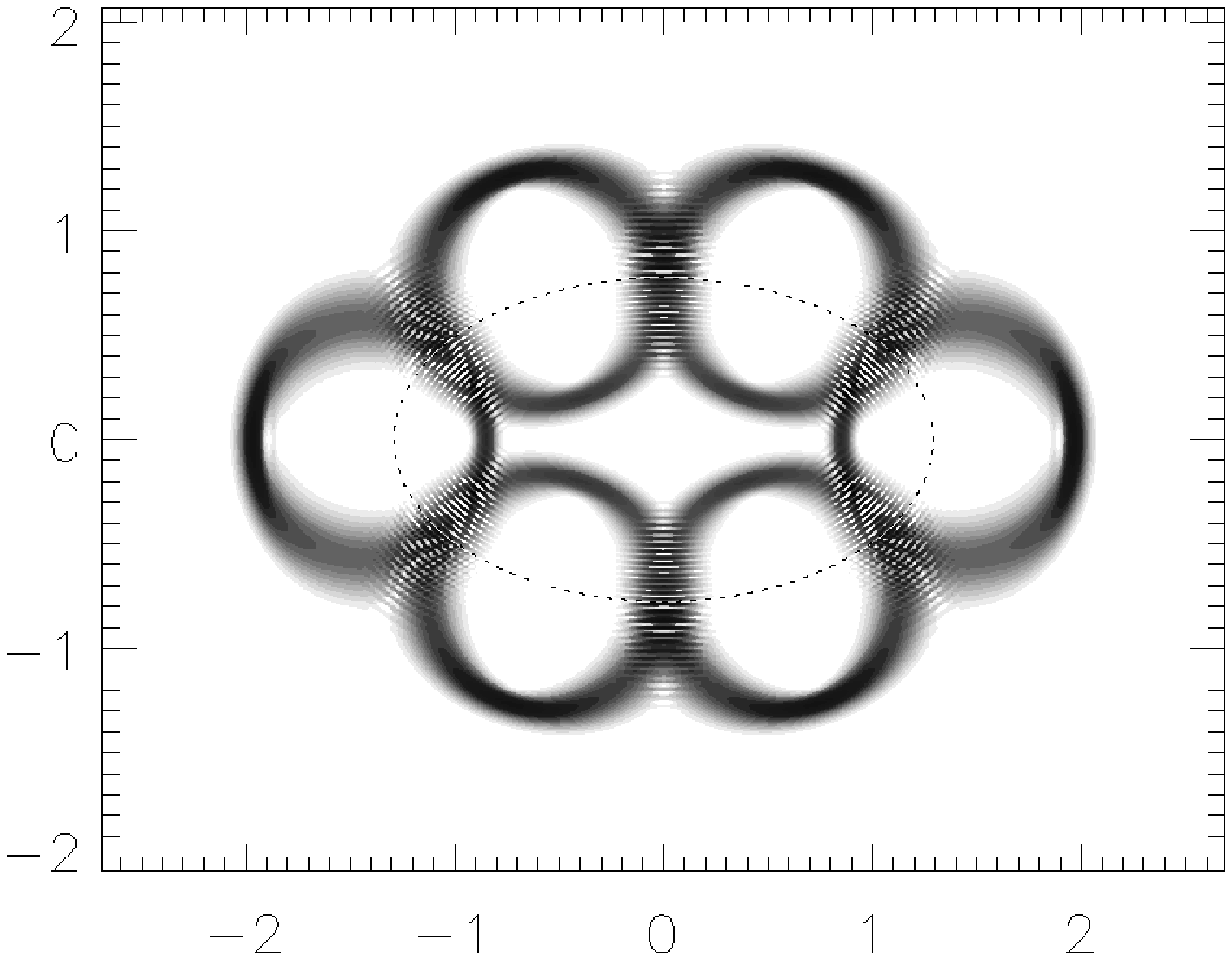}
    \includegraphics[bb=138 272 474 512,clip,keepaspectratio,
    width=0.68\linewidth] {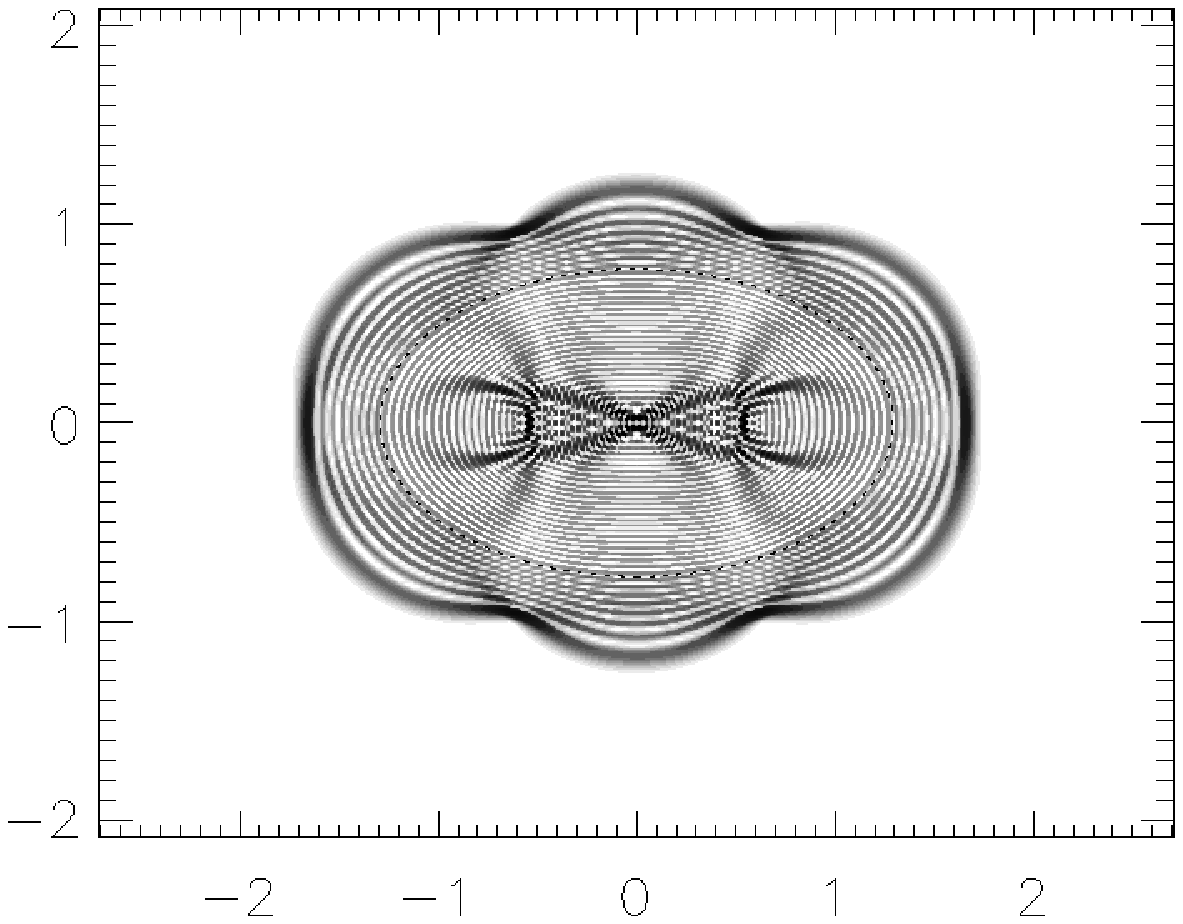}
    \includegraphics[bb=137 250 474 512,clip,keepaspectratio,
    width=0.68\linewidth] {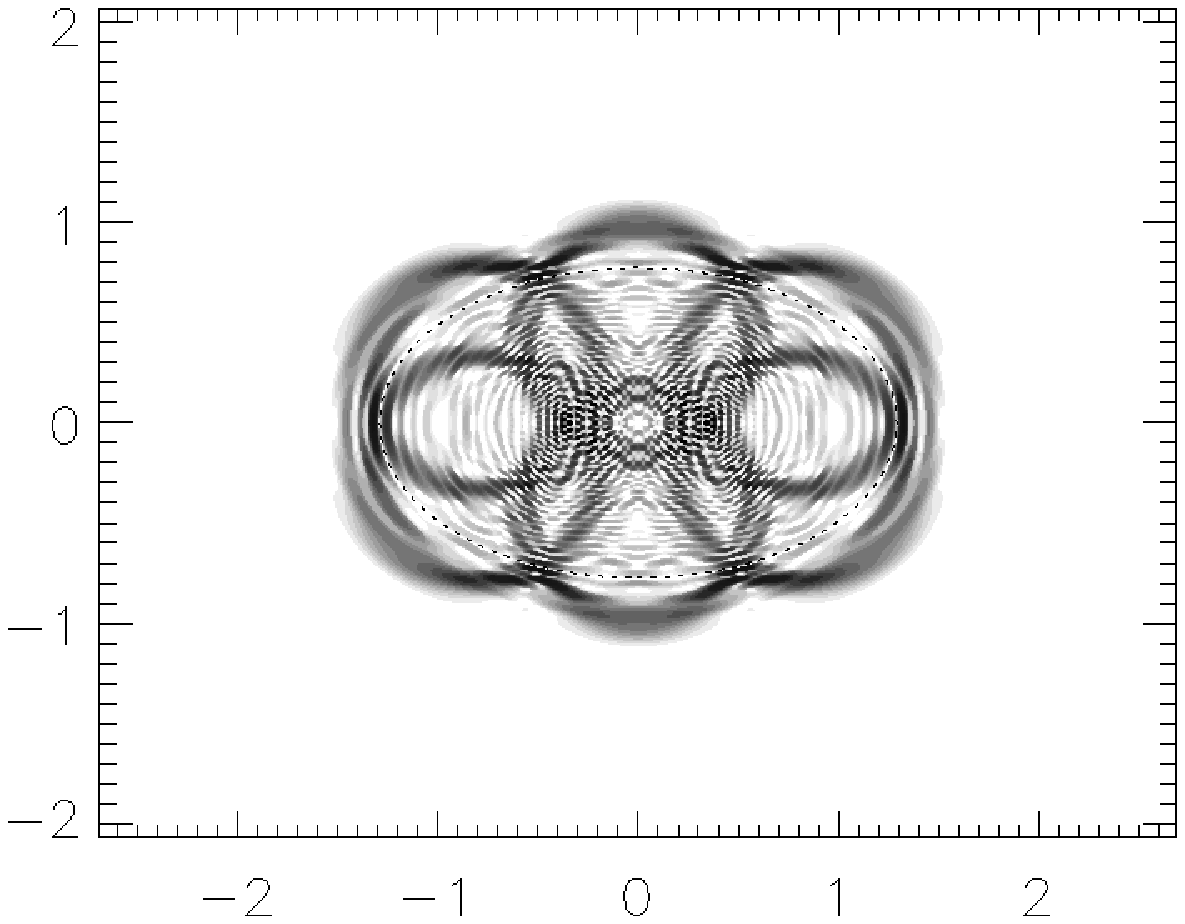}
    \figurecaption{%
      Pairs of correlated interior and exterior wave functions.  The
      energies and weights are given Table \ref{tab:cpair}.  (Ellipse billiard
      at $b=0.1$; the shading is  proportional to the modulus of the
      wave function, and the boundary is indicated by a dotted line.)
      [figure quality reduced]}
\label{fig:cpair}%
\end{center}%
\end{figure}

\begin{figure}[p]%
  \begin{center}%
    \includegraphics[bb= 75 83 509 396,clip,width=0.68\linewidth] 
    {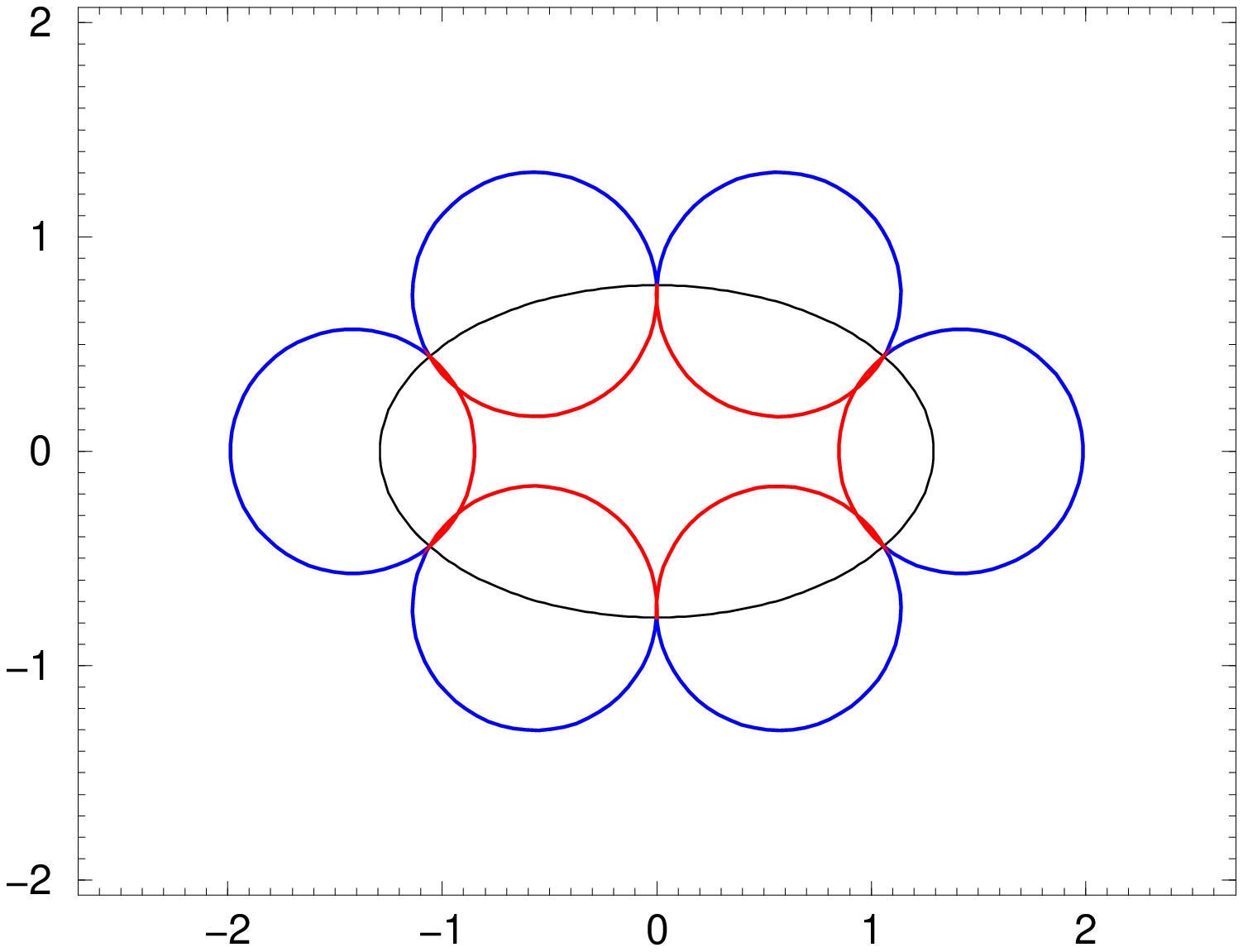}
    \includegraphics[bb= 75 83 509 396,clip,width=0.68\linewidth] 
    {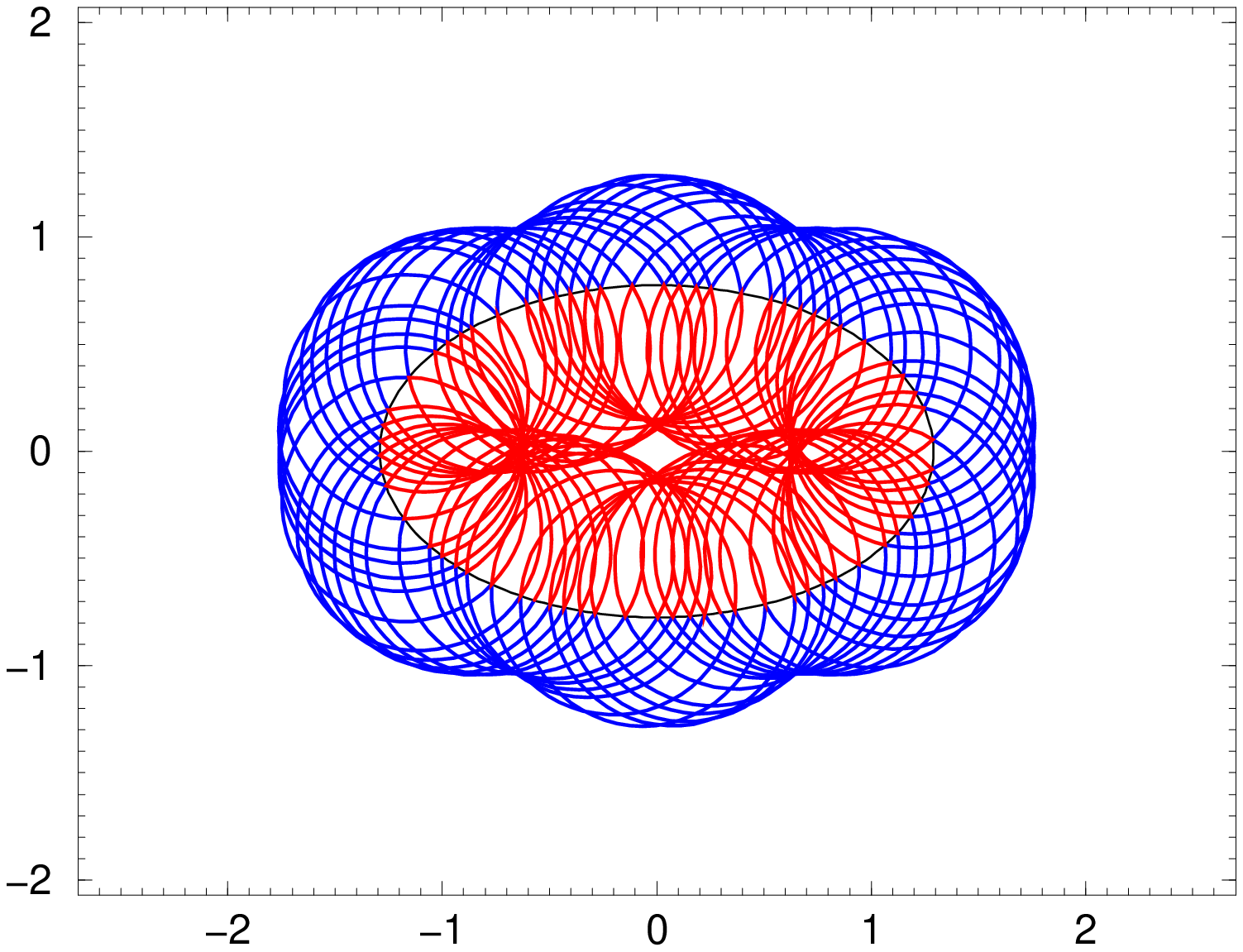}
    \includegraphics[bb= 75 66 509 396,clip,width=0.68\linewidth] 
    {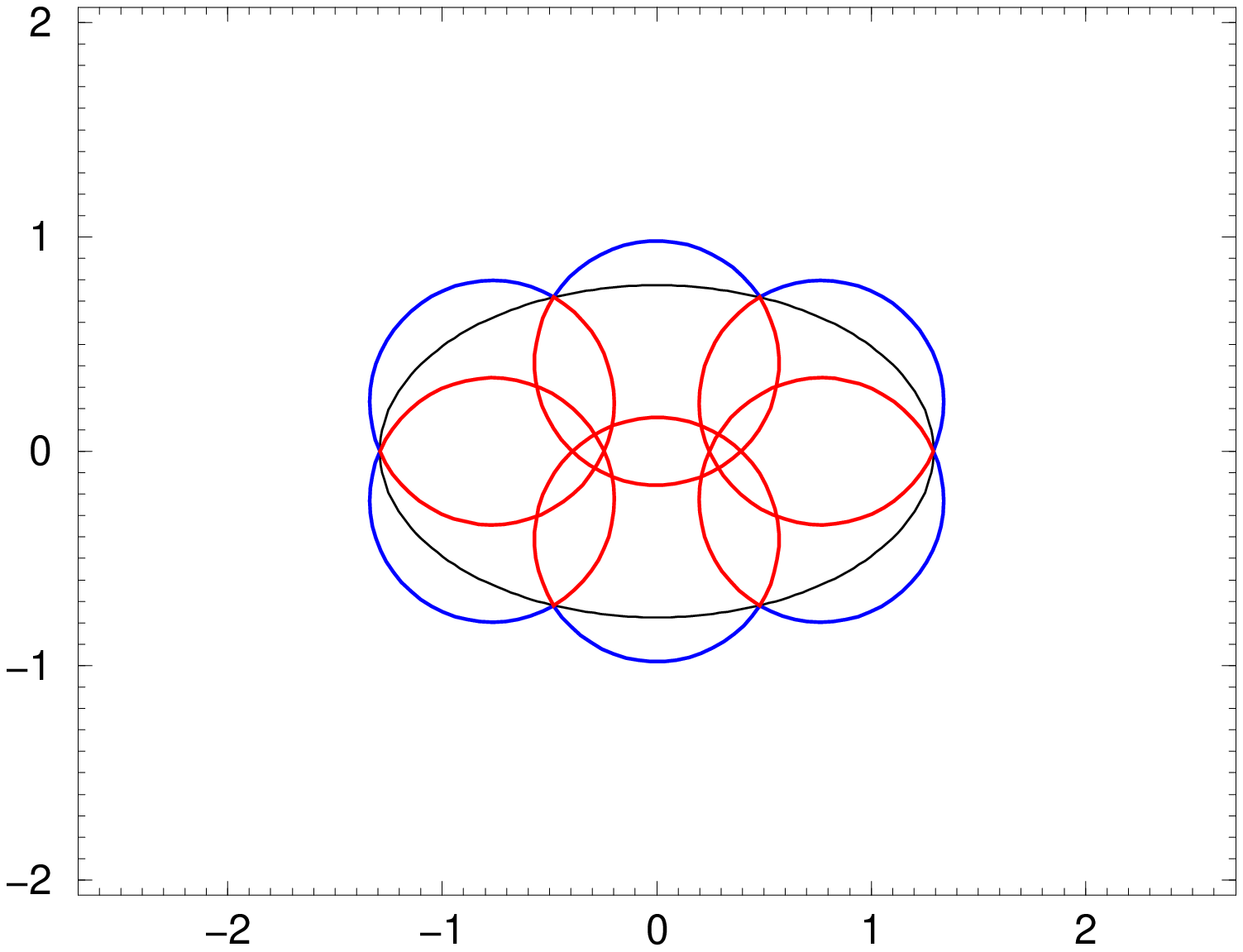}
    \figurecaption{%
      Dual pairs of classical periodic orbits in the ellipse billiard,
      at $\rho=0.57$.  The top and bottom orbits are stable, while the
      middle one in unstable.  Their classical weights \eref{eq:defwp}
      correspond to the quantum weights \eref{eq:defweight} of the
      states in Fig.~\ref{fig:cpair}. }
\label{fig:copair}%
\end{center}%
\end{figure}

\myparagraph{Correlated wave functions}

We proceed to present three pairs of correlated wave functions of the
ellipse billiard. The interior states were chosen to have different
locations in Figure \ref{fig:qmclswt} which displays the distribution
of quantum weights on page \pageref{fig:qmclswt} (top part).  At
energies corresponding to $\rho\approx0.57$ we took states with
weights lying in the top branch of the rightmost bifurcation
structure, in the middle, and in the bottom branch, respectively.  The
respective correlated exterior states were identified using the pair
relation \eref{eq:qmrel}.  Table \ref{tab:cpair} lists the data for
the three pairs.

Figure~\ref{fig:cpair} displays superimpositions of the interior and
the exterior wave functions.  One clearly observes that the top and
bottom wave functions are localized on dual periodic orbits.  In
agreement with this observation one finds that the structures of
increased density of classical weights in Fig.~\ref{fig:qmclswt} (page
\pageref{fig:qmclswt}, bottom part) may be attributed to periodic
orbits which bifurcate as the cyclotron radius $\rho$ is increased.
The top and bottom wave functions were taken from a fork which belongs
to orbits with period 6. Two of these orbits are shown in
Fig.~\ref{fig:copair} along with their dual partners.

The middle wave functions in Figure \ref{fig:cpair}, in contrast, are
localized on a chaotic region in phase space confined by un-destroyed
invariant tori. For comparison, an unstable pair of dual classical
orbits from this region is given in the middle part of
Fig.~\ref{fig:copair}. Note that it exhibits the same spatial
extension as the wave functions.  Here, the correlation of interior
and exterior wave functions is not evident from the visual inspection.

\begin{figure}[t]%
  \begin{center}%
    \psfrag{y}{\hspace*{-12mm}$|\dnb\psi-\rmi\At_n\psi|$}
    \psfrag{s}{$s$}
    \includegraphics[width=\linewidth] 
    {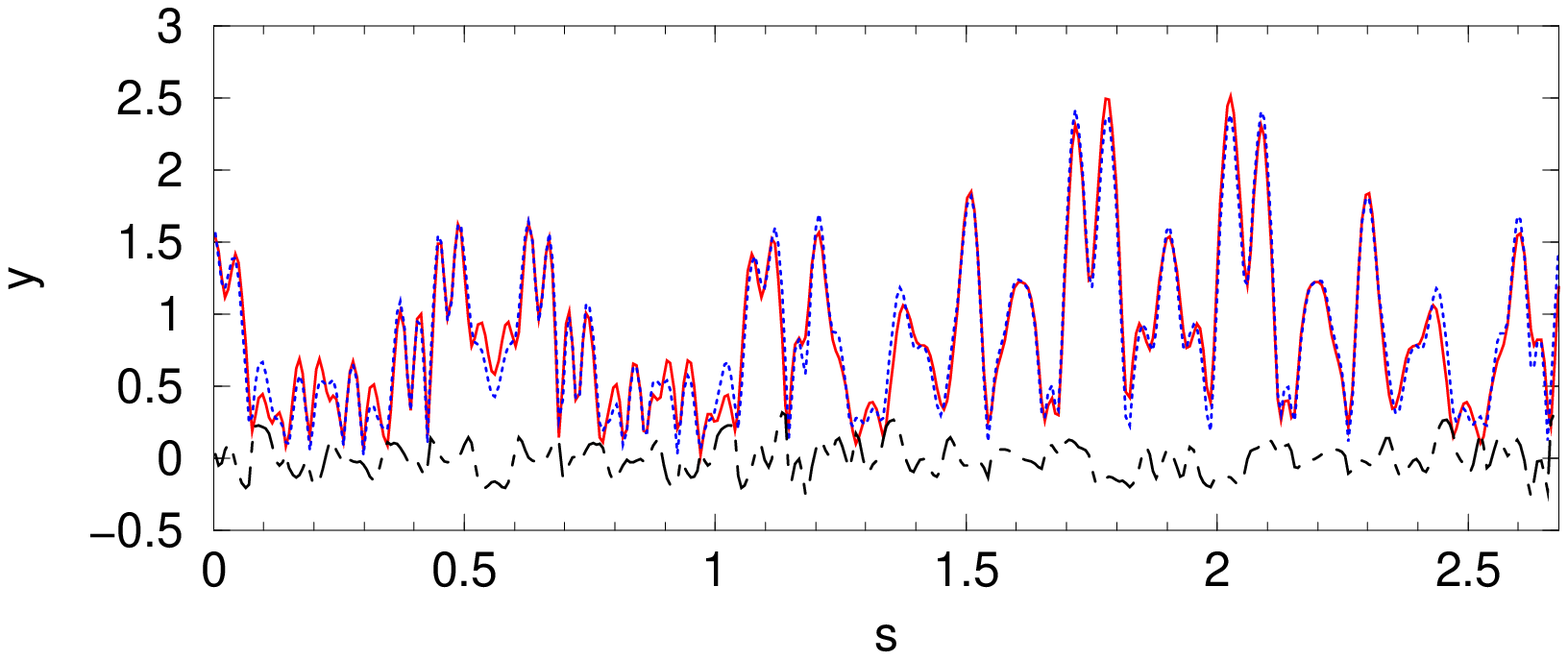}
    \figurecaption{%
      Boundary functions $|u|=|\dnb\psi-\rmi\At_n\psi|$ of the
      correlated wave functions depicted in Fig.~\ref{fig:mpair1} (along
      one half of the boundary). Solid line: interior ($\nu=
      110.6567$), dotted line: exterior ($\nu'=110.4841$). The
      difference is given as a dashed line.}
    \label{fig:pairs1boundary}%
  \end{center}%
\end{figure}

\begin{figure}[tbp]%
  \begin{center}%
    \psfrag{-4}{$-4$}
    \psfrag{-2}{$-2$}
    \psfrag{0}{$0$}
    \psfrag{2}{$2$}
    \psfrag{4}{$4$}
    \includegraphics[width=\linewidth] {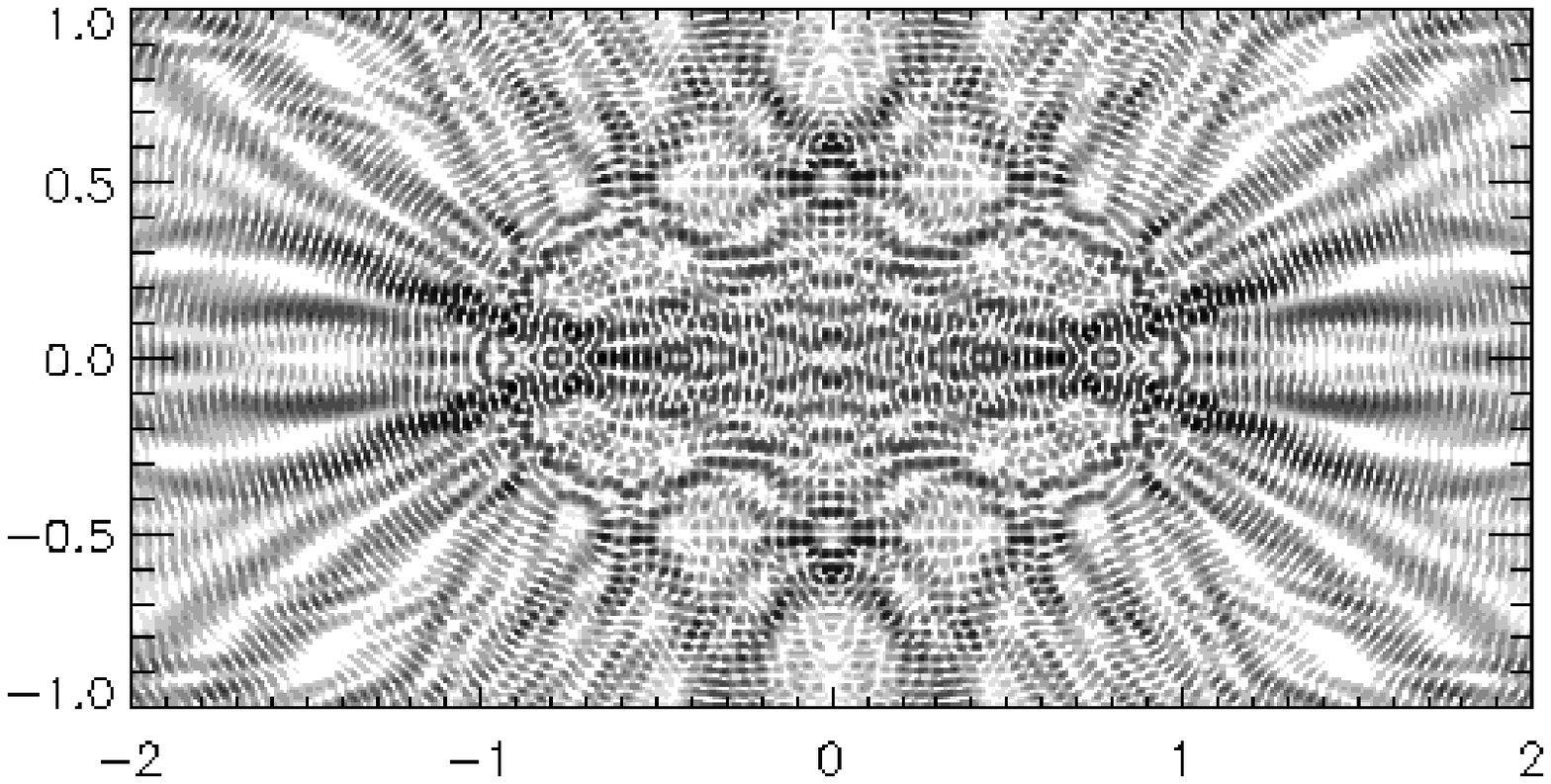} 
    \figurecaption{%
      A typical pair of correlated wave functions taken from the
      spectrum displayed in Fig.~\ref{fig:s14magspec} (superimposed;
      $\nu=110.6567$, $\nu'=110.4841$, $b=0.2$). The stadium-shaped
      boundary is not drawn but visible as a regular nodal line.  
      Figure \ref{fig:mpair1ext} shows the pair on a larger
      scale. [figure quality reduced]
    }
\label{fig:mpair1}%
\end{center}%
\end{figure}

\begin{figure}[tbp]%
  \begin{center}%
    \includegraphics[width=0.8\linewidth] 
{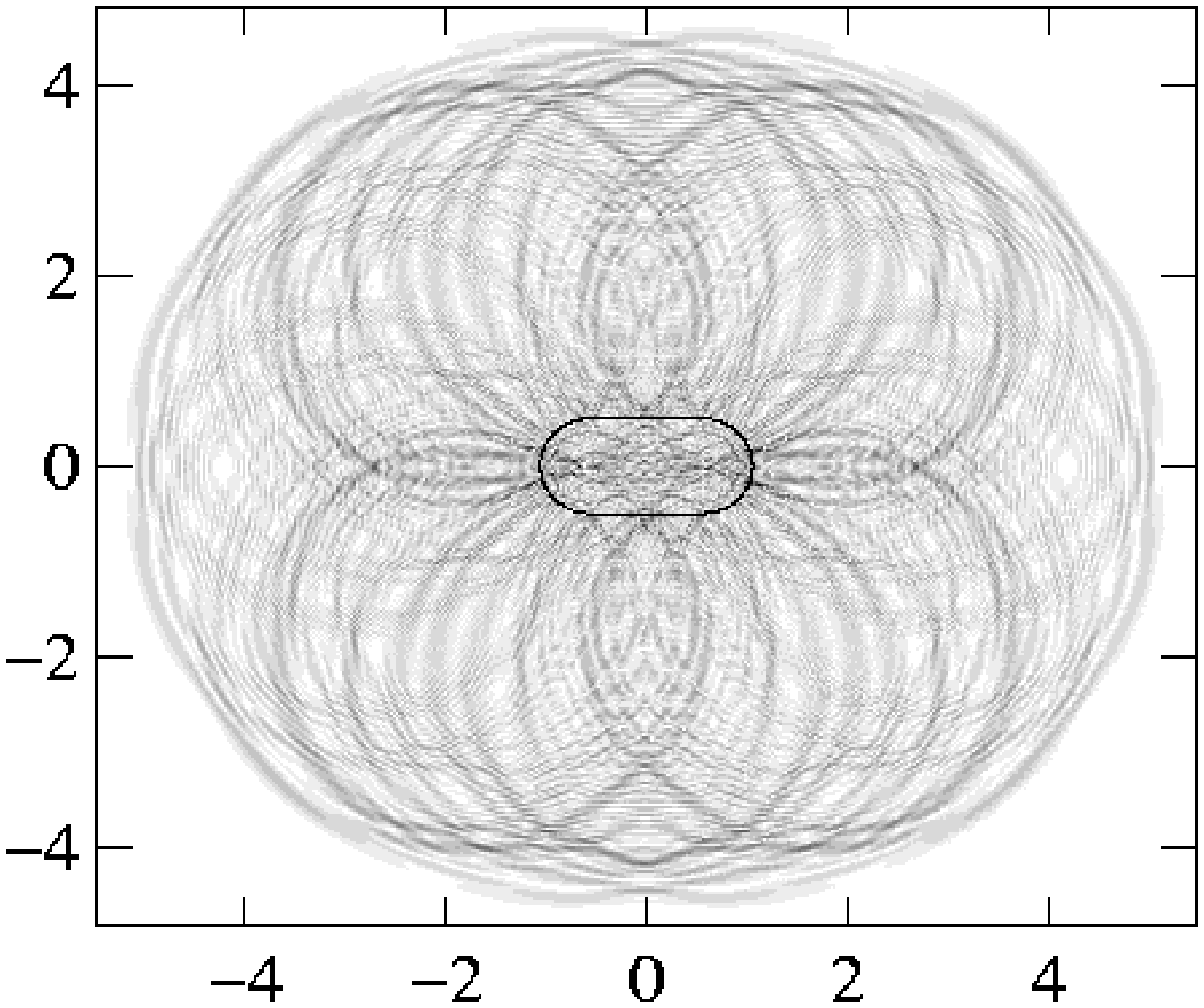}
    \figurecaption{%
      The pair of correlated interior and exterior wave functions from
      Fig.~\ref{fig:mpair1} on a larger scale. The black line
      indicates the stadium-shaped  boundary.  The circular
      scars in the exterior wave function match the classical
      cyclotron radius $\rho\simeq 2.10\,$.  [figure quality reduced]  }
\label{fig:mpair1ext}%
\end{center}%
\end{figure}

The most convincing evidence for the correlation between interior and
exterior eigenfunctions comes from a striking similarity of their respective
normal derivatives at the boundary.
This is a feature shared
by all pairs of correlated wave functions, including those which are
based on a chaotic part of the phase space.  To show this clearly we
compare in Fig.~\ref{fig:pairs1boundary} the normal derivatives of a
typical pair of correlated states taken from the \emph{stadium}
billiard.  One observes that the moduli resemble each other, even
though they are rather irregular. The difference of the interior and
the exterior values is indicated by the dashed line.  As a consequence
of \eref{eq:qmrel} one expects in general that the difference is the
smaller the closer two edge states are in energy. If the edge energies
happened to coincide this would have to take place on a Landau level
according to \eref{eq:qmrel} and the continuation of one wave function
would simply yield the other, ie the normal derivatives would
coincide.
 
In Figure \ref{fig:mpair1} we show the wave functions of the pair from
the stadium billiard for which the boundary functions are shown in
Fig.~\ref{fig:pairs1boundary}.
Although they exhibit the typical irregular pattern of wave function
based on a chaotic part of phase space, one can notice that the
interior and exterior structures match.  In the exterior wave
function scars of periodic orbits may be discerned if viewed with
some measure of imagination, see Fig.~\ref{fig:mpair1ext}. Here, the
circular structures match the classical cyclotron radius $\rho\cong
2.10$.

\section{Conclusions}

The main goal of this review was to present a practical and coherent
theoretical framework for the study of magnetic edge states allowing
in particular the investigation of their semiclassical properties.  An
important step was to set up a quantitative and physically meaningful
definition for the notion of edge states. It became the crucial
ingredient for identifying the quantum analogue of the classical
interior-exterior duality.  In the remainder we discuss a few of the
unsettled problems and comment on possible new directions of research
motivated by the progress made so far.

The boundary integral method presented in Chapter \ref{chap:bim} was
formulated only for smooth boundaries.  This restriction was crucial
for the regularization of the hypersingular integral operators, which
is necessary at non-Dirichlet boundary conditions, and a general
treatment of corners or cusps is still missing.  Also the effect of
the general boundary conditions on the smooth number counting function
was computed indirectly only. Balian and Bloch and others developed a
systematic method for the computation of these quantities
\cite{BB70,SPSUS95} which should be extended to the magnetic case. The
implementation of this program is technically far from being simple
and awaits a proper treatment.

As a natural direction of further research one should ask whether the
proposed spectral measure is applicable and useful in other areas,
specifically for the physics of the (fractional) Quantum Hall effect,
where the concept of edge states is frequently employed without a
clear definition.  The introduction of confining walls which extend to
infinity adds an essentially new dimension to the study of edge
states. It requires a scattering theory appropriate to the continuous
spectrum encountered in such systems. The models investigated so far
dealt with the simplest geometries\cite{LS90,LS91}.  Once an obstacle
(ie, a billiard) is placed near a wall or between parallel walls,
scattering resonances will appear.  Their analysis is important from
an experimental point of view and theoretically challenging since they
introduce resonant transitions between counter-propagating modes along
adjacent leads.  In the field-free case the scattering matrix is often
discussed in terms of boundary integrals, and we expect that the
techniques developed in Chapter \ref{chap:bim} may provide convenient
tools for the corresponding magnetic scattering theory --- the same
holds for the development a semiclassical scattering theory.  These
problems are bound to find applications for magneto-transport experiments
involving ballistic mesoscopic devices.

The quantum interior-exterior duality belongs to a more general class
of problems emerging in the field of quantum chaos. Here, one
considers systems whose classical dynamics are related in some way and
asks for the quantum correlations implied by the classical
correlations and how they emerge in the semiclassical limit.  Recently
B. Gutkin analyzed the boundary operators for a line partitioning the
$2d$-sphere \cite{GutkinThesis} and found that the corresponding
semiclassical map operators are related by an analogue of Equation
\eref{eq:durel}. More complicated surfaces or the consequences of
other classical relations were not yet analyzed but are a natural
direction of further research.

\begin{ack}
  We thank B. Gutkin for helpful discussions and A. Buchleitner for
  many helpful comments on the manuscript at an early stage.  The work
  was partially supported by a Minerva fellowship for KH and by the
  Minerva Center for Nonlinear Physics at the Weizmann Institute.
\end{ack}

\begin{appendix}
\section{Appendices}

\subsection{Green function in angular momentum representation}
\label{app:Gang}

In Section \ref{sec:freeGreen} the magnetic Green function  was
obtained by a direct evaluation of the Fourier integral. 
In the present appendix we derive its angular momentum decomposition.
This lets us correct some erroneous results in the literature
and discuss the irregular Green function. Moreover,
the solutions of the radial Schr{\"o}dinger equation will be needed
below, in App.~\ref{app:diskexact}.

The symmetric gauge must be employed since only this choice renders
the angular momentum a constant of the motion.  In polar coordinates
the inhomogeneous Schr{\"o}dinger equation \eref{eq:seconvinh} then
assumes the form
\begin{gather}
  \label{eq:SEsyminh}
  \Big[
  -\frac{1}{4}(\partial^2_{\rt}+\frac{1}{\rt}\partial_{\rt})
  +\frac{1}{4}(\rt+\rmi\frac{\partial_\vartheta}{\rt})^2-\nu
  \Big]
  \Gnu
  = -\frac{1}{4}\,\delta(\rvect-\rvect_0)
\PO
\end{gather}
An ansatz in terms of the difference of polar angles,
\begin{gather}
  \Gnu(\rvec,\rvec_0)=
  \frac{1}{2\pi}\sum_{m=-\infty}^\infty \rme^{\rmi m(\vartheta-\vartheta_0)}
  G_m(\rt,\rt_0)
\CO
\end{gather}
which cannot be justified a priori, separates the radial and the
angular coordinates.  For $\rt\neq\rt_0$ the functions $G_m$ solve the
radial Schr{\"o}dinger equation in the plane
\begin{gather} 
  \label{eq:radSG}
  \Big[
  \partial^2_{\rt}+\frac{1}{\rt}\partial_{\rt}
  -\frac{(\rt^2-m)^2}{\rt^2}+4\nu
  \Big]
  G_m(\rt,\rt_0)= 0
\PO
\end{gather}
The definition
\begin{gather}
   G_m(\rt,\rt_0) = \rt^{|m|}\, \rme^{-\rt^2/2}\, g_m(\rt^2)
\end{gather}
leads to an equation for $g_m$,
\begin{gather}
  z  g_m''(z) + (1+|m|-z) g_m'(z) - \Big(\oh-\nu+\frac{|m|-m}{2}\Big) 
  g_m(z) =0
\CO
\end{gather}
which is known as Kummer's differential equation and is solved by
the regular and irregular hypergeometric function, $\oFo$ and
${\rm U}$, respectively  \cite{AS65}.
For energies different from the Landau levels it follows that 
a pair of independent
solutions $u_1, u_2$ of the radial Schr{\"o}dinger equation
\eref{eq:radSG} 
is given by
\begin{align}
\label{eq:u1}
  u_1(\rt) &=  \rt^{|m|}\, \rme^{-\rt^2/2}\,
  \oFo\Big(\oh-\nu+\frac{|m|-m}{2}, 1+|m|, \rt^2\Big)
\intertext{and}
\label{eq:u2}
  u_2(\rt) &=  \rt^{|m|}\, \rme^{-\rt^2/2}\,
  {\rm U}\Big(\oh-\nu+\frac{|m|-m}{2}, 1+|m|,  \rt^2\Big)
  \PO
\end{align}
Both are real valued solutions. $u_1$ is bounded at
$\rt=0$ and diverges as $\rt\to \infty$. The function
$u_2$, on the other hand, decays like a Gaussian in this limit but
displays a (logarithmic) singularity as $\rt\to 0$.

Another fundamental system of equation \eref{eq:radSG} is obtained
if one replaces $u_2$ by
\begin{align}
  \label{eq:irru}
  u^{\rm irr}_2(\rt) &= \rt^{|m|}\, \rme^{+\rt^2/2}\,
  {\rm U}\Big(\oh+\nu+\frac{|m|+m}{2}, 1+|m|, - \rt^2\Big)
\PO
\end{align}
This is a \emph{complex} valued solution \cite{Buchholz69} which we call ``irregular''. Apart
from its logarithmic singularity at $\rt\to 0$, it diverges
exponentially as $\rt\to \infty$.

Both $u_1$ and $u_2$ are needed to form a solution $G_m$ of the
inhomogeneous equation \eref{eq:SEsyminh} since the $\delta$-function
implies a discontinuity of the derivative,
\begin{align}
  \partial_1 G_m(\rt_0+0,\rt_0) -   
  \partial_1 G_m(\rt_0-0,\rt_0) = \frac{1}{\rt_0}
  \PO
\end{align}
The requirement that the Green function must vanish as $\rt\to\infty$,
together with its
continuity at $\rt=\rt_0$,
 leads necessarily to the form
\begin{align}
  G_m(\rt,\rt_0)
  = \frac{1}{\rt_0 W(\rt_0)}
  \begin{cases}
    u_1(\rt) u_2(\rt_0)  &\text{if $r<r_0$}
    \\
    u_2(\rt) u_1(\rt_0) &\text{if $r>r_0$\CO}
  \end{cases}
\end{align}
with Wronskian $W=u_1 u_2'-u_1' u_2$. In total, the Green function in
angular momentum decomposition and symmetric gauge is given by
\begin{align}
\label{eq:Gangular}
  {\rm G}_\nu(\rvec;\rvec_0)
  &=
  \frac{-1}{4\pi}
  \sum_{m=-\infty}^\infty\!\!
  \rme^{\rmi m(\vartheta-\vartheta_0)}\,
 \frac{\Gamma\Big(\oh-\nu+\frac{|m|-m}{2}\Big)}{|m|!}\,
  \Big(\frac{r r_0}{b^2}\Big)^{|m|}
  \exp\Big(-\frac{r^2+r_0^2}{2b^2}\Big)
  \nnn
  &\phantom{\frac{-1}{4\pi}\sum_{m=-\infty}^\infty}
  \times
  \oFo\Big(\oh-\nu+\frac{|m|-m}{2},1+|m|,z_<\Big)
  \nnn
  &\phantom{\frac{-1}{4\pi}\sum_{m=-\infty}^\infty}
  \times
  {\rm U}\Big(\oh-\nu+\frac{|m|-m}{2},1+|m|,z_>\Big)
\\
  &=
  \frac{-1}{4\pi}
  \sum_{m=-\infty}^\infty\!\!
  \rme^{\rmi m(\vartheta-\vartheta_0)}\,
 \frac{\Gamma\Big(\oh-\nu+\frac{|m|-m}{2}\Big)}{|m|!}\,
   \Big(\frac{r r_0}{b^2}\Big)^{-1}\,
  M_{\nu+\frac{m}{2},\frac{|m|}{2}}(z_<)\;
  \nnn
  &\phantom{\frac{-1}{4\pi}\sum_{m=-\infty}^\infty}
  \times
    W_{\nu+\frac{m}{2},\frac{|m|}{2}}(z_>)
\end{align}
with $M_{k,\mu}(z)$ and $W_{k,\mu}(z)$ the regular and irregular
Whittaker functions \cite{AS65}, and
\begin{align}
  z_< &\defas \min\left(\frac{\rvec^2}{b^2},\frac{\rvec_0^2}{b^2}\right)
  \q\text{and}\q
  z_> \defas \max\left(\frac{\rvec^2}{b^2},\frac{\rvec_0^2}{b^2}\right)
\PO
\end{align}
Note that this expression differs slightly from the (incorrect) expressions in
\cite{KR92} and \cite[eq (6.2.26)]{GS98}.

An independent solution to the inhomogeneous problem
\eref{eq:SEsyminh} may obtained if one drops the requirement that the
Green function should vanish as $\rt\to\infty$. It involves the
irregular solution \eref{eq:irru} and leads to the  Green
function
\begin{align}
\label{eq:Girrangular}
  {\rm G}^{\rm (irr)}_\nu(\rvec;\rvec_0)
  &=
  \frac{-1}{4\pi}
  \sum_{m=-\infty}^\infty\!\!
  \rme^{\rmi(\vartheta-\vartheta_0+\pi)m}\,
  \frac{\Gamma\Big(\oh+\nu+\frac{|m|+m}{2}\Big)}{|m|!}\,
  \Big(\frac{r r_0}{b^2}\Big)^{|m|}
  \exp\Big(\frac{r^2+r_0^2}{2b^2}\Big)
  \nnn
  &\phantom{\frac{-1}{4\pi}\sum_{m=-\infty}^\infty}
  \times
  \oFo\Big(\oh+\nu+\frac{|m|+m}{2},1+|m|,-z_<\Big)
  \nnn
  &\phantom{\frac{-1}{4\pi}\sum_{m=-\infty}^\infty}
  \times
  {\rm U}\Big(\oh+\nu+\frac{|m|+m}{2},1+|m|,-z_>\Big)
\end{align}
which we call ``irregular''.  This expression was derived by Tiago \etal
\cite{TCA97}. Unlike the regular Green function \eref{eq:Gangular},
this one diverges exponentially once the distance between initial and
final point exceeds one cyclotron diameter.  This property renders the
irregular Green function impractical for most  purposes.

\subsection{The null field method}
\label{app:nullfield}

The null field method is an alternative scheme to quantize magnetic
billiards in the interior \cite{TCA97}. We include it for
completeness although its practical use is limited.

Let us start with equation \eref{eq:split}.  In terms of the
irregular Green function it reads
\begin{gather}
\label{eq:nullfield1}
\int_{\Boundary}
{\rm G}^{\rm (irr)}_\nu(\rvec;\rvec_0)
\dnb\psi^* \frac{\rmd\Boundary}{b}
=0
\CO
\end{gather}
where we chose $\rvec_0\in\Rtwo\setminus\Domain$, Dirichlet boundary
conditions, and the symmetric gauge.  Rather than transforming this
into an integral equation we place $\rvec_0$ on a (large) circle with
radius $R_p$ which is centered at the origin and surrounds the
billiard domain.

Now assume that the billiard boundary is given as a function
$r(\theta)$ of the polar angle and expand the unknown boundary
function in a Fourier series,
\begin{align}
  \dnb\psi^*\big(r(\theta)\big) 
=
   \sum_\ell \rme^{\rmi \theta\ell} c_\ell
\PO
\end{align}
Using the angular momentum decomposition \eref{eq:Girrangular} of the
irregular Green function equation \eref{eq:nullfield1} assumes the form
\begin{align}
\label{eq:null1}
  &\sum_{\ell,m=-\infty}^\infty
  \rme^{-\rmi\theta_0 m} a_m  B_{m\ell} c_\ell
  =0
\CO
\end{align}
{with}
\begin{align}
\label{eq:nullA}
  a_m =&\,(-)^m
  \frac{\Gamma\Big(\oh+\nu+\frac{|m|+m}{2}\Big)}{|m|!}\,
  \Big(\frac{R_p}{b^2}\Big)^{|m|}
  \exp\Big(\frac{R_p^2}{2b^2}\Big)
  \nnn
  &\,\times
  {\rm U}\Big(\oh+\nu+\frac{|m|+m}{2},1+|m|,-\frac{R_p^2}{b^2}\Big)
\intertext{and}
\label{eq:nullB}
  B_{m\ell} =&\, \int_0^{2\pi}\!\!
  \rme^{\rmi(m+\ell)\theta}\,
 \Big(\frac{r(\theta)}{b^2}\Big)^{|m|}
  \exp\Big(\frac{r^2(\theta)}{2b^2}\Big)
  \nnn
  &\,\times
  \oFo\Big(\oh+\nu+\frac{|m|+m}{2},1+|m|,-\frac{r^2(\theta)}{b^2}\Big)
  \,\rmd\theta
\\
  =&\, \int_0^{2\pi}\!\!
  \rme^{\rmi(m+\ell)\theta}\,
 \Big(\frac{r(\theta)}{b^2}\Big)^{|m|}
  \exp\Big(-\frac{r^2(\theta)}{2b^2}\Big)
  \nnn 
\label{eq:nullBa}
\tag{\ref{eq:nullB}a}
  &\,\times
  \oFo\Big(\oh-\nu+\frac{|m|-m}{2},1+|m|,\frac{r^2(\theta)}{b^2}\Big)
  \,\rmd\theta
  \PO
\end{align}
Equation \eref{eq:null1}  holds for all polar angles $\theta_0$.
For negative arguments the function $ {\rm U}$ is known to be complex
and non-zero \cite{Buchholz69}. Therefore we can divide by $a_m$ for
all $R_p$ which leaves the condition for the existence of a
nontrivial solution $c_\ell$ to
\begin{align}
  \det( B_{m\ell})=0
\PO
\end{align}
This is a spectral equation which was derived by Tiago \etal\ 
\cite{TCA97} (except for a misprint in their paper).

\subsection{Exact quantization of the magnetic disk}
\label{app:diskexact}

We briefly  describe how to
quantize the interior and the exterior of the magnetic disk.  As discussed in
Sect.~\ref{sec:disksep}, the problem is separable in the symmetric
gauge.  For this choice the exact solutions of the free Schr\"odinger
equation are given
above, see \eref{eq:u1} and
\eref{eq:u2}. 
It follows that the interior and the exterior eigenfunctions of the disk
are specified uniquely by their behavior at the origin and at
infinity, respectively, and by the angular momentum quantum number $m$.
Since the interior wave function (at energy $\nu$) must be regular at
the origin
it has the form
\begin{align}
  \psi_m(r,\vartheta) = \mathcal{N}_{\rm int}\, \rme^{\rmi m
    \vartheta}\, \Big(\frac{r}{b}\Big)^{|m|} \,
  \rme^{-\tfrac{r^2}{2b^2}}\,
  \oFo\!\Big(\oh-\nu+\frac{|m|-m}{2},1+|m|;\frac{r^2}{b^2}\Big) \PO
\end{align}
For the exterior wave function, which must vanish at infinity,
we have
\begin{align}
  \psi_m(r,\vartheta) = \mathcal{N}_{\rm ext}\, \rme^{\rmi m
    \vartheta}\, \Big(\frac{r}{b}\Big)^{|m|} \,
  \rme^{-\tfrac{r^2}{2b^2}}\,
  {\rm U}\Big(\oh-\nu+\frac{|m|-m}{2},1+|m|;\frac{r^2}{b^2}\Big) \PO
\end{align}
Upon applying the general boundary conditions \eref{eq:bcond} at the
disk radius $r=R$
one obtains the spectral functions
\begin{align}
  \label{eq:xidisk}
  \xi_{\rm disk}\Big(\nu;m,\Lambda,\frac{R}{b}\Big) 
  &= \Big[
    \sqrt{\nu}+(\pm\Lambda)\Big(\frac{R}{2 b}-\frac{|m| b}{2R}\Big) \Big]
    K\Big(\oh-\nu+\frac{|m|-m}{2},1+|m|;\frac{R^2}{b^2}\Big) \nnn
    &-(\pm\Lambda)\frac{R}{b}\, \partial_3
    K\Big(\oh-\nu+\frac{|m|-m}{2},1+|m|;\frac{R^2}{b^2}\Big)
\intertext{with}
\label{eq:defK}
  K(a,b;z)&=
  \begin{cases}
    \oFo(a,b;z) & \text{for the interior problem}
    \\
    {\rm U}(a,b;z) & \text{for the exterior problem.}
  \end{cases}
\end{align}
($\partial_j K$ indicates partial derivation with respect to the $j$th
argument.)  Unlike the semiclassical case \eref{eq:xidisksc}, one
cannot predetermine the radial quantum number here but has to search
for all zeros at given angular quantum number $m$.  The derivatives of
the energies with respect to external parameters
are given explicitely by derivatives of the spectral function like in
the semiclassical case. For variations in the boundary condition we
find
\begin{align}
\label{eq:wdisk}
  \left.\frac{\rmd \nu}{\rmd \Lambda}\right|_{\Lambda=0} = \mp
  \frac{R}{\sqrt{\nu} b} \, \frac{\partial_3 K}{\partial_1 K}
\end{align}
with the arguments of $K$ like above.  Similarly, the derivatives with
respect to the magnetic length are given by the quotient
\begin{align}
  b^2\frac{\rmd \nu}{\rmd b^2} = -\frac{R^2}{b^2}\, \frac{\partial_3
    K}{\partial_1 K}
\PO
\end{align}
We note the relation
\begin{align}
\label{eq:wmagrel}
  b^2\frac{\rmd \nu}{\rmd b^2} = \pm \sqrt{\nu}\,\frac{R}{b}
  \left.\frac{\rmd \nu}{\rmd \Lambda}\right|_{\Lambda=0}
\end{align}
which holds in the semiclassical case as well.

\subsection{The stationary phase approximation}
\label{app:statphase}

The method of the stationary phase yields asymptotic expansions of
integrals with rapidly oscillating integrands like $ \int \! g({x})
\rme^{2\pi\rmi \nu f({x})} \rmd {x}$. One can show that for large
$\nu$ the leading order contribution stems from the stationary points
of the phase $f$. After an expansion of the phase to second order
around these points and the use of the Gaussian integral
\begin{gather}
  \label{eq:Gaussianint}
  \int_{-\infty}^{\infty}
  \rme^{\ts\rmi a x^2} \rmd x
  = \left(\frac{\pi}{|a|}\right)^\oh \rme^{\ts\rmi\sgn(a)\piof}
\end{gather}
one finds that
for functions $f,g\in C^\infty(\mathbb{R})$ where $f$ has a
finite number of non-degenerate stationary points ${x}_j$, ie,
$f'({x}_j)=0$,
the asymptotic expansion reads  \cite{Erdelyi56}
\begin{gather}
  \label{eq:stphase1d}
  \int \!\! g({x}) \rme^{2\pi\rmi \nu f({x})} \rmd {x}
  =
  \frac{1}{\sqrt{\nu}}
  \sum_{ {x}_j}
  \frac{  g({x}_j)}{|f''({x}_j)|^\oh}
  \rme^{2\pi\rmi \nu f({x}_j)+\rmi\piof\sgn(f''({x}_j))}
  \,
  \big(1+\Or({\nu}^{-1})\big)
\CO
\end{gather}
as $\nu\to\infty$. 
For functions of an $N$-dimensional argument, $f,g\in
C^\infty(\mathbb{R}^N)$, an analogous form can be found \cite{BH75}:
\begin{gather}
  \label{eq:stphaseNd}
  \int \!\! g(\mathbf{x}) \rme^{2\pi\rmi \nu f(\mathbf{x})} \rmd^N \mathbf{x}
  \sim
  \left(\frac{\rmi}{\nu}\right)^{\frac{N}{2}}
  \sum_{ \mathbf{x}_j}
  \frac{  g(\mathbf{x}_j)}{|\det f''(\mathbf{x}_j)|^\oh}
  \rme^{2\pi\rmi \nu f(\mathbf{x}_j)-\rmi\nu_j\piot}
\PO
\end{gather}
Here, $\nu_j$ gives the number of negative eigenvalues of the matrix
$f''(\mathbf{x}_j)$.

\subsubsection*{A peculiar $\boldsymbol{\delta}$-function}

As an immediate application the stationary phase approximation allows
to show that the complex function
\begin{align}
  \label{eq:deltaedef}
  &\delta_\epsilon(\xi) \defas \frac{1}{(2\pi\rmi)^\oh} 
  \frac{\exp\left({\rmi\, \frac{\xi^2}{2\epsilon}}\right)}{\sqrt{\epsilon}}
\intertext{has the property of a one-dimensional Dirac $\delta$-function,}
 \label{eq:dprop}
  &\int  \delta_\epsilon(\xi)\, \rmd \xi =1
  \\
 \tag{\ref{eq:dprop}a}     
  &\int g(\xi)\,  \delta_\epsilon(\xi) \,\rmd \xi = g(0) \, (1+\Or(\epsilon)) 
  \qq\text{as $\epsilon\to 0$.}
\end{align}
This follows from
\eref{eq:Gaussianint} and \eref{eq:stphase1d}, with $\nu=1/\epsilon$
and $f=x^2/2$, and is not easily proven otherwise. 
The product of \eref{eq:deltaedef} for the two Cartesian components
of the vector $\rvec$ yields the two dimensional $\delta$-function
\begin{align}
  \label{eq:delta2d}
  \lim_{\epsilon\to 0}
  \frac{1}{2\pi\rmi b^2}\,\frac{1}{\epsilon}
  \exp\left[\rmi\frac{(\rvec-\rvec_0)^2}{2\epsilon b^2}\right] 
  = \delta(\rvec-\rvec_0)
  \CO
\end{align}
which shows up in \eref{eq:deltaprop}.

\subsection{The product relation of the map operators}
\label{app:prod}

We show that the relations \eref{eq:Prel}
for the products of the interior and exterior map operators
\eref{eq:pkern} hold semiclassically.  They were needed to prove the
factorization of the spectral function.  Since possible saddle point
contributions are excluded by the vanishing prefactors and the duality
condition the only relevant contribution to the product stems from
regions where the initial and the final point are close. Hence, we are
allowed to replace the boundary locally by a circular arc.  It follows
that the expressions \eref{eq:pdisk1} -- \eref{eq:pdisk4} derived for
the disk billiards may be employed to show that the kernel of the
product \eref{eq:Prel} acts like a $\delta$-function.  Assuming the
angle between the initial and the final point $\delta\phi=(s-s_0)/R$
to be small we find
\begin{align}
  \label{eq:dp1}
  &\left({\rm p}_{\rm S}^{\rm int} {\rm p}_{\rm L}^{\rm ext}\right)
  ( s,s_0)=
\nnn
  =&
  \frac{1}{2\pi\rmi}\,\frac{b}{R}
  \int \rmd\phi'
  \left(
    \frac{\rmd^2 (2\pi\nu\ga_{\rm S})}
    {\rmd\phi\,\rmd \phi_0}(\phi'-\phi_0) \,
    \frac{\rmd^2 (2\pi\nu\ga_{\rm L})}
    {\rmd\phi_0\,\rmd \phi'}(\phi_0+\delta\phi-\phi')
  \right)^\oh
  \nnn
  &
  \times
  \rme^{ 2\pi\rmi\nu  
    (\ga_{\rm S}(\phi'-\phi_0)+\ga_{\rm L}(\phi_0+\delta\phi-\phi'))}
  \begin{cases}
    -\Theta(\phi'-\phi_0)\Theta(\phi'-\phi_0+\delta) & \text{if
      $\Gam>1$}
    \\
    -1 &\text{if      $\Gam<1$}
  \end{cases}
  \nnn
  \simeq&
  \frac{1}{\pi}\,\frac{ b\nu}{R}
  \int \rmd\phi'
  \left|
    \frac{\rmd^2 (\pi\ga_{\rm L})}
    {\rmd\phi'\,\rmd \phi_0}(\phi_0-\phi')
  \right|
  \rme^{ 2\pi\rmi\nu  +  2\pi\rmi\nu  
    \partial_{\phi_0}\ga_{\rm L}(\phi_0-\phi')\,\delta\phi}
  \begin{cases}
    -\Theta(\phi'-\phi_0) 
    \\
    -1 
  \end{cases}
  \nnn
  =&
  \frac{1}{\pi}\,\frac{ b\nu}{R}\,\rme^{ 2\pi\rmi\nu }
  \int \rmd\phi'\,
  \frac{\rmd u_{\rm L}}{\rmd\phi'} \,\rme^{ 2\rmi\nu u_{\rm L}\,\delta\phi}
  \begin{cases}
    \Theta(\phi'-\phi_0)     & \text{if      $\Gam>1$}
    \\
    1  &\text{if      $\Gam<1$\PO}
  \end{cases}
\end{align}
with $\Gam=R/(b\nu^\oh)$, see \eref{eq:Gamdef}.
The dependence on $\delta\phi$ was expanded linearly in the
phase and neglected in the prefactor. The latter is cancelled by the
change of the integration variable to $u_{\rm L}\defas \pi
\partial_{\phi_0}\ga_{\rm L}(\phi_0-\phi')$. Likewise, one
finds for the second combination of interior and exterior operators:
\begin{align}
  \label{eq:dp2}
  &\left({\rm p}_{\rm L}^{\rm int} {\rm p}_{\rm S}^{\rm ext}\right)
  ( s,s_0)=
  \frac{1}{\pi}\,\frac{ b\nu}{R}\,\rme^{ 2\pi\rmi\nu }
  \int \rmd\phi'\,
  \frac{\rmd u_{\rm S}}{\rmd\phi'} \,\rme^{ 2\rmi\nu u_{\rm S}\,\delta\phi}
  \begin{cases}
    -\Theta(\phi'-\phi_0)     & \text{if      $\Gam>1$}
    \\
    0  &\text{if      $\Gam<1$\CO}
  \end{cases}
  \nnn
\end{align}
with $u_{\rm S}\defas \pi \partial_{\phi_0}\ga_{\rm S}(\phi_0-\phi')$.
The sum of the kernels  assumes the form of a semiclassical
$\delta$-function once the integration is carried out. The ranges of
integration differ for weak and strong fields. They can be found in
Table \ref{tab:int}. Setting ${\overline{\phi}}\equiv2\arcsin(1/\Gam)$
one gets
\begin{align}
  \label{eq:dp3}
  \big({\rm p}_{\rm S}^{\rm int} &   {\rm p}_{\rm L}^{\rm ext}
    +{\rm p}_{\rm L}^{\rm int} {\rm p}_{\rm S}^{\rm ext}
  \big)
  (s,s_0)=
  \nnn
&=  \frac{1}{\pi}\,\frac{ b\nu}{R}\,\rme^{ 2\pi\rmi\nu }
  \begin{cases}
    \ds
    \int_{\Gam+\oh\Gam^2}^{\oh\Gam^2\cos({\overline{\phi}})}
    \hspace{-2em}
    \rmd u_{\rm L} \,\rme^{2\rmi\nu  u_{\rm L} \delta\phi  }
    -
    \int_{-\Gam+\oh\Gam^2}^{\oh\Gam^2\cos({\overline{\phi}})}
    \hspace{-2em}
    \rmd u_{\rm S} \,\rme^{2\rmi\nu  u_{\rm S} \delta\phi  }
    &  \text{if      $\Gam>1$}
    \\[2.5ex]
    \ds
    \int_{\Gam+\oh\Gam^2}^{-\Gam+\oh\Gam^2}
    \hspace{-2em}
    \rmd u_{\rm L} \,\rme^{2\rmi\nu  u_{\rm L} \delta\phi  }
    &  \text{if      $\Gam<1$}
  \end{cases}
  \nnn
&  =   
  - \,\rme^{ 2\pi\rmi\nu }\,
   \frac{1}{\pi}\,
   \frac{\sin\left(2\sqrt{\nu}\,\frac{s-s_0}{b}\right)}
   {\frac{s-s_0}{b}}
   \,   \rme^{ \rmi\pi\sqrt{\nu}\Gam(s-s_0)/b }
   \begin{CD} @>{2\sqrt{\nu}\to\infty}>> \end{CD}
   - \,\rme^{ 2\pi\rmi\nu }\,
   {\delta}\!\left(\frac{s- s_0}{b}\right)
\PO
\end{align}
This proves the identity \eref{eq:Prel}.  In a similar fashion one
finds that the product \eref{eq:Prel2} does not contribute
semiclassically.

\begin{table}[tbp]
  \begin{center}%
    \small%
    \fbox{\begin{minipage}{0.9\linewidth}%
    \vspace*{-5ex}
    \begin{alignat*}{5}
       \intertext{If $\Gam>1$:}
      \phi'&:& \q
      \phi_0-{\overline{\phi}} & \llongrightarrow   \phi_0-0  &
      {\q}
      \phi_0+0 & \llongrightarrow   \phi_0+{\overline{\phi}}
      \\
      u_{\rm S}&:&\q
      \oh\Gam^2\cos({\overline{\phi}})  & \llongrightarrow  \Gam+\oh\Gam^2  &
      {\q}
      -\Gam+\oh\Gam^2   & \llongrightarrow   \oh\Gam^2\cos({\overline{\phi}})
      \\
      u_{\rm L}&:&\q
      \oh\Gam^2\cos({\overline{\phi}})  & \llongrightarrow  -\Gam+\oh\Gam^2  &
      {\q}
      \Gam+\oh\Gam^2   & \llongrightarrow   \oh\Gam^2\cos({\overline{\phi}})
      \\[1mm]
      \intertext{If $\Gam<1$:}
      \phi'&:& \q
      \phi_0-\pi & \llongrightarrow   \phi_0-0  &
      {\q}
      \phi_0+0 & \llongrightarrow   \phi_0+\pi 
      \\
      u_{\rm S}&:&\q
      -\oh\Gam^2  & \llongrightarrow  \Gam+\oh\Gam^2  &
      {\q}
      -\Gam+\oh\Gam^2   & \llongrightarrow   -\oh\Gam^2
      \\
      u_{\rm L}&:&\q
      -\oh\Gam^2  & \llongrightarrow  -\Gam+\oh\Gam^2  &
      {\q}
      \Gam+\oh\Gam^2   & \llongrightarrow   -\oh\Gam^2
    \end{alignat*}
    \end{minipage}}
    \vspace*{\baselineskip}
\end{center}
\figurecaption{Ranges of the integrations in Equation \eref{eq:dp3}.}
\label{tab:int}
\end{table}

To show the semiclassical unitarity of the interior map operator,
${\sf P}^{\rm int} ({\sf P}^{\rm int})^\dagger={\sf id}$, we note the
kernel of its adjoint explicity. Using \eref{eq:trel},
\eref{eq:vnadj}, \eref{eq:psk}, \eref{eq:plk}, and \eref{eq:pkern} we
find
\begin{align}
\label{eq:intadj}
  &\left({\rm p}^{\rm int}\right)^\dagger(s;s_0)
  = 
  \left({\rm p}^{\rm int}\right)^*(s_0;s)
\\
 &=
  -\rme^{-2\pi\rmi\nu}
  \frac{1}{(2\pi\rmi)^\oh}\,
\,
  \Bigg\{
  \frac{-\big(\vvech_{\rm S}\,\nvec\big)(s;s_0)}
       {(\sin(\alpha)\cos(\alpha))^\oh}
  \,
  \Theta(-\vvech_{\rm S}\,\nvec)
  \,
  \rme^{2\pi\rmi\nu\ga_{\rm S}(s;s_0)}
  \,
  \rme^{\rmi\chit-\rmi\chit_0}
\nnn
&\phantom{=-\rme^{2\pi\rmi\nu}\frac{1}{(2\pi\rmi)^\oh}\,}
 +
  \frac{-\big(\vvech_{\rm L}\,\nvec\big)(s;s_0)}
  {(\sin(\alpha)\cos(\alpha))^\oh}
  \,
  \Theta(-\vvech_{\rm L}\,\nvec)
  \,
  \rme^{-\rmi\piot}\,
  \rme^{2\pi\rmi\nu\ga_{\rm L}(s;s_0)}
  \,
  \rme^{\rmi\chit-\rmi\chit_0}
  \Bigg\}
\PO
\nn
\end{align}
If the integral corresponding to the operator multiplication ${\sf
  P}^{\rm int} ({\sf P}^{\rm int})^\dagger$ is evaluated
semiclassically one obtains a finite contribution only if the initial
and the final points
are close, like in \eref{eq:dp1} above. We may again replace the
boundary locally by an arc of constant curvature and we have in this
case $-\vvech_{\rm S/L}\nvec(s;s_0)=\vvech^0_{\rm S/L}\nvec_0(s;s_0)$.
Comparison with \eref{eq:pkern} shows that the operator corresponding
to \eref{eq:intadj} assumes the form $({\sf P}^{\rm int})^\dagger=
-\rme^{-2\pi\rmi\nu}{\sf P}^{\rm ext}$. The unitarity of ${\sf P}^{\rm
  int}$ follows now immediately with \eref{eq:durel}, and the same
holds for ${\sf P}^{\rm ext}$.

\subsection{The straight line with periodic boundary conditions}
\label{app:line}
In this appendix we 
discuss a model system which allows studying the transition from edge
states to bulk states asymptotically.
In order to remove the effects of a a finite curvature
we deform the boundary $\Gamma$ of a billiard to a straight line
of length $\Len$. In addition to the (mixed) boundary conditions along
the straight line we prescribe periodic boundary conditions at the
end points of the line and perpendicular to $\Gamma$.
This is clearly no longer a billiard problem in its proper sense and
there is no distinction between an interior and an exterior.
Nonetheless, the classical and the quantum problem is well-defined,
with a discrete quantum spectrum.  This simple system permits
discussing of the asymptotics of bulk and edge states in a
straightforward fashion, see Sect.~\ref{sec:asymp}.

The problem is separable in the Landau gauge \eref{eq:AconvL} and
may be solved analogous to the disk problem in Section
\ref{sec:disksep}. Now it is the
\emph{longitudinal} canonical momentum (ie, the transverse component
of the scaled center of motion) which is the second constant of the
motion. Due to the periodic boundary conditions it is quantized,
\begin{align}
\label{eq:defmline}
\frac{c_y }{ b} = \frac{\pi b}{\Len} m
\CO
\end{align}
with integer $m$ (here we put the boundary on the $x$-axis).
The transverse part $\phi$ of the wave function
obeys 
\begin{align}
\label{eq:psiline}
\phi''(z)+\big(\nu-\frac{1}{4}z^2\big)\phi(z)=0 \CO
\end{align}
with $z\defas 2(y-c_y)/b$.
The semiclassical and exact solutions of this equation yield spectral
functions like in the case of the disk, see Section \ref{sec:disksep}
and Appendix \ref{app:diskexact}, respectively.  We report only the
results.

\subsubsection{Semiclassical  quantization}

For given longitudinal and transverse quantum numbers, $m$ and $n$,
the semiclassical energies of skipping states are determined by the roots
of the spectral function
\begin{align}
  \label{eq:xilinesc}
  \xi_{\rm line}^{\rm (sc)}\Big(\nu;n,m,\Lambda,\frac{\Len}{b}\Big) =&\,
  \nu \left[ \piot+\arcsin\Big(\frac{\pi
      m}{\sqrt{\nu}}\frac{b}{\Len}\Big) + \Big(\frac{\pi
      m}{\sqrt{\nu}}\frac{b}{\Len}\Big) \Big[{1-\Big(\frac{\pi
        m}{\sqrt{\nu}}\frac{b}{\Len}\Big)^2}\Big]^\oh \right] 
\nnn
  &-\alpha_\Lambda^{\rm line}  \Big(\nu,m,\frac{\Len}{b}\Big)
  -\pi \Big(n+\frac{3}{4}\Big)
\PO
\intertext{The phase shift} 
  \label{eq:alphalinesc}
  \alpha_\Lambda^{\rm line}\Big(\nu,m,\widetilde{\Len}\Big)
  =&\, \arctan\left(\Lambda
    \Big[{1-\Big(\frac{\pi m}{\sqrt{\nu}\widetilde{\Len}}\Big)^2}\Big]^\oh
  \right) 
\end{align}
is determined by the boundary condition $\Lambda$.

\subsubsection{Exact quantization}

Equation \eref{eq:psiline} is solved by the parabolic cylinder
functions.
It follows that the exact spectral function has the form
\begin{align}
\label{eq:xiline}
\xi_{\rm line}\Big(\nu;m,\Lambda,\frac{\Len}{b}\Big) 
 =\,  
 {\rm  D}_{\nu-\oh}\Big(-2\pi m \frac{b}{\Len} \Big)
 &+\Lambda \Big[ \pi m \frac{b}{\Len} 
 {\rm D}_{\nu-\oh}\Big(-2\pi m \frac{b}{\Len}\Big) 
 \nnn
 & +\frac{1}{\sqrt{\nu}}\, {\rm D}_{\nu+\oh}
 \Big(-2\pi m \frac{b}{\Len}\Big) \Big]
 \CO
\end{align}
where ${\rm D}_k$ is Whittaker's form of the regular parabolic cylinder
function \cite{AS65}.

\subsubsection{The uniform approximation}
\label{app:uniform}

As the most important point, we are interested in a semiclassical
description of the situation when the corresponding classical
trajectory is just detached from the boundary.  Since the WKB
approximation of the wave function fails close to the classical
turning points we have to resort to a uniform approximation, see eg
\cite{BM72}. It yields the asymptotic wave function in the whole
region around one classical turning point, $z_\nu=-2\sqrt{\nu}$, in
terms of the (action) integral
\begin{align}
  &w(z)\defas \left| \int_{z_\nu}^z \Big|\nu-\frac{1}{4}{z'}^2\Big|^\oh
    \, \rmd z' \right|
  \\
  &=
 \begin{cases}
   \ds 
   \nu \left[
     \oh\sinh\!\Big(2\arccos\!\Big(\frac{z}{z_\nu}\Big)\Big) -
     \arccos\!\Big(\frac{z}{z_\nu}\Big)
   \right] 
   &\text{if     $z<-2\sqrt{\nu}$}
   \\[1.8ex]
   \ds
   \nu\left[
    \piot+\arcsin\Big(\frac{z}{2\sqrt{\nu}}\Big)
    + \Big(\frac{z}{2\sqrt{\nu}}\Big)\sqrt{1-\Big(\frac{z}{2\sqrt{\nu}}\Big)^2}
    \,\right]
  &\text{if $-2\sqrt{\nu}<z<2\sqrt{\nu}$,}
 \end{cases}
  \nn
\end{align}
which we define to be positive for any $z$.  In the uniform approximation
the two independent solutions of \eref{eq:psiline} are given (for
$-\infty<z<2\sqrt{\nu}$) in terms of the Airy functions \cite{AS65}
\begin{align}
  \label{eq:psilineuni1}
  \phi_1(z)
  &=
  \mathcal{N}\,
  \frac{\big(w(z)\big)^{\frac{1}{6}}}{|\nu-\frac{1}{4}z^2|^\frac{1}{4}} 
  \Ai\!\bigg(\!\!-\sgn(z-z_\nu)
  \Big(\frac{3}{2}w(z)\Big)^\frac{2}{3}\bigg) 
\intertext{and} 
\phi_2(z)&=
\mathcal{N}\,
  \frac{\big(w(z)\big)^{\frac{1}{6}}}{|\nu-\frac{1}{4}z^2|^\frac{1}{4}} 
  \Bi\!\bigg(\!\!-\sgn(z-z_\nu)
  \Big(\frac{3}{2}w(z)\Big)^\frac{2}{3}\bigg)
\PO
\end{align}
The general solution may be parametrized by an angle $\au\in[-\piot;\piot]$.
\begin{align}
\label{eq:psilineuni2}
\phi(z)&=\cos(\au)\, \phi_1(z) - \sin(\au)\, \phi_2(z)
\end{align}
This form is particularly convenient. By virtue of the asymptotic
expansions of the Airy functions \cite{AS65} we regain the WKB wave
functions in both the energetically forbidden   region,
\begin{align}
\label{eq:wkbline1}
  \psi(z)
  &\sim \frac{1}{(\frac{1}{4}{z}^2-\nu)^\oh} \left( \oh
    \cos(\au)\, \rme^{\ts-w(z)} - \sin(\au)\, \rme^{\ts w(z)}
  \right) \q&(z\ll z_\nu,)
\intertext{and in the energetically allowed one,} 
\label{eq:wkbline2}
  \psi(z) &\sim \frac{1}{(\nu-\frac{1}{4}{z}^2)^\oh} \,
  \cos\!\Big(w(z)-\piof-\au\Big) \q&(z\gg z_\nu.) 
\PO
\end{align}
Note the factor one-half in \eref{eq:wkbline1} which arises in a
non-trivial fashion when connecting the WKB solutions of the two
regions \cite{BM72}.

\begin{figure}[tb]%
\begin{center}%
  \psfrag{x}{\hspace*{-1.6em}$\arctan(\Lambda)$}
  \psfrag{y}{$\Delta\nu$}
    \includegraphics[clip,width=0.8\linewidth] 
    {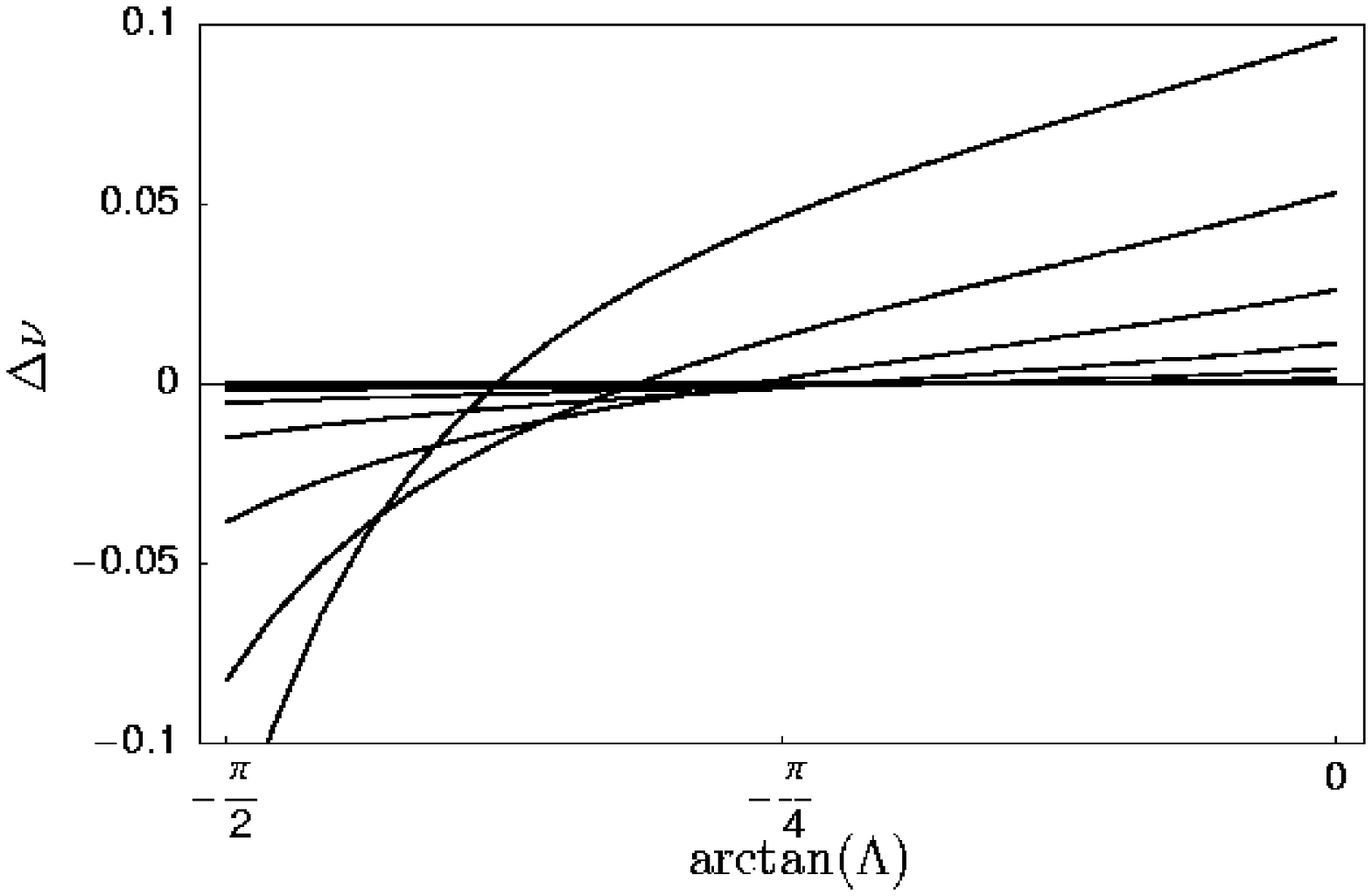}
    \figurecaption{%
      Energy shifts of the bulk states \eref{eq:bulkshiftuni} for the
      $4$th Landau level and parameter values ($\Len=5.39724$,
      $b=0.25$, $m=14\ \ldots\ 20$) which allow the comparison with
      the right part of Fig.~\ref{fig:lambdadyn}. As $m$ is increased
      the energy shifts $\Delta\nu$ get exponentially small
      \eref{eq:bulkshiftwkb} and the boundary mixing parameter for
      which there is no energy shift approaches the Neumann
      condition.  }
    \label{fig:uniform}
  \end{center}%
\end{figure}

The eigenfunctions turn into bulk states once the longitudinal quantum
number $m$ is large enough to leave the boundary
in the energetically forbidden region, 
\begin{align}
\label{eq:qdef}
   \qm&\defas\frac{\pi b m}{\Len\sqrt{\nu}}
   =\frac{z}{z_\nu} > 1
\PO
\end{align}
From the uniform approximation \eref{eq:psilineuni1},
\eref{eq:psilineuni2} we find that in this case the angle $\au$ is
determined by the ratio
\begin{align}
  \tan(\au) &=
  \frac{\Ai\big((\frac{3}{2}w)^\frac{2}{3}\big)
        -\Lambda\sqrt{\qm^2-1}\Ai'\big((\frac{3}{2}w)^\frac{2}{3}\big)}
  {\Bi\big((\frac{3}{2}w)^\frac{2}{3}\big)
        -\Lambda\sqrt{\qm^2-1}\Bi'\big((\frac{3}{2}w)^\frac{2}{3}\big)}
\PO
\end{align}
By comparing the asymptotic expression \eref{eq:wkbline2} of the wave
function in the allowed region with that of a Landau state (which has
no phase shift) one obtains the energy shift $\Delta\nu$ compared to
the Landau energy,
\begin{align}
  \label{eq:bulkshiftuni}
  \Delta\nu(m,\Lambda) = \frac{1}{\pi}
  \arctan\left(
  \frac{\Ai\big((\frac{3}{2}w)^\frac{2}{3}\big)
        -\Lambda\sqrt{\qm^2-1}\Ai'\big((\frac{3}{2}w)^\frac{2}{3}\big)}
  {\Bi\big((\frac{3}{2}w)^\frac{2}{3}\big)
        -\Lambda\sqrt{\qm^2-1}\Bi'\big((\frac{3}{2}w)^\frac{2}{3}\big)}
      \right)
\PO
\end{align}
Figure \ref{fig:uniform} shows the energy shifts for the fourth Landau
level as a function of the boundary mixing parameter. Here, the value
of $\Len/b$ was chosen to correspond to the situation of
Fig.~\ref{fig:lambdadyn} in Chap.~\ref{chap:numres}. We observe that
the bulk state behavior is reproduced qualitatively even at this low
Landau level.  A quantitative comparison of the bulk energy shifts
\eref{eq:bulkshiftuni} with a quantum spectrum is given in
Sect.~\ref{sec:bulkasymp}.

For quantum numbers $m$ which put the boundary into the energetically
allowed region ($|\qm|<1$) the angle $\au$ is semiclassically given
by the phase shift \eref{eq:alphalinesc} obtained above. For these
states
the energy derivative with respect to $\Lambda$ 
reads
\begin{align}
  \left.\frac{\rmd \nu}{\rmd \Lambda}\right|_{\Lambda=0} 
  =
  \frac{\ds\sqrt{1-\qm^2}} {    \piot+\arcsin(\qm) }
  \PO
\end{align}
It coincides with
the limiting expression of a large disk
if we set $|\cvec|=R+c_y$ in \eref{eq:dnudLdisksc2},
\eref{eq:epsilondef} and let $R\to\infty$.

\subsection{Scaled spectra}
\label{app:rho}

We collect a number of formulas for spectra defined
in the semiclassical direction. As discussed in Sections
\ref{sec:scaling} and \ref{sec:qspec} those spectra are obtained by
decreasing the magnetic length $b$ at fixed cyclotron radius $\rho$
(unlike conventional spectra where $\rho$ is increased at fixed $b$).
Since in both cases the spectra are noted in terms of the scaled energy
$\nu=\rho^2/b^2$  the superscript-$(\rho)$ is used to
indicate spectra taken at fixed $\rho$.

Scaled spectroscopy \cite{MWHW86,BD95} has the advantage that the
classical dynamics remains fixed as the spectral variable is
increased. This allows to ensure that the underlying classical motion
is chaotic throughout the spectral interval (Sect.~\ref{sec:specstat})
and to extract classical actions easily by Fourier transformation
(Sect.~\ref{sec:acspec}).

However, one should be aware of the fact that the spectrum obtained
this way does not belong to a single self-adjoint operator. Rather, a
sweep through a family of operators (parametrized by an effective
Planck's constant) is performed as the spectral
variable is increased.  Clearly, the energies are real and the
eigenvectors are still proper solutions of the Schr{\"o}dinger
equation but the latter are not orthogonal.  Moreover, it may happen
that two energies coalesce and vanish as an external parameter is
varied.

Most of the formulas in the main part of this article hold as well for
spectra at fixed $\rho$ after the substitution $b\to \rho/\sqrt{\nu}$.
In particular, this is the case for the spectral functions and the
trace formulas which are to leading order in $\nu$ The smooth number
counting function \eref{eq:Nsmooth}, for example, reads
\begin{align}
  \label{eq:Nrhosm}
  \Nsm^{(\rho)}(\nu)
  =& \frac{\Area}{\rho^2\pi} \, \nu^2
  - \frac{\Len}{2\pi \rho} \, \nu + \frac{1}{6}
\PO
\end{align}
However, care is needed in the case of the spectral densities. The
density of edge states is now given as
\begin{align}
  \label{eq:dedgerho}
  \dedge^{^{(\rho)}\!}(\nu) 
  &= \sum_{n=1}^\infty w_n^{(\rho)} \delta(\nu-\nu_n^{(\rho)})
\CO
\intertext{with the weights now defined at constant $\rho$,}
   w_n^{(\rho)} 
   &\defas
      \frac {\rmd\nu_n^{(\rho)}(\Lambda)}{\rmd
      \Lambda}
  \Big|_{\Lambda=0}
\PO
\end{align}
Here, we obtain the mean edge counting function
\begin{align}
  \label{eq:Nedgerho}
  \Nedgesm^{\rm (\rho)}(\nu)
  = \oh \,\frac{\Len}{2\pi \rho}\, \nu^2 
  \mp\oh\nu
\PO
\end{align}

\end{appendix}

\addcontentsline{toc}{section}{List of Symbols}
\section*{List of important Symbols}

\setlength{\tabcolsep}{0pt}
\setlength{\parindent}{0pt}
\newcommand{\RAGRIGHT}{\let\temp=\\\raggedright\let\\=\temp\hspace{0pt}}

\subsubsection*{Most important:}
\begin{align*}
  \begin{aligned}
    \text{cyclotron radius:\q}&  \rho 
    \\
    \text{magnetic length:\q} & b
  \end{aligned}
  \qq\q
  \text{scaled energy:\q} &  \nu =\frac{\rho^2}{b^2} 
\end{align*}

\subsubsection*{Latin symbols:}
\begin{longtable}{
    >{$}
    p{.33\linewidth}
    <{$\hspace*{.3em}\dotfill\hspace*{.3em}}
    >{\RAGRIGHT}
    p{.65\linewidth}
    }
  \Area
  & area of the billiard domain ($\Area=|\Domain|$)
  \\
  \Area_\po
  & area enclosed by the trajectory of $\po$ \eref{eq:areapodef}
  \\
  \Area_{\rm skip}
  & area determining the phase space of skipping orbits, Fig.~\ref{fig:area}
  \\
  {\rm Ai}(z)
  & Airy function  \cite{AS65}
  \\
  \Ga(\po)  
  & geometric part of the action of the periodic orbit $\po$ \eref{eq:Gadef}
  \\
  \Avec(\rvec)
  & vector potential at arbitrary gauge \eref{eq:Aconv}
  \\
  \Avect(\rvec)
  & scaled vector potential at arbitrary gauge
  (${\Avect(\rvect)={2}\Avec(b\rvect)/({B b})}$) 
  \\
  \Avec_{\rm Lan}(\rvec),\   \Avec_{\rm sym}(\rvec)
  & vector potential in Landau gauge \eref{eq:AconvL} (symmetric
  gauge \eref{eq:Aconvsym})
  \\
  \ga_{\rm S}(\rvec;\rvec_0),\ \ga_{\rm L}(\rvec;\rvec_0)
  & geometric part of the action for the short (long) arc
  \eref{eq:deftsl}, \eref{eq:tdef}, \eref{eq:trel}
  \\
  \aop_{\rm R},\ \aop_{\rm L}
  & annihilation operator of right (left) circular quanta
  \eref{eq:adef}
  \\
  B
  & magnetic induction ($B=\grad\times\Avec$)
  \\
  \mathcal{B} 
  & billiard bounce map \eref{eq:Birmap}
  \\
  {\rm Bi}(z)
  & Airy function \cite{AS65}
  \\
  b
  & magnetic length \eref{eq:bdef}, \eref{eq:rhobdef}
  \\
  C(\nu_0)
  & cross correlation function \eref{eq:Cdef}, \eref{eq:qmcorr}
  \\
  \cvec,\ \cvect
  & (scaled) center of cyclotron motion ($\cvec\in\Rtwo$)
  \eref{eq:rhocqm}
  \\
  \Domain
  & domain of the interior billiard  ($\Domain\subset\Rtwo$)
  \\
  D(t)
  & Fourier transform of $C(\nu_0)$ \eref{eq:corrft}
  \\
  D_k(z)
  & parabolic cylinder function (Whittaker's form) \cite{AS65}
  \\
  d(\nu)
  & standard spectral density \eref{eq:ddef}
  \\
  \dedge(\nu),\  \dedge^\magn(\nu)
  & spectral density of edge states  \eref{eq:dedgedef},
  \eref{eq:dedgedef2}, \eref{eq:dmagdef}
  \\
  \dedgesm(\nu),\ \dedgesm^\magn(\nu)
  & smooth spectral density of edge states  \eref{eq:dedgesm}, \eref{eq:dmagsm}
  \\
  \dedge^{\rm osc}(\nu)
  & fluctuating part of the spectral density of edge states  
\eref{eq:dedgeosc}, \eref{eq:dedgeosc2}
  \\
  \dedge^{(\rho)\!}(\nu)
  & spectral density of edge states in the semiclassical direction
  \eref{eq:dedgerho} 
  \\
  \overline{d}_{\rm skip}(\nu)
  & smooth spectral density of skipping states \eref{eq:dskip}
  \\
  {d}^{\rm osc}_{\rm skip}(\nu)
  & fluctuating part of the spectral density of skipping states 
  \eref{eq:dskiposc}
  \\
  {E}
  & (kinetic) energy
  \\
  \widetilde{E}
  & proper scaled energy ($\widetilde{E}=E/(\hbar\omega)=2\nu$), 
  page \pageref{eq:defnu}
  \\
  f(n)
  & weighted classical sum over $n$-orbits $\pon$ \eref{eq:fn}
  \\
  \oFo(a,b;z)
  & regular confluent hypergeometric function \cite{AS65}
  \\
  \GF
  & generating function of the billiard bounce map \eref{eq:dGF}
  \\
  \Gnu(\rvec;\rvec_0)
  & free Green function at energy $\nu$, with $\rvec_0$
  the initial point \eref{eq:Gintexp}
  \\
  \Gnu^{\rm (sc)}(\rvec;\rvec_0)
  & semiclassical free Green function at energy $\nu$, 
  \eref{eq:Gsctsl}, \eref{eq:Gsccos}
  \\
  \Gn_\nu(z),\  \Gtn_\nu(z)
  & gauge independent part of the (regularized) free Green function at
  energy $\nu$  
  \eref{eq:GreenPhase}, \eref{eq:Gregdef}, \eref{eq:Gregdef2} 
  \\
  \Gnsc_\nu(z)
  & gauge independent part of the semiclassical free Green 
  function at energy $\nu$ 
  \eref{eq:Gscn}
  \\
  g(z)
  & normalized Gaussian window function,
  $g(z)\equiv$ $(2\pi\sigma_g^2)^{-\oh}\exp\big(-z^2/(2\sigma^2_g)\big)$, with
  ``small'' $\sigma_g$
  \\
  \hat{g}(t)
  & Fourier transform of g(z)
  \\
  \Ham
  & magnetic Hamiltonian \eref{eq:Hamconv}
  \\
  \Hamt
  & scaled Hamiltonian \eref{eq:Hamtdef}
  \\
  h(z),\ \hat{h}(t)
  & normalized Gaussian window function, cf
  $g(z)$, with 
  ``large'' width $\sigma_h$, (and its Fourier transform)
  \\
  \jvec(\rvec)
  & probability current density \eref{eq:defj}
  \\
  K(\tau)
  & form factor \eref{eq:formfactordef}, \eref{eq:formfactorqm}
  \\
  {K}(a,b;z)
  &  confluent hypergeometric function \eref{eq:defK}
  \\
  \Len
  & circumference of the billiard domain ($\Len=|\Boundary|$)
  \\
  \Len_\po
  & length of the trajectory of $\po$ \eref{eq:lenpodef}
  \\
  L
  & canonical angular momentum \eref{eq:Aconvsym}
  \\
  \Lag
  & magnetic Lagrangian \eref{eq:Lagrangian}
  \\
  \Lagt
  & scaled Lagrangian \eref{eq:Lagtdef}
  \\
  \Mag(\nu)
  & scaled magnetization  \eref{eq:Mdef}
  \\
  \Mag_{\rm edge}(\nu)
  & scaled edge magnetization  \eref{eq:Medge}, \eref{eq:Mag2}
  \\
  \overline{\Mag}_{\rm edge}  (\nu)
  & smooth edge magnetization \eref{eq:Magsmboth}
  \\
  M_{k,\mu}(z)
  & Whittaker function \cite{AS65}
  \\
  \m_{\rm max}
  & maximum winding number in magnetic disk  \eref{eq:Mmax}
  \\
  {\rm M}(\po)
  & stability matrix of $\po$ \eref{eq:Mrel}
  \\
  m
  & angular (or longitudinal) momentum quantum number
  \eref{eq:defmdisk}, \eref{eq:defmline} (or else integer)
  \\
  m_{\rm max},\ m_{\rm min}
  & maximum (minimum) angular momentum quantum number corresponding to
  skipping motion in the disk \eref{eq:minmax}
  \\
  \mass
  &  particle mass
  \\
  \mmt(\nu)
  & scaled magnetization density \eref{eq:magdef}
  \\
  \mmt_{\rm edge}(\nu)
  & edge magnetization density \eref{eq:mmedge}, \eref{eq:mmedge2}
  \\
  \overline{\mm}_{\rm edge}(\nu)
  & smooth edge magnetization density \eref{eq:mmedgesm}
  \\
  {\mmt}^{\rm osc}(\nu)
  & fluctuating part of the scaled  magnetization density \eref{eq:mmosc}
  \\
  \mathcal{N}
  & normalization constant of the wave function
  \\
  \N(\nu)
  & spectral number counting function (spectral staircase) \eref{eq:Ndef}
  \\
  \Nsm(\nu)
  & smooth  number counting function \eref{eq:Nsmooth}
  \\
  \Nedge(\nu)
  &  edge state counting function  \eref{eq:Nedgedef}
  \\
  \Nedgesm(\nu)
  &  smooth part of the edge state counting function  \eref{eq:Nedgesm}
  \\
  \Nedge^{\rm osc}(\nu)
  &  fluctuating part of the edge state counting function \eref{eq:defNosc}
  \\
  \Nsm_{\rm skip}(\nu)
  &  smooth counting function for skipping states \eref{eq:Nsmskip}
  \\
  \N_{\rm osc}^{\rm skip}(\nu)
  & fluctuating part of the counting  function  for skipping states
  \eref{eq:Nskipint}, \eref{eq:Nskiposc},
  \eref{eq:Ndiskint}, \eref{eq:Ndiskext}, \eref{eq:Nosc}
  \\
  n_\po
  & number of reflections in $\po$, page \pageref{pag:nrdef}
  \\
  \nvec
  & normal vector of billiard boundary, pointing \emph{outwards}
  \eref{eq:parametr2}
  \\
  \PSet_{\rm int}^\n,\   \PSet_{\rm ext}^\n
  & angular increment of $N$-orbit in interior (exterior) disk
  \eref{eq:Mint}, \eref{eq:Mext}
  \\
  \mathsf{P}
  & semiclassical map operator \eref{eq:defP}
  \\
  p(s,s_0)
  & kernel of  semiclassical map operator \eref{eq:psk},
  \eref{eq:plk}, \eref{eq:pkern}
  \\
  p_s
  & Birkhoff coordinate conjugate to $s$, page \pageref{eq:Birmap}
  \\
  \pvec,\  \pvect
  & (scaled) canonical momentum vector \eref{eq:pdef}, \eref{eq:defrtpt}
  \\
  \mathsf{Q}
  & boundary integral operator \eref{eq:Opdef} -- \eref{eq:Opdef4}, 
  \eref{eq:combdef}
  \\
  q
  & particle charge ($q\,B>0$)
  \\
  q_m
  & quantized relative distance of the center of motion from the
  boundary \eref{eq:qdef} 
  \\
  {\rm q}(\rvec;\rvec_0)
  & boundary integral kernel \eref{eq:QexpslD}  -- \eref{eq:QexpdlN}
  \\
  {\rm q}^{\rm (sc)}(\rvec;\rvec_0)
  & semiclassical boundary integral kernel 
  \eref{eq:bopsc2}, \eref{eq:bopsc3}, \eref{eq:bopsc4}
  \\
  R
  & disk radius 
  \\
  \Rt
  & scaled disk radius ($\Rt\equiv R/b$)
  \\
  r_\po
  & number of repetitions in $\po$, page \pageref{pag:nrdef}
  \\
  \rvec,  \rvect
  & (scaled) particle position vector ($\rvec\in\Rtwo$),  \eref{eq:defrtpt}
  \\
  s
  & curvilinear coordinate on boundary ($s_j\equiv s(\rvec_j)$), 
  \eref{eq:parametr}
  \\
  T
  & Larmor period ($T= 2\pi/\omega$)
  \\
  T_{\rm cyc}
  & cyclotron period ($T_{\rm cyc} = \oh T$)
  \\
  \tit
  & scaled time ($\tit=\omega t$)
  \\
  \tvec
  & tangent vector of billiard boundary \eref{eq:parametr2}
  \\
  {\rm U}(a,b;z)
  & irregular confluent hypergeometric function \cite{AS65}
  \\
  {\rm U}(\rvec;\rvec_0)
  & free quantum propagator \eref{eq:propagator}
  \\
  \tilde{v}
  & scaled velocity ($\tilde{v} = v/(\omega b)= 2\rho/b$)
  \\
  \vvec
  & velocity vector \eref{eq:vorrn}
  \\
  \vvech_{\rm S},\  \vvech_{\rm L}
  & normalized velocity vector at point of incidence for short (long)
  arc\eref{eq:vn},
  Fig.~\ref{fig:alphabeta} 
  \\
  \vvech^0_{\rm S},\  \vvech^0_{\rm L}
  & normalized velocity vector after reflection  for short (long) arc 
  \eref{eq:vn},
  Fig.~\ref{fig:alphabeta} 
  \\
  \W
  & classical action (time domain) \eref{eq:acq}, \eref{eq:Wt}
  \\
  W_{k,\mu}(z)
  & Whittaker function \cite{AS65}
  \\
  w_n,\ w_n^\magn
  & quantum weight of state $\ket{\psi_n}$ \eref{eq:defweight}, 
   \eref{eq:wnmagdef}
  \\
  w_\po,\ w_\po^\magn
  & classical weight of orbit $\po$ \eref{eq:defwp},  \eref{eq:wpomagdef}
\addtocounter{table}{-1}%
\end{longtable}

\subsubsection*{Greek symbols:}

\begin{longtable}{
    >{$}
    p{.33\linewidth}
    <{$\hspace*{.3em}\dotfill\hspace*{.3em}}
    >{\RAGRIGHT}
    p{.65\linewidth}
    }
  \alpha(\rvec;\rvec_0)
  & relative distance of the initial and the final point
      ($0\le\alpha\le\piot$) \eref{eq:alphadef}, Fig.~\ref{fig:alphabeta}
  \\
  \alpha_j
  & $\alpha(\rvec_j;\rvec_{j+1})$ \eref{eq:ajdef}, Fig.~\ref{fig:saddle}
  \\
  \pshift
  & phase shift (depending on boundary condition $\Lambda$)
  \eref{eq:pshiftdisk}, \eref{eq:alphalinesc}, \eref{eq:ps}
  \\
  \beta(\rvec;\rvec_0),\ \beta^0(\rvec;\rvec_0)
  & relative direction of the normal vector at incidence (reflection)
  \eref{eq:betadef}, Fig.~\ref{fig:alphabeta}
  \\
  \beta_j,\ \beta^0_j
  & angles \eref{eq:bjdef}, Fig.~\ref{fig:saddle}
  \\
  \Boundary
  & billiard boundary ($\Boundary=\partial\Domain$) \eref{eq:parametr}
  \\
  \Gam
  & relative radius of the magnetic disk \eref{eq:Gamdef}
  \\
  \po,\ \pon
  & physical periodic orbit (with $n$ reflections),  page \pageref{pag:podef}
  \\
  \epsilon
  & parameterization of the angular momentum ($\-1<\epsilon<1$)
  \eref{eq:epsilondef} 
  \\
  \wasgamma
  & relative distance \eref{eq:wasgammadef}
  \\
  \SoL
  & index for type of arc, $\SoL\in\{ {\rm S, L}\}$
  \\
  \Theta(x)
  & Heaviside step function 
  \\
  \theta
  & polar angle
  \\
  \kappa(s)
  & curvature of the billiard boundary at the point $s$ \eref{eq:parametr3}
  \\
  \Lambda
  & dimensionless boundary mixing parameter \eref{eq:Lambdadef}
  \\
  \lambda
  & boundary mixing parameter \eref{eq:bcond}
  \\
  \mu_\po
  & Maslov index (number of conjugate points in $\po$), 
  page \pageref{pag:mudef}
  \\
  \nu
  & scaled energy \eref{eq:defnu}
  \\
  \xi(\nu)
  & spectral function \eref{eq:xibim}, \eref{eq:xidisk}, \eref{eq:xiline}
  \\
  \xi^{\rm (sc)}(\nu)
  & semiclassical spectral function \eref{eq:xidisksc}, \eref{eq:xilinesc}
  \\
  \Phi
  & radial WKB phase in the disk \eref{eq:Phiint}, \eref{eq:Phiext}
  \\
  \phi
  & polar angle in the disk ($\phi\equiv s/R$)
  \\
  \rho
  & cyclotron radius \eref{eq:rhobdef}
  \\
  \rhovec
  & radius vector ($\rhovec=\rvec-\cvec\in\Rtwo$) \eref{eq:rhovecdef}
  \\
  \sigma_j
  & arc parameterization ($-1<\sigma_j<1$) \eref{eq:sigmadef}
  \\
  \sigma_g,\ \sigma_h
  & width of the normalized Gaussians $g(\nu),\ h(\nu)$
  \\
  \tau_\po
  & scaled time of flight of $\po$ \eref{eq:deftau}
  \\
  \Chi(\rvec)
  & gauge field, page \pageref{eq:Aconv}
  \\
  \Chit(\rvect)
  & scaled gauge field ($\Chit\equiv \Chit(\rvect)$, $\Chit_0\equiv
  \Chit(\rvect_0)$)   \eref{eq:Chitdef}
  \\
  \Psi(z)
  & digamma function \cite{AS65}
  \\
  \psi
  & stationary wave function
  \\
  \oc
  & cyclotron frequency ($\oc=2\omega>0$)
  \\
  \omega
  & Larmor frequency \eref{eq:larmordef}
\addtocounter{table}{-1}%
\end{longtable}

{ Some of the figures in this article were reduced in quality. 
A high quality version is available at \tt http:www.klaus-hornberger.de}


\begin{thebibliography}{100}
\raggedright

\bibitem{Halperin82}
B.~I. Halperin,
\newblock Quantized {H}all conductance, current-carrying edge states, and the
  existence of extended states in a two-dimensioinal disordered potential,
\newblock Phys. Rev. B {\bf 25}(4), 2185--2190 (1982).

\bibitem{OzoriodeAlmeida88}
A.~M. {Ozorio de Almeida},
\newblock {\em Hamiltonian Systems: Chaos and Quantization},
\newblock Cambridge University Press, 1988.

\bibitem{Gutzwiller90}
M.~C. Gutzwiller,
\newblock {\em Chaos in Classical and Quantum Mechanics},
\newblock Springer-Verlag, Berlin, 1990.

\bibitem{Haake90}
F.~Haake,
\newblock {\em Quantum Signatures of Chaos},
\newblock Springer-Verlag, Berlin, 1990.

\bibitem{LesHouches91}
M.-J. Giannoni, A.~Voros, and J.~Zinn-Justin, editors,
\newblock {\em Proceedings of the 1989 Les Houches Summer School on ``Chaos and
  Quantum Physics''}, North-Holland, Amsterdam, 1991.

\bibitem{BB97b}
M.~Brack and R.~K. Bhaduri,
\newblock {\em Semiclassical Physics}, volume~96 of {\em Frontiers in Physics},
\newblock Addison-Wesley, Reading, 1997,
\newblock Errata in
  http://www.physik.uni-regensburg.de/$\sim$brm04014/Publications.html.

\bibitem{Gutzwiller98}
M.~C. Gutzwiller,
\newblock The interplay between classical and quantum chaos {(Resource Letter
  ICQM-1)},
\newblock Am. J. Phys. {\bf 66}(4), 304--324 (1998).

\bibitem{Tabachnikov95}
S.~Tabachnikov,
\newblock {\em Billiards}, volume 289 of {\em Panoramas et Synth\`eses},
\newblock Soci\'et\'e Math\'ematique de France, 1995.

\bibitem{DS93}
B.~Dietz and U.~Smilansky,
\newblock A scattering approach to the quantization of billiards -- The
  inside-outside duality,
\newblock Chaos {\bf 3}, 581--590 (1993).

\bibitem{EP95}
J.-P. Eckmann and C.-A. Pillet,
\newblock Spectral duality for planar billiards,
\newblock Commun. Math. Phys. {\bf 170}, 283--313 (1995).

\bibitem{Comtet87}
A.~Comtet,
\newblock On the {L}andau levels on the hyperbolic plane,
\newblock Ann. Phys. (N.Y.) {\bf 173}, 185--209 (1987).

\bibitem{CGO93}
A.~Comtet, B.~Georgeot, and S.~Ouvry,
\newblock Trace formula for Riemann surfaces with magnetic field,
\newblock Phys. Rev. Lett. {\bf 71}, 3786--3789 (1993).

\bibitem{Tasnadi98}
T.~Tasn{\'a}di,
\newblock Hard chaos in magnetic billiards (on the hyperbolic plane),
\newblock J. Math. Phys. {\bf 39}(7), 3783--3804 (1998).

\bibitem{Gutkin01}
B.~Gutkin,
\newblock Hyperbolic magnetic billiards on surfaces of constant curvature,
\newblock Commun. Math. Phys. {\bf 217}, 33--53 (2001).

\bibitem{AKP92}
J.~E. Avron, M.~Klein, and A.~Pnueli,
\newblock Hall conductance and adiabatic charge transport of leaky tori,
\newblock Phys. Rev. Lett. {\bf 69}, 128--131 (1992).

\bibitem{Dunne92}
G.~V. Dunne,
\newblock Hilbert space for charged particles in perpendicular magnetic fields,
\newblock Ann. Phys. (N.Y.) {\bf 215}, 233--263 (1992).

\bibitem{FV01}
E.~V. Ferapontov and A.~P. Veselov,
\newblock Integrable {S}chr{\"o}dinger operators with magnetic fields:
  Factorization method on curved surfaces,
\newblock J. Math. Phys. {\bf 42}, 590--607 (2001).

\bibitem{LaLi2}
L.~D. Landau and E.~M. Lifshitz,
\newblock {\em The classical theory of fields}, volume~2 of {\em Course of
  theoretical physics},
\newblock Butterworth-Heinenann, Oxford, 1975.

\bibitem{Landau30}
L.~Landau,
\newblock Diamagnetismus der {M}etalle,
\newblock Z. Phys. {\bf 64}, 629--637 (1930).

\bibitem{CTDLqm1}
C.~Cohen-Tannoudji, B.~Diu, and F.~Laloe,
\newblock {\em Quantum mechanics}, volume~1,
\newblock Wiley-Interscience, New York, 1977.

\bibitem{AHS78}
J.~Avron, I.~Herbst, and B.~Simon,
\newblock Schr{\"o}dinger operators with magnetic fields. I. General
  interactions,
\newblock Duke Math. J. {\bf 45}, 847883 (1978).

\bibitem{Helffer94}
B.~Helffer,
\newblock On spectral theory for {S}chr{\"o}dinger operators with magnetic
  fields,
\newblock Advanced Studies in Pure Mathematics {\bf 23}, 113--141 (1994).

\bibitem{Schroedinger26}
E.~Schr{\"o}dinger,
\newblock Der stetige {\"U}bergang von der {M}akro- zur {M}ikromechanik,
\newblock Die Naturwissenschaften {\bf 14}, 664 (1926).

\bibitem{FH65}
R.~P. Feynman and A.~R. Hibbs,
\newblock {\em Quantum Mechanics and Path Integrals},
\newblock McGraw-Hill, New York, 1965.

\bibitem{Schulman81}
L.~S. Schulman,
\newblock {\em Techniques and Applications of path integration},
\newblock John Wiley \& Sons, New York, 1981.

\bibitem{GS98}
C.~Grosche and F.~Steiner,
\newblock {\em Handbook of {F}eynman integrals}, volume 145 of {\em Springer
  {T}racts in {M}odern {P}hysics},
\newblock Springer-Verlag, Berlin, 1998.

\bibitem{Gutzwiller67}
M.~C. Gutzwiller,
\newblock Phase-integral approximation in momentum space and the bound states
  of an atom,
\newblock J. Math. Phys. {\bf 8}, 1979--2000 (1967).

\bibitem{Morse73}
M.~Morse,
\newblock {\em Variational Analysis},
\newblock Wiley, New York, 1973.

\bibitem{Glasser64}
K.~L. Glasser,
\newblock Summation over {F}eynman histories: {C}harged particle in a uniform
  magnetic field,
\newblock Phys. Rev. {\bf 133}(3), B831--834 (1964).

\bibitem{LS77}
S.~Levit and U.~Smilansky,
\newblock A new approach to {G}aussian path integrals and the evaluation of the
  semiclassical propagator,
\newblock Ann. Phys. (N.Y.) {\bf 103}, 198--207 (1977).

\bibitem{Cheng84}
B.~K. Cheng,
\newblock Exact evaluation of the propagator for a charged particle in a
  constant magnetic field,
\newblock Physica Scripta {\bf 29}, 351--352 (1984).

\bibitem{KR92}
S.~Klama and U.~R\"ossler,
\newblock The {G}reen's function of confined electrons in an external magnetic
  field: analytical formulation,
\newblock Ann. Phys. (Leipzig) {\bf 1}(6), 460--466 (1992).

\bibitem{TCA97}
M.~L. Tiago, T.~O. de~Carvalho, and M.~A.~M. de~Aguiar,
\newblock Boundary integral method for quantum billiards in a constant magnetic
  field,
\newblock Phys. Rev. A {\bf 55}(1), 65--70 (1997).

\bibitem{BM72}
M.~V. Berry and K.~E. Mount,
\newblock Semiclassical approximations in wave mechanics,
\newblock Rep. Prog. Phys. {\bf 35}, 315--397 (1972).

\bibitem{HS00a}
K.~Hornberger and U.~Smilansky,
\newblock The boundary integral method for magnetic billiards,
\newblock J. Phys. A {\bf 33}, 2829--2855 (2000).

\bibitem{AS65}
M.~Abramowitz and I.~Stegun,
\newblock {\em Handbook of Mathematical Functions},
\newblock Dover Publications, New York, 1965.

\bibitem{Ueta92}
T.~Ueta,
\newblock Green's function of a charged particle in magnetic fields,
\newblock J. Phys. Soc. Jpn. {\bf 61}, 4314--4324 (1992).

\bibitem{MF53}
P.~M. Morse and H.~Feshbach,
\newblock {\em Methods of Theoretical Physics},
\newblock McGraw-Hill, New York, 1953.

\bibitem{NP67}
T.~W. Nee and R.~E. Prange,
\newblock Quantum spectroscopy of the low field oscillations of the surface
  impedance,
\newblock Phys. Lett. {\bf 25A}(8), 582--583 (1967).

\bibitem{KMF69}
E.~A. Kaner, N.~M. Makarov, and I.~M. Fuks,
\newblock The spectrum and damping of surface electron states in a magnetic
  field,
\newblock Sov. Phys. - JETP {\bf 28}(3), 483--488 (1969).

\bibitem{KDP80}
K.~von Klitzing, G.~Dorda, and M.~Petter,
\newblock New method for high-accuracy determination of the fine-structure
  constant based on quantized {Ha}ll resistance,
\newblock Phys. Rev. Lett. {\bf 45}(6), 494--497 (1980).

\bibitem{Laughlin81}
R.~B. Laughlin,
\newblock Quantized {H}all conductivity in 2 dimensions,
\newblock Phys. Rev. B {\bf 23}(10), 5632--5633 (1981).

\bibitem{Buttiker88}
M.~B{\"u}ttiker,
\newblock Absence of backscattering in the quantum Hall effect in multiprobe
  conductors,
\newblock Phys. Rev. B {\bf 38}, 9375--9389 (1988).

\bibitem{Shizuya94}
K.~Shizuya,
\newblock Edge current in the {Quantum Hall Effect},
\newblock Phys. Rev. Lett. {\bf 73}(21), 2907--2910 (1994).

\bibitem{MMP99}
N.~Macris, P.~A. Martin, and J.~V. Pul\'e,
\newblock On edge states in semi-infinite quantum {H}all systems,
\newblock J. Phys. A {\bf 33}, 1985--1996 (1999).

\bibitem{FGW00}
J.~Fr{\"o}hlich, G.~M. Graf, and J.~Walcher,
\newblock On the extended nature of edge states of {Quantum Hall Hamiltonians},
\newblock Ann. Henri Poincar\'e {\bf 1}(3), 405--442 (2000).

\bibitem{SBKR00}
H.~Schulz-Bades, J.~Kellendonk, and T.~Richter,
\newblock Simultaneous quantization of edge and bulk {H}all conductivity,
\newblock J. Phys. A {\bf 33}, L27--L32 (2000).

\bibitem{HS02a}
K.~Hornberger and U.~Smilansky,
\newblock Spectral cross correlations of magnetic edge states,
\newblock Phys. Rev. Lett. {\bf 88}, 024101 (2002),
\newblock in press.

\bibitem{RB85}
M.~Robnik and M.~V. Berry,
\newblock Classical billiards in magnetic fields,
\newblock J. Phys. A {\bf 18}, 1361--1378 (1985).

\bibitem{MBG93}
O.~Meplan, F.~Brut, and C.~Gignoux,
\newblock Tangent map for classical billiards in magnetic fields,
\newblock J. Phys. A {\bf 26}, 237--246 (1993).

\bibitem{Kleberetal96}
X.~Kleber et~al.,
\newblock Chaotic electron dynamics around a single elliptically shaped
  antidot,
\newblock Phys. Rev. B {\bf 54}, 13859--13867 (1996).

\bibitem{BK96}
N.~Berglund and H.~Kunz,
\newblock Integrability and ergodicity of classical billiards in a magnetic
  field,
\newblock J. Stat. Phys. {\bf 83}, 81--126 (1996).

\bibitem{Tasnadi96}
T.~Tasn{\'a}di,
\newblock The behavior of nearby trajectories in magnetic billiards,
\newblock J. Math. Phys. {\bf 37}(11), 5577--5598 (1996).

\bibitem{Tasnadi97}
T.~Tasn{\'a}di,
\newblock Hard chaos in magnetic billiards (On the {E}uclidean plane),
\newblock Commun. Math. Phys. {\bf 187}, 597--621 (1997).

\bibitem{Kovacs97}
Z.~Kov{\'a}cs,
\newblock Orbit stability in billiards in magnetic field,
\newblock Phys. Rep. {\bf 290}, 49--66 (1997).

\bibitem{AL97}
A.~V. Aivazyan and M.~L. Lyubimova,
\newblock The Spectrum of stochastic motion in near-circular magnetic
  billiards.,
\newblock Physics--Doklady {\bf 42}(12), 641--645 (1997).

\bibitem{DA00}
L.~G. G.~V. {Dias da Silva} and M.~A.~M. {de Aguiar},
\newblock Periodic orbits in magnetic billiards,
\newblock Europ. Phys. J. B {\bf 16}(4), 719--728 (2000).

\bibitem{Sinai76}
Y.~G. Sinai,
\newblock {\em Introduction to Ergodic Theory},
\newblock Princeton Univ. Press, 1976.

\bibitem{Bunimovich74}
L.~Bunimovich,
\newblock On the ergodic properties of certain billiards,
\newblock Anal. Appl. {\bf 8}, 254--255 (1974).

\bibitem{Smilansky95}
U.~Smilansky,
\newblock Semiclassical quantization of chaotic billiards -- a scattering
  approach,
\newblock in {\em Proceedings of the 1994 Les Houches Summer School on
  ``Mesoscopic Quantum Physics''}, edited by E.~Akkermans, G.~Montambaux, J.-L.
  Pichard, and J.~Zinn-Justin, volume LXI, Elsevier, 1995.

\bibitem{Berry78}
M.~V. Berry,
\newblock Regular and Irregular Motion,
\newblock in {\em Topics in Nonlinear Mechanics}, edited by S.~Jorna, number~46
  in Am. Inst. Ph. Conf. Proc., pages 16--120, 1978.

\bibitem{LL83}
A.~J. Lichtenberg and M.~A. Lieberman,
\newblock {\em Regular and Stochastic Motion}, volume~38 of {\em Applied
  Mathematical Sciences},
\newblock Springer-Verlag, Berlin, 1983.

\bibitem{Reichl92}
L.~E. Reichl,
\newblock {\em The Transition to Chaos (in Conservative Classical Systems:
  Quantum Manifestations)},
\newblock Springer-Verlag, Berlin, 1992.

\bibitem{GSG99}
B.~Gutkin, U.~Smilansky, and E.~Gutkin,
\newblock Hyperbolic billiards on surfaces of constant curvature,
\newblock Commun. Math. Phys. {\bf 208}(1), 65--90 (1999).

\bibitem{NT88}
N.~Nakamura and H.~Thomas,
\newblock Quantum billiard in a magnetic field: {C}haos and diamagnetism,
\newblock Phys. Rev. Lett. {\bf 61}(3), 247--250 (1988).

\bibitem{PAKGEC94}
S.~D. Prado, M.~A.~M. {de Aguiar}, J.~P. Keating, and R.~{Egydio de Carvalho},
\newblock Semiclassical theory of magnetization for a two-dimensional
  non-interacting electron gas,
\newblock J. Phys. A {\bf 27}, 6091--6106 (1994).

\bibitem{BGOdAS95}
O.~Bohigas, M.-J. Giannoni, A.~M. {Ozorio de Almeida}, and C.~Schmit,
\newblock Chaotic dynamics and the GOE-GUE transition,
\newblock Nonlinearity {\bf 8}, 203--221 (1995).

\bibitem{YH95}
Z.~Yan and R.~Harris,
\newblock Stadium in a magnetic field: Time-reversal invariance symmetry
  breaking and energy level statistics,
\newblock Europhys. Lett. {\bf 32}(5), 437--442 (1995).

\bibitem{JB95}
Z.-L. Ji and K.-F. Berggren,
\newblock Transition from chaotic to regular behavior of electrons in a
  stadium-shaped quantum dot in a perpendicular magnetic field,
\newblock Phys. Rev. B {\bf 52}, 1745--1750 (1995).

\bibitem{Rensink69}
M.~E. Rensink,
\newblock Electron eigenstates in uniform magnetic fields,
\newblock Am. J. Phys. {\bf 37}(9), 900--904 (1969).

\bibitem{BB97a}
J.~Blaschke and M.~Brack,
\newblock Periodic orbit theory of a circular billiard in homogeneous magnetic
  fields,
\newblock Phys. Rev. A {\bf 56}(1), 182--194 (1997),
\newblock Erratum in Phys. Rev. A {\bf 57}, 3136 (1997).

\bibitem{RUJ96}
K.~Richter, D.~Ullmo, and R.~A. Jalabert,
\newblock Orbital magnetism in the ballistic regime: geometrical effects,
\newblock Phys. Rep. {\bf 276}, 1--83 (1996).

\bibitem{NPZ00}
R.~Narevich, R.~E. Prange, and O.~Zaitsev,
\newblock Square billiard with a magnetic flux,
\newblock Phys. Rev. E {\bf 62}(2), 2046--2059 (2000).

\bibitem{KSG64}
N.~S. Koshlyakov, M.~M. Smirnov, and E.~B. Gliner,
\newblock {\em Differential Equations of Mathematical Physics},
\newblock North-Holland, Amsterdam, 1964.

\bibitem{BB70}
R.~Balian and C.~Bloch,
\newblock Distribution of eigenfrequencies for the wave equation in a finite
  domain: I. Three-dimensional problem with smooth boundary surface,
\newblock Ann. Phys. (N.Y.) {\bf 60}, 401--447 (1970),
\newblock Erratum in Ann. Phys. {\bf 84}, 559-563 (1974).

\bibitem{SPSUS95}
M.~Sieber, H.~Primack, U.~Smilansky, I.~Ussishkin, and H.~Schanz,
\newblock Semiclassical quantization of billiards with mixed boundary
  conditions,
\newblock J. Phys. A {\bf 28}, 5041--5078 (1995).

\bibitem{AANS98}
E.~Akkermans, J.~E. Avron, R.~Narevich, and R.~Seiler,
\newblock Boundary conditions for bulk and edge states in {Q}uantum {H}all
  systems,
\newblock Europ. Phys. J. B {\bf 1}, 117--121 (1998).

\bibitem{BH78}
H.~P. Baltes and E.~R. Hilf,
\newblock {\em Spectra of Finite Systems},
\newblock B.I.-{Wissen\-schafts\-verlag}, Mannheim, 1978.

\bibitem{MMP97}
N.~Macris, P.~A. Martin, and J.~V. Pul\'e,
\newblock Large volume asymptotics of {B}rownian integrals and orbital
  magnetism,
\newblock Ann. Inst. Henri Poincar\'e Physique th\'eorique {\bf 66}(2),
  147--183 (1997).

\bibitem{NS99}
R.~Narevich and D.~Spehner,
\newblock Weyl expansion of a circle billiard in a magnetic field,
\newblock J. Phys. A {\bf 32}(19), L227--L230 (1999).

\bibitem{BH94}
M.~V. Berry and C.~J. Howls,
\newblock High orders of the {W}eyl expansion for quantum billiards: resurgence
  of periodic orbits, and the {S}tokes phenomenon,
\newblock Proc. Roy. Soc. Lond. A {\bf 447}, 527--555 (1994).

\bibitem{Peierls79}
R.~Peierls,
\newblock {\em Surprises in Theoretical Physics},
\newblock Princeton University, New Jersey, 1970.

\bibitem{Leeuwen21}
H.-J. van Leeuwen,
\newblock Problem{\`e}s de la th{\'e}orie {\'e}lectronique du magn{\'e}tisme,
\newblock Le Journal de Physique el le Radium {\bf S{\'e}rie VI}(12), 261--377
  (1921).

\bibitem{AM76}
N.~W. Ashcroft and N.~D. Mermin,
\newblock {\em Solid state physics},
\newblock Saunders College, Fort Worth, 1976.

\bibitem{vRvL93}
J.~van Ruitenbeek and D.~A. van Leeuwen,
\newblock Size effects in orbital magnetism,
\newblock Mod. Phys. Lett. B {\bf 7}(16), 1053--1069 (1993).

\bibitem{Kunz94}
H.~Kunz,
\newblock Surface Orbital Magnetism,
\newblock J. Stat. Phys. {\bf 76}, 183--207 (1994).

\bibitem{JS95}
P.~John and L.~G. Suttorp,
\newblock Boundary effects in a magnetized free-electron gas: {G}reen function
  approach,
\newblock J. Phys. A {\bf 28}(21), 6087--6097 (1995).

\bibitem{NSA98}
R.~Narevich, D.~Spehner, and A.~Akkermans,
\newblock Heat kernel of integrable billiards in a magnetic field,
\newblock J. Phys. A {\bf 31}, 4277--4287 (1998).

\bibitem{vonOppen94}
F.~von Oppen,
\newblock Magnetic susceptibility of ballistic microstructures,
\newblock Phys. Rev. B {\bf 50}(23), 17151--17161 (1994).

\bibitem{Agam94}
O.~Agam,
\newblock The magnetic response of chaotic mesoscopic systems,
\newblock J. Phys. I France {\bf 4}, 697--730 (1994).

\bibitem{URJ95}
D.~Ullmo, K.~Richter, and R.~A. Jalabert,
\newblock Orbital magnetism in ensembles of ballistic billiards,
\newblock Phys. Rev. Lett. {\bf 74}(3), 383--386 (1995).

\bibitem{RM98}
K.~Richter and B.~Mehlig,
\newblock Orbital magnetism of classically chaotic quantum systems,
\newblock Europhys. Lett. {\bf 41}(6), 587--592 (1998).

\bibitem{Tanaka98}
K.~Tanaka,
\newblock Semiclassical study of the magnetization of a quantum dot,
\newblock Ann. Phys. (N.Y.) {\bf 268}, 31--60 (1998).

\bibitem{Richter00}
K.~Richter,
\newblock {\em Semiclassical theory of mesoscopic quantum systems}, volume 161
  of {\em Springer Tracts in Modern Physics},
\newblock Springer-Verlag, Berlin, 2000.

\bibitem{deAguiar96}
M.~A.~M. de~Aguiar,
\newblock Eigenvalues and eigenfunctions of billiards in a constant magnetic
  field,
\newblock Phys. Rev. E {\bf 55}(5), 4555--4561 (1996).

\bibitem{KR74}
R.~E. Kleinman and G.~F. Roach,
\newblock Boundary integral equations for the three-dimensional {H}elmholtz
  equation,
\newblock SIAM Review {\bf 16}(2), 214--236 (1974).

\bibitem{Riddell79}
R.~J. Riddell,
\newblock Boundary-distribution solution of the {H}elmholtz equation for a
  region with corners,
\newblock J. Comp. Phys. {\bf 31}, 21 (1979).

\bibitem{BW84}
M.~V. Berry and M.~Wilkinson,
\newblock Diabolical points in the spectra of triangles,
\newblock Proc. Roy. Soc. Lond. A {\bf 392}, 15--43 (1984).

\bibitem{Boasman94}
P.~A. Boasman,
\newblock Semiclassical accuracy for billiards,
\newblock Nonlinearity {\bf 7}, 485--537 (1994).

\bibitem{KS97}
I.~Kosztin and K.~Schulten,
\newblock Boundary integral method for stationary states of two-dimensional
  quantum systems,
\newblock Int. J. Mod Phys. C {\bf 8}(2), 293--325 (1997).

\bibitem{LRH98}
B.~W. Li, M.~Robnik, and B.~Hu,
\newblock Relevance of chaos in numerical solutions of quantum billiards,
\newblock Phys. Rev. E {\bf 57}(4), 4095--4105 (1998).

\bibitem{MK88}
S.~W. McDonald and A.~N. Kaufman,
\newblock Wave chaos in the stadium: {S}tatistical properties of short-wave
  solutions of the {H}elmholtz equation,
\newblock Phys. Rev. A {\bf 37}(8), 3067--3086 (1988).

\bibitem{CLH01}
D.~Cohen, N.~Lepore, and E.~J. Heller,
\newblock Unified framework for finding eigenstates of Helmholtz equation using
  boundary methods,
\newblock submitted to Phys. Rev. E  (2001),
\newblock nlin.CD/0108014.

\bibitem{Martin82}
P.~A. Martin,
\newblock Acoustic scattering and radiation problems, and the null-field
  method,
\newblock Wave Motion {\bf 4}, 391--408 (1982).

\bibitem{Guiggiani98}
M.~Guiggiani,
\newblock Formulation and numerical treatment of boundary integral equations
  with hypersingular kernels,
\newblock in {\em Singular Integrals in Boundary Element Methods}, edited by
  V.~Sladek and J.~Sladek, Computational Mechanics Publications, Billerica,
  1998.

\bibitem{Hornberger01}
K.~Hornberger,
\newblock {\em Spectral properties of magnetic edge states},
\newblock PhD thesis, Ludwig-Maximilians-Uni\-versi\-t{\"a}t M{\"u}n\-chen,
  2001.

\bibitem{Bohigas91}
O.~Bohigas,
\newblock Random matrices and chaotic dynamics,
\newblock In Giannoni et~al. \cite{LesHouches91}.

\bibitem{Berry85}
M.~Berry,
\newblock Semiclassical theory of spectral rigidity,
\newblock Proc. Roy. Soc. Lond. A {\bf 400}, 229--251 (1985).

\bibitem{AIS93}
N.~Argaman, Y.~Imry, and U.~Smilansky,
\newblock Semiclassical analysis of spectral correlations in mesoscopic
  systems,
\newblock Phys. Rev. B {\bf 47}(8), 4440--4457 (1993).

\bibitem{HM01}
B.~Helffer and A.~Morame,
\newblock Magnetic bottles in connection with superconductivity,
\newblock J. Funct. Anal. {\bf 185}, 604--680 (2001).

\bibitem{Gutzwiller71}
M.~C. Gutzwiller,
\newblock Periodic orbits and classical quantization conditions,
\newblock J. Math. Phys. {\bf 12}, 343--358 (1971).

\bibitem{BT76}
M.~V. Berry and M.~Tabor,
\newblock Closed orbits and regular bound spectrum,
\newblock Proc. Roy. Soc. Lond. A {\bf 349}, 101 (1976).

\bibitem{BT77}
M.~V. Berry and M.~Tabor,
\newblock Calculating bound spectrum by path summation in action-angle
  variables,
\newblock J. Phys. A {\bf 10}, 371--379 (1977).

\bibitem{BB72}
R.~Balian and C.~Bloch,
\newblock Distribution of eigenfrequencies for the wave equation in a finite
  domain: {III}. Eigenfrequency density oscillations,
\newblock Ann. Phys. (N.Y.) {\bf 69}, 76--160 (1972).

\bibitem{HS92}
T.~Harayama and A.~Shudo,
\newblock Zeta function derived from the boundary element method,
\newblock Phys. Lett. A {\bf 165}, 417--426 (1992).

\bibitem{Bogomolny92}
E.~B. Bogomolny,
\newblock Semiclassical quantization of multidimensional systems,
\newblock Nonlinearity {\bf 5}, 805--866 (1992).

\bibitem{DS92}
E.~Doron and U.~Smilansky,
\newblock Semiclassical quantization of chaotic billiards: a scattering theory
  approach,
\newblock Nonlinearity {\bf 5}, 1055--1084 (1992).

\bibitem{GP95a}
B.~Georgeot and R.~E. Prange,
\newblock Exact and semiclassical {F}redholm solutions of quantum billiards,
\newblock Phys. Rev. Lett. {\bf 74}, 2851--1854 (1995).

\bibitem{THS97}
S.~Tasaki, T.~Harayama, and A.~Shudo,
\newblock Interior {D}irichlet eigenvalue problem, exterior {N}eumann
  scattering problem, and boundary element method for quantum billiards,
\newblock Phys. Rev. E {\bf 56}(1), R13--R16 (1997).

\bibitem{HST99}
T.~Harayama, A.~Shudo, and S.~Tasaki,
\newblock Semiclassical {F}redholm determinant for strongly chaotic billiards,
\newblock Nonlinearity {\bf 12}, 1113--1149 (1999).

\bibitem{MM92}
M.~Marcus and H.~Minc,
\newblock {\em A Survey of Matrix Theory and Matrix Inequalities},
\newblock Dover Publishing, New York, 1992.

\bibitem{SPS97}
M.~Sieber, N.~Pavloff, and C.~Schmit,
\newblock Uniform approximation for diffractive contributions to the trace
  formula in billiard systems,
\newblock Phys. Rev. E {\bf 55}(3), 2279--2299 (1997).

\bibitem{Sieber98}
M.~Sieber,
\newblock Billiard systems in three dimensions: the boundary integral equation
  and the trace formula,
\newblock Nonlinearity {\bf 11}, 1607--1623 (1998).

\bibitem{BB97c}
J.~Blaschke and M.~Brack,
\newblock Quantum corrections to the semiclassicall level density of the
  circular disk in homogeneous magnetic fields,
\newblock Physica E {\bf 1}, 288--291 (1997).

\bibitem{CL91}
S.~C. Creagh and R.~G. Littlejohn,
\newblock Semiclassical trace formulas in the presence of continuous
  symmetries,
\newblock Phys. Rev. A {\bf 44}, 836--850 (1991).

\bibitem{LaLi3}
L.~D. Landau and E.~M. Lifshitz,
\newblock {\em Quantum mechanics}, volume~3 of {\em Course of theoretical
  physics},
\newblock Butterworth-Heinenann, Oxford, 1975.

\bibitem{Titchmarsch48}
E.~C. Titchmarsch,
\newblock {\em Introduction to the theory of Fourier integrals},
\newblock Clarendon Press, Oxford, 1948.

\bibitem{HS01a}
K.~Hornberger and U.~Smilansky,
\newblock The exterior and interior edge states of magnetic billiards: Spectral
  statistics and correlations,
\newblock Physica Scripta {\bf T90}, 64--74 (2001).

\bibitem{BFFMPW81}
T.~A. Brody, J.~Flores, J.~B. French, P.~Mello, A.~Pandey, and S.~Wong,
\newblock Random-matrix physics -- Spectrum and strength fluctuations,
\newblock Rev. Mod. Phys. {\bf 53}(3), 385--479 (1981).

\bibitem{Robbins89}
J.~M. Robbins,
\newblock Discrete symmetries in periodic-orbit theory,
\newblock Phys. Rev. A {\bf 40}(4), 2128--2136 (1989).

\bibitem{LS90}
Y.~B. Levinson and E.~V. Sukhorukov,
\newblock Scattering of electron edge states in a magnetic field by small
  irregularities of the boundary,
\newblock Physics Letters A {\bf 149}, 167--171 (1990).

\bibitem{LS91}
Y.~B. Levinson and E.~V. Sukhorukov,
\newblock Bending of electron edge states in a magnetic field,
\newblock J. Phys.: Condens. Matter {\bf 3}, 7291--7306 (1991).

\bibitem{GutkinThesis}
B.~Gutkin,
\newblock PhD thesis, Weizmann Institute of Science, Israel, 2001,
\newblock submitted.

\bibitem{Buchholz69}
H.~Buchholz,
\newblock {\em The Confluent Hypergeometric Function},
\newblock Springer-Verlag, Berlin, 1969.

\bibitem{Erdelyi56}
A.~Erd{\'e}lyi,
\newblock {\em Asymptotic Expansions},
\newblock Dover, New York, 1956.

\bibitem{BH75}
N.~Bleistein and R.~A. Handelsman,
\newblock {\em Asymptotic expansions of integrals},
\newblock Holt, Rinehart and Watson, New York, 1975.

\bibitem{MWHW86}
J.~Main, G.~Wiebusch, A.~Holle, and K.~H. Welge,
\newblock New quasi-Landau structure of highly excited atoms: The hydrogen
  atom,
\newblock Phys. Rev. Lett. {\bf 57}, 2789--2792 (1986).

\bibitem{BD95}
A.~Buchleitner and D.~Delande,
\newblock Spectral aspects of the microwave ionization of atomic {R}ydberg
  states,
\newblock Chaos, Solitons \& Fractals {\bf 5}(7), 1125--1141 (1995).

\end{thebibliography}
\end{document}